\documentclass[a4paper,usenames,dvipsnames]{article}
\pdfoutput=1

\usepackage{jheppub}
\usepackage{soul}
\usepackage[utf8]{inputenc}
\usepackage{bbold}
\usepackage{multirow}
\usepackage{subcaption}
\usepackage{amssymb}
\usepackage{graphicx}
\usepackage[dvipsnames]{xcolor}
    \definecolor{darkgreen}{rgb}{0,0.5,0}
    \definecolor{darkblue}{rgb}{0,0,0.6}
    \definecolor{purple}{rgb}{0.4,.2,0.7}
\usepackage{slashed,cancel}    
\newcommand{\be}{\begin{equation}}
\newcommand{\ee}{\end{equation}}
\newcommand{\ba}{\begin{eqnarray}}
\newcommand{\ea}{\end{eqnarray}}
\newcommand{\bea}{\begin{eqnarray}}
\newcommand{\eea}{\end{eqnarray}}

\newcommand{\bmat}{\begin{pmatrix}}
\newcommand{\emat}{\end{pmatrix}}

	\newcommand{\bes}{\begin{equation} \begin{split} }	
	\newcommand{\ees}{\end{split} \end{equation} }

	\usepackage{physics}

    \def\nn{\nonumber}

\def\Tr{\,{\rm Tr}\,}

\def\PhiT{ \tilde \Phi }
\def\PhiV{ \bar \Phi }

\def\Pbar{ \overline {\cal P} }

\def\r{\rho}

\def\b{\beta}
\def\g{\gamma}

\def\d{\delta}

\def\e{\epsilon}

\def\m{\mu}
\def\n{\nu}

\def\l{\lambda}

\def\s{\sigma}

\def\ket{\rangle}
\def\bra{\langle}
\def\ketbra{\rangle \langle}

\title{Quantum properties of heavy-fermion pairs at a lepton collider with polarised beams}
\author[a]{Mohammad Mahdi Altakach,}
\author[b,c]{Priyanka Lamba,}
\author[b,c,d,e]{Fabio Maltoni,}
\author[f]{Kazuki Sakurai}
\affiliation[a]{University of Sciences and Arts in Lebanon, Beirut 1102 2801, Lebanon}
\affiliation[b]{Dipartimento di Fisica e Astronomia, Universit\`{a} di Bologna, Via Irnerio 46, 40126 Bologna, Italy}
\affiliation[c]{INFN, Sezione di Bologna, Via Irnerio 46, 40126 Bologna, Italy}
\affiliation[d]{Centre for Cosmology, Particle Physics and Phenomenology (CP3), Universit\'{e} Catholique de Louvain, B-1348 Louvain-la-Neuve, Belgium}
\affiliation[e]{European Organisation for Nuclear Research (CERN), Geneva, Switzerland}
\affiliation[f]{Institute of Theoretical Physics, Faculty of Physics, University of Warsaw, Pasteura 5, 02-093, Warsaw, Poland}

\abstract{
We investigate the quantum properties of heavy-fermion pairs, such as $t\bar t$ or $\tau^+\tau^-$, 
produced in lepton-lepton collisions with polarised beams. Focusing on spin correlations, entanglement, 
Bell-inequality violation, and quantum-information--theoretic measures such as purity and magic, 
we analyse how beam polarisation shapes the structure of the spin-density matrix. We derive analytic 
expressions for a wide range of helicity configurations, including both Standard Model contributions 
and generic new-physics effects parametrised by scalar, vector, and tensor four-fermion operators 
within an effective field theory framework. We show that beam polarisation unlocks a substantially 
richer set of spin configurations and significantly enhances sensitivity to non-standard interactions. 
As a phenomenological application, we study $t\bar t$ production at a future linear collider and 
demonstrate that quantum observables provide a comprehensive and complementary probe of top-quark interactions and stronger constraints on the scale of new physics. 
}
\begin{document}
\maketitle

\section{Introduction}
Quantum information concepts provide a powerful and unifying language to describe the spin
structure of states produced in high-energy collisions. When particles are created in pairs,
their spins generally form a mixed quantum state, fully characterised by a spin-density matrix.
Beyond single-particle polarisations, this object encodes genuine two-particle quantum
correlations, including spin correlations, entanglement, Bell-inequality violation, and other
markers of non-classical behaviour. These observables offer a complementary probe of
fundamental interactions, as they are directly sensitive to the Lorentz and chiral structure
of the underlying dynamics. As a result, the density-matrix formalism provides a natural framework to investigate
quantum aspects of fundamental interactions at present and future colliders within and beyond the Standard Model; see Ref.~\cite{Afik:2025ejh} for a general overview of the research goals and Ref.~\cite{Durupt:2025wuk} for a range of applications. 

The first arena in which these ideas have been extensively explored is $t \bar t$ production
at the LHC. The top quark, as the heaviest elementary particle in the Standard Model (SM),
plays a special role due to its very short lifetime, which prevents spin decorrelation before
decay and allows its spin information to be directly imprinted in the kinematic distributions of
its decay products. This feature enables direct access to the spin configuration of the
$t\bar t$ system and has motivated a large body of theoretical work on spin correlations,
entanglement, and related quantum observables~\cite{2003.02280,Fabbrichesi:2021npl,Severi:2021cnj,Severi:2022qjy,Aoude:2022imd,Afik:2022kwm,Aguilar-Saavedra:2022uye, Afik:2022dgh, Cheng:2023qmz, Han:2023fci, Dong:2023xiw, Cheng:2024btk, Aguilar-Saavedra:2024hwd,Aguilar-Saavedra:2023lwb,Aguilar-Saavedra:2024fig,Cheng:2024rxi, Aguilar-Saavedra:2024vpd, Han:2024ugl, Maltoni:2024tul, Maltoni:2024csn, Aoude:2025jzc, Fabbrichesi:2025psr},
together with the first experimental studies by the ATLAS and CMS collaborations~\cite{ATLAS:2014aus, CMS:2015cal, ATLAS:2016bac, CMS:2016piu, CMS:2019nrx, ATLAS:2019zrq, CMS:2024pts, CMS:2024zkc}.
Closely related investigations have also been carried out for $\tau^+\tau^-$ production,
either directly or as final states in $Z$- and $H$-boson decays~\cite{Altakach:2022ywa,Fabbrichesi:2022ovb,Ehataht:2023zzt,Fabbrichesi:2024xtq,Fabbrichesi:2024wcd,Zhang:2025mmm,Han:2025ewp}.

A particularly clean realisation of this programme could be provided by the study of the production of massive
fermion pairs in lepton collisions, $l \bar l \to F\bar F$. In this environment, the well-defined centre-of-mass energy, the absence
of QCD backgrounds and the well--defined initial state allow for precise control over the spin
degrees of freedom. In this case, the top quark remains a prominent example at high-energies, the same theoretical
framework applies equally well to other fermions such as $\tau$ leptons or muons produced at lower energies. In all cases,
the spin state of the $F\bar F$ system can be analysed in full generality using its density
matrix, enabling a systematic study of quantum correlations and their dependence on kinematics,
interaction structure, and experimental conditions.

So far, the production of heavy fermion pairs $F\bar F$ ($F=t,\tau$) at lepton colliders has been
primarily considered in the context of unpolarised circular machines~\cite{Altakach:2022ywa, Maltoni:2024csn}, such as the FCC-ee. These studies have shown that
$F\bar F$ pairs produced in such environments can exhibit strong spin entanglement, opening
new avenues to probe fundamental interactions at the TeV scale. However, the physics reach of
these machines is intrinsically constrained by their beam characteristics: while hadron
colliders suffer from limited control over the partonic initial state, circular lepton
colliders typically operate with unpolarised beams, restricting access to specific helicity
configurations.

Lepton colliders offering highly polarised initial beams, such as linear colliders, could provide a
qualitatively new handle on the spin structure of the produced states. Beam polarisation can be
exploited to engineer specific initial helicity configurations, thereby unlocking a much richer
landscape of quantum states in $F\bar F$ production. This capability not only enhances sensitivity
to SM parameters, but can also amplify the effects of new physics operators in an Effective Field
Theory (EFT) framework, significantly improving the sensitivity of quantum observables as probes
of physics beyond the SM. The impact of beam polarisation on the final-state density matrix has been studied for $e^+ e^- \to ZH$ \cite{Rao:2019hsp,Rao:2021eer} and in electron–ion collisions \cite{Cheng:2025zaw}.

In this work, we present a detailed theoretical analysis of $F\bar F$ production with polarised
lepton beams at leading order in the electroweak interactions. We first study heavy fermion pair
production in full generality, $l \bar l \to F\bar F$, within a quantum field theory framework
with polarised initial states, assuming a single four--fermion operator of scalar, vector, or
tensor type. From the $F\bar F$ spin density matrix, we derive closed--form expressions for key
quantum observables, including the purity, concurrence, Bell-inequality violation, and stabiliser
R\'enyi entropies, both in the helicity and beam bases. These results make explicit how quantum
properties depend on kinematic, experimental, and theoretical inputs, in particular on beam
polarisations and the Lorentz structure of the interaction. We stress that since our derivations are fully general, they can be directly applied to any massive fermion pair final states, such as $t \bar t$,
$\tau^+\tau^-$ or $\mu^+\mu^-$.

As a leading application in this work, we consider the process $e^+e^- \to t\bar t$ by identifying $F=t$ and $\ell=e$,
and map out regions of parameter space in which specific quantum features are enhanced. Our main goal is to address  the central question: \emph{what does a beam--polarised lepton collider offer
to quantum--information--driven searches for new physics?} To answer this, we compare the
polarisation response of the quantum observables for $e^+e^- \to t\bar t$ in the SM and in a general EFT benchmark scenario, featuring new four-fermion contact interactions.  We demonstrate that beam polarisation, together with kinematic
selection, enables the isolation of specific $t\bar t$ final states, featuring maximal or
vanishing entanglement, Bell-inequality violation, and stabiliser versus high--magic
behaviour, in a manner that is diagnostic of the underlying operator structure. This capability
not only enhances sensitivity to new physics, but also provides discriminating power to infer
the nature of the interaction responsible for any observed deviation.

The paper is organised as follows. Section \ref{sec:spin_density_matrix} sets up the kinematics and defines the initial and final spin density matrices for $l\bar l\!\to\! F \bar F$. 
Section \ref{sec:QO} introduces the quantum observables used in this study.  
Section \ref{sec:QOI} analyses the quantum properties of the polarised initial state. 
Section \ref{sec:EFT} presents closed–form results (and numerical maps where needed) for the quantum observables in the $l\bar l\!\to\! F \bar F$ process within an EFT framework, assuming a single four-fermion operator of scalar, vector or tensor type. 
Section \ref{sec:sm} specialises to the SM process $e^+e^-\!\to t \bar t$ and investigates the kinematic and polarisation dependences of the observables.  
Section \ref{sec:sm+eft} compares the polarisation response of the observables across SM and benchmark EFT scenarios, highlighting patterns that enable discrimination among operator structures. 
Section \ref{sec:concl} summarises our findings. Several appendices collect further information and studies. In particular, in Appendix A we include the derivation of a key formula we use throughout, in Appendix B we include the analytical expressions for the density matrices and then in Appendices C and D we provide additional plots at $\sqrt s=500~\mathrm{GeV}$ and $1~\mathrm{TeV}$. In Appendix E we show that requiring the EFT-truncated spin-density matrix to remain positive semi-definite leads to constraints on linear EFT corrections that are stronger than those derived from cross sections alone.

\section{Spin density matrices} 
\label{sec:spin_density_matrix}

We consider the production of a massive spin-$\frac{1}{2}$ fermion $F$ and its antiparticle $\bar F$, by the collision of an almost massless spin-$\frac{1}{2}$ fermion $l$ and its antiparticle $\bar l$:
\be
l \bar l \,\to\, F \bar F
\ee
In Section \ref{sec:sm}, $l$ and $F$ will be specifically identified with the electron $e^-$ and the top quark $t$, respectively. Prior to that section, however, we refrain from making this identification to study the quantum information in the general QFT framework.

\subsection{The initial spin state}
The spins of the incoming particles $l$ and $\bar l$ are assumed to be in a separable (non-entangled) state and (imperfectly) polarised.
Following the standard convention,
the polarised $l$ beam is described by the spin-density operator
\ba
\hat \rho_l &=& {\cal P} \cdot | + \ketbra + |_l + (1 - {\cal P}) \cdot \rho_l^{\rm mix}
\nn \\
&=& 
\frac{1}{2} (1+{\cal P}) \cdot | + \ketbra + |_l \,+\, \frac{1}{2} (1-{\cal P}) \cdot | - \ketbra - |_l\,,
\ea
where ${\cal P} \in [-1,1]$ denotes the beam polarisation parameter for $l$,
and $\rho_l^{\rm mix} \equiv \tfrac{1}{2}[ |+ \ketbra +|_l + |- \ketbra -|_l ]$ represents the maximally mixed (unpolarised) spin state~\cite{Adolphsen:2013kya}.
Here, $|\lambda_l \rangle$ ($\lambda_l = \pm$) denotes the pure state, in which $l$'s spin is aligned ($\lambda_l = +$) or anti-aligned ($\lambda_l = -$) with its momentum direction. In other words, $\lambda_l$ denotes the lepton helicity.

A similar expression applies to the antiparticle $\bar l$, characterised by its own beam polarisation parameter $\overline{\cal P} \in [-1,1]$.
For the antiparticle, however, we quantise the spin opposite to its momentum direction:
the positive ($\lambda_{\bar l} = +$) and negative ($\lambda_{\bar l} = -$) states $|\lambda_{\bar l}\rangle$ correspond to the spin being anti-aligned and aligned with the $\bar l$ momentum, respectively.
In the standard convention, a positive beam polarisation ($\overline{\cal P} > 0$) indicates that the $\bar l$ beam contains a larger fraction of particles whose spins are aligned with their momenta.
To maintain consistency with this convention, the spin density operator for the $\bar l$ beam is written as
\ba
\hat \rho_{\bar l} &=& \Pbar \cdot | - \ketbra - |_{\bar l} + (1 - \Pbar) \cdot \rho_{\bar l}^{\rm mix}
\nn \\
&=& 
\frac{1}{2} (1-\Pbar) \cdot | + \ketbra + |_{\bar l} \,+\, \frac{1}{2} (1+\Pbar) \cdot | - \ketbra - |_{\bar l}\,.
\ea
In this convention, the same spin quantisation axis is adopted for both $l$ and $\bar l$, namely along the $l$–momentum direction.

Combining the two beams, the total initial spin state is described by the density operator of the product (separable) state 
\ba
\hat \rho_{\rm in} \,=\, \hat \rho_l \otimes \hat \rho_{\bar l}  
\,=\,
\sum_{ \{ \l_{l \bar l} \}  }
\rho^{\rm in}_{ (\l'_l, \l'_{\bar l}),(\l_l, \l_{\bar l})  }
| \l'_l, \l'_{\bar l} \ketbra \l_l, \l_{\bar l} |\,,
\label{hat_rho}
\ea
where the summation $\sum_{ \{ \l_{l \bar l} \} }$ is taken over all possible configurations of the indices 
$\l'_l, \l'_{\bar l}, \l_l, \l_{\bar l} \in \{ +, -\}$
and
$\rho^{\rm in}_{ (\l'_l, \l'_{\bar l}),(\l_l, \l_{\bar l})  }$ 
is the $4 \times 4$ density matrix of the diagonal form
\ba
&& \rho^{\rm in}_{ (\l'_l, \l'_{\bar l}),(\l_l, \l_{\bar l})  }
\,=\,
\delta_{\l'_l,\l_l}
\delta_{\l'_{\bar l},\l_{\bar l}}
\rho^{\rm in}_{ (\l_l, \l_{\bar l})  } \,, \nn \\
&&
\rho^{\rm in}_{ (\l_l, \l_{\bar l})  } 
\,=\,
\left\{ 
\tfrac{1}{4}(1 + {\cal P})(1 - \Pbar), \,
\tfrac{1}{4}(1 + {\cal P})(1 + \Pbar), \,
\tfrac{1}{4}(1 - {\cal P})(1 - \Pbar), \,
\tfrac{1}{4}(1 - {\cal P})(1 + \Pbar) 
\right\} , 
\label{rho^in}
\ea
where the basis representing the matrix is ordered as $(\l_l, \l_{\bar l}) = (++), (+-), (-+), (--)$.

\subsection{The final spin state}
\label{sec:final_state}

We are interested in the final $F \bar{F}$ spin state, $\hat{\rho}_{\mathrm{f}}$, resulting from the collision of $l \bar{l}$ with the beam polarisation parameters ${\cal P}$ and $\Pbar$.
In particular, we aim to retain the full kinematical dependence of the final spin state.
To describe the $F\bar{F}$ kinematics, we adopt the following coordinate system.
In the centre-of-mass frame, the $\hat z$-axis is defined along the direction of the incoming $l$-beam.
The production plane of the $F\bar{F}$ pair is chosen to coincide with the $(\hat x, \hat z)$ plane, with the $\hat x$-component of the $F$-momentum taken to be positive.
The $\hat y$-axis is then fixed so that the ($\hat x, \hat y, \hat z$) axes form a right-handed coordinate system.
In this setup, 
the $l$($\bar l$)-momentum is given by
$p^\mu = (E, 0,0,E)$ ($\bar p^\mu = (E, 0,0,-E)$)
and 
the $F$($\bar F$)-momentum is parametrised as $k^{\mu} = (E, k \sin \Theta, 0, k\cos \Theta)$
($\bar k^{\mu} = (E, -k \sin \Theta, 0, -k\cos \Theta)$)
with $E = \sqrt{s}/2$, $k = \sqrt{E^2 - m_F^2}$ and $\Theta \in[0, \pi]$, where $\sqrt{s}$ is the $l \bar l$ collision energy (which is equal to the $F \bar F$ invariant mass $M_{F \bar F}$) and $\Theta$ is the opening angle between the $l$-beam direction and the $F$-momentum direction. 
Our goal is to describe the final spin state, $\hat{\rho}_{\mathrm{f}}$, as a function of $\sqrt{s}$ and $\Theta$, as well as ${\cal P}$ and $\Pbar$.

The description of the final density matrix requires specifying the spin quantisation axes for $F$ and $\bar F$.
Following the same convention as for $l$ and $\bar l$, we quantise the spins of $F$ and $\bar F$ along the $F$–momentum direction, $\hat{k}$.
When the state is expressed as $|\lambda_F, \lambda_{\bar F}\rangle$ with $\lambda_F = \pm$ and $\lambda_{\bar F} = \pm$, the label $\lambda_F$ denotes the helicity of $F$, whereas $\lambda_{\bar F}$ corresponds to the negative of the helicity of $\bar F$ as the $\bar F$-momentum direction is $- \hat k$.

Expanding the final spin state in terms of the basis operators as
\be
\hat \rho_{\rm f} \,=\, 
\sum_{ \{ \l_{F \bar F} \}  }
\rho^{\rm f}_{ (\l'_F, \l'_{\bar F}),(\l_F, \l_{\bar F})}(\sqrt{s}, \Theta, {\cal P}, \Pbar) \cdot 
| \l'_F, \l'_{\bar F} \ketbra \l_F, \l_{\bar F} |\,,
\ee
the $4 \times 4$ spin density matrix $\rho^{\rm f}_{ (\l'_F, \l'_{\bar F}),(\l_F, \l_{\bar F})}$ is defined (the representation basis is ordered as $| \l_F, \l_{\bar F} \ket = |++\ket, |+-\ket, |-+\ket, |--\ket $), as a function of the beam collision energy $\sqrt{s}$, the production angle $\Theta$ and the beam polarisation parameters ${\cal P}$ and $\Pbar$.

From an explicit calculation in Appendix \ref{derivation}, the final spin density matrix is given by
\be
\rho^{\rm f}_{ (\l'_F, \l'_{\bar F}),(\l_F, \l_{\bar F})}
\,=\,
\frac{1}{\cal N} \sum_{ \l_l, \l_{\bar l} } 
\rho^{\rm in}_{(\l_l, \l_{\bar l})}({\cal P},\Pbar)
{\cal M}^{\l_l, \l_{\bar l}}_{\l'_F, \l'_{\bar F}}(\sqrt{s},\Theta)
\left[ {\cal M}^{\l_l, \l_{\bar l}}_{\l_F, \l_{\bar F}}(\sqrt{s},\Theta) \right]^*,
\label{rho_formula}
\ee
where ${\cal M}^{\l_l, \l_{\bar l}}_{\l'_F, \l'_{\bar F}}(\sqrt{s},\Theta)$ is the transition amplitude between the pure initial state $| l( p, \l_l ), \bar l( \bar p, \l_{\bar l} ) \ket$
and the pure final state $| F( k, \l_F ), \bar F( \bar k, \l_{\bar F} ) \ket$. 
The normalisation factor $1/\mathcal{N}$ ensures that the density matrix satisfies
$ 
{\rm Tr} \left( \rho^{\rm f}  \right)
= \sum_{ \{ \l_{F \bar F}\} } \rho^{\rm f}_{ (\l'_F, \l'_{\bar F}),(\l_F, \l_{\bar F})} = 1,
$
where the summation $\sum_{ \{\lambda_{F \bar F} \} }$ runs over all helicity indices
$\lambda’_F, \lambda’_{\bar F}, \lambda_F, \lambda_{\bar F} \in \{ +, - \}$.

\section{Quantum observables} 
\label{sec:QO}

In this section we introduce the set of quantum observables that will be used throughout this work to characterise the spin state of the $F\bar F$ system produced in polarised lepton collisions. These observables are constructed from the spin density matrix and are designed to capture complementary aspects of the underlying quantum correlations, ranging from classical mixedness to genuinely non-classical features. In particular, we consider the purity, as a measure of how close the state is to being pure; the concurrence and the Bell-inequality violation, which quantify bipartite entanglement; and the stabiliser R\'enyi entropy, which probes non-stabiliserness or ``magic'', a quantum computing resource. Together, these quantities provide a systematic and basis-controlled framework to analyse how beam polarisation, kinematics, and the Lorentz structure of the interaction shape the quantum properties of the $F\bar F$ final state, both within and Beyond the SM (BSM).

In this paper we study the following quantum observables. 
\begin{itemize}

\item 
{\bf Purity}, defined as
\be
\Gamma[\rho] \,\equiv\, {\rm Tr} \left( \rho^2 \right)\,,
\ee
quantifies how pure a quantum state is.
A state is pure when $\Gamma = 1$, and maximally mixed when $\Gamma = \frac{1}{D}$, where $D$ is the dimension of the Hilbert space.
For the density matrix $\rho^{\rm f}$ describing the $F\bar{F}$ spin state, we have $\frac{1}{4} \leq \Gamma \leq 1$.

\item 
{\bf Concurrence} quantifies the degree of entanglement between two qubits, that is, within a bipartite quantum system of dimension $2 \otimes 2$. 
For a given $4 \times 4$ density matrix $\rho$, the concurrence $\mathcal{C}$ is defined as~\cite{Wootters:1997id}
\begin{equation}
\mathcal{C}[\rho] = \max(0,\, \lambda_1 - \lambda_2 - \lambda_3 - \lambda_4)\,,
\end{equation}
where $\lambda_1, \lambda_2, \lambda_3, \lambda_4$ are the square roots of the eigenvalues (arranged in decreasing order) of the non-Hermitian matrix
$R = \rho (\sigma_y \otimes \sigma_y) \rho^* (\sigma_y \otimes \sigma_y)$.
Here, $\rho^*$ denotes the complex conjugate of $\rho$ in the computational basis, and $\sigma_y$ is the Pauli-$y$ matrix. 
The concurrence satisfies $0 \leq \mathcal{C}[\rho] \leq 1$: 
$\mathcal{C}[\rho] = 0$ indicates a separable (non-entangled) state, 
while $\mathcal{C}[\rho] = 1$ corresponds to a maximally entangled state.

There is a simpler way to compute the concurrence when the rank of the density matrix is 2~\cite{Hill:1997pfa}, which is the case we will encounter in section \ref{sec:EFT}.
A rank-2 density matrix can be written as a mixture of two pure states:
\be
\rho = p_1 | \psi_1 \ketbra \psi_1 | \,+\, p_2
| \psi_2 \ketbra \psi_2 |\,,
\ee
with $0 < p_1, p_2 < 1$ and $p_1 + p_2 = 1$.
In this case, the concurrence can be obtained by
\be
{\cal C}[\rho] = s_1 - s_2\,,
\ee
where $s_1$ and $s_2$ are the singular values ($s_1 > s_2$) of the matrix 
\be
\tau_{ij} = \sqrt{p_i p_j} 
\bra \psi_i | (\sigma_y \otimes \sigma_y) | \psi_j \ket\,.
\ee

\item
{\bf Bell-CHSH observable}~\cite{PhysicsPhysiqueFizika.1.195, PhysRevLett.23.880} is defined as the following combination of the spin correlations among two qubits: 
\be
{\cal B}_{\rm CHSH} \equiv 
\max_{ {\bf n},{\bf n}',{\bf m},{\bf m}' }
\left|
\bra s_{\bf n} s_{\bf m} \ket 
+ \bra s_{\bf n} s_{{\bf m}'} \ket
+ \bra s_{{\bf n}'} s_{\bf m} \ket
- \bra s_{{\bf n}'} s_{{\bf m}'} \ket 
\right|\,,
\ee
where $\bra s_{\bf n} s_{\bf m} \ket$ denotes the correlation (i.e., the expectation value of the product of spin measurement outcomes) of the first spin measured along ${\bf n}$ and the second along ${\bf m}$.
The maximum is taken over all choices of measurement directions ${\bf n},{\bf n}',{\bf m},{\bf m}'$.
If we assume that the two spins (the observables that take the binary outcome $\pm 1$) are described by local hidden variable theories (LHVTs), it follows~\cite{PhysRevLett.23.880}
\be
{\cal B}_{\rm CHSH} \leq 2 \quad (\rm LHVTs)\,.
\label{CHSH}
\ee

In quantum mechanics, for a given spin density matrix $\rho$, the spin correlation is computed as
$
\bra s_{\bf n} s_{\bf m} \ket
= {\rm Tr} \left[ 
\left(
{\bf n} \cdot {\boldsymbol \sigma} \otimes {\bf m} \cdot {\boldsymbol \sigma} \right)
\rho
\right],
$
and the Bell-CHSH observable is obtained as~\cite{Horodecki:1995nsk}
\be
{\cal B}_{\rm CHSH}[\rho] = 2 \sqrt{ \mu_1 + \mu_2 }\,,
\label{Horodecki}
\ee
where $\mu_1$ and $\mu_2$ are the two largest eigenvalues of the matrix $C^T C$ with $C$ being the spin correlation matrix 
$
C_{ij} \,\equiv\, {\rm Tr} \left[ \left( 
{\boldsymbol x}_i \cdot {\boldsymbol \sigma} \otimes {\boldsymbol x}_j \cdot {\boldsymbol \sigma} \right) \rho \right],
$
where ${\boldsymbol x}_i$ ($i=1,2,3$) are the three orthonormal vectors.\footnote{ The Bell-CHSH observable is independent of the choice of the orthonormal basis ${\boldsymbol x}_i$.
When computing the Bell-CHSH observable for the $F \bar F$ spin state, we adopt the helicity-basis ${\boldsymbol x}_i = ({\hat n},{\hat r},{\hat k})$. }
As the eigenvalues of the $C^T C$ matrix can range from 0 to 1, the CHSH inequality \eqref{CHSH} can be violated in quantum mechanics.
If the state violates the CHSH bound, 
${\cal B}_{\rm CHSH}[\rho] > 2$, then $\rho$ must be entangled. 
From the formula \eqref{Horodecki}, we see that the Bell-CHSH observable cannot exceed $2 \sqrt{2}$ in quantum mechanics, which is known as the Tsirelson bound~\cite{Cirelson:1980ry}.

\item
{\bf Magic} quantifies the non-stabiliser resource required for universal quantum computation. 
The Gottesman–Knill theorem~\cite{Gottesman:1998hu,Gottesman:1999tea,Aaronson:2004xuh,Knill:1996dv} shows that any quantum computation restricted to stabiliser states~\cite{Gottesman:1997zz}, which are states prepared from computational basis states using Clifford gates, can be efficiently simulated on a classical computer.
Since stabiliser states can already contain a large amount of entanglement, this result shows that entanglement alone does not provide a quantum computational advantage. 
The additional resource enabling quantum computational advantage is non-stabiliserness or ``magic'', which quantifies the nonclassical features required for universal quantum computation~\cite{Emerson:2013zse,Howard:2014zwm}.

The stabiliser R{\'e}nyi entropy (SRE) of order two is a commonly used measure to quantify the amount of magic or non-stabiliserness in a quantum state.
For a mixed two-qubit state, it is defined as~\cite{Leone:2021rzd,Wang:2023sre}
\be
M_2[\rho]
\,=\,
 -\log_2
 \left[ \frac{\sum_{P_1, P_2 \in \tilde {\cal P}} \left( \Tr [(P_1 \otimes P_2) \rho ] \right)^4 }{ \sum_{P_1, P_2 \in \tilde {\cal P}} \left( \Tr [(P_1 \otimes P_2) \rho ] \right)^2 } \right]\,,
 \label{M2rho}
\ee
with $\tilde {\cal P} = \{ 1, \sigma_x, \sigma_y, \sigma_z \}$.
With this definition, $M_2[\rho]$ is invariant under the Clifford unitaries~\cite{Leone:2021rzd}. 
The SRE $M_2[\rho]$ vanishes if and only if $\rho$ is a stabiliser state, that is, a convex mixture of pure stabiliser states~\cite{Leone:2021rzd,Wang:2023sre}.
For the two-qubit system, $M_2[\rho]$ attains its maximum value
$M_2^{\rm max} = - \log_2 \frac{7}{16} \simeq 1.1926$~\cite{Liu:2025frx} for maximally magical states, such as
$| \psi \ket = \tfrac{1}{2}( |+ - \ket + |-+ \ket + \sqrt{2} e^{-i\frac{\pi}{4}} |-- \ket)$~\cite{Ohta:2025utz}.

A general two-qubit density matrix can be expanded in the tensor-product basis of Pauli operators as
\be
\rho \,=\, \frac{1}{4} \left[  {\mathbb 1} + B_i \sigma_i \otimes {\mathbb 1} + \overline B_j {\mathbb 1} \otimes \sigma_j
+ C_{ij} \sigma_i \otimes \sigma_j \right]\,,
\label{BBC}
\ee
where the coefficients $B_i$, $\overline B_j$ and $C_{ij}$ ($i,j=1,2,3$) are real. 
Here $B_i = \mathrm{Tr}[(\sigma_i \otimes \mathbb{1}) \rho]$ and $\overline{B}_j = \mathrm{Tr}[(\mathbb{1} \otimes \sigma_j) \rho]$ describe the Bloch vectors (spin polarisations) of the first and second qubits, respectively, while $C_{ij} = \mathrm{Tr}[ (\sigma_i \otimes \sigma_j) \rho]$ quantifies the spin correlations between the two qubits.
Substituting Eq.\ \eqref{BBC} into Eq.\ \eqref{M2rho}, one obtains 
\be
M_2[\rho]
\,=\,
 -\log_2
\left[ 
\frac{1+\sum_{i=1}^3 (B_i)^4+\sum_{j=1}^3 (\overline B_j)^4+\sum_{i,j=1}^3 (C_{ij})^4}{1+\sum_{i=1}^3 (B_i)^2+\sum_{j=1}^3 (\overline B_j)^2+\sum_{i,j=1}^3 (C_{ij})^2}
\right]\,.
\label{M2rho_2}
\ee

As is evident from Eqs.~\eqref{M2rho} and \eqref{M2rho_2}, $M_2[\rho]$ is not invariant under local basis transformations.
In general,
$M_2[\rho] \neq M_2[\rho’]$ for
$\rho’ = (U_1 \otimes U_2)\, \rho\, (U_1^\dagger \otimes U_2^\dagger)$,
where $U_1$ and $U_2$ are local unitary operators acting on the individual qubit subsystems.
This behaviour contrasts with the other quantum observables defined above, $\Gamma[\rho]$, $\mathcal{C}[\rho]$ and $\mathcal{B}_{\mathrm{CHSH}}[\rho]$, all of which are invariant under local unitaries.

When calculating the SRE for the initial $l\bar l$ state, we adopt the beam-basis, i.e., the $(\hat x,\hat y,\hat z)$ coordinate system, defined in Sec.~\ref{sec:final_state}, which serves as the natural reference frame for this system.
This quantity is denoted by $M_2^{(\hat z)}[\rho^{\rm in}]$.

For the final $F \bar F$ spin state, two natural coordinate systems can be considered.
The first is the helicity-basis~\cite{Baumgart:2012ay}, defined by the orthonormal triad $({\hat r}, {\hat n}, {\hat k})$, where ${\hat k}$ is the $F$-momentum direction, ${\hat n} = \hat y$, and ${\hat r} \equiv {\hat n} \times \hat k$.
Since we expressed the density matrix $\rho^{\rm f}$ in Eq.~\eqref{rho_formula} using the quantisation axis $\hat k$, the SRE can be computed directly from Eqs.~\eqref{BBC} and~\eqref{M2rho_2}.
The resulting value is denoted by $M_2[\rho^{\rm f}]$.

The second natural coordinate system is the beam-basis, $(\hat x,\hat y,\hat z)$, introduced earlier.
To compute the SRE in the beam-basis, denoted by $M_2^{(\hat z)}[\rho^{\rm f}]$, the density matrix $\rho^{\rm f}$ in Eq.~\eqref{rho_formula} must be transformed such that the axes $({\hat r}, {\hat n}, {\hat k})$ are rotated to $({\hat x}, {\hat y}, {\hat z})$.
This transformation is achieved by rotating the Bloch vectors and the spin correlation matrix with the rotation matrix
\be
R \,=\,
\bmat
\cos \Theta & 0 & \sin \Theta \\
0 & 1 & 0 \\
-\sin \Theta & 0 & \cos \Theta 
\emat\,,
\ee
so that $B \to R B$, $\overline{B} \to R \overline{B}$, and $C \to R C R^{\mathrm{T}}$,
before applying Eq.~\eqref{M2rho_2}.

\end{itemize}

\section{Quantum properties of the initial state} 
\label{sec:QOI}
For the initial spin state of the $l\bar l$ system defined in Eq.~\eqref{rho^in}, it is straightforward to evaluate the four quantum observables introduced in the previous section.
The results are
\ba
\Gamma[ \rho^{\rm in} ] &=& \frac{ 1  }{4}
(1 + {\cal P}^2 ) (1 + \Pbar^2 )
\,,
\label{pure_in} \\
{\cal C}[ \rho^{\rm in} ] &=& 0
\,,
\label{conc_in}
 \\
{\cal B}_{\rm CHSH}[ \rho^{\rm in} ] &=& 2 | {\cal P} \Pbar |
\label{chsh_in}
\,,
 \\
M_2^{(\hat z)}[ \rho^{\rm in} ] &=& - \log_2 \left[ 
\frac{ (1 + {\cal P}^4 ) (1 + \Pbar^4 ) }{ (1 + {\cal P}^2 ) (1 + \Pbar^2 ) }
\right]\,.
\label{M2_in}
\ea

We see from Eq.~\eqref{pure_in} that the initial spin state is maximally mixed,
$\Gamma[\rho^{\mathrm{in}}] = \tfrac{1}{4}$, when both beams are unpolarised,
${\cal P} = \overline{\cal P} = 0$,
and becomes a pure state,
$\Gamma[\rho^{\mathrm{in}}] = 1$,
when both beams are perfectly polarised,
$|{\cal P}| = |\overline{\cal P}| = 1$.

Since the $l$- and $\bar l$-beams are prepared independently,
the initial state is a product state (see Eq.\ \eqref{hat_rho}),
and therefore contains no entanglement between the two subsystems.
Consequently, the concurrence vanishes for all polarisation configurations,
${\cal C}[\rho^{\mathrm{in}}] = 0$.

As shown in Eq.\ \eqref{chsh_in}, the Bell-CHSH observable ${\cal B}_{\mathrm{CHSH}}$
is zero for the maximally mixed state,
${\cal P} = \overline{\cal P} = 0$,
and reaches its maximum value ${\cal B}_{\mathrm{CHSH}}[\rho^{\mathrm{in}}] = 2$
in the pure state limit $|{\cal P}| = |\overline{\cal P}| = 1$.
This value still satisfies the CHSH inequality \eqref{CHSH},
indicating that no Bell-inequality violation can occur in the initial state
due to the absence of entanglement.

As discussed earlier, the SRE for the initial state is most naturally defined in the $(\hat x, \hat y, \hat z)$ coordinate system, and its analytic expression is given in Eq.~\eqref{M2_in}.
The quantity $M_2^{(\hat z)}[\rho^{\rm in}]$ vanishes when $|{\cal P}|$ and  $|\overline{\cal P}|$ are each either 0 or 1.
In contrast, it attains its maximum value, 
\be
M_2^{(\hat z)}[\rho^{\rm in}]_{\rm max} \,=\, -2-\log_2[3-2\sqrt{2}] \,\simeq\, 0.543,
\ee
at $|{\cal P}| = |\overline{\cal P}| = \sqrt{-1+\sqrt{2}} \simeq 0.644$.
This peak value is roughly half of the global maximum for a general two–qubit system,
$M_2^{\mathrm{max}} = -\log_2 \tfrac{7}{16} \simeq 1.193$.

\begin{figure}[h!]
\centering
\includegraphics[scale=0.335]{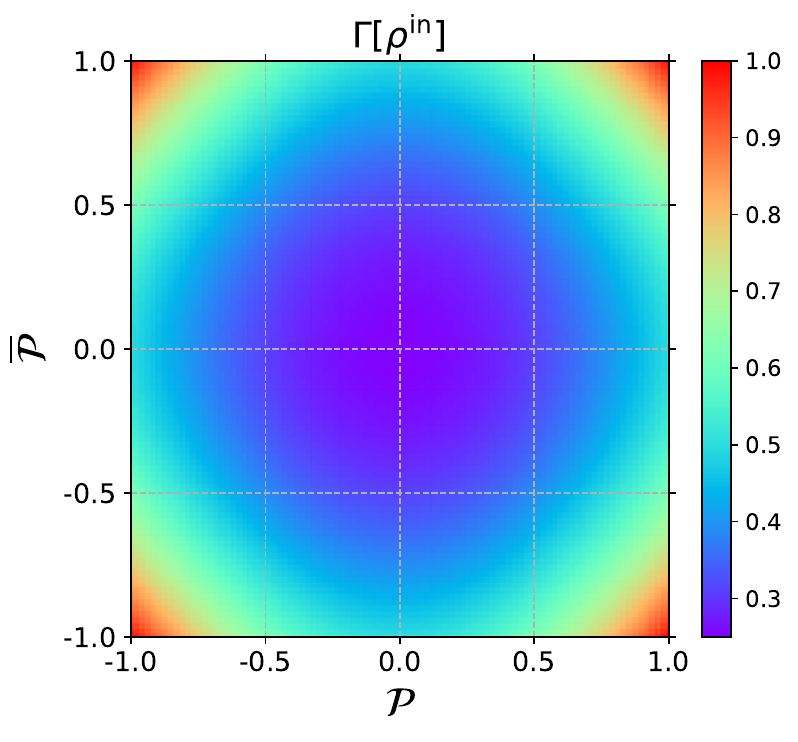}
\includegraphics[scale=0.335]{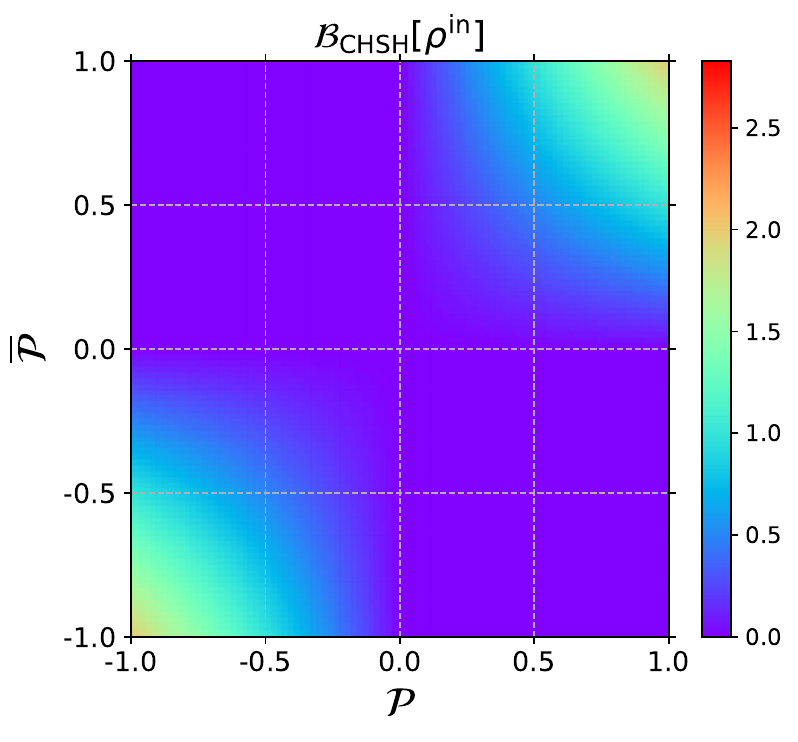}
\includegraphics[scale=0.335]{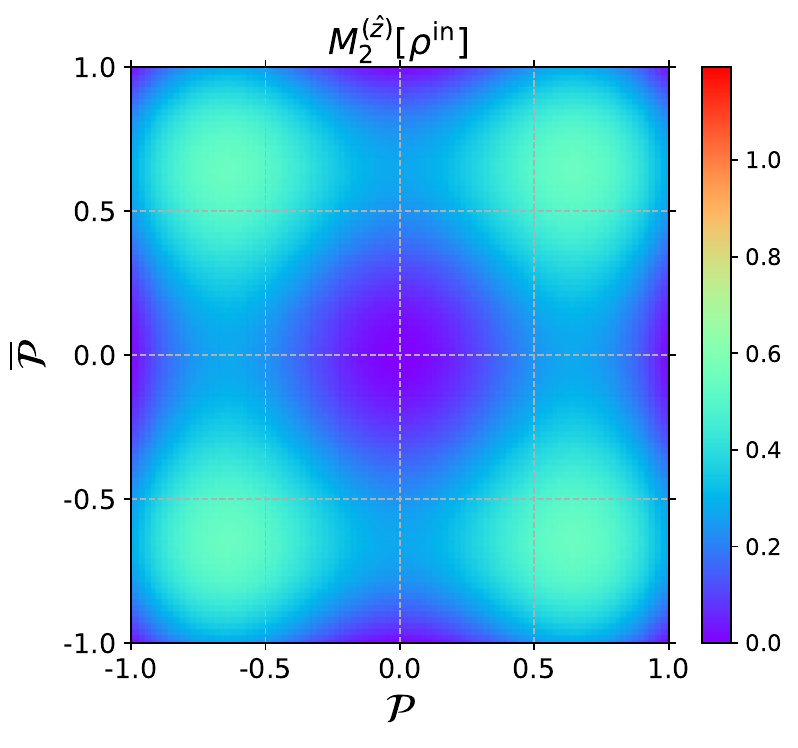}
\caption{\label{fig:QOI}
\small
The purity $\Gamma$ (left), 
the Bell-CHSH observable ${\cal B}_{\rm CHSH}$ (middle) 
and the SRE in the beam-basis $M_2^{(\hat z)}$ (right) 
are computed for the initial state $\rho^{\rm in}$ and shown in the $({\cal P}, \overline{\cal P})$ plane.
}
\end{figure}

The quantum observables that take non-trivial values for the initial $l\bar l$ state, namely, $\Gamma[\rho^{\rm in}]$, ${\cal B}_{\mathrm{CHSH}}[\rho^{\rm in}]$ and $M_2^{(\hat z)}[\rho^{\rm in}]$, are shown in the $({\cal P}, \overline{\cal P})$ plane in Fig.~\ref{fig:QOI}.

\section{Quantum properties in $l \bar l \to F \bar F$ } 
\label{sec:EFT}

In this section, we investigate the quantum properties of the $F \bar F$ spin state, produced by the collision of polarised $l \bar l$ beams. 
We examine this in a general EFT framework, assuming the following four-fermion interactions:
\ba
{\bf \rm Scalar:} &~&
~~~
 \frac{c_S}{\Lambda^2}
[ \bar \psi_l (\cos \xi_S + i \sin \xi_S \g_5) \psi_l ]
[ \bar \psi_F (\cos \eta_S + i \sin \eta_S \g_5) \psi_F ]\,,
\label{Lscalar}
\\
{\bf \rm Vector:} &~&
~~~
\frac{c_V}{\Lambda^2}
[ \bar \psi_l \g^\m (\cos \xi_R P_L + \sin \xi_R P_R)  \psi_l ]
[ \bar \psi_F \g_\m (\cos \eta_A + \sin \eta_A \gamma_5)  \psi_F ]\,,
\label{Lvector}
\\
{\bf \rm Tensor:} &~&
~~~
\frac{c_T}{\Lambda^2}
[ \bar \psi_l (\cos \xi_T + i \sin \xi_T \g_5) \s^{\m \n} \psi_l ]
[ \bar \psi_F (\cos \eta_T + i \sin \eta_T \g_5) \s_{\m \n} \psi_F ]\,.
\label{Ltensor}
\ea
Here, $P_{R,L}=\tfrac12(1\pm\gamma_5)$ are the chiral projection operators. 
The $c_S$, $c_V$ and $c_T$ are the dimensionless Wilson coefficients and $\Lambda$ is the cutoff scale of the EFT.
The real angles $\xi_S,\xi_R,\xi_T$ parametrise the $l\bar l$ spin structure in the interaction, while $\eta_S,\eta_R,\eta_T$ parametrise the $F \bar F$ spin structure.
Note that for the vector operator the spin decomposition is arranged asymmetrically: we use the chiral projectors $P_{L,R}$ for the $l\bar l$ current, while the $F \bar F$ current is expanded in ${\mathbb 1}$ and $\gamma_5$. With this convention the vector--EFT transition amplitudes take a particularly compact form, see Eq.~\eqref{amp_vec}.

In the tensor interaction, we include the fermion bilinear with $\gamma_5\sigma^{\mu\nu}$.
Using the identity $i\gamma_5\sigma^{\mu\nu}=- \frac{1}{2} \,\epsilon^{\mu\nu\rho\sigma}\sigma_{\rho\sigma}$
the above operator can be rewritten as
\ba
&&
[ \bar \psi_l (\cos \xi_T + i \sin \xi_T \g_5) \s^{\m \n} \psi_l ]
[ \bar \psi_F (\cos \eta_T + i \sin \eta_T \g_5) \s_{\m \n} \psi_F ]
\nn \\
&&~~~
=\,
\cos(\xi_T + \eta_T)
[ \bar \psi_l \s^{\m \n} \psi_l ]
[ \bar \psi_F \s_{\m \n} \psi_F ]
\,-\,
\frac{1}{2}\sin(\xi_T + \eta_T)
\e^{\m \n \r \s} 
[ \bar \psi_l \s_{\m \n} \psi_l ]
[ \bar \psi_F \s_{\r \s} \psi_F ]\,.
\ea
Therefore, only the combination $\xi_T + \eta_T$ is physical. 
We, however, retain the redundant parametrisation in Eq.\ \eqref{Ltensor} since it aligns our coupling conventions with the scalar and vector cases.

\subsection{Scalar interaction} 
\label{sec:scalar}

We first consider the scattering process $l \bar l \to F \bar F$ with the scalar-type four-fermion interaction \eqref{Lscalar}.
Due to the scalar nature of the interaction, the transition amplitudes are independent of the production angle $\Theta$ and 
only four of them, ${\cal M}^{+-}_{+-}, {\cal M}^{+-}_{-+}, {\cal M}^{-+}_{+-}, {\cal M}^{-+}_{-+}$, are non-vanishing in the $m_l^2/s \to 0$ limit.
From explicit calculations, we find
\ba
{\cal M}^{\l_l,\l_{\bar l}}_{\l_F,\l_{\bar F}}
\,=\, \delta_{\l_l,-\l_{\bar l}} \delta_{\l_F,-\l_{\bar F}} \cdot 
\frac{c_S}{\Lambda^2} 
\cdot s \cdot r \cdot 
e^{ i ( \l_l \xi_S - \l_F \delta_S) },
\label{M_S}
\ea
where 
\be
r \equiv \sqrt{ \beta^2 \cos^2 \eta_S  + \sin^2 \eta_S },~~~
\delta_S \equiv \arctan \left( \frac{\sin \eta_S}{\beta \cos \eta_S} \right)\,,
\label{dS}
\ee
being $\sqrt{s}$ the collision energy and $\beta \equiv \sqrt{1 - 4 m^2_F/s}$ the relative velocity of $F$ (and $\bar F$) in their centre-of-mass frame. 
Substituting the amplitudes \eqref{M_S} into the formula \eqref{rho_formula}, we obtain
\be
\rho^{\rm f}
\,=\,
\frac{1}{2}
\bmat
0 & 0 & 0 & 0\\
0 & 1 &  e^{-i2 \delta_S}  & 0 \\
0 & e^{i2 \delta_S} & 1 & 0 \\
0 & 0 & 0 & 0
\emat\,.
\label{rho_S}
\ee
Interestingly, the final spin density matrix $\rho^{\rm f}$ is independent of the $l$ CP parameter $\xi_S$ and of the beam polarisations ${\cal P}$ and $\overline{\cal P}$.
It depends only on the $F$ properties, i.e. on the  CP parameter $\eta_S$ and the velocity $\beta$, through $\delta_S$.
Moreover, $\rho^{\rm f}$ possesses a single non-zero eigenvalue equal to 1, indicating that the state is pure:
\be
\Gamma[\rho^{\rm f}] = 1\,,
\ee
independently of the beam polarisations and the initial state purity $\Gamma[\rho^{\rm in}]$.
This is because the two initial spin configurations $(\lambda_l, \lambda_{\bar l}) = (+,-)$ and $(-,+)$ evolve into the phase-equivalent pure final states: 
\ba
| + - \ket_{l \bar l}, ~ | - + \ket_{l \bar l} \,\longrightarrow\, \frac{1}{\sqrt{2}} \left[  | +- \ket_{F \bar F} + e^{ i 2 \d_S} | - + \ket_{F \bar F}  \right]  \,.
\label{scalar_state}
\ea
This final state is maximally entangled for arbitrary $\delta_S$, highlighting the fact that $CP$ conservation/violation and entanglement are independent. 
Accordingly, both the concurrence and the Bell-CHSH observable attain their maximal values:
\be
{\cal C} = 1,~~~~~
{\cal B}_{\rm CHSH} = 2 \sqrt{2}\,.
\ee
Unlike the $\Gamma$, ${\cal C}$ and ${\cal B}_{\rm CHSH}$ observables, $M_2$ exhibits a sensitivity to $\delta_S$. From the expression \eqref{rho_S}, one finds vanishing Bloch vectors
$B_i = \overline B_j = 0$ ($i,j=1,2,3$) and the spin correlation matrix
\be
C \,=\,
\bmat
\cos 2 \delta_S & -\sin 2 \delta_S & 0 \\
\sin 2 \delta_S & \cos 2 \delta_S & 0 \\
0 & 0 & -1
\emat\,,
\ee
which leads to the stabiliser R{\' e}nyi entropy  
\be
M_2 \,=\, - \log_2 \Xi_S \,~~~~{\rm with}~~~~\Xi_S 
\,=\, 1 - \frac{\sin^2 4 \delta_S}{4}
\,.
\label{M2_scalar}
\ee

As is evident from this expression, the $F \bar F$ final state is a stabiliser state, $M_2 = 0$, for $\delta_S = 0, \pm \frac{\pi}{4}$ and $\pm \frac{\pi}{2}$.
The case $\delta_S = 0$ corresponds to $\eta_S = 0$, whereas $\delta_S = \pm \frac{\pi}{4}$ implies $\beta = |\tan \eta_S|$ with ${\rm sign}(\delta_S) = {\rm sign}(\eta_S)$.
For $\delta_S = \pm \frac{\pi}{2}$, one obtains $\eta_S = \pm \frac{\pi}{2}$.

On the other hand, $M_2$ takes the maximum value $M^{\rm max}_2 = - \log_2 \frac{3}{4} \simeq 0.4150$ only when the $F$'s CP parameter $\eta_S$ and the beam collision energy conspire in the specific relations $\delta_S = \arctan( \frac{\sin \eta_S}{\beta \cos \eta_S} ) = 
\pm \frac{\pi}{8}$ and $\pm \frac{3\pi}{8}$.
Note that this maximum value is only about 1/3 of the theoretical maximum value of the two-qubit SRE $M_2^{\rm max} = - \log_2 \frac{7}{16} \simeq 1.1926$.

\begin{figure}[t!]
\centering
\includegraphics[scale=0.46]{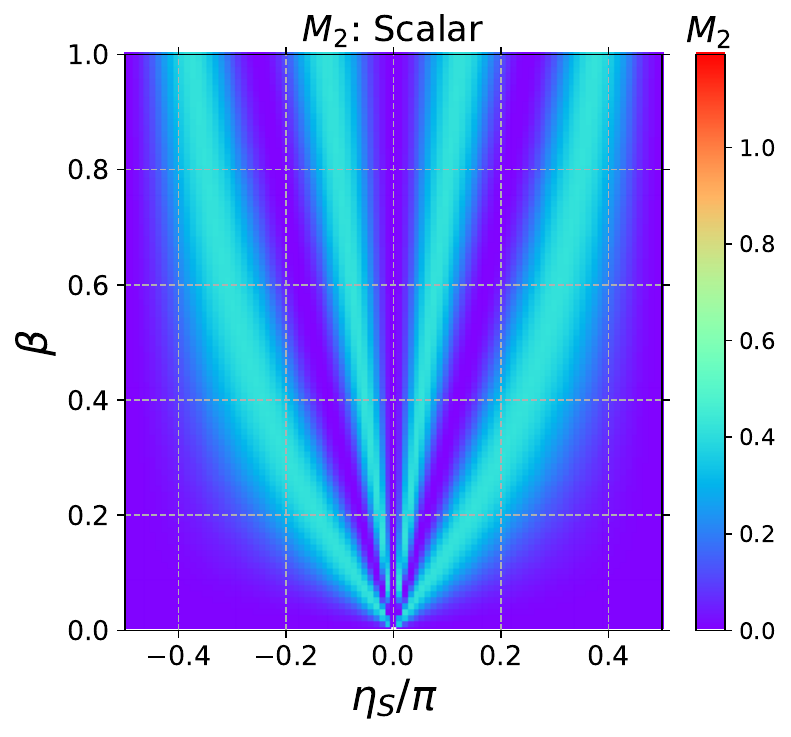}
\includegraphics[scale=0.46]{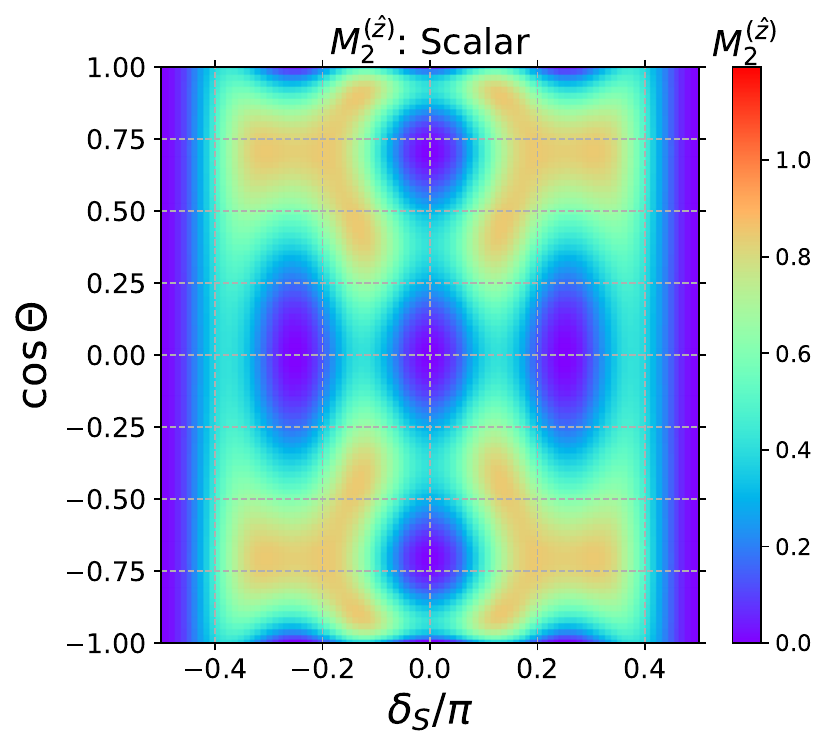}
\caption{\label{fig:M2_S}
\small
The stabiliser R{\' e}nyi entropies (SREs) for the scalar four-fermion interaction.
The left panel shows the SRE in the helicity-basis, $M_2$, over the $(\eta_S, \beta)$ plane.
The right panel shows the SRE in the beam-basis, $M_2^{(\hat{z})}$, over the $(\d_S, \cos \Theta)$ plane.
}
\end{figure}

In the left panel of Fig.\ \ref{fig:M2_S}, we present the behaviour of $M_2$ in the $(\eta_S, \beta)$ plane.
The five contour lines corresponding to the stabiliser states at $\delta_S = 0, \pm \frac{\pi}{4}, \pm \frac{\pi}{2}$, as well as the four contours representing the maximal values at $\delta_S = \pm \frac{\pi}{8}, \pm \frac{3\pi}{8}$, are clearly visible.

As mentioned earlier, the SRE is basis dependent. 
For the SRE in the beam-basis, we find 
\be
M_2^{(\hat{z})} \,= - \log_2 \left(  \Xi_S + \Delta_S \right)
~~~~{\rm with}~~~~
\Delta_S \,=\,
-\frac{\sin^2  2\Theta}{16}
\left[
7\,\sin^{4} 2\delta_S 
+ 16\,\cos^{8}\!\delta_S\,\cos^{2} 2\Theta 
\right]\,.
\label{M2z_scalar}
\ee

The right panel of Fig.\ \ref{fig:M2_S} displays $M_2^{(\hat z)}$ in the $(\d_S, \cos \Theta)$ plane.
$M_2^{(\hat z)}$ vanishes identically for $\d_S = \pm \frac{\pi}{2}$.
There are additional eleven discrete zeros in the plane:
nine are located at $\delta_S \in \{0, \pm \frac{\pi}{4} \}$
and $\cos \Theta \in \{0, \pm 1 \}$;
and two are located at $\delta_S = 0$ and $\cos \Theta = \pm \frac{1}{\sqrt{2}} \simeq \pm 0.707$.

There are four maximum magic points in the plane, which are located at
$\cos \Theta = \pm \frac{1}{\sqrt{5}} \simeq \pm 0.447$
and 
$\delta_S = \pm \arctan \frac{1}{\sqrt{5}} \simeq \pm 0.134 \pi$.
At these points, the SRE takes the maximum value 
$M_2^{(\hat z)} = - \log_2 \frac{5}{9} \simeq 0.848$.

\subsection{Vector interaction} 
\label{sec:vector}
For the vector-type interaction \eqref{Lvector}, 
the transition amplitudes are given by

\bea
{\cal M}^{++}_{++} &=&
- \frac{c_V}{\Lambda^2} \cdot s \cdot  \sin \xi_R ( \cos \eta_A + \beta \sin \eta_A) \cdot (1 + \cos \Theta) \,,
\nonumber \\
{\cal M}^{++}_{+-} \,=\, {\cal M}^{++}_{-+} &=&  \frac{c_V}{\Lambda^2} \cdot s \cdot \gamma^{-1} \cdot \sin \xi_R \cos \eta_A \cdot \sin \Theta\,,
\nonumber \\
{\cal M}^{++}_{--} &=& 
- \frac{c_V}{\Lambda^2} \cdot s \cdot  \sin \xi_R ( \cos \eta_A - \beta \sin \eta_A) \cdot (1 - \cos \Theta) \,.
\nn \\
{\cal M}^{--}_{++} &=&
- \frac{c_V}{\Lambda^2} \cdot s \cdot  \cos \xi_R ( \cos \eta_A + \beta \sin \eta_A) \cdot (1 - \cos \Theta) \,,
\nonumber \\
{\cal M}^{--}_{+-} \,=\, {\cal M}^{--}_{-+} &=&  - \frac{c_V}{\Lambda^2} \cdot s \cdot \gamma^{-1} \cdot \cos \xi_R \cos \eta_A \cdot \sin \Theta\,,
\nonumber \\
{\cal M}^{--}_{--} &=& 
- \frac{c_V}{\Lambda^2} \cdot s \cdot  \cos \xi_R ( \cos \eta_A - \beta \sin \eta_A) \cdot (1 + \cos \Theta) \,.
\label{amp_vec}
\eea
The amplitudes ${\cal M}_{\l_F,\l_{\bar F}}^{+-}$
and ${\cal M}_{\l_F,\l_{\bar F}}^{-+}$ vanish in the limit $m_l^2/s \to 0$ due to the vector nature of the interaction. 
As before, $\sqrt{s}$ represents the collision energy, $\beta \in (0,1)$ is the velocity of $F$ (and $\bar F$) and $\gamma = 1/\sqrt{1 - \b^2}$ is the Lorentz boost factor. 

Since the amplitudes are non-zero only for $(\l_l,\l_{\bar l}) = (+,+)$ and $(-,-)$ and $\rho^{\rm f}$ is normalised as ${\rm Tr} \rho^{\rm f} = 1$, the dependence of $\rho^{\rm f}$ on the beam polarisations arises solely through the ratio
\be
\frac{ \rho^{\rm in}_{+,+} \cdot \sin^2 \xi_R }{\rho^{\rm in}_{-,-} \cdot \cos^2 \xi_R } 
\,=\,
\frac{ (1 + {\cal P})(1 - \Pbar)\cdot \sin^2 \xi_R }{ (1 - {\cal P})(1 + \Pbar)\cdot \cos^2 \xi_R }
\,\equiv\, \tan^2 \PhiV \,,
~~~~~\PhiV \in \left[ 0, \frac{\pi}{2}  \right]
\label{PhiV}.
\ee
Namely, the two $F \bar F$ pure states that originated from the $|+ + \ket_{l \bar l}$ and $|-- \ket_{l \bar l}$ initial states are mixed with the corresponding weights $\omega_{++} \propto \sin^2 \PhiV$ and $\omega_{--} \propto \cos^2 \PhiV$, respectively.
The final spin state is therefore pure at $\PhiV = 0$ (either the interacting $l$ is purely left-handed $(\sin \xi_R = 0)$ or one of the beams is perfectly polarised with ${\cal P} = -1$ or $\overline {\cal P} = 1$)
and 
$\PhiV = \frac{\pi}{2}$
(either the interacting $l$ is purely right-handed $(\cos \xi_R = 0)$ or one of the beams is perfectly polarised with ${\cal P} = 1$ or $\overline {\cal P} = -1$).
We observe that 
$\PhiV$ becomes independent of the values of ${\cal P}$ and $\overline {\cal P}$
if ${\cal P} = \overline {\cal P}$.
This means the beam is effectively ``unpolarised'' in this limit 
regardless of the ${\cal P}$ values.

The $F \bar F$ spin density matrix is then parametrised by the effective beam polarisation angle $\PhiV$, the production angle $\Theta$, $F$'s velocity $\beta$ and $F$'s axial coupling angle $\eta_A$.
The explicit form of $\rho^{\rm f}$ is given in Eq.\ \eqref{eq:rho_vec} (Appendix \ref{app:rho}).

Unlike the scalar case, quantum observables do not admit simple forms for the general case of the vector interaction.  
Below, we consider two special cases:
({\it i}) the vector-like $F \bar F$ interaction, characterised by $\sin \eta_A = 0$; and
({\it ii}) the axial-vector-like $F \bar F$ interaction with $\cos \eta_A = 0$.
We also consider ({\it iii}) an example parameter point interpolating between ({\it i}) and ({\it ii}).

\subsubsection*{ $(i)$ The vector-like ($\sin \eta_A = 0$) interaction}

The quantum observables are evaluated under the assumption $\sin\eta_A = 0$ (vector-like). 
From a direct calculation we find the following analytic expressions:
\ba
&&
\quad \quad \quad \quad \quad \quad \quad \quad 
\quad \quad \quad
\Gamma \,=\, 1 - \frac{ 2 (1 - B) }{ (2 - B)^2 }
\sin^2 2 \PhiV
\,,
\label{pure_v}
\\
&&
\quad \quad \quad \quad \quad \quad \quad \quad 
{\cal C} \,=\, \frac{ B }{2 - B} \,,
\quad \quad
{\cal B}_{\rm CHSH} \,=\, 2 \sqrt{1 + {\cal C}^2}\,,
\label{conc_v}
\\
&& 
M_2 \,=\, -\log_2 \left[ \frac{F_V + G_V \cos^4 2 \PhiV }{ 4 \Gamma (2 - B)^4 } \right]\,,~~~~~
M_2^{(\hat z)} \,=\, -\log_2
\left[ \frac{F_V^{(\hat z)} + G_V^{(\hat z)} \cos^4 2 \PhiV }{ 4 \Gamma (2 - B)^4 } \right] \,,
\label{magic_v}
\ea
where $B \equiv \beta^2 \sin^2 \Theta$ and
\ba
F_V &\equiv& 
( 2 - B )^{4}
+\bigl( B + 2 \cos^2 \Theta \bigr)^{4}
+2 \gamma^{-4} \sin^{4} 2\Theta
+ \left[ \beta^{8}+(2 - \beta^{2})^{4}\right] \sin^{8}\Theta
\,,
\nn \\
G_V &\equiv&
32 \left[ \cos^4 \Theta + \gamma^{-4} \sin^4 \Theta \right]\,,
\nn \\
G_V^{(\hat z)}
&\equiv&
32\big[\big(\cos^{2}\Theta+\gamma^{-1} \sin^{2}\Theta\big)^{4}
 + 2 \big( 1-\gamma^{-1} \big)^{4} \sin^{4}2\Theta\big]\,,
 \nn \\
F_V^{(\hat z)} &\equiv& (2-B)^{4}+\frac{1}{256}(-6+\beta^{2}-2\gamma^{-1}-2\beta^{2}\cos 2\Theta+H\cos 4\Theta)^{4} \nn \\
&&  +\sin^{4}\Theta \cos^{4}\Theta\!\left[2(\beta^{2}-H\cos 2\Theta)^{4}\right] +\sin^{8}\Theta\!\left[\beta^8+(2(-1+\gamma^{-1})+H\cos 2\Theta)^{4}\right]\,,
 \label{FV}
\ea
with $\gamma^{-1} = \sqrt{1 - \beta^2}$ and $H \equiv \b^2 + 2(\g^{-1} -1)$.

\begin{figure}[t!]
\centering
\includegraphics[scale=0.46]{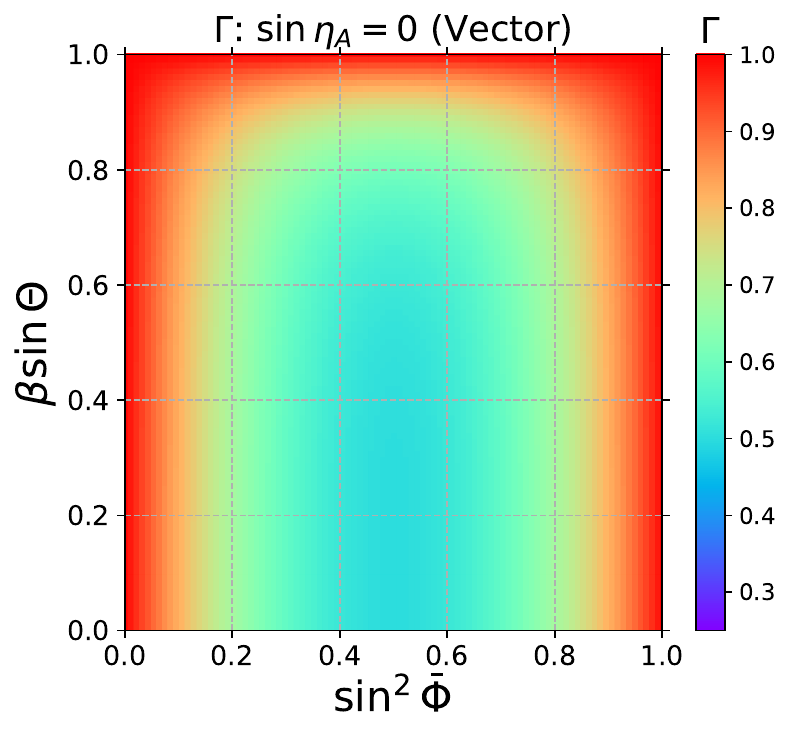}
\hspace{3mm}
\includegraphics[scale=0.46]{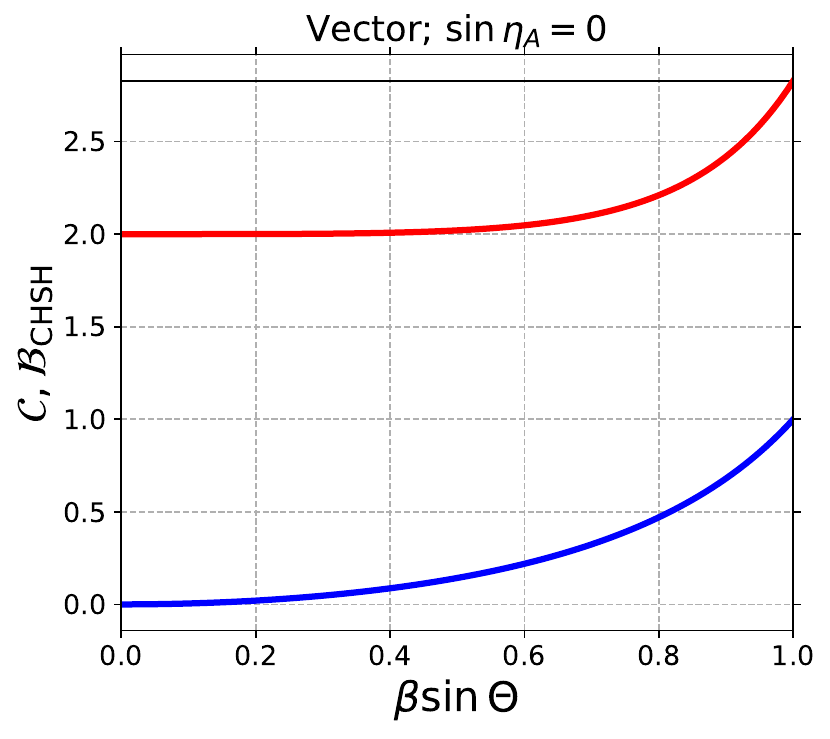}
\caption{\label{fig:vec_v}
\small
The left panel displays the purity $\Gamma$ for the vector interaction ($\sin \eta_A = 0$) over the ($\sin^2 \PhiV, \b \sin \Theta$) plane. 
The right panel shows the concurrence ${\cal C}$ (blue) and the Bell-CHSH observable ${\cal B}_{\rm CHSH}$ (red) as functions of $\beta \sin\Theta$. 
}
\end{figure}

We observe that the purity $\Gamma$ depends on $\PhiV$ and on $\beta$ and $\Theta$ only through the combination $\beta \sin \Theta$.
The expression manifests the fact that the final spin state is pure at $\PhiV=0$ and $\PhiV=\pi/2$.
It also reveals that the state is pure for $\beta \sin \Theta = 1$ for any polarisation angle $\PhiV$.
This is because in this limit, the final states arising from the initial states $|++\rangle_{l\bar l}$ and $|–-\rangle_{l\bar l}$ approach a Bell state
\be
|++\rangle_{l\bar l},\,
|--\rangle_{l\bar l}\,\longrightarrow\,
\frac{1}{\sqrt{2}} \left[\, |++\rangle_{F \bar F}
+ |--\rangle_{F \bar F} \right]\,.
\label{triplet}
\ee
The state is never maximally mixed ($\Gamma_{\rm max\,mix} = \frac{1}{4}$). 
The minimal purity, $\Gamma = \frac{1}{2}$, is achieved when $\beta \sin \Theta = 0$ and $\PhiV=\pi/4$.
These behaviours are visible in the left panel of Fig.\ \ref{fig:vec_v}, which shows the purity $\Gamma$ over the $(\sin^2 \PhiV, \, \beta\sin\Theta)$ plane.

The analytical expression for the concurrence ${\cal C}$ is found in Eq.\ \eqref{conc_v}.
Remarkably, in the vector-like case ($\sin \eta_A =0$), the concurrence is independent of the effective polarisation angle $\PhiV$ and depends on $\beta$ and $\Theta$ only through the combination $\beta \sin \Theta$.
We see that the $F \bar F$ spin is maximally entangled, ${\cal C} = 1$, when $\beta \sin \Theta = 1$.
This can be understood as the state approaches the Bell state \eqref{triplet} in this limit.

One can also see that the $F \bar F$ state always possesses non-zero entanglement except for the special case, $\beta \sin \Theta = 0$.
In fact, {\it for any axial coupling angle $\eta_A$}, the final states arising from the two pure initial states are the following product (separable) states
\ba
| ++ \ket_{l \bar l} &\,\longrightarrow\,&
 \left(  \cos \tfrac{\Theta}{2} |+ \ket_F - \sin \tfrac{\Theta}{2} | - \ket_F  \right)
\otimes \left( \cos \tfrac{\Theta}{2} |+ \ket_{\bar F} - \sin \tfrac{\Theta}{2} | - \ket_{\bar F} \right)
\,=\, |+_z\ket_F \otimes |+_z\ket_{\bar F}
\,,
\nn \\
| -- \ket_{l \bar l} &\,\longrightarrow\,&
 \left( \sin \tfrac{\Theta}{2} |+ \ket_F + \cos \tfrac{\Theta}{2} | - \ket_F \right)
\otimes \left( \sin \tfrac{\Theta}{2} |+ \ket_{\bar F} + \cos \tfrac{\Theta}{2} | - \ket_{\bar F} \right) 
\,=\, |-_z\ket_F \otimes |-_z\ket_{\bar F}\,,
\label{b=0state}
\ea
in the threshold $\beta \to 0$ limit.
In this limit, $F$ and $\bar F$ are at rest and $\Theta$ simply indicates the spin quantisation axis.  
If the spin quantisation axis is taken along in the $z$-direction, the $F \bar F$ spins are $++$ or $--$ as indicated in the right-hand-side of Eq.\ \eqref{b=0state}. 

In the forward/backward limit $\sin \Theta \to 0$, we instead have
\be
\begin{tabular}{@{}l@{\hspace{1.6em}}l@{}}
\hspace{24mm} $\Theta=0$ & $\Theta=\pi$\\[2pt]
$\begin{aligned}
|++ \ket_{l \bar l} &~~~\longrightarrow~~ |++ \ket_{F \bar F}\\
|-- \ket_{l \bar l} &~~~\longrightarrow~~ |-- \ket_{F \bar F}
\end{aligned}$ &
$\begin{aligned}
|--\ket_{F \bar F}\\
|++\ket_{F \bar F}
\end{aligned}$
\end{tabular}
\ee
Since any convex mixture of pure product states is separable, the entanglement vanishes in the above limits and the concurrence is zero, ${\cal C}=0$, for all $\PhiV$ and $\eta_A$. The Bell-CHSH observable in the vector-like case $\sin \eta_A = 0$ is monotonically related to the concurrence ${\cal C}$ as indicated in Eq.\ \eqref{conc_v}.
The expression implies a CHSH violation for all non-zero $\b \sin \Theta$ and saturation of the Tsirelson bound, ${\cal B}_{\rm CHSH} = 2 \sqrt{2}$, at $\b \sin \Theta = 1$. The right panel of Fig.\ \ref{fig:vec_v} illustrates the concurrence (blue) and the Bell-CHSH observable (red) as functions of $\beta \sin\Theta$.

\begin{figure}[t!]
\centering
\includegraphics[scale=0.335]{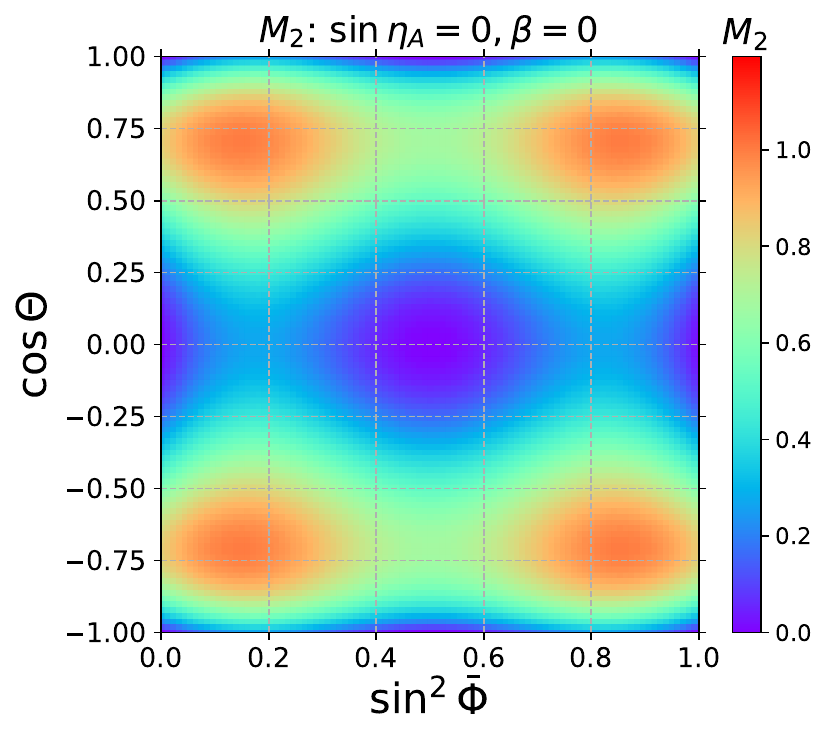}
\includegraphics[scale=0.335]{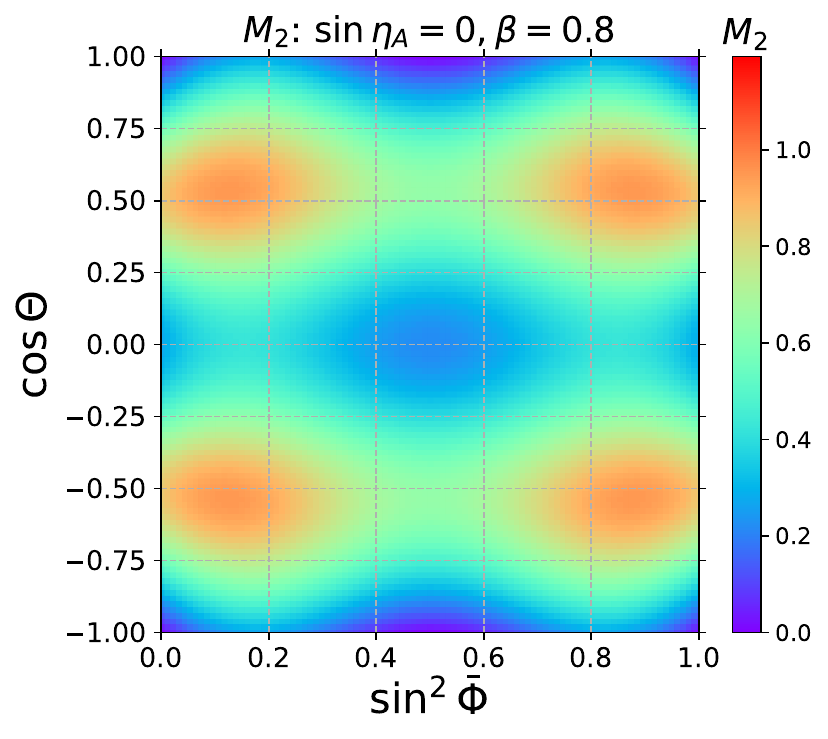}
\includegraphics[scale=0.335]{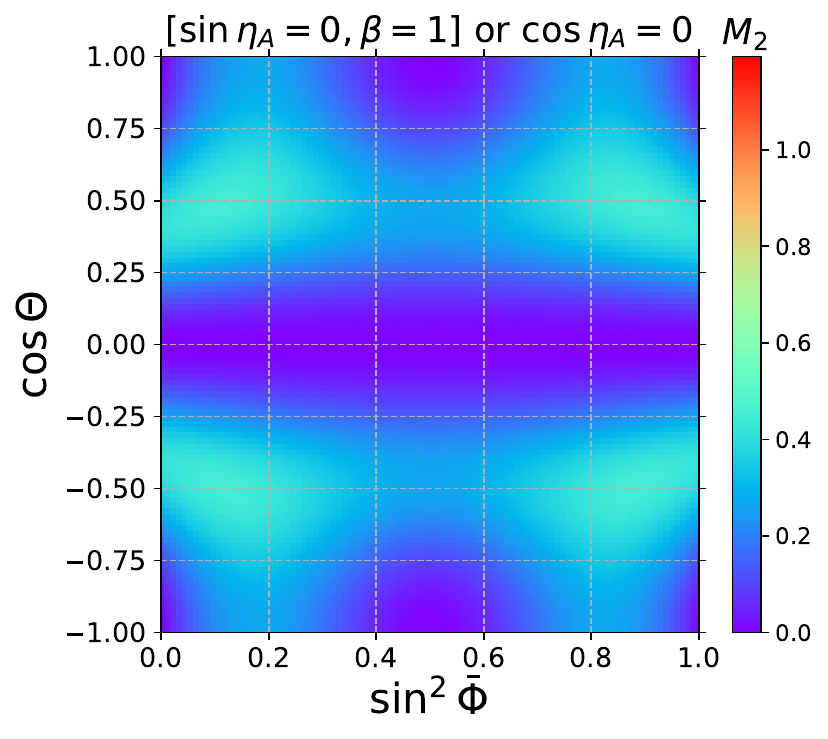}
\includegraphics[scale=0.335]{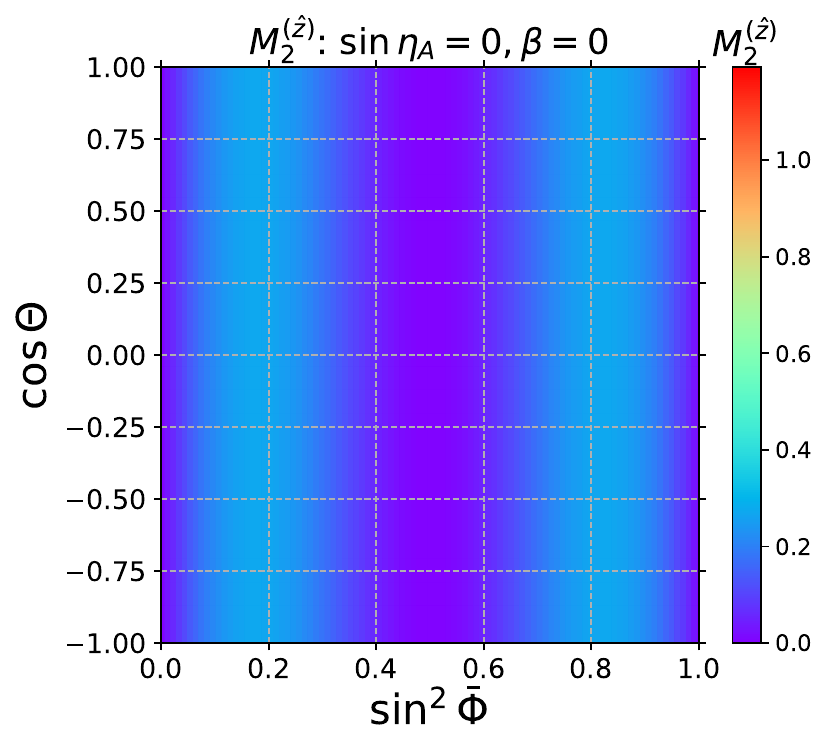}
\includegraphics[scale=0.335]{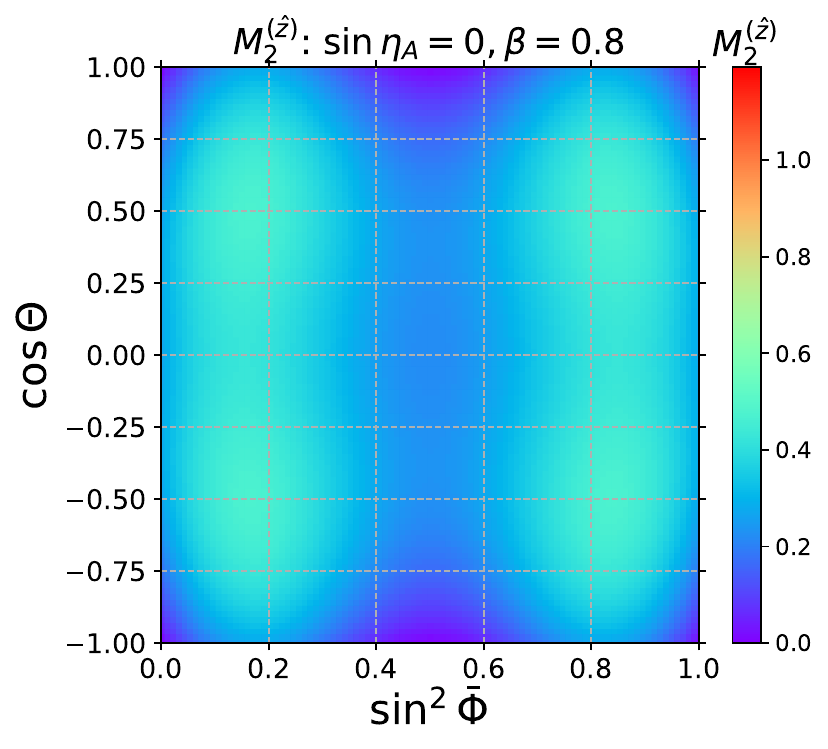}
\includegraphics[scale=0.335]{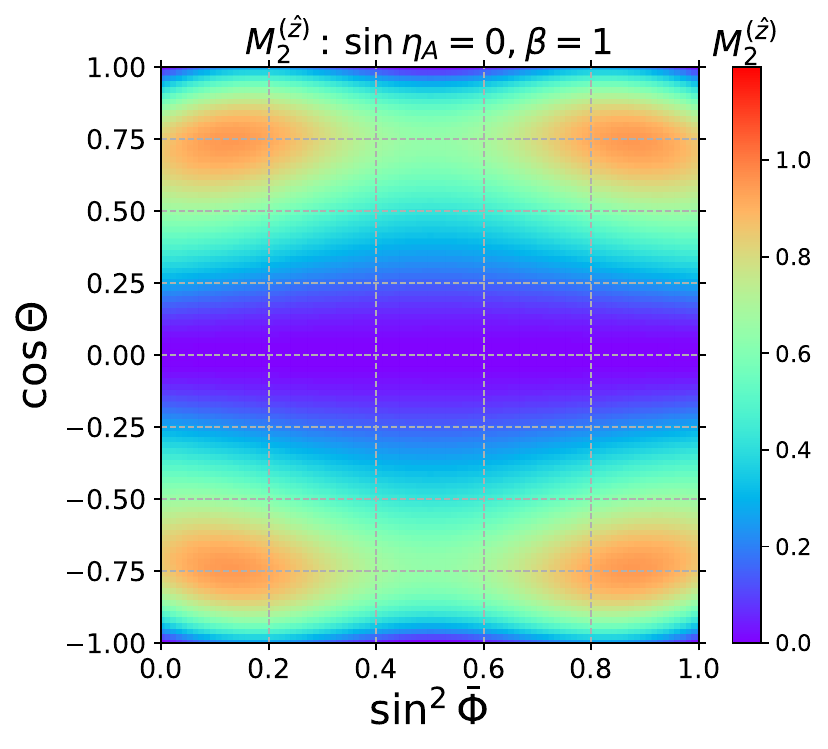}
\caption{\label{fig:m2_v}
\small
The upper (lower) panel shows the helicity (beam) basis stabiliser R{\' e}nyi entropy $M_2$ ($M_2^{(\hat z)}$) for the vector-like case $(\sin \eta_A = 0)$ over the $(\sin^2 \PhiV, \cos \Theta)$ plane, respectively. 
The left, middle and right panels correspond to different velocities: $\b = 0$ (threshold); $\b = 0.8$ (intermediate);
and $\b = 1$ (ultra-relativistic).
The upper-right plot also represents the $M_2$ for the axial-vector-like interaction ($\cos \eta_A = 0$).
}
\end{figure}

The analytical expressions for the stabiliser R{\' e}nyi entropies in the helicity and beam bases, $M_2$ and $M_2^{(\hat z)}$, are given in Eq.\ \eqref{magic_v} through the functions $F_V$, $G_V$, $F_V^{(\hat z)}$, $G_V^{(\hat z)}$ defined in Eq.\ \eqref{FV}. 
In these expressions, $\Gamma$ represents the purity function defined in \eqref{pure_v}.

Fig.\ \ref{fig:m2_v} illustrates $M_2$ (upper panel) and $M_2^{(\hat z)}$ (lower panel) over the ($\sin^2 \PhiV$, $\cos \Theta$) plane for three representative $\beta$ values: $\beta = 0$ (left panel), 0.8 (middle panel) and 1 (right panel).  
We observe that there are two independent symmetry transformations 
\begin{align}
  {\rm Sym\,1}:~~ \PhiV &\to \pi/2 - \PhiV     \nn \\
  {\rm Sym\,2}:~~\Theta &\to \pi - \Theta 
  \label{sym_v}
\end{align}
that leave $M_2$ and $M_2^{(\hat z)}$ invariant.

For $\beta = 0$, $\rho^{\rm f}$ is a classical mixture of two $\Theta$-independent product states $|+_z\ket_T \otimes |+_z\ket_{\bar T}$ and $|-_z\ket_T \otimes |-_z\ket_{\bar T}$ (see Eq.\ \eqref{b=0state}) with the weights $\sin^2 \PhiV$ and $\cos^2 \PhiV$, respectively.  
The upper-left panel of Fig.\ \ref{fig:m2_v} shows the SRE $M_2$ computed in the $\Theta$-dependent helicity-basis.   
We see that although the quantum state is $\Theta$-independent, the resulting SRE has a $\Theta$-dependence.
We observe that $M_2$ has the maximum value $M_2 = 1$ at $(\PhiV, \Theta) = (\frac{\pi}{8}, \frac{\pi}{4})$ in the fundamental region $0 \leq \PhiV \leq \frac{\pi}{4}$ and $0 \leq \Theta \leq \frac{\pi}{2}$.  
The other three maximal points can be obtained via the symmetries \eqref{sym_v}.
One can also see that $M_2$ possesses nine minima with $M_2 = 0$ at $\sin^2 \PhiV \in \{ 0, \,0.5,\, 1 \}$ and $\cos \Theta \in \{ -1, 0, 1\}$.

The lower-left plot of Fig.\ \ref{fig:m2_v} shows the SRE in the threshold limit $\beta \to 0$ computed in the beam-basis, $M_2^{(\hat z)}$.
One can see that $M_2^{(\hat z)}$ is $\Theta$-independent, since neither the quantum state nor the basis depends on $\Theta$.
The $M_2^{(\hat z)}$ vanishes at $\sin^2 \PhiV \in \{ 0,\, 0.5,\, 1\}$, while the maximal value $M_2^{(\hat z)} = -1 - \log_2 \left[ \sqrt{2} - 1 \right] \simeq 0.272$ is found at $\sin^2\PhiV = \frac{1}{2} \left( 1 \pm \sqrt{ \tan \frac{\pi}{8} } \right) \in \{ 0.178, 0.822 \}$.

We now turn our attention to the right panel of Fig.\ \ref{fig:m2_v} which represents $M_2$ and $M_2^{(\hat z)}$ in the ultra-relativistic limit $\beta \to 1$.
We see that both $M_2$ and $M_2^{(\hat z)}$ present the new continuous minima at $\cos \Theta = 0$.
Along this line, $\rho^{\rm f}$ is the Bell state \eqref{triplet}.    
This reflects the fact that the maximally entangled Bell state has the vanishing magic. 
Compared to the $\b = 0$ case, the maximal value of $M_2$ is reduced, $M_2=-\log_2[-1+\sqrt{3}]\simeq 0.450$, and given at $\cos \Theta = \pm \frac{1}{2}$ and $\sin^2 \PhiV = \{ \sin^2 \frac{\pi}{9}, \sin^2 \frac{7 \pi}{18} \} \simeq \{ 0.117, 0.883\}$.
On the other hand, the maximum of $M_2^{(\hat z)}$ is substantially larger, $M_2^{(\hat z)} \simeq 0.956$, attained at $\cos \Theta \simeq \pm 0.765$ and $\sin^2 \PhiV \simeq \{ 0.133, 0.866 \}$.

The SREs $M_2$ and $M_2^{(\hat z)}$ between the two extremal limits $\b =0$ and $\b=1$ are shown in the middle panel of Fig.\ \ref{fig:m2_v} by fixing $\b = 0.8$.
Although $\beta$ is rather large, the plots are more similar to the corresponding plots for $\b = 0$.
The $\b$-dependence on the SREs is rather weak for small $\b$.

\subsubsection*{ $(ii)$ The axial-vector-like ($\cos \eta_A = 0$) $F \bar F$ interaction}

Substituting $\cos \eta_A \to 0$ in the formula \eqref{amp_vec}, we have ${\cal M}_{+-}^{\l_l, \l_{\bar l}} = {\cal M}_{-+}^{\l_l, \l_{\bar l}} = 0$.
As a consequence, the $F \bar F$ spin states arising from the two initial states are the superposition of $|++ \ket_{F \bar F}$ and $|-- \ket_{F \bar F}$ only.  
They are given by
\ba
| ++ \ket_{l \bar l} &\,\longrightarrow\,&
\frac{1}{N}
\left[ \cos^2 \tfrac{\Theta}{2} | ++ \ket_{F \bar F}
+ h \cdot  \sin^2 \tfrac{\Theta}{2} | -- \ket_{F \bar F} \right]
\,,
\nn \\
| -- \ket_{l \bar l} &\,\longrightarrow\,&
\frac{1}{N}
\left[
\sin^2 \tfrac{\Theta}{2} | ++ \ket_{F \bar F}
+ h \cdot  \cos^2 \tfrac{\Theta}{2} | -- \ket_{F \bar F}
\right]
\,.
\label{b=1state}
\ea
with $N = \sqrt{\cos^4 \tfrac{\Theta}{2} + \sin^4 \tfrac{\Theta}{2}}$ and $h = -1$.
As can be seen, the final state is independent of $\beta$.

The situation is similar for the ultra-relativistic vector-like case with $\sin \eta_A = 0$ and $\beta =1$.
In this case, we also have ${\cal M}_{+-}^{\l_l, \l_{\bar l}} = {\cal M}_{-+}^{\l_l, \l_{\bar l}} = 0$ due to $\gamma^{-1} = 0$,
and the final state is given by the same formula \eqref{b=1state} with the different sign factor $h = +1$.

By mixing the two final states in Eq.\ \eqref{b=1state} with the probabilities $\sin^2 \PhiV$ and $\cos^2 \PhiV$, the density matrix is obtained as 
\be
\rho^{\rm f} \,=\,
\bmat
\frac{1}{2} - a & 0 & 0 & h \cdot b \\
0 & 0 & 0 & 0 \\ 
0 & 0 & 0 & 0 \\ 
h \cdot b & 0 & 0 & \frac{1}{2} + a 
\emat\,,
\label{rhof_ax}
\ee
with $a \equiv 2 \cos \Theta \cos 2 \PhiV / D$,
$b \equiv \sin^2 \Theta / D$
and 
$D \equiv 3 + \cos 2 \Theta$.

It is straightforward to see that the quantum observables listed in section \ref{sec:QO}, other than $M_2^{(\hat z)}$, are not sensitive to the sign factor $h$; these observables (except for $M_2^{(\hat z)}$) depend quadratically on $h$.
As a consequence, the expressions for the purity $\Gamma$, concurrence ${\cal C}$, the Bell-CHSH observable ${\cal B}_{\rm CHSH}$ for the axial-vector-like case are obtained by substituting $\beta = 1$ in the corresponding formulae given in Eqs.\ \eqref{pure_v}, \eqref{conc_v}. 
Their response to the other parameters, $\PhiV$ and $\Theta$, can also be seen in Fig.\ \ref{fig:vec_v} by setting $\beta = 1$.    

\begin{figure}[t!]
\centering
\includegraphics[scale=0.46]{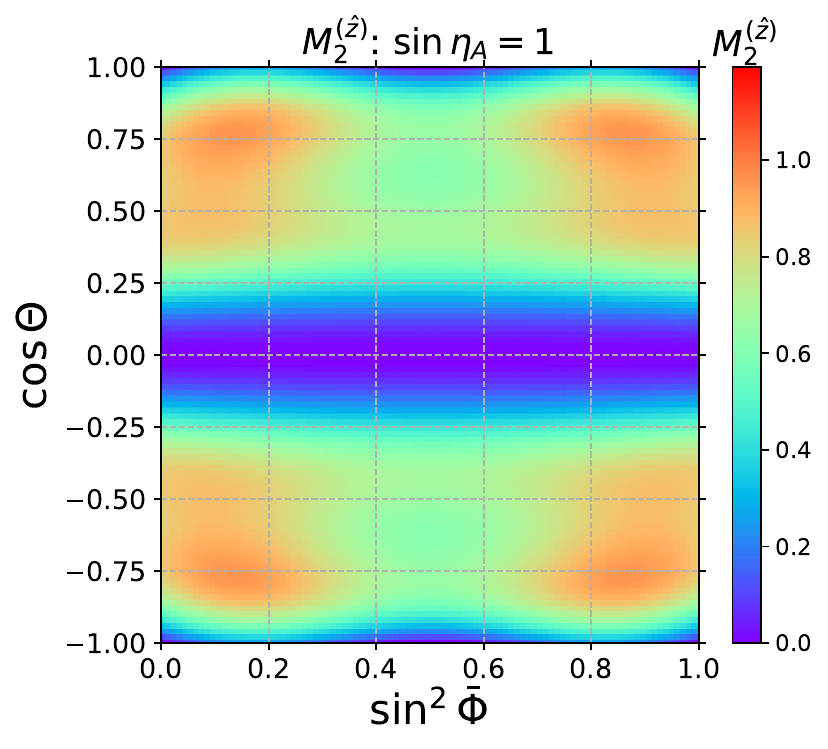}
\caption{\label{fig:M2z_axial}
\small The beam-basis stabiliser R{\' e}nyi entropy $M_2^{(\hat z)}$ for the axial-vector-like interaction ($\cos \eta_A = 0$) presented over the $(\sin^2 \PhiV, \cos \Theta)$ plane.
}
\end{figure}

Similarly, the SRE in the helicity-basis $M_2$ can be obtained by setting $\beta = 1$ in the expressions \eqref{magic_v} and \eqref{FV}.
The beam-basis SRE $M_2^{(\hat z)}$, on the other hand, cannot be obtained by sending $\beta = 1$ in the corresponding vector-like formula, as it carries linear dependence on the sign factor $h$.  
The analytical expressions for $M_2$ and $M_2^{({\hat z})}$ for the axial-vector-like case are given by
\ba
M_2 =
- \log_2 \frac{F_A}{2 \Gamma D^4}  \,,
~~~
M_2^{({\hat z})} 
=
-\log_2 \frac{F_A + \Delta_A}{2 \Gamma D^4} 
\,,
\label{magic_a}
\ea
with
\ba
F_A &=& D^4
+ (D-4)^4 + 64 (D-2)^2 \cos^4 2 \PhiV  
\,,
\nn \\
\Delta_A &=& 
- 8(4-D) (D-2) \left[
7 D^2 - 36 D + 48 + 4 (D - 2)^2 \cos^4 2 \PhiV 
\right]
\,.
\label{FA}
\ea

The $M_2$ distribution of the axial-vector-like case ($\cos \eta_A = 0$) in the $(\sin^2\!\PhiV,\cos\Theta)$ plane is shown in the upper-right panel of Fig.~\ref{fig:m2_v}. 
The corresponding maximal and minimal values of $M_2$ are reported above.

Fig.\ \ref{fig:M2z_axial} shows the beam-basis SRE, $M_2^{(\hat z)}$, for the axial–vector–like case ($\cos\eta_A=0$) over the $(\sin^2\!\PhiV,\cos\Theta)$ plane. 
Its pattern closely mirrors the vector–like case in the ultra–relativistic limit $\beta\to1$ (lower–right panel of Fig.~\ref{fig:m2_v}). 
Remarkably, the extrema and their locations coincide in the two cases. 
Specifically:
(i) $M_2^{(\hat z)}=0$ at $\cos\Theta=0$ for all polarisations, and it also vanishes at the six discrete points $\cos\Theta=\pm1$ with $\sin^2\!\PhiV\in\{0,\,\tfrac12,\,1\}$;
(ii) the maximum $M_2^{(\hat z)}\simeq0.956$ occurs at $\cos\Theta\simeq\pm0.765$ with $\sin^2\!\PhiV\simeq\{0.133,\,0.867\}$.

\subsubsection*{ $(iii)$ An intermediate case: $\eta_A = \frac{2}{5} \pi$ and $\beta=0.8$}

\begin{figure}[t!]
\centering
\includegraphics[scale=0.335]{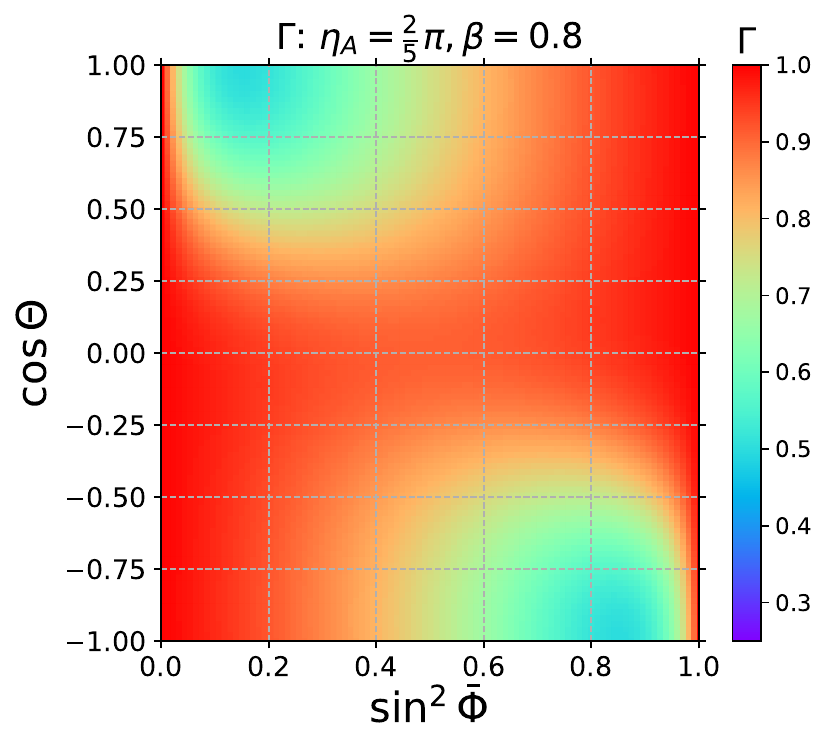}
\includegraphics[scale=0.335]{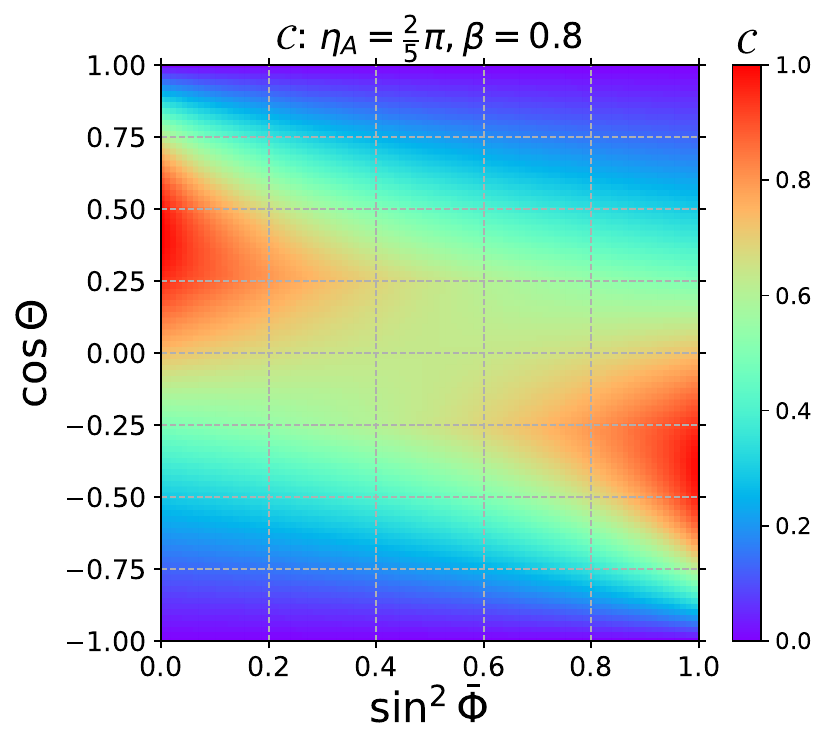}
\includegraphics[scale=0.335]{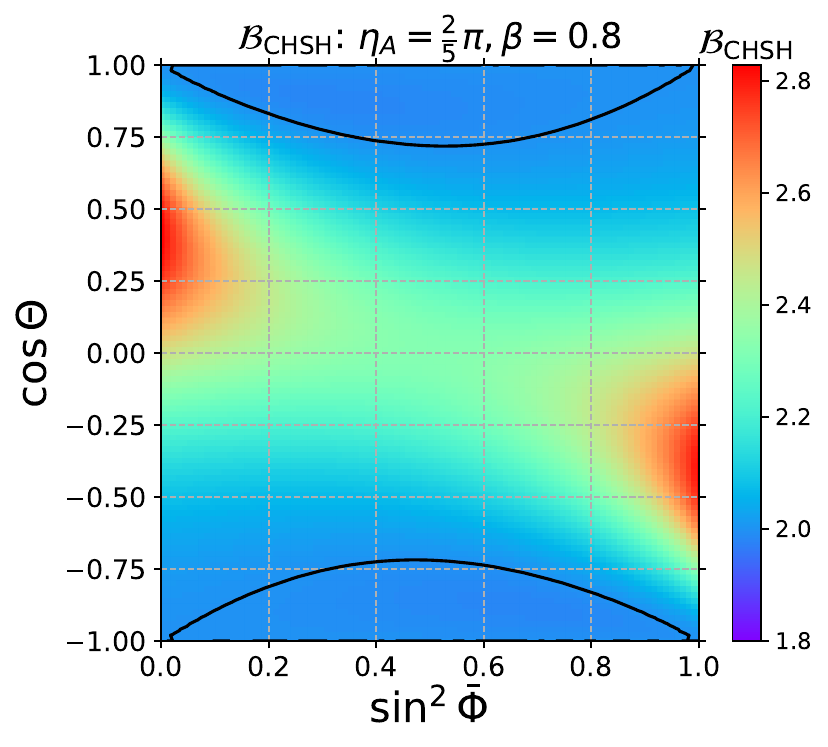}
\includegraphics[scale=0.335]{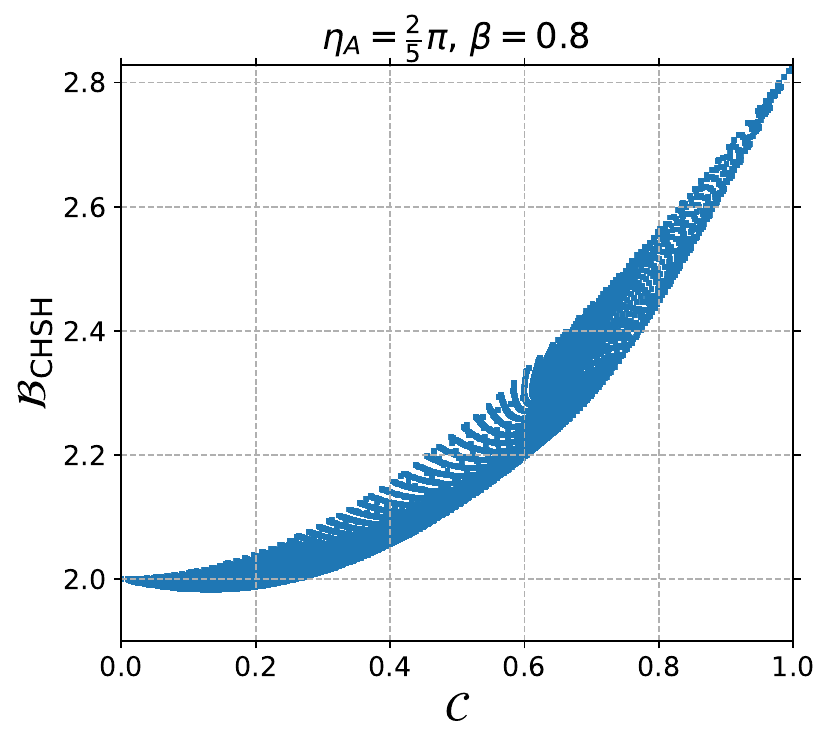}
\includegraphics[scale=0.335]{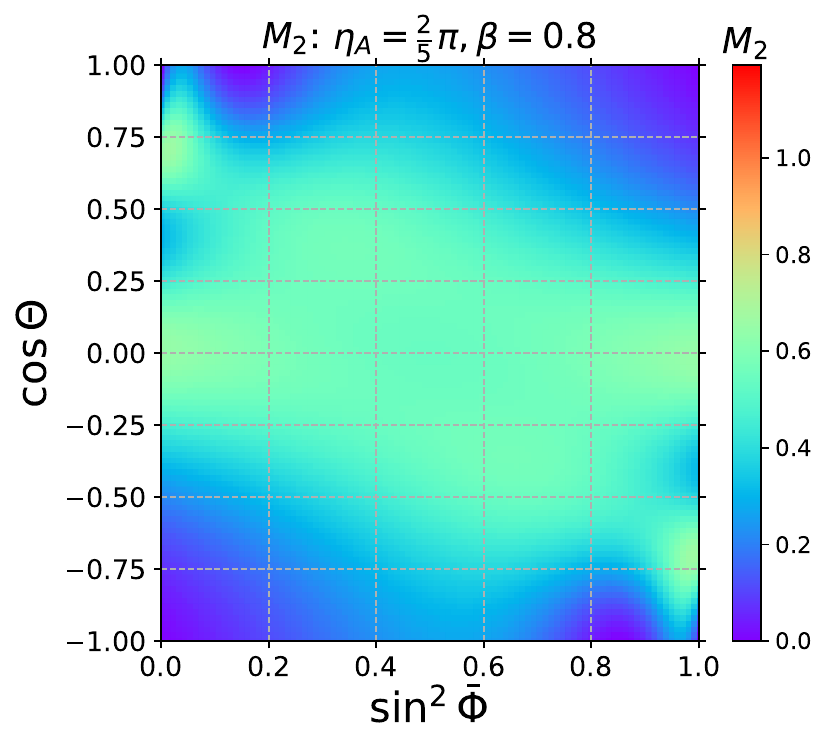}
\includegraphics[scale=0.335]{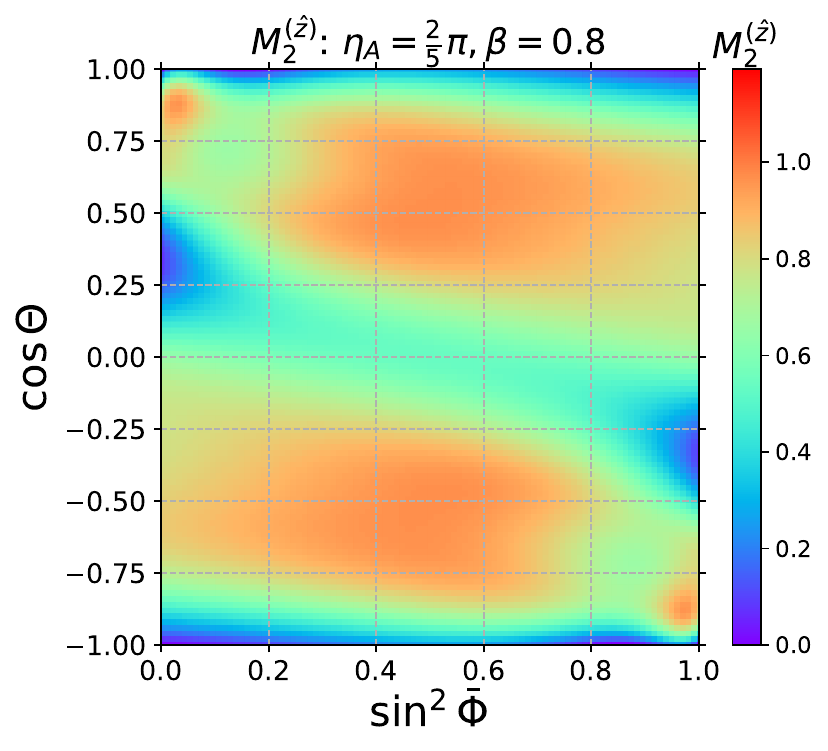}
\caption{\label{fig:iii}
\small
The quantum observables for the vector interaction with $\eta_A = \frac{2}{5} \pi$ and $\b = 0.8$ are shown.
In the upper panel, the purity $\Gamma$ (left), the concurrence ${\cal C}$ (middle) and the Bell-CHSH observable ${\cal B}_{\rm CHSH}$ (right) are shown in the ($\sin^2 \PhiV, \cos \Theta$) plane. 
The lower-left panel shows the correlation between the concurrence and the Bell-CHSH observable. 
The lower-middle and lower-right panels represent $M_2$ and $M_2^{(\hat z)}$ over the ($\sin^2 \PhiV, \cos \Theta$) plane. 
}
\end{figure}

In this case, we examine an intermediate regime interpolating between {\it (i)} $\sin\eta_A=0$ and {\it (ii)} $\cos\eta_A=0$. 
For a generic chiral coupling angle $\eta_A$, the observables depend on $\beta$, $\Theta$, and $\PhiV$ and do not admit compact closed-form expressions. 
For definiteness, we set $\eta_A=\frac{2}{5}\pi$ and $\beta=0.8$ and evaluate the observables numerically over the $(\sin^2 \PhiV, \cos \Theta)$ plane. The results are shown in Fig.\ \ref{fig:iii}.

We see that all observables are symmetric under the discrete transformation: 
\be
(\Theta,\,\PhiV) \to (\Theta',\,\PhiV') \,=\, (\pi - \Theta, \, \pi/2 - \PhiV)
\label{eq:sym}
\,.
\ee
In fact, $\rho^{\rm f}(\Theta, \PhiV)$ and $\rho^{\rm f}(\Theta', \PhiV')$ are related by a local Clifford unitary
\be
\rho^{\rm f}(\Theta', \PhiV') = 
U \rho^{\rm f}(\Theta, \PhiV) U^\dagger
~~~{\rm with}~~~
U = \sigma_z \otimes \sigma_z \,,
\ee
for general $\eta_A$ and $\beta$.
As discussed earlier, the observables $\Gamma$, ${\cal C}$ and ${\cal B}_{\rm CHSH}$ are invariant under local unitaries, while the SREs are invariant under Clifford unitaries. 

The purity $\Gamma$ is shown in the upper–left panel of Fig.\ \ref{fig:iii}.
The state is pure only at $\sin^2 \PhiV = 0$ and 1, for any production angle $\Theta$.
The minimum purity, $\Gamma=\tfrac{1}{2}$, is attained at $\cos \Theta=1$ with $\sin^2 \PhiV \simeq 0.151$ and,
by symmetry, at $\cos \Theta=-1$ with $\sin^2 \PhiV \simeq 0.849$. 


The upper–middle panel of Fig.\ \ref{fig:iii} shows the concurrence $\mathcal{C}$.
The state is separable at $\cos \Theta = \pm 1$ for all $\PhiV$.
Maximal entanglement, $\mathcal{C}=1$, is reached on the pure-state lines at $\sin^2 \PhiV=0$ and $\sin^2 \PhiV= 1$, at the specific production angles $\cos \Theta \simeq 0.406$
and $\simeq -0.406$, respectively. 

The upper-right panel of Fig.\ \ref{fig:iii} shows the Bell-CHSH observable $\mathcal{B}_{\rm CHSH}$. 
Black contours delineate the boundary $\mathcal{B}_{\rm CHSH}=2$, separating CHSH-violating from non-violating regions. 
Non-violation appears in the forward/backward regimes, $\cos\Theta \sim \pm1$, most prominently for the effectively unpolarised configuration $\sin^2 \PhiV \sim 0.5$.
The Tsirelson bound, $\mathcal{B}_{\rm CHSH}=2\sqrt{2}$, is attained precisely at the points where the concurrence reaches unity, $\mathcal{C}=1$.
The minimum value observed is $\mathcal{B}_{\rm CHSH}\simeq 1.98$, occurring at 
$(\sin^2 \PhiV, \cos \Theta) \simeq (0.322,\, 0.878)$
and
$(0.678,\, -0.878)$
Note that ${\cal B}_{\rm CHSH} < 2$ is never realised for the pure vector-like and axial-vector-like cases.

For the pure vector-like and axial–vector–like cases, the concurrence and the Bell-CHSH observable satisfy
${\cal B}_{\rm CHSH}=2\sqrt{1+{\cal C}^2}$ (see Eq.~\eqref{conc_v}). 
This relation breaks down for generic chiral admixtures and/or finite velocity. 
In the lower-left panel of Fig.~\ref{fig:iii}, we illustrate the ${\cal C}$–${\cal B}_{\rm CHSH}$ correlation for $\eta_A=2\pi/5$ and $\beta=0.8$: ${\cal C}$ and ${\cal B}_{\rm CHSH}$ are no longer in one-to-one correspondence, except at the endpoints ${\cal C}=0 \Leftrightarrow {\cal B}_{\rm CHSH}=2$ and ${\cal C}=1 \Leftrightarrow {\cal B}_{\rm CHSH}=2\sqrt{2}$.

The lower–middle and lower–right panels of Fig.\ \ref{fig:iii} display $M_2$ and $M_2^{(\hat z)}$, respectively.
The SRE landscape is non-trivial. 
Both measures vanish along the pure-state lines $\sin^2\PhiV \in \{0, 1 \}$ and at forward/backward production $\cos \Theta = \pm1$.
In addition, $M_2$ exhibits two further zeros at $\cos \Theta = \pm 1$ for intermediate $\PhiV$, while $M_2^{(\hat z)}$ shows four additional zeros at $\cos \Theta =\pm 1$ for intermediate $\sin^2 \PhiV$, and at $\sin^2 \PhiV \in \{0, 1\}$ for intermediate $\cos \Theta$.
The global maximum of the helicity-basis SRE is $M_2\simeq 0.667$, attained at $(\sin^2 \PhiV, \cos\Theta) \simeq(0.0134,\, 0.705)$ and $(0.987,\, -0.705)$.
For $M_2^{(\hat z)}$, the maximal value is slightly larger, $M_2^{(\hat z)}\simeq 0.968$, occurring at $(\sin^2 \PhiV, \cos \Theta) \simeq (0.484,\, 0.484)$ and 
$(0.516,\, -0.484)$.

\subsection{Tensor interaction } 
\label{sec:tensor}
We consider the scattering process $l \bar l \to F \bar F$ with the tensor-type interaction \eqref{Ltensor}.
The non-zero transition amplitudes are given by
\bea
{\cal M}^{+-}_{++} \,=\, - {\cal M}^{+-}_{--} &=&  - e^{i(\xi_T + \eta_T)} \frac{2 c_T}{\Lambda^2} \cdot s \cdot \gamma^{-1} \cdot  \sin \Theta \,,
\nonumber \\
{\cal M}^{+-}_{+-} &=&
- e^{i(\xi_T + \eta_T)} \frac{2 c_T}{\Lambda^2} \cdot s \cdot 
(1 - \beta) \cdot \cos \Theta
 \,,
\nonumber \\
{\cal M}^{+-}_{-+} &=&
- e^{i(\xi_T + \eta_T)} \frac{2 c_T}{\Lambda^2} \cdot s \cdot 
(1 + \beta) \cdot \cos \Theta
\,.
\nn \\
{\cal M}^{-+}_{++} \,=\, - {\cal M}^{-+}_{--} 
&=&  
- e^{-i(\xi_T + \eta_T)} \frac{2 c_T}{\Lambda^2} \cdot s \cdot \gamma^{-1} \cdot \sin \Theta 
\,,
\nonumber \\
{\cal M}^{-+}_{+-} &=&
- e^{-i(\xi_T + \eta_T)} \frac{2 c_T}{\Lambda^2} \cdot s \cdot 
(1 + \beta) \cdot \cos \Theta
 \,,
\nonumber \\
{\cal M}^{-+}_{-+} &=& 
- e^{-i(\xi_T + \eta_T)} \frac{2 c_T}{\Lambda^2} \cdot s \cdot 
(1 - \beta) \cdot \cos \Theta \,.
\label{amp_t}
\eea
In the $m_l^2/s \to 0$ limit, ${\cal M}_{\l_F, \l_{\bar F}}^{++} = {\cal M}_{\l_F, \l_{\bar F}}^{--} = 0$. 
Note that all amplitudes with $(\l_l,\l_{\bar l}) = (+,-)$
and $(-,+)$ are proportional to the phase factor $e^{i (\xi_T + \eta_T)}$
and $e^{-i (\xi_T + \eta_T)}$, respectively.
These phase factors are cancelled in the construction of $\rho^{\rm f}$ (see Eq.\ \eqref{rho_formula}).
As a result, the spin density matrix is independent of the coupling angles $\xi_T$ and $\eta_T$.
Since only $(\l_l,\l_{\bar l}) =(+,-)$ and $(-,+)$ amplitudes contribute, the density matrix depends on the beam polarisation parameters only through the ratio 
\be
\frac{ \rho^{\rm in}_{+,-}  }{\rho^{\rm in}_{-,+} } 
\,=\,
\frac{ (1 + {\cal P})(1 + \Pbar) }{ (1 - {\cal P})(1 - \Pbar) }
\,\equiv\, \tan^2 \PhiT \,, 
~~~~~\PhiT \in \left[ 0, \frac{\pi}{2}  \right]\,.
\label{PhiT}
\ee
As discussed in the previous section, the final state is manifestly pure at $\PhiT = 0$ and $\pi/2$ as only one initial spin configuration contributes. 
The angle $\PhiT$ is symmetric under exchange of the beam polarisations ${\cal P} \leftrightarrow \overline {\cal P}$.
The beams are effectively unpolarised, i.e., $\PhiT = \pi/4$, whenever ${\cal P} = - \overline {\cal P}$, independent of their magnitudes.

The $F \bar F$ spin density matrix $\rho^{\rm f}$ is obtained as a classical mixture of the two pure states 
\be
\rho^{\rm f} \,=\, \sin^2 \PhiT | \psi^{+-} \ketbra \psi^{+-} |
\,+\,
\cos^2 \PhiT | \psi^{-+} \ketbra \psi^{-+} |\,,
\label{rho_T}
\ee
where 
\ba
| \psi^{+-} \ket &=& \frac{1}{N}
\left[ \sqrt{1 - \b^2} \sin \Theta ( | ++ \ket_{F \bar F} - | -- \ket_{F \bar F} ) 
+ \cos \Theta ( 
[1 - \beta] | + - \ket_{F \bar F} 
+ [1 + \beta] |- + \ket_{F \bar F} 
)
\right]\,,
\nn \\
| \psi^{-+} \ket &=& \frac{1}{N}
\left[ \sqrt{1 - \b^2} \sin \Theta ( | ++ \ket_{F \bar F} - | -- \ket_{F \bar F} ) 
+ \cos \Theta ( 
[1 + \beta] | + - \ket_{F \bar F} 
+ [1 - \beta] | - + \ket_{F \bar F}
)
\right]\,,
\nn \\
\ea
are the $F \bar F$ spin states associated with the $(\l_l,\l_{\bar l}) =(+,-)$ and $(-,+)$ initial states, respectively.
Here, $N = \sqrt{2(1+\b^2 \cos 2 \Theta})$ is the normalisation factor.

Note that in the threshold limit $\b \to 0$ the two pure final states coincide, $| \psi^{+-} \ket = | \psi^{-+} \ket$.
In this limit, $\rho^{\rm f}$ becomes pure regardless of the beam polarisation angle $\PhiT$.
On the other hand, in the ultra-relativistic limit $\b \to 1$, the two states become trivial product states:
$| \psi^{+-} \ket = | +- \ket_{ F \bar F}$ and
$| \psi^{-+} \ket = | -+ \ket_{ F \bar F}$.
In this case, $\rho^{\rm f}$ becomes independent of the production angle $\Theta$.
Another interesting limit is the central production, $\Theta \to \frac{\pi}{2}$. 
In that case, $| \psi^{+-} \ket = | \psi^{-+} \ket$ and they coincide with the maximally entangled singlet state $| \psi_{0} \ket = \frac{1}{\sqrt{2}} [ | ++ \ket_{F \bar F} - | -- \ket_{F \bar F}]$\,. 
The state is again pure, $\rho^{\rm f} = | \psi_0 \ketbra \psi_0|$ regardless of the polarisation angle $\PhiT$.

With the density matrix of Eq.\ \eqref{rho_T} in hand, one can directly compute the quantum observables. The results are: 
\ba
&&
\hspace{28mm}
\Gamma \,=\, 
1 - \frac{ \b^2 \cos^2 \Theta ( \gamma^{-2}  + V) }{V^2} 
\sin^2 2 \PhiT
\,,
\label{pure_t}
\\
&&
\hspace{28mm}
{\cal C} \,=\, \gamma^{-2} V^{-1}\,,
~~~~~
{\cal B}_{\rm CHSH} \,=\, 
2\sqrt{1 + {\cal C}^2 } \,,
\label{conc_t}
\\
&&
M_2 \,=\, - \log_2 
\left[
\frac{ F_T + G_T \cos^4 2 \PhiT }{4 \Gamma V^{4}}
\right]
,~~~
M_2^{(\hat z)} \,=\, - \log_2 
\left[
\frac{ F_T^{(\hat z)} + G_T^{(\hat z)} \cos^4 2 \PhiT }{4 \Gamma V^{4}}
\right]
\,,
\label{M2_t}
\ea
where $\gamma^{-1} =\sqrt{1 - \b^2}$, $V \equiv 1 + \b^2 \cos 2 \Theta$ and
\ba
F_T &\equiv&
V^4 +
(\b^2 + \cos 2 \Theta)^4 +  2 \g^{-4} \sin^4 2 \Theta
+ \g^{-8} (1 + \cos^4 2 \Theta) 
\,,
\nn \\
G_T &\equiv&
2 \b^4 ( 16 \cos^8 \Theta + \g^{-4} \sin^4 2 \Theta )
\,, 
\nn \\
F_T^{(\hat z)} &\equiv&
V^4 + \g^{-8} 
+ \frac{1}{256} \left[ 2 \b^2 + U_- + U_+ \cos 4 \Theta \right]^4
+ \frac{\sin^4 2 \Theta}{8} \left[ \b^2 - U_+ \cos 2 \Theta \right]^4
\nn \\
&& + \left[
(\beta^2+\cos 2 \Theta) \cos^2 \Theta - \g^{-2} \cos 2 \Theta \sin^2 \Theta + \g^{-1} \sin^2 2 \Theta
\right]^4
\,,
\nn \\
G_T^{(\hat z)} &\equiv&
2 \b^4 \left[ 16 (1-\g^{-1})^4 \sin^4 \Theta \cos^8 \Theta  + 
(2 \cos^3 \Theta + \g^{-1} \sin \Theta \sin 2 \Theta)^4
\right]
\,, 
\label{FDT}
\ea
with $U_\pm = \pm 2\g^{-1} +\b^2 -2$.

\begin{figure}[t!]
\centering
\includegraphics[scale=0.335]{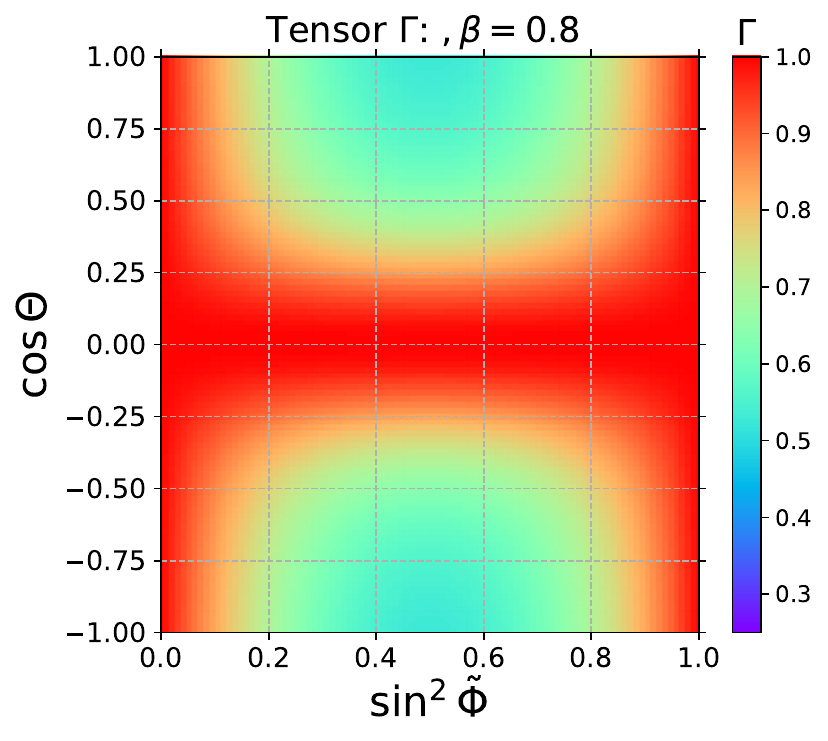}
\includegraphics[scale=0.335]{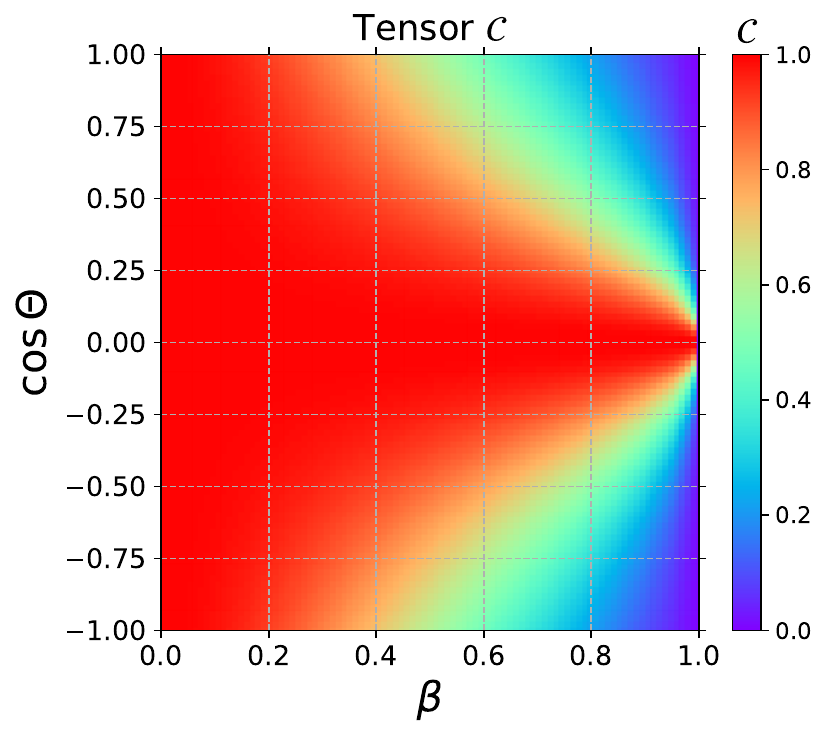}
\includegraphics[scale=0.335]{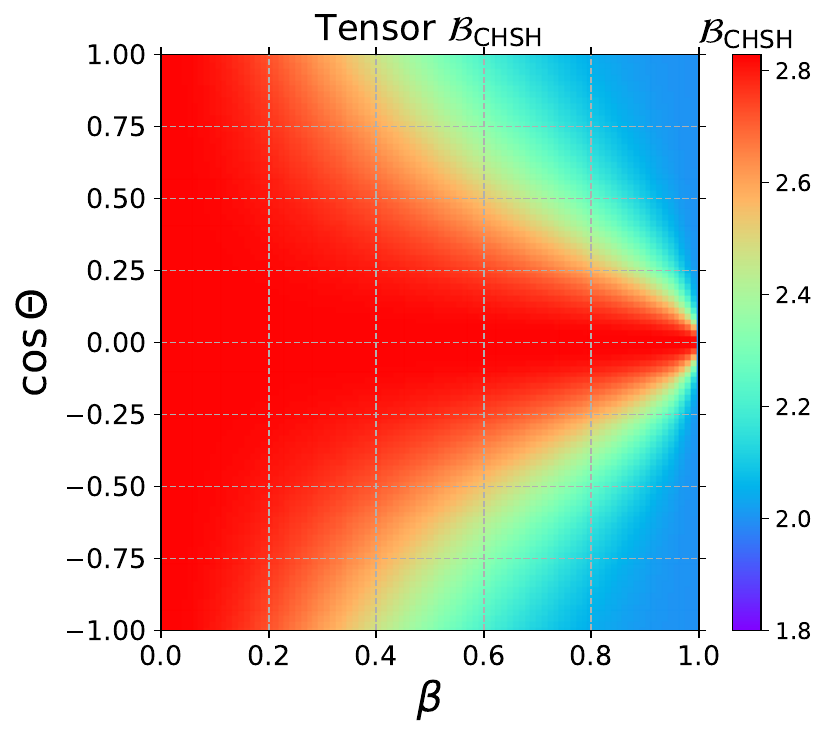}
\caption{\label{fig:tens}
\small
The purity $\Gamma$ (left), the concurrence ${\cal C}$ (middle) and the Bell-CHSH observable ${\cal B}_{\rm CHSH}$ (right) for the tensor interaction are shown.
The purity is displayed in the $(\sin^2 \PhiT, \cos \Theta)$ plane, whereas the concurrence and the Bell-CHSH observable are presented in the $(\beta, \cos \Theta)$ plane.
}
\end{figure}

In the first expression, the purity $\Gamma$ depends on $\PhiT$, $\beta$ and $\Theta$.
As discussed above, the state is pure, $\Gamma=1$, at $\PhiT \in\{0,\pi/2\}$ or $\beta = 0$ or $\Theta = \frac{\pi}{2}$.
For a given $\beta$, the lowest purity is given by $\Gamma = 1 - \frac{2 \b^2}{(1+\b^2)^2}$ at $\PhiT = \frac{\pi}{4}$ and $\Theta \in \{ 0, \pi \}$.
The global minimum $\Gamma =\frac{1}{2}$ is attained in the $\beta \to 1$ limit.
The left panel of Fig.\ \ref{fig:tens} illustrates 
the purity at $\b = 0.8$ over the $(\sin^2 \PhiT, \cos \Theta)$ plane. 

Eq.\ \eqref{conc_t} shows the closed-form expressions for the concurrence $\mathcal{C}$ and the Bell-CHSH observable $\mathcal{B}_{\rm CHSH}$. 
We see that they are independent of the polarisation angle $\PhiT$ and depend only on $\beta$ and $\Theta$.
The state is maximally entangled, ${\cal C} = 1$, for $\b \to 0$ or $\Theta = \frac{\pi}{2}$,
while it is separable, ${\cal C} = 0$, for $\b \to 1$ except at $\Theta \to \frac{\pi}{2}$.

We also see in Eq.\ \eqref{conc_t} that the Bell-CHSH observable is monotonically related to the concurrence in the same way as in the pure vector-like or pure axial-vector-like cases discussed in the previous subsection.  
The Bell-CHSH observable reaches  
the Tsirelson bound, $\mathcal{B}_{\rm CHSH}=2\sqrt{2}$, when $\b = 0$ or $\Theta = \frac{\pi}{2}$.
The state is always Bell-inequality violating, $\mathcal{B}_{\rm CHSH} > 2$, except at $\b = 1$ and $\Theta \neq \frac{\pi}{2}$.

The middle and right panels of Fig.\ \ref{fig:tens} display the concurrence and Bell-CHSH observable over the $(\beta,\, \cos \Theta)$ plane, respectively. 

Analytical expressions for the SREs $M_2$ and $M_2^{(\hat z)}$ are given in Eq.\ \eqref{M2_t}. 
Here $\Gamma$ is the purity function in Eq.\ \eqref{pure_t}.
Since both $M_2$ and $M_2^{(\hat z)}$ depend non-trivially on $\beta$, $\Theta$, and $\PhiT$, below we consider three representative values of $\beta$: 0 (threshold), 
$1$ (ultra-relativistic)
and
$0.8$ (intermediate).

In the threshold limit $\beta \to 0$, the state becomes pure and independent of $\PhiT$.
The expressions of $M_2$ and $M_2^{(\hat z)}$ in this limit are drastically simplified to
\be
M_2\,=\,-\log_2 \left[ \frac{7+\cos(8\Theta)}{8} \right],\qquad
M_2^{(\hat z)}\,=\,0.
\ee
Remarkably, $M_2^{(\hat z)}$ vanishes independently of $\Theta$. 
On the other hand, $M_2$ is an oscillating function of $\Theta$ with the period $\frac{\pi}{4}$.
The $M_2$ vanishes when $\cos 8 \Theta = 1$, while it takes the maximal value $M_2 = - \log_2 \frac{3}{4} \simeq 0.415$ for $\cos 8 \Theta = -1$.

\begin{figure}[t!]
\centering
\includegraphics[scale=0.335]{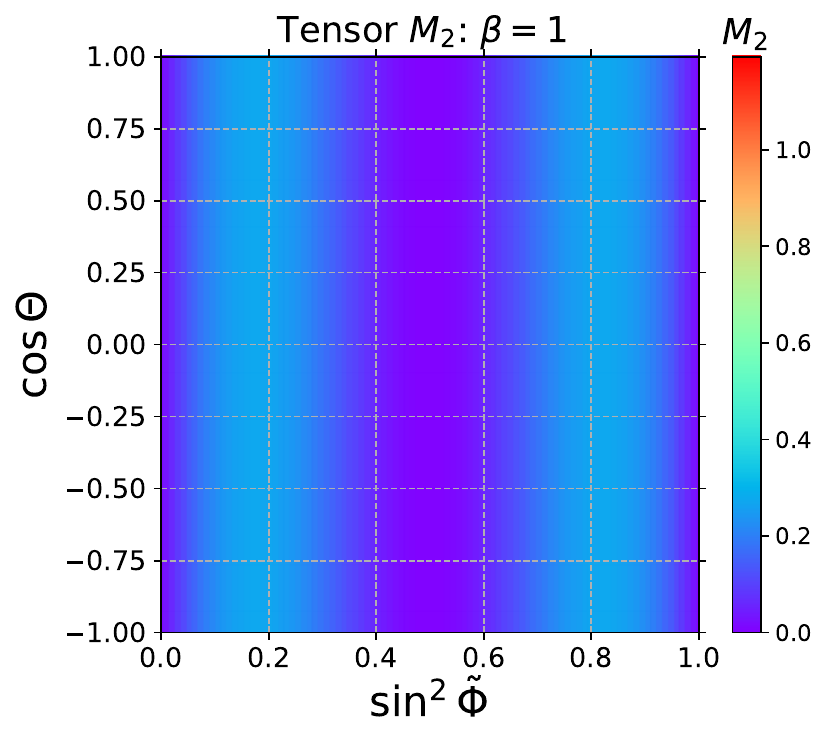}
\hspace{3mm}
\includegraphics[scale=0.335]{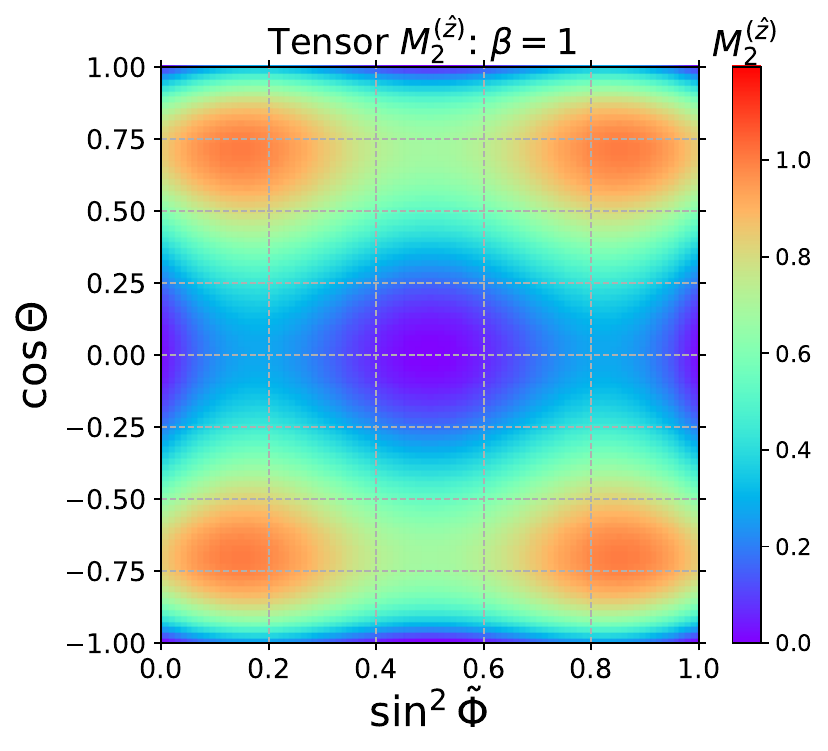}
\\
\includegraphics[scale=0.335]{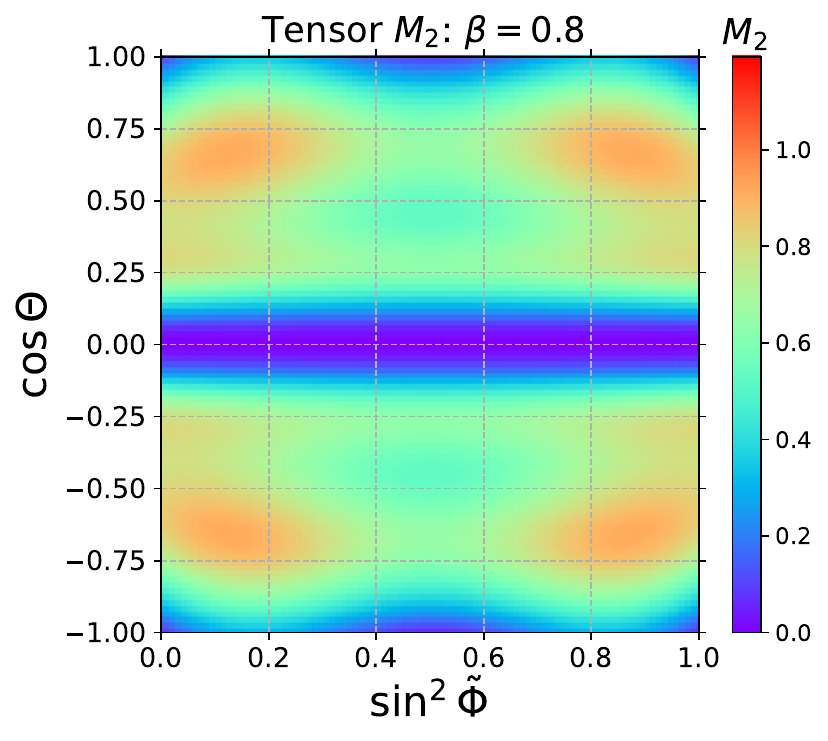}
\hspace{3mm}
\includegraphics[scale=0.335]{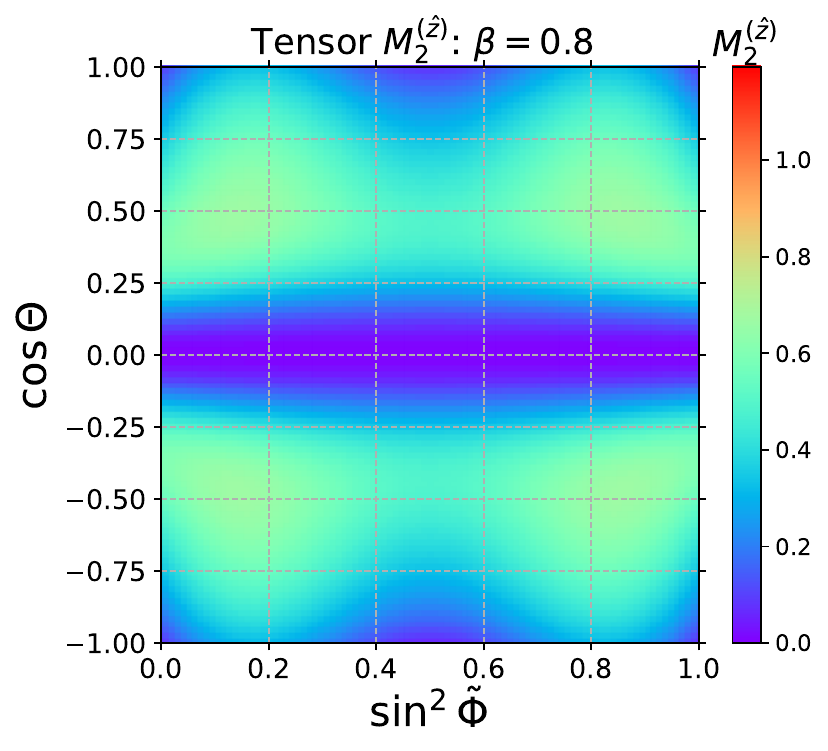} 
\caption{\label{fig:M2tens}
\small
The left and right panels show $M_2$ and $M_2^{(\hat z)}$ for the tensor interaction, respectively, over the $(\sin \PhiT, \cos \Theta)$ plane.
The upper and lower panels correspond to $\beta = 1$ and $\beta = 0.8$, respectively. 
}
\end{figure}

In the ultra-relativistic limit $\beta \to 1$, 
$\rho^{\rm f} = \sin^2 \PhiT |+- \ketbra +-| + \cos^2 \PhiT |-+ \ketbra -+|$, as discussed above. 
In this expression, the basis states $|\l_F, \l_{\bar F} \ket$ are quantised along the $F$'s momentum direction $\Theta$.
Computing the SRE in the helicity-basis therefore gives the $\Theta$-independent result, while the SRE in the beam-basis is $\Theta$-dependent:
\ba
M_2 &=& - \log_2 \left[ \frac{8 - (1 - \cos 4 \PhiT)(3 + \cos 4 \PhiT) }{2(3 + \cos 4 \PhiT)}
\right]
\,,
\label{M2_tens_B1}
\nn \\
M_2^{(\hat z)} &=&
- \log_2 \left[
\frac{ \cos 8 \Theta + 12 \cos 4 \Theta + 51  
+ 4 (3 + \cos 4 \Theta) (1 + \cos 4 \PhiT)^2 }{32(3 + \cos 4 \PhiT)}
\right]\,.
\label{M2z_tens_B1}
\ea
The situation is almost parallel to the vector-like interaction case in the $\beta \to 0$ limit, where the final state is given by $\rho^{\rm f} = \sin^2 \PhiT |++ \ketbra ++|_{z} + \cos^2 \PhiT |-- \ketbra --|_{z}$, as discussed in the previous section. 
In this expression, the basis kets are quantised in the $z$-direction.
Therefore, $M_2^{(\hat z)}$ for this case is given by the $\Theta$-independent function \eqref{M2_tens_B1}, while $M_2$ is given by the right-hand-side of Eq.\ \eqref{M2z_tens_B1} and carries the $\Theta$-dependence.
The upper panel of Fig.\ \ref{fig:M2tens} shows the $M_2$ (left) and $M_2^{(\hat z)}$ (right) for the tensor-type interaction in the $\b \to 1$ limit. 
The similarity between these plots and the plots in the left panel of Fig.\ \ref{fig:m2_v} is obvious.

For the intermediate $\beta$, both $M_2$ and $M_2^{(\hat z)}$ depend non-trivially on $\Theta$ and $\PhiT$.
The lower panels of Fig.~\ref{fig:M2tens} display $M_2$ (left) and $M_2^{(\hat z)}$ (right) for $\beta=0.8$.

One observes that both SREs are symmetric under the independent discrete transformations $\PhiT\to\frac{\pi}{2}-\PhiT$ and $\Theta\to\pi-\Theta$.
The effect of the former transformation is to switch the beam polarisation, $\sin^2 \PhiT \leftrightarrow \cos^2 \PhiT$ in Eq.\ \eqref{rho_T}.
In terms of the density matrix, this can be implemented by the SWAP operation: 
\be
\rho^{\rm f} ~\xrightarrow{\PhiT \to \frac{\pi}{2} - \PhiT}~ (\rho^{\rm f})' \,=\, U_{\rm sw} \rho^{\rm f} U_{\rm sw}^\dagger\,,
~~~~~~
U_{\rm sw} =
\bmat
~1\, & 0 & 0 & 0 \\
0 & 0 & \,1\, & 0 \\
0 & \,1\, & 0 & 0 \\
0 & 0 & 0 & \,1~ 
\emat\,.
\ee
It is straightforward to check that this operation sends a Pauli element to another Pauli element, meaning that it is an element of the Clifford group.

By inspecting the density matrix \eqref{eq:rho_tens} (Appendix \ref{app:rho}), one can see that the latter operation can be implemented by another Clifford operation:
\be
\rho^{\rm f} ~\xrightarrow{\Theta \to \pi - \Theta}~ (\rho^{\rm f})' \,=\, U_{z} \rho^{\rm f} U_{z}^\dagger\,,
~~~~~~
U_{z} = \sigma_z \otimes \sigma_z\,.
\ee
As the magic is invariant under the Clifford group, these operations leave $M_2$ and $M_2^{(\hat z)}$ invariant. 

In the lower two plots in Fig.\ \ref{fig:M2tens}, both $M_2$ and $M_2^{(\hat z)}$ vanish at $\cos \Theta = \pm 1$ with $\sin^2 \PhiT \in \{ 0, \tfrac12, 1 \}$.
These SREs vanish also at $\cos \Theta = 0$ for all $\PhiT$.
As mentioned above, the final state in this configuration is the pure singlet state $| \psi_0 \ket$ for all $\PhiT$.

The maxima of the SREs can be obtained numerically.
We find $M_2$ takes the maximal value $M_2 \simeq 0.914$ at $(\sin^2 \PhiT, \cos \Theta) \simeq (0.129, 0.664)$ in the fundamental region $0 \leq \sin^2\PhiT \leq \tfrac12$ and $0 \leq \cos \Theta \leq 1$.  
Similarly, the maximal value $M_2^{(\hat z)} \simeq 
0.665$ is obtained at $(\sin^2 \PhiT, \cos \Theta) \simeq (0.145, 0.480)$ in the fundamental region. 
The corresponding three other maxima can be obtained by the symmetries, $\sin^2\PhiT \to 1 - \sin^2\PhiT$ and $\cos\Theta \to - \cos \Theta$, discussed above.

\section{$e^+ e^- \to t \bar t$ in the SM} 
\label{sec:sm}

We extend the previous analysis to the SM process
$e^+e^-\to F \bar F$, with the identifications $F = t$ and $l = e$. 
The leading order SM amplitudes follow from the
effective four–fermion interaction
\be
{\cal L}_{\rm int} \,=\, \sum_{i=A,Z}
\frac{1}{\Lambda^2_i} [\bar \psi_e \gamma^{\mu}(c_L^i P_L + c_R^i P_R) \psi_e]
[\bar \psi_t \gamma_{\mu}(d_L^i P_L + d_R^i P_R) \psi_t]\,,
\label{Lsm}
\ee
where the sum runs over $s$-channel photon ($A$) and $Z$-boson exchange.
The propagator factors $\Lambda_i^2$ and chiral couplings,
$c_{L/R}^{\,i}$ and $d_{L/R}^{\,i}$, are listed in Table~\ref{tab:coup}.
\begin{table}[t]
\centering
\renewcommand{\arraystretch}{1.3}
\begin{tabular}[t!]{ c | c c c c c}
$i$ & $\Lambda^2_i$ & $c_L^i$ & $c_R^i$ & $d_L^i$ & $d_R^i$ \\ 
\hline
$A$ & $s$ & $-e$ & $-e$ & $\tfrac{2}{3} e$ & $\tfrac{2}{3} e$  \\ 
$Z$ & $s-m_Z^2 + i m_Z \Gamma_Z$ 
& $g_Z \left( -\tfrac{1}{2} + \sin^2 \theta_w \right)$ 
& $g_Z \sin^2 \theta_w$
& $g_Z \left( \tfrac{1}{2} - \tfrac{2}{3} \sin^2 \theta_w \right)$
& $g_Z \left( - \tfrac{2}{3} \sin^2 \theta_w \right)$
\end{tabular}
\caption{\label{tab:coup}
\small The propagator factor $\Lambda^2_i$ and the chiral couplings $c_{L/R}^i$ and $d_{L/R}^i$ for the photon ($i=A$) and the $Z$ boson ($i=Z$). }
\end{table}
In the table, $\theta_w$ is the weak mixing angle,
$m_Z$ and $\Gamma_Z$ are the $Z$-boson mass and width, and
$e$ and $g_Z=e/(\sin\theta_w\cos\theta_w)$ are the electromagnetic and the effective $Z$-boson couplings, respectively.

The leading order amplitudes are obtained by superposing the photon and $Z$ contributions in the form of Eq.~\eqref{amp_vec}.
For instance, the first line of Eq.~\eqref{amp_vec} generalises to
\be
{\cal M}^{++}_{++} \,=\,
- \sum_{i=A,Z} \frac{c_V^i}{\Lambda^2_i} \cdot s \cdot  \sin \xi_R^i ( \cos \eta_A^i + \beta \sin \eta_A^i) \cdot (1 + \cos \Theta) \,,
\ee
where the parameters are defined in terms of the chiral couplings in
Eq.~\eqref{Lsm} as
\be
c_V^i = \frac{L_c^i L_d^i}{\sqrt{2}},
\quad
\cos \xi_R^i = \frac{c_L^i}{L_c^i} , \quad
\sin \xi_R^i = \frac{c_R^i}{L_c^i} , \quad
\cos \eta_A^i = \frac{d_R^i + d_L^i}{\sqrt{2} L_d^i} , \quad
\sin \eta_A^i = \frac{d_R^i - d_L^i}{\sqrt{2} L_d^i} , 
\ee
with $L_c^i \equiv \sqrt{ (c_L^i)^2 + (c_R^i)^2 }$
and $L_d^i \equiv \sqrt{ (d_L^i)^2 + (d_R^i)^2 }$.
The top-quark velocity, $\beta = \sqrt{1 - 4 m_t^2/s}$, is fixed by the centre-of-mass energy $\sqrt{s}$, where $m_t = 173$ GeV is the top-quark mass.
Amplitudes for the remaining spin configurations follow analogously
from Eq.~\eqref{amp_vec}.

Due to the $s$-channel gauge boson exchange, the leading order SM amplitudes are non-zero only for $(\l_{e^-}, \l_{e^+}) = (+,+)$ and $(-,-)$ configurations. 
Therefore, the density matrix depends on the beam polarisation parameters ${\cal P}$ and $\overline {\cal P}$ through the ratio 
\be
\frac{ \rho^{\rm in}_{+,+} }{\rho^{\rm in}_{-,-} } 
\,=\,
\frac{ (1 + {\cal P})(1 - \Pbar) }{ (1 - {\cal P})(1 + \Pbar) }
\,\equiv\, \tan^2 \Phi \,,
~~~~~\Phi \in \left[ 0, \frac{\pi}{2}  \right].
\label{Phi}
\ee
As in the vector-type case, discussed in subsection \ref{sec:vector}, 
the beams are effectively unpolarised if the two polarisations are equal, ${\cal P} = \overline {\cal P}$, irrespective of their value.

In the above setup, the $F \bar F$ spin density matrix and the associated quantum observables are parametrised by the three parameters: the beam collision energy $\sqrt{s}$, the production angle $\Theta$ and the effective polarisation angle $\Phi$.

\begin{figure}[t!]
\centering
\includegraphics[scale=0.335]{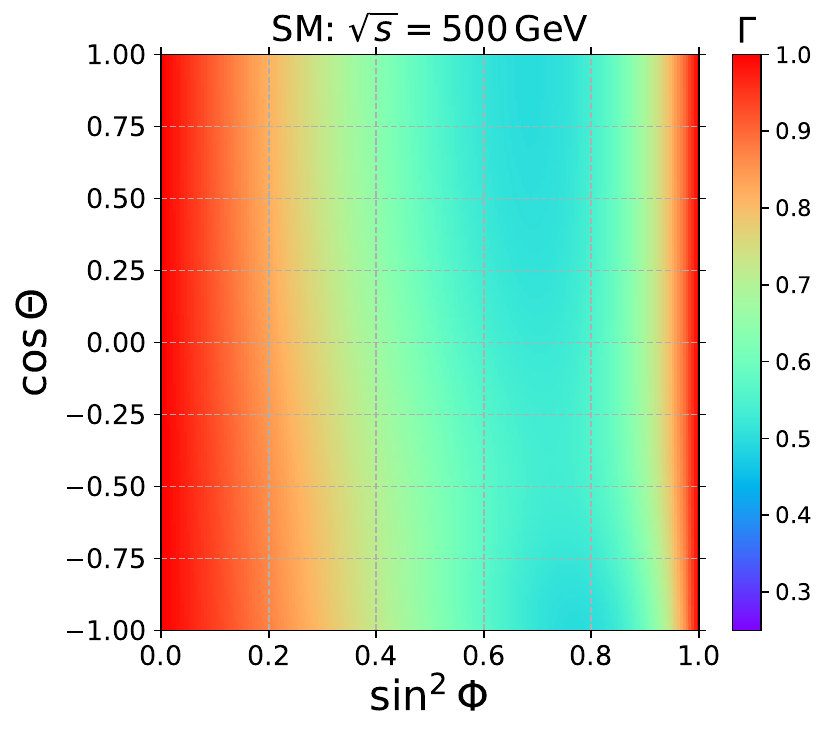}
\includegraphics[scale=0.335]{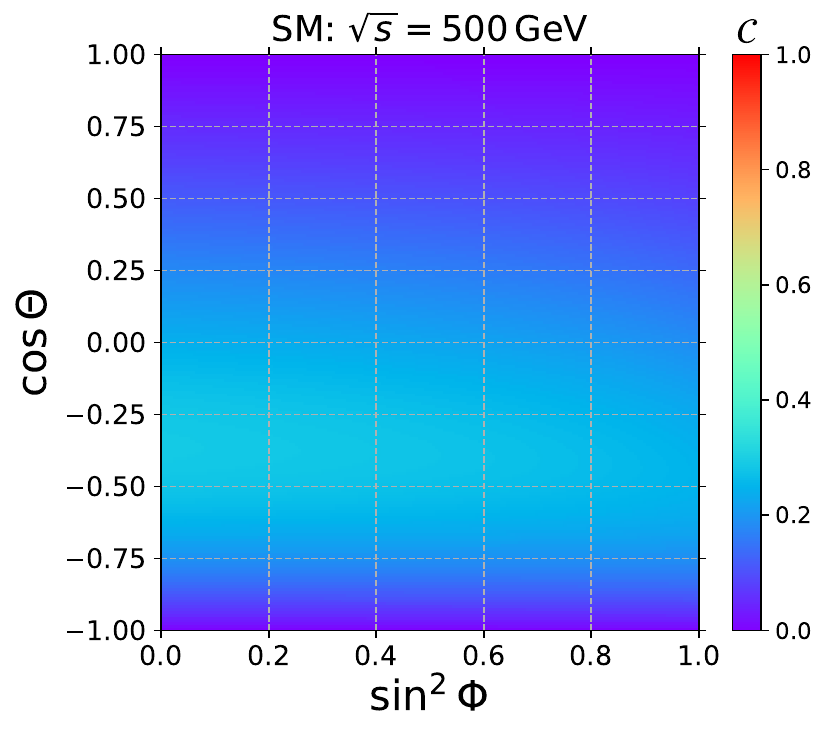}
\includegraphics[scale=0.335]{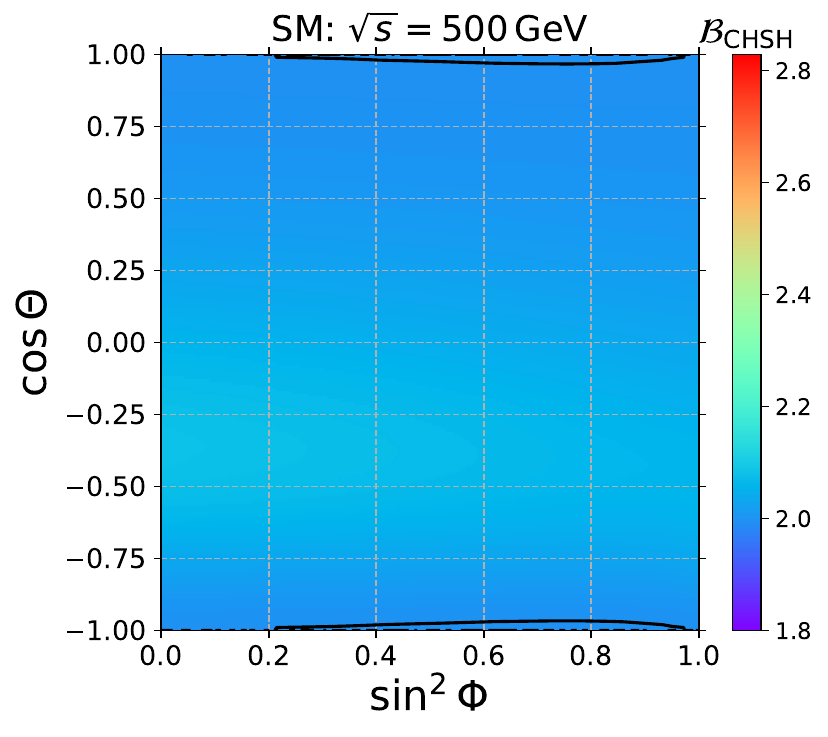}
\includegraphics[scale=0.335]{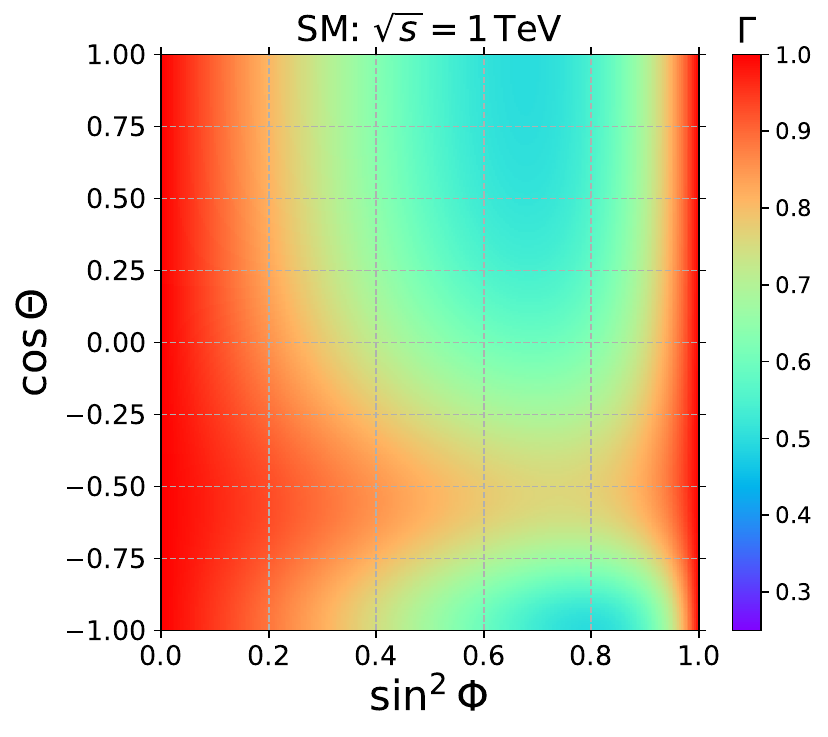}
\includegraphics[scale=0.335]{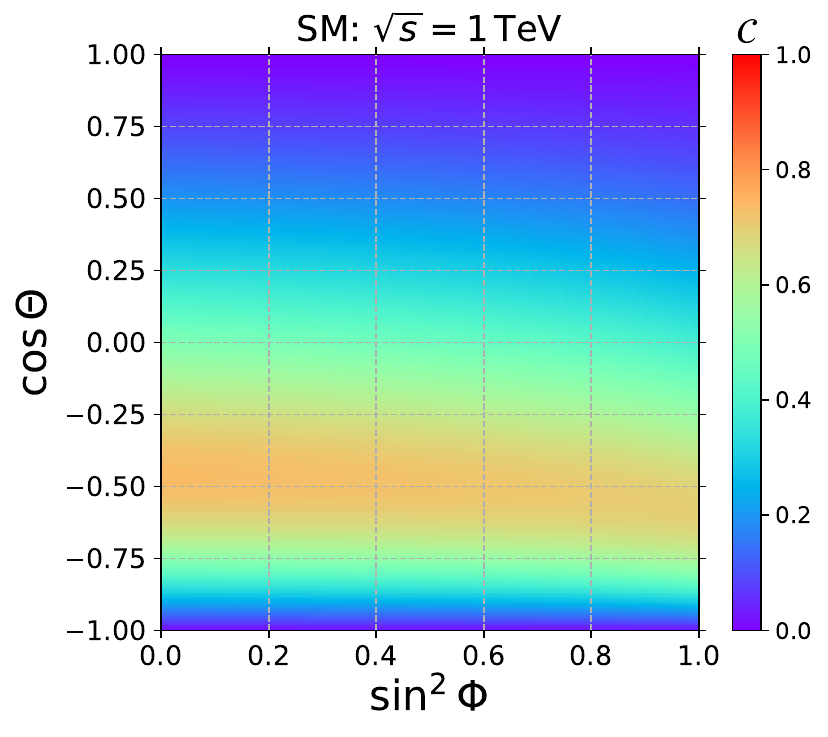}
\includegraphics[scale=0.335]{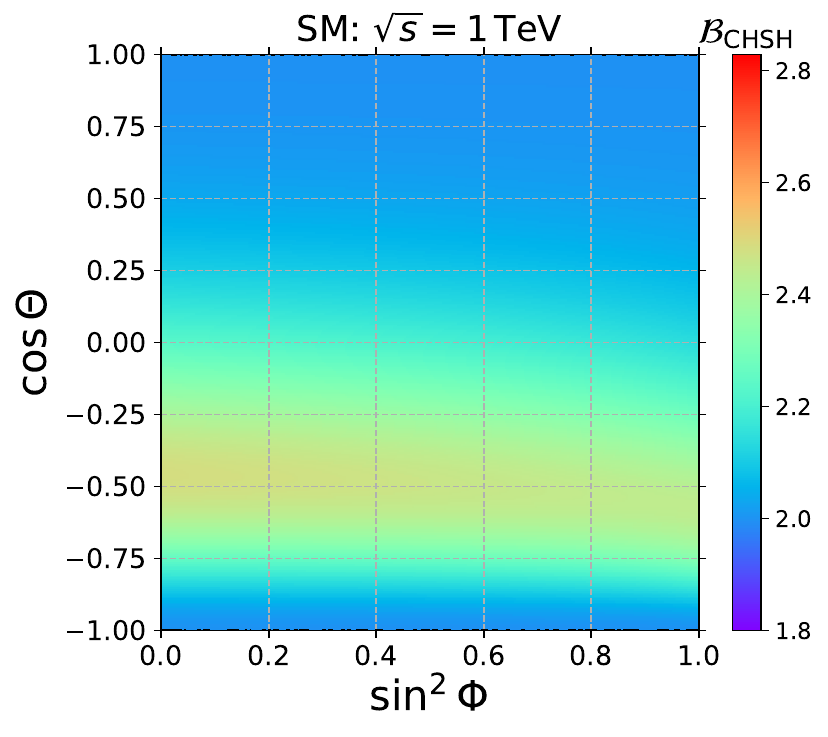}
\caption{\label{fig:sm}
\small 
The purity $\Gamma$ (left panel), the concurrence ${\cal C}$ (middle panel) and the Bell-CHSH observable ${\cal B}_{\rm CHSH}$ (right panel) computed for the SM are shown over the ($\sin^2 \Phi$, $\cos \Theta$) plane.   
The upper and lower panels correspond to the $e^+ e^-$ collision energies $\sqrt{s} = 500$ GeV and 1 TeV, respectively.
}
\end{figure}

Fig.\ \ref{fig:sm} displays the SM predictions for the purity $\Gamma$ (left), the concurrence ${\cal C}$ (middle), and the Bell-CHSH observable ${\cal B}_{\rm CHSH}$ (right) over the $(\sin^2\!\Phi,\cos\Theta)$ plane. The upper and lower rows correspond to $e^+e^-$ collision energies of $\sqrt{s}=500$ GeV and $1$ TeV, respectively.

The resulting patterns differ markedly from those obtained for the single vector--operator case in section \ref{sec:vector}\,(iii) (see Fig.\ \ref{fig:iii}). 
In particular, the symmetry in Eq.\ \eqref{eq:sym} is absent in the SM because the photon- and $Z$-exchange amplitudes contribute coherently, breaking the simple structure that holds for a pure vector interaction.

In the left panel, we see that the state is pure, $\Gamma=1$, for perfectly polarised beams, i.e. at $\sin^2 \Phi=0$ or $1$. 
The purity falls off more steeply near $\sin^2\Phi=1$ than near $\sin^2\Phi=0$. 
For $\sqrt{s}=500~\mathrm{GeV}$, it attains the minimum $\Gamma=\tfrac12$ at $(\sin^2\!\Phi,\cos\Theta)\simeq(0.684,\,1)$ and $(0.772,\,-1)$, with only mild $\Theta$ dependence overall. 
At $\sqrt{s}=1~\mathrm{TeV}$ the $\Theta$ dependence becomes stronger, especially for $\sin^2\!\Phi\simeq0.5$–$0.8$. 
The minimum remains $\Gamma=\tfrac12$, occurring at $(\sin^2\!\Phi,\cos\Theta)\simeq(0.676,\,1)$ and $(0.793,\,-1)$. 
Within this $\sin^2\!\Phi$ window, $\Gamma$ can reach $\simeq0.765$ around $(\sin^2\!\Phi,\cos\Theta)\simeq(0.676,\,-0.522)$.

In the middle panel, we see that the concurrence is almost independent of the polarisation angle $\Phi$.
We observe that ${\cal C} = 0$ in the forward/backward limits, $\cos \Theta = \pm 1$.
For $\sqrt{s} = 500$ GeV, the concurrence is rather small; the maximal value is ${\cal C} \simeq 0.291$ at $(\sin^2 \Phi,\, \cos \Theta) \simeq (0,\,-0.355)$.
For higher collision energies, the entanglement becomes stronger.
For $\sqrt{s} = 1$ TeV, we find the maximum
${\cal C} \simeq 0.737$ at $(\sin^2 \Phi,\, \cos \Theta) \simeq (0,\,-0.457)$.

The behaviour of the Bell-CHSH observable mirrors that of the concurrence, as seen in the right panel of Fig.~\ref{fig:sm}. In general, ${\cal B}_{\rm CHSH}$ is suppressed in the forward/backward regions $\cos\Theta\simeq\pm1$ and enhanced near $\cos\Theta\simeq-0.5$.

For $\sqrt{s} = 500$ GeV, the Bell-CHSH observable is almost constant, ${\cal B}_{\rm CHSH} \simeq 2$, and depends weakly on $\Phi$ and $\Theta$.
In almost the entire plane, ${\cal B}_{\rm CHSH}$ is slightly larger than 2; the maximum ${\cal B}_{\rm CHSH} \simeq 2.08$ is found at $(\sin^2 \Phi,\, \cos \Theta) \simeq (0,\,-0.359)$.
In the region $\sin^2 \Phi \simeq 0.2-0.9$ and $\cos \Theta \simeq \pm 1$, ${\cal B}_{\rm CHSH}$ can go lower than the LHVT threshold 2. 
We detected the minimum $1.99$ around $(\cos \Theta,\,\sin^2 \Phi) \simeq (\pm1,\,0.76)$.
The contour of ${\cal B}_{\rm CHSH} = 2$ is shown in the upper-right plot.

For $\sqrt{s} = 1$ TeV, the Bell-CHSH observable is generally larger than for $\sqrt{s} = 500$ GeV.
In the lower-right plot, the ${\cal B}_{\rm CHSH} = 2$ contour is no longer visible.
The maximal value ${\cal B}_{\rm CHSH}  \simeq 2.48$ is found at $(\cos \Theta,\,\sin^2 \Phi) \simeq (-0.457,\,0)$.

\begin{figure}[t!]
\centering
\includegraphics[scale=0.335]{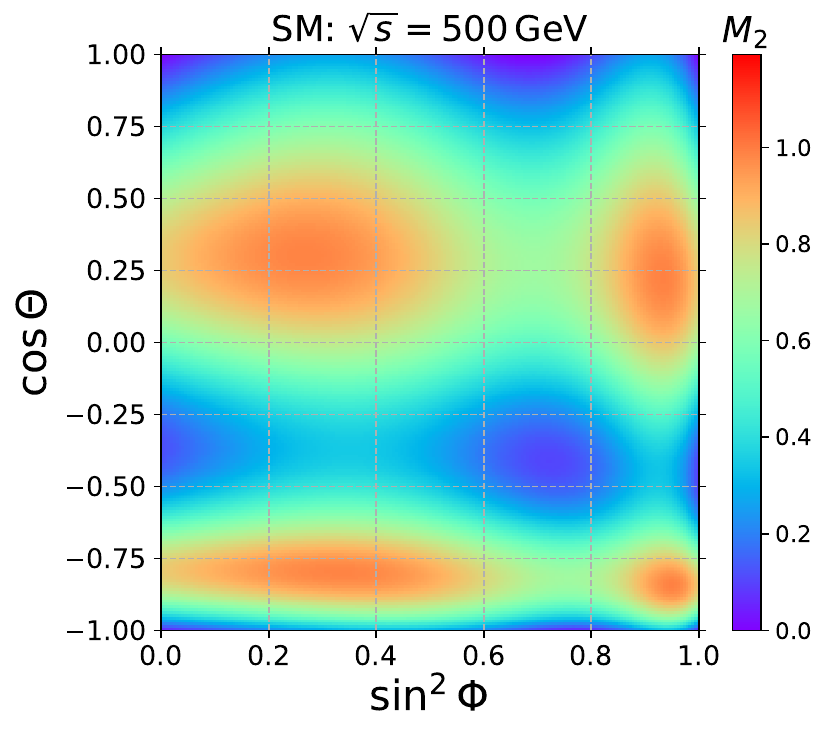}
\hspace{5mm}
\includegraphics[scale=0.335]{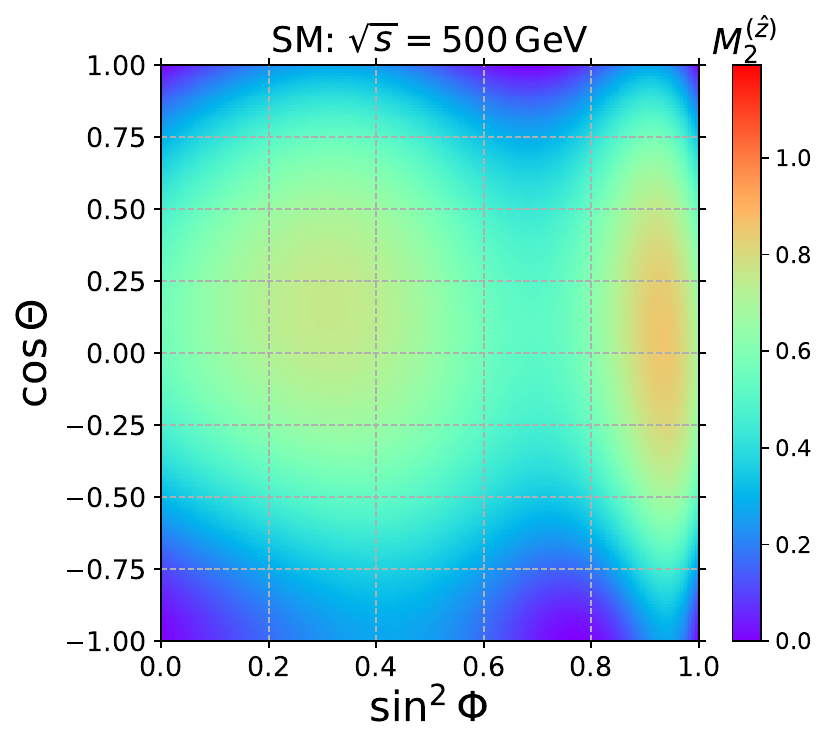}
\\
\includegraphics[scale=0.335]{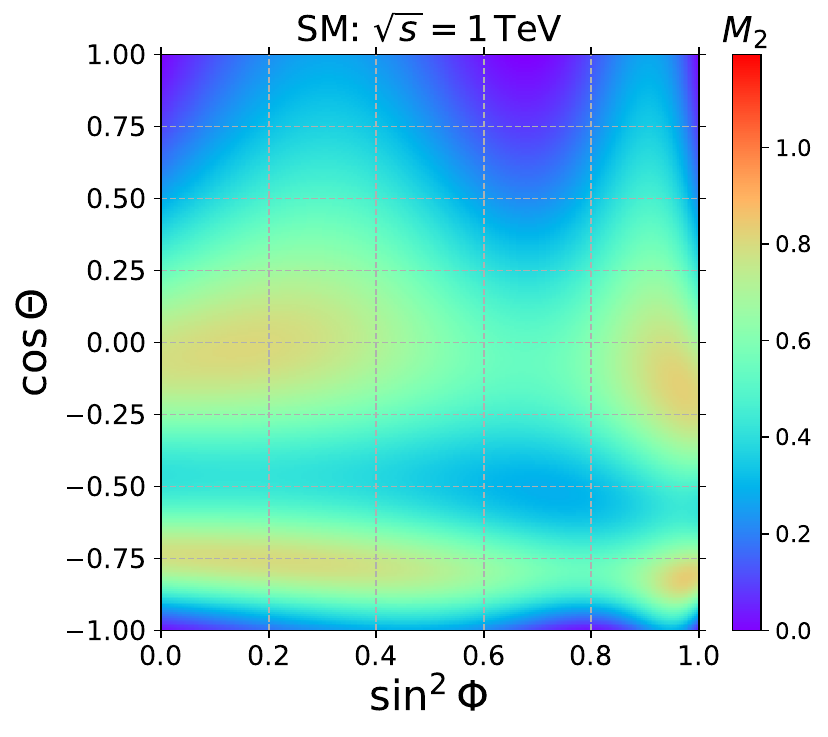}
\hspace{5mm}
\includegraphics[scale=0.335]{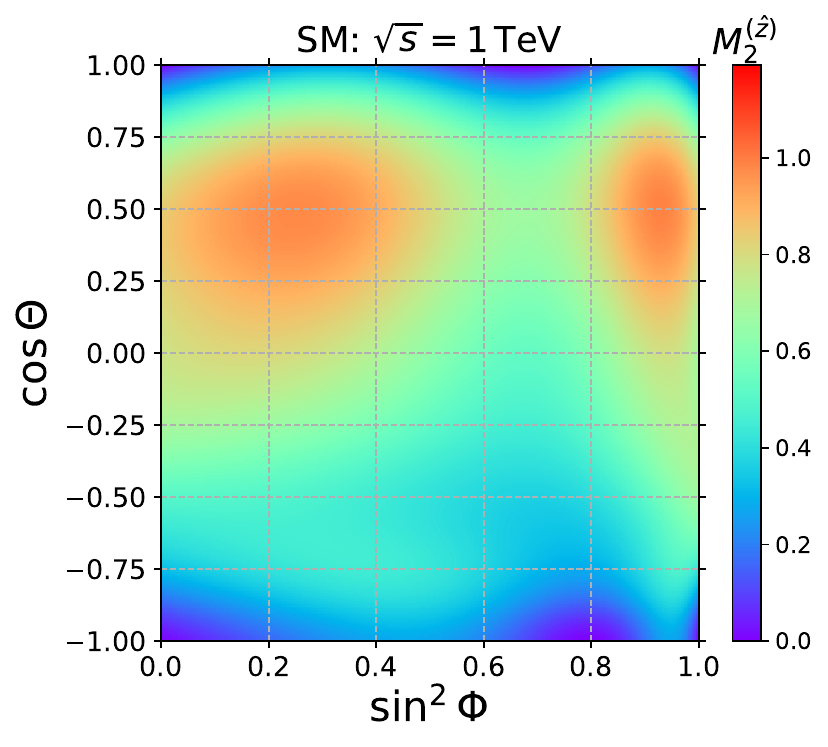}
\caption{\label{fig:sm_SRE}
\small The left (right) panel shows the helicity (beam) basis stabiliser R{\' e}nyi entropy $M_2$ ($M_2^{(\hat z)}$) predicted by the SM over the $(\sin^2 \Phi,  \cos\Theta)$ plane.  
The upper and lower panels correspond to the beam collision energy $\sqrt{s} = 500$ GeV and 1 TeV, respectively.
}
\end{figure}

Fig.\ \ref{fig:sm_SRE} illustrates $M_2$ (left panel) and $M_2^{(\hat z)}$ (right panel) predicted by the SM.
The upper and lower panels represent the collision energies $\sqrt{s} = 500$ GeV and 1 TeV, respectively. 
We observe that both SREs vanish at the corner of the plot, i.e., $(\sin^2 \Phi,\, \cos\Theta) = (0, \pm 1)$ and $(1, \pm 1)$.
In addition to these zeros, SREs also vanish at intermediate $\sin^2 \Phi$ values when $\cos\Theta = \pm 1$.
The locations of those non-trivial zeros are listed in Table \ref{tab:M2sm}.
\begin{table}[t!]
\centering
\begin{minipage}{0.48\linewidth}
\centering
\renewcommand{\arraystretch}{1.1}
\setlength{\tabcolsep}{4pt}
\begin{tabular}{|c|c|c|}
  \hline
  $\sqrt{s}$ & $M_2$ & $(\sin^2 \Phi,\, \cos\Theta)$ \\ \hline
  \multirow{5}{*}{$0.5\,\mathrm{TeV}$} 
    & 0     & $(0.684,\, 1),\,(0.772,\, -1)$ \\ \cline{2-3}
    & 0.986 & $(0.268,\, 0.306)$ \\ \cline{2-3}
    & 0.984 & $(0.333,\, -0.792)$ \\ \cline{2-3}
    & 0.989 & $(0.932,\, 0.233)$ \\ \cline{2-3}
    & 0.989 & $(0.950,\, -0.834)$ \\ \hline
  \multirow{5}{*}{$1\,\mathrm{TeV}$} 
    & 0     & $(0.676,\, 1),\,(0.793,\, -1)$ \\ \cline{2-3}
    & 0.809 & $(0.151,\, -0.017)$ \\ \cline{2-3}
    & 0.800 & $(0.191,\, -0.748)$ \\ \cline{2-3}
    & 0.829 & $(0.97,\, -0.154)$ \\ \cline{2-3}
    & 0.838 & $(0.976,\, -0.812)$ \\ \hline
\end{tabular}
\end{minipage}
\begin{minipage}{0.48\linewidth}
\centering
\renewcommand{\arraystretch}{1.1}
\setlength{\tabcolsep}{4pt}
\begin{tabular}{|c|c|c|}
  \hline
  $\sqrt{s}$ & $M_{2}^{(\hat z)}$ & $(\sin^2 \Phi,\, \cos\Theta)$ \\ \hline
  \multirow{3}{*}{$0.5\,\mathrm{TeV}$}
    & 0     & $(0.684,\, 1),\,(0.772,\, -1)$ \\ \cline{2-3}
    & 0.758 & $(0.311,\, 0.163)$ \\ \cline{2-3}
    & 0.854 & $(0.933,\, 0.051)$ \\ \hline
  \multirow{3}{*}{$1\,\mathrm{TeV}$}
    & 0     & $(0.676,\, 1),\,(0.793,\, -1)$ \\ \cline{2-3}
    & 0.977 & $(0.250,\, 0.480)$ \\ \cline{2-3}
    & 0.988 & $(0.928,\, 0.501)$ \\ \hline
\end{tabular}
\end{minipage}
\caption{\small The information about the zeros and local maxima of $M_2$ (left) and $M_2^{(\hat z)}$ (right) in the SM.
In addition to the non-trivial zeros shown in the table, $M_2$ and $M_2^{(\hat z)}$ vanish also at $(\sin^2 \Phi,\, \cos\Theta) = (0, \pm 1)$ and $(1, \pm 1)$. 
\label{tab:M2sm}}
\end{table}

We further observe that $M_2$ exhibits four local maxima, while $M_2^{(\hat z)}$ shows two.
The local maxima of $M_2$ decrease, whereas those of $M_2^{(\hat z)}$ increase, as $\sqrt{s}$ rises from $500$ GeV to 1 TeV. 
The values and locations of those local maxima are listed in Table \ref{tab:M2sm}.



\section{$e^+ e^- \to t \bar t$ in an EFT} 
\label{sec:sm+eft}

We now consider the $t\bar t$ final state in an EFT obtained by augmenting the SM Lagrangian with the generic four-fermion dimension-six operators listed in Eqs.~(\ref{Lscalar})–(\ref{Ltensor}).
The aim of this section is to present the response of quantum observables to different types of new physics contributions and beam polarisations.
By adding the scalar or the tensor four-fermion operator to the SM, all initial spin configurations, $(\l_l, \l_{\bar l}) = (\pm,\pm)$ and $(\pm, \mp)$, provide non-zero amplitudes.
Hence, the final state density matrix cannot be parametrised by a single effective polarisation angle.
We describe $\rho^{\rm f}$ with the original polarisation parameters, ${\cal P}$ and $\overline {\cal P}$.

Throughout this section we consider the beam collision energy $\sqrt{s} = 500$ GeV.
The quantum observables for $\sqrt{s} = 1$ TeV are presented in Appendix \ref{app:eft_1tev_plots}.
The Wilson coefficient of the EFT operator under consideration is conveniently set to one, $c_X = 1$ ($X = S,\, V$ or $T$). 
The cut-off scale of the operator is fixed to three times the centre-of-mass energy, $\Lambda = 3 \cdot \sqrt{s}$. 
Note that the SM amplitudes scale as $\sim 1/s$, whereas the new physics contributions scale as $\sim s/\Lambda^2$.  
Therefore, in this setup the relative importance of the new physics to the SM is independent of $\sqrt{s}$.

\begin{table}[t]
\centering
\renewcommand{\arraystretch}{1.1}
\begin{tabular}[t!]{| c | c |}
\hline
Benchmark scenario & Parameters  \\ 
\hline 
SM & ---    
\\ 
\hline
Scalar (S) & $\eta_S = 0$ 
\\ 
\hline
Scalar (P) & $\eta_S = \frac{\pi}{2}$  
\\ 
\hline
Scalar (CPV) & $\eta_S = \frac{\pi}{4}$  
\\ 
\hline
Vector (V,V) & $(\xi_R,\eta_A) = (\frac{\pi}{4}, 0)$ 
\\ 
\hline
Vector (A,A) & $(\xi_R,\eta_A) = (\frac{3\pi}{4}, \frac{\pi}{2})$ 
\\ 
\hline
Vector (R,R) & $(\xi_R,\eta_A) = (\frac{\pi}{2}, \frac{\pi}{4})$  
\\
\hline
Tensor & --- 
 \\ 
\hline
\end{tabular}
\caption{\label{tab:scenarios}
\small 
Definition of benchmark scenarios. 
}
\end{table}

The benchmark scenarios studied in this paper are summarised in Table~\ref{tab:scenarios}. 
Each scenario is specified by the Lorentz structure of the EFT operator (scalar, vector or tensor) and by fixing the relevant coupling angles (e.g.\ $\eta_S$ and $\xi_R$). 
Not all coupling angles are physical, however. 
At leading order, the SM has non-vanishing amplitudes only for initial spin configurations $(\lambda_{e^-},\lambda_{e^+})=(\pm,\pm)$, whereas the scalar and tensor operators contribute only for $(\pm,\mp)$. 
As a result, there is no SM--EFT interference for scalar or tensor operators. 
The overall phases $e^{\pm i\xi_S}$ (scalar) and $e^{\pm i(\xi_T+\eta_T)}$ (tensor) therefore drop out of the final $F \bar F$ spin density matrix, and all quantum observables are independent of $\xi_S$ in the scalar case and of $\xi_T,\eta_T$ in the tensor case.

As indicated in Table~\ref{tab:scenarios}, we select three scalar benchmarks: (S), (P) and (CPV), defined by $\eta_S=0$, $\pi/2$, and $\pi/4$, respectively. 
For vectors, we consider three cases:
(V,V) with $(\xi_R,\eta_A)=(\pi/4,0)$, 
(A,A) with $(\xi_R,\eta_A)=(3\pi/4,\pi/2)$ and
(R,R) with $(\xi_R,\eta_A)=(\pi/2,\pi/4)$. 
For the tensor operator, the $t \bar t$ spin density matrix is independent of $\xi_T$ and $\eta_T$, yielding a single benchmark.

In what follows we focus on scalar (S) and (P), vector (V,V) and (A,A) and tensor benchmarks. 
The scalar (CPV) and vector (R,R) scenarios are presented in Appendix~\ref{app:eft_500gev_plots}.

\begin{figure}[t!]
\centering
\fbox{\includegraphics[scale=0.335]{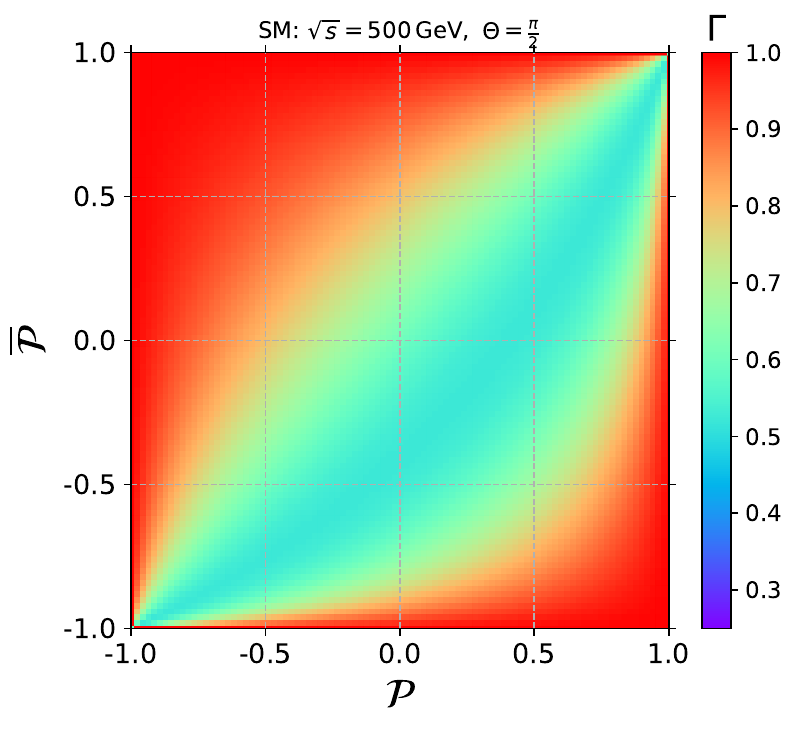}}
\includegraphics[scale=0.335]{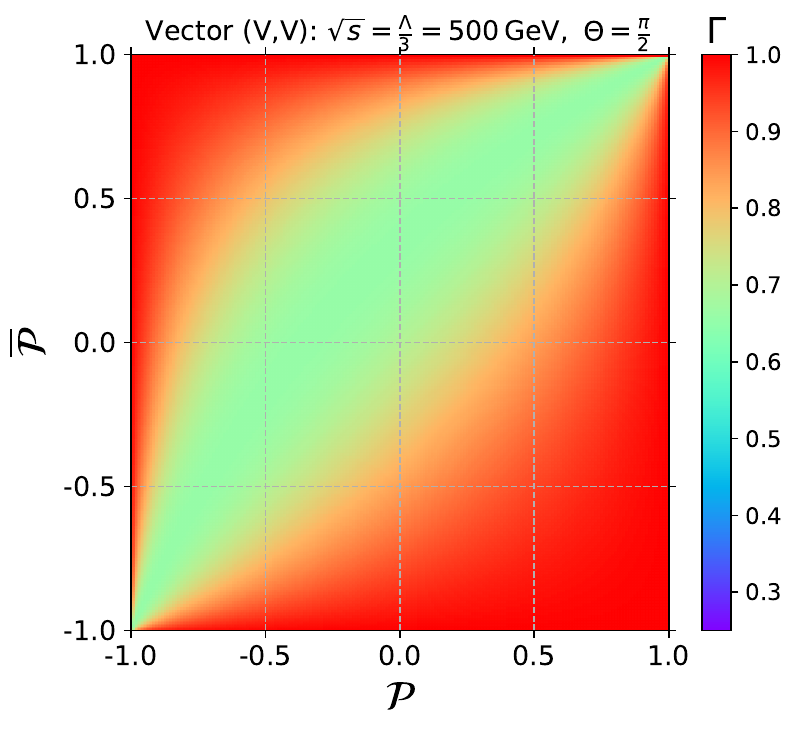}
\includegraphics[scale=0.335]{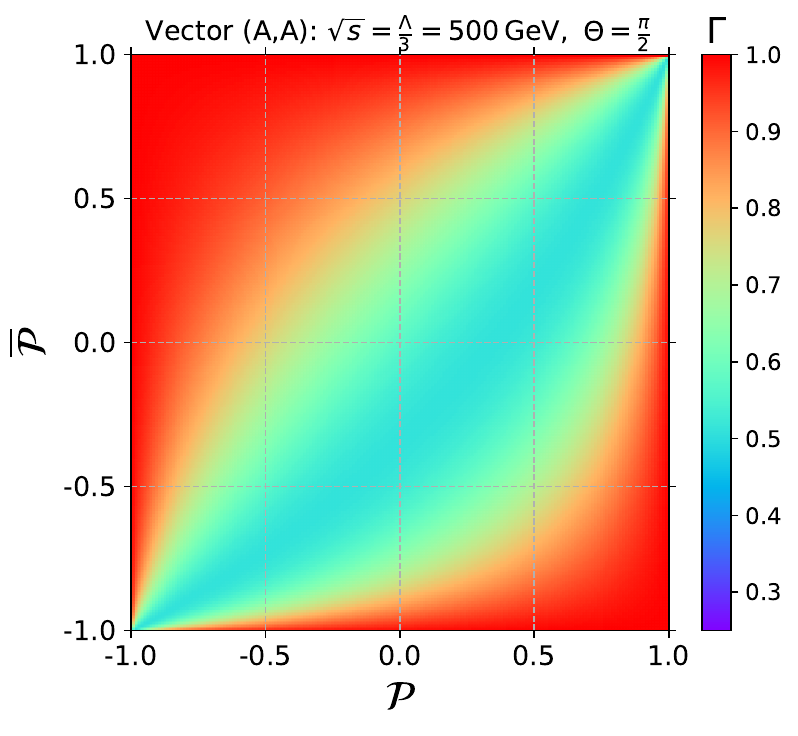}
\\
\includegraphics[scale=0.335]{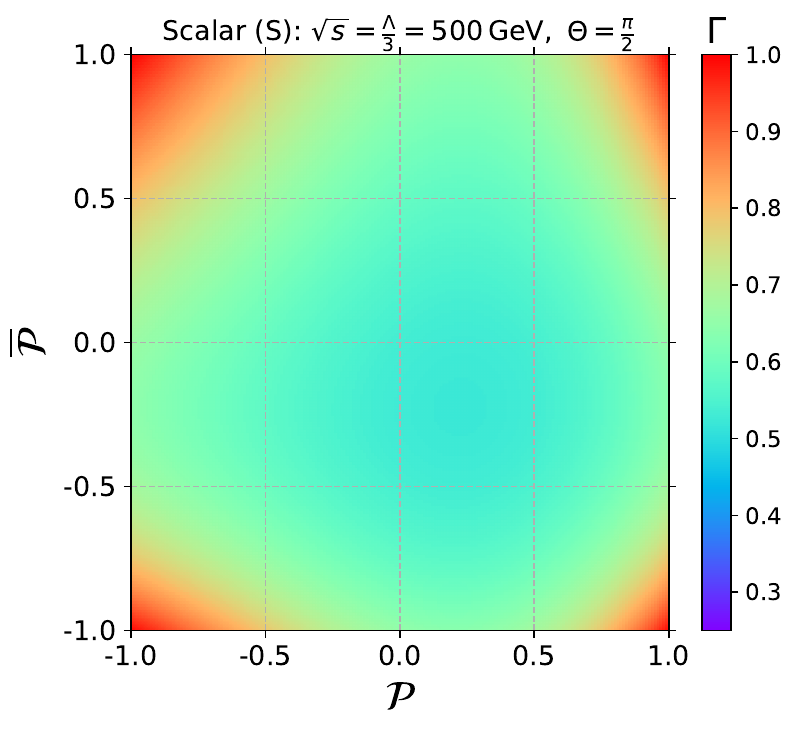}
\includegraphics[scale=0.335]{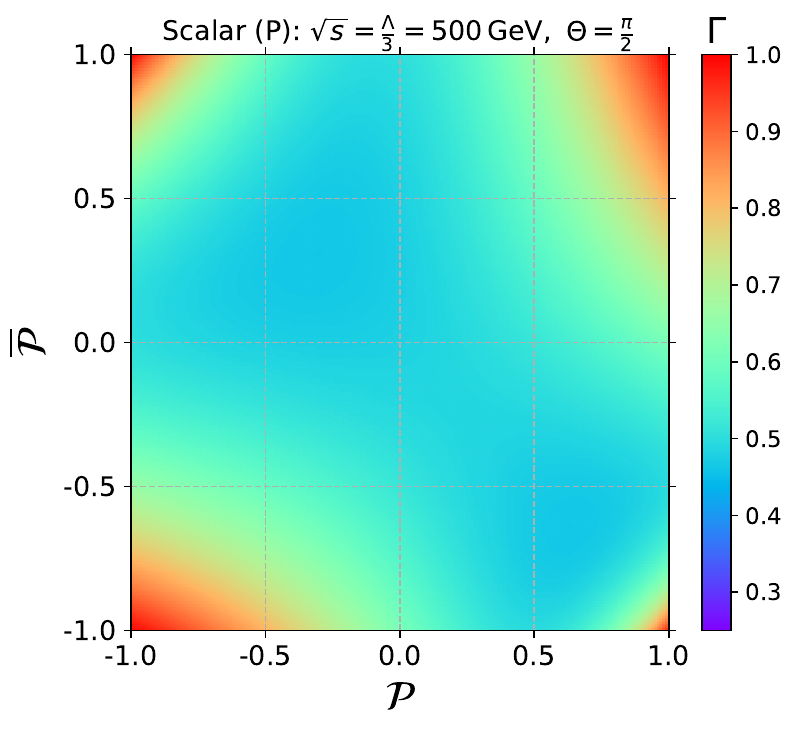}
\includegraphics[scale=0.335]{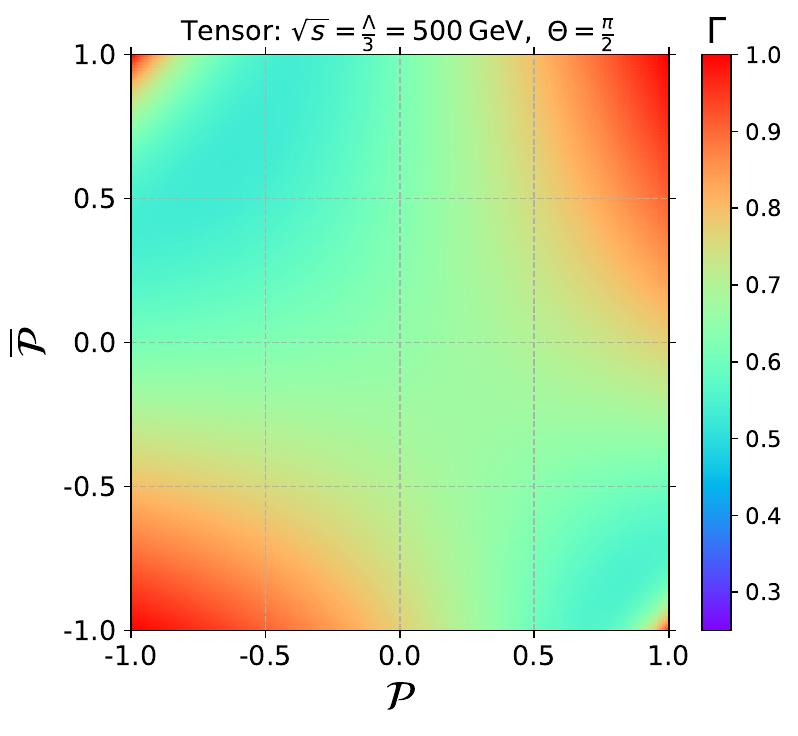}
\caption{\label{fig:pure_500gev}
\small 
Purity $\Gamma$ over the $(\cal P,\overline {\cal P})$ plane. 
The centre-of-mass energy is $\sqrt{s}=500$ GeV and the production angle is fixed to $\Theta=\pi/2$. 
Top row: SM (left; outlined in black), vector (V,V) (middle), and axial–vector (A,A) (right). 
Bottom row: scalar (S) (left), scalar (P) (middle) and tensor (right).
}
\end{figure}

\begin{figure}[t!]
\centering
\fbox{\includegraphics[scale=0.335]{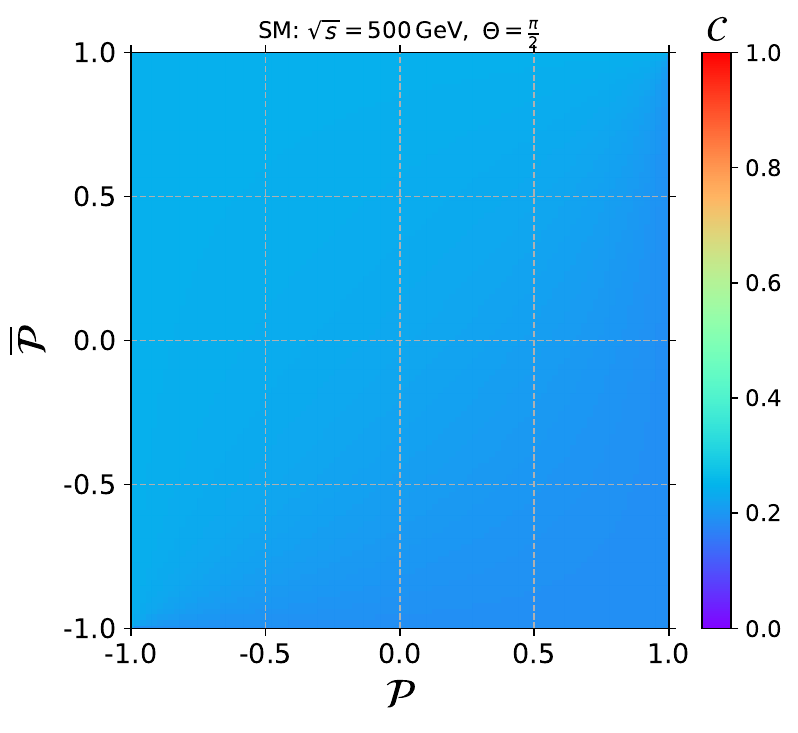}}
\includegraphics[scale=0.335]{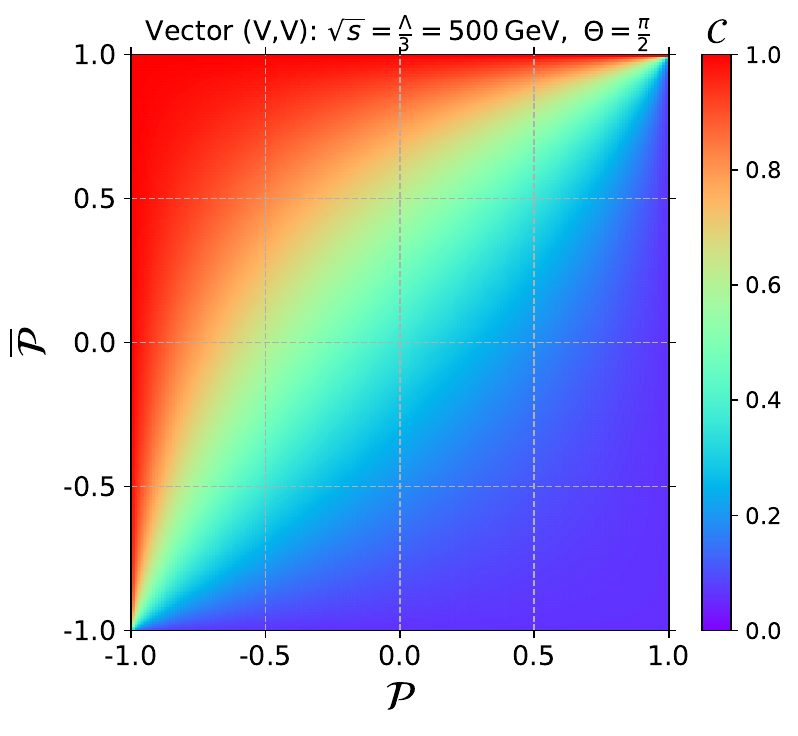}
\includegraphics[scale=0.335]{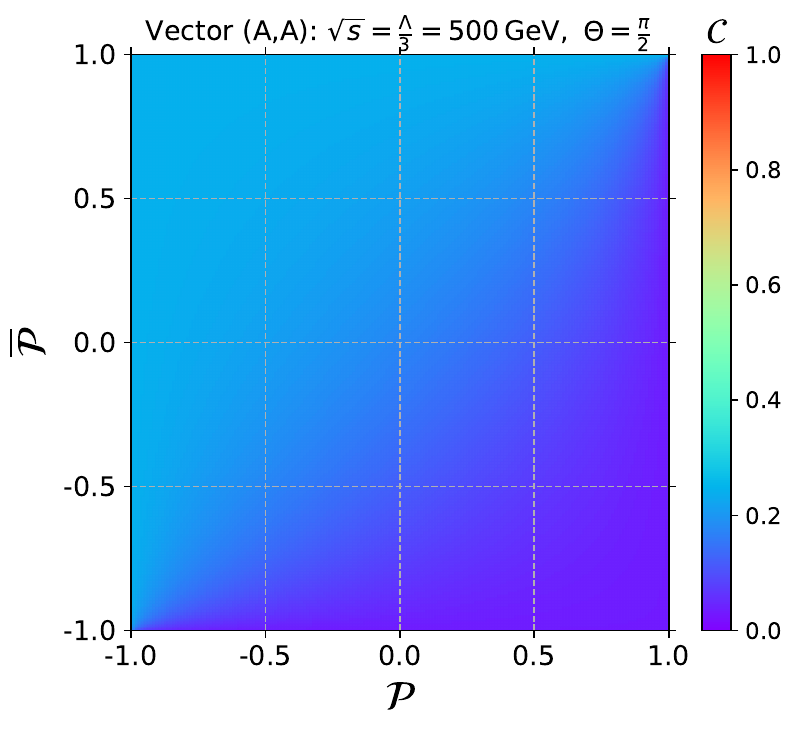}
\\
\includegraphics[scale=0.335]{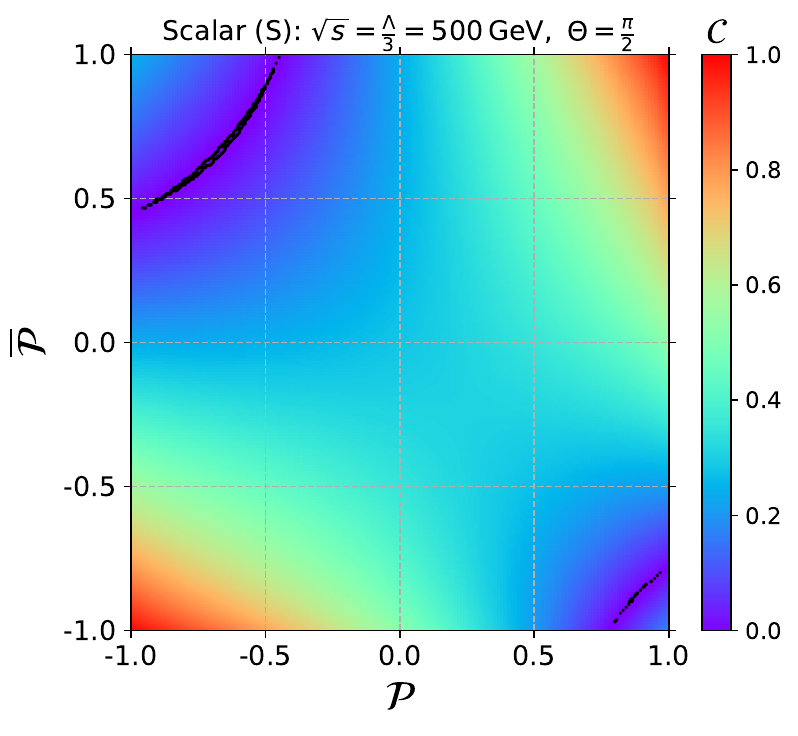}
\includegraphics[scale=0.335]{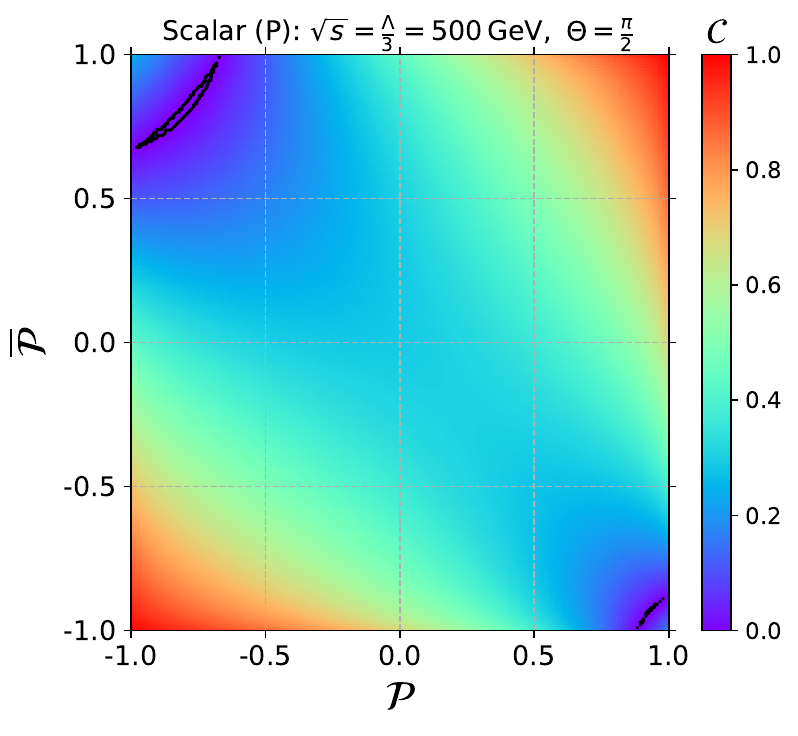}
\includegraphics[scale=0.335]{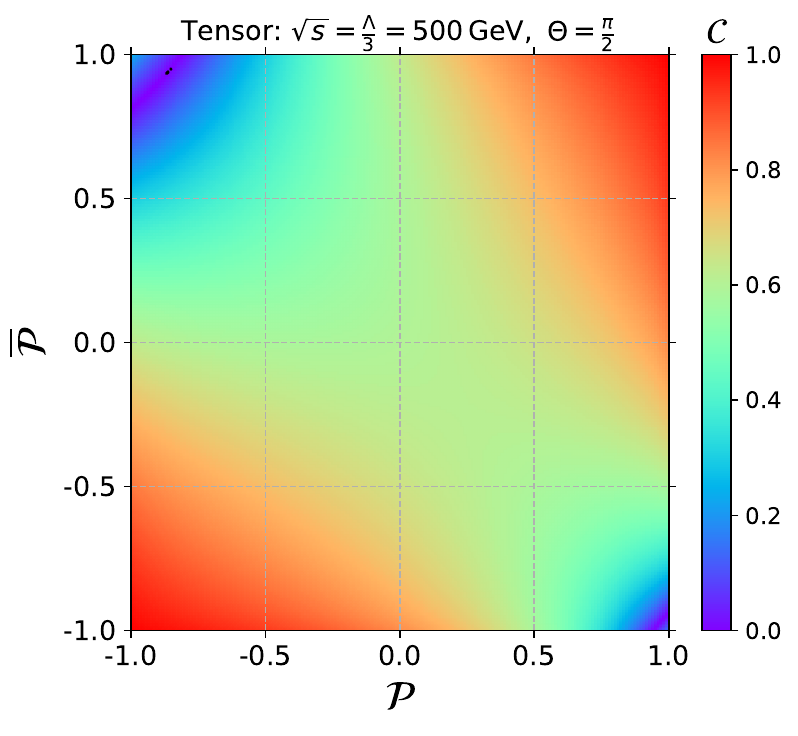}
\caption{\label{fig:conc_500gev}
\small Concurrence ${\cal C}$ over the $({\cal P},\overline {\cal P})$ plane. 
Contours of ${\cal C}=10^{-4}$ are overlaid in black.
The centre-of-mass energy is $\sqrt{s}=500$ GeV and the production angle is fixed to $\Theta=\pi/2$. 
Top row: SM (left; outlined in black), vector (V,V) (middle), and axial–vector (A,A) (right). 
Bottom row: scalar (S) (left), scalar (P) (middle) and tensor (right).
}
\end{figure}

\begin{figure}[t!]
\centering
\fbox{\includegraphics[scale=0.335]{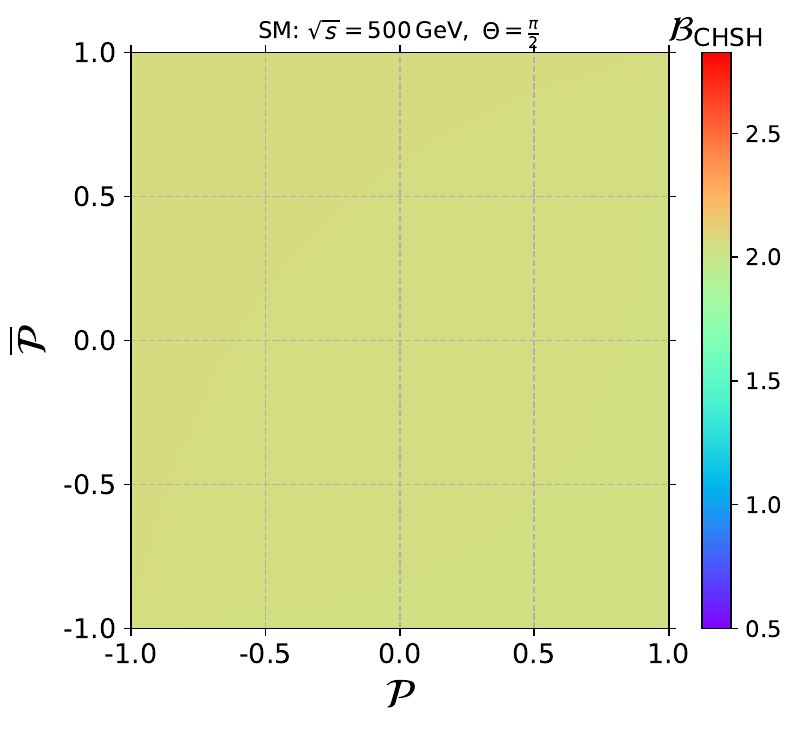}}
\includegraphics[scale=0.335]{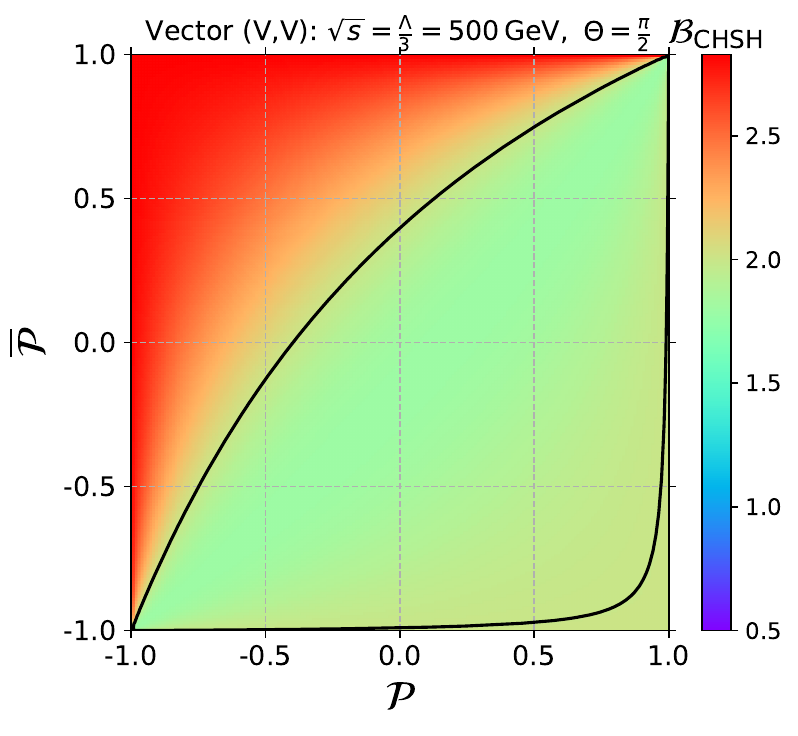}
\includegraphics[scale=0.335]{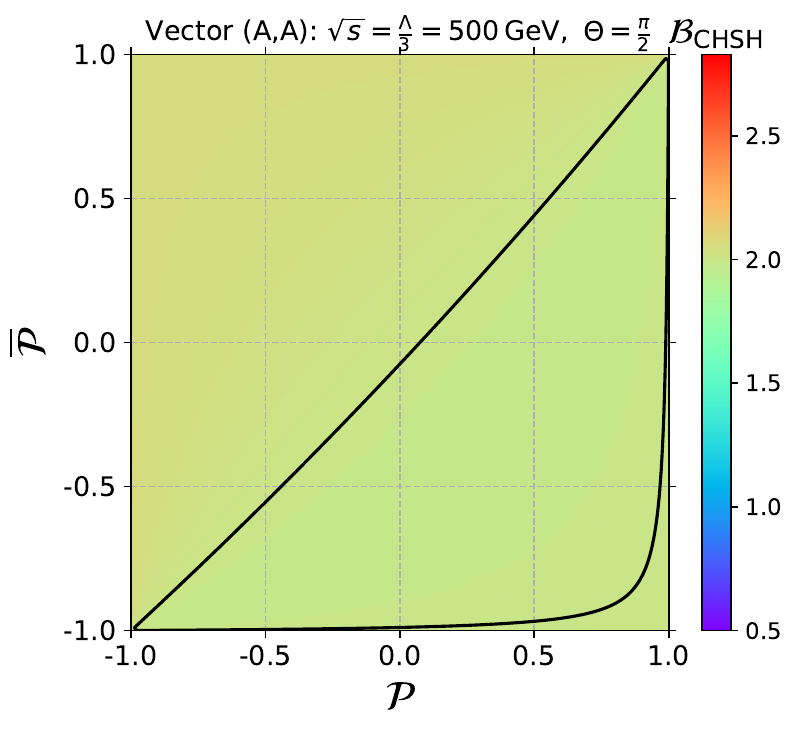}
\\
\includegraphics[scale=0.335]{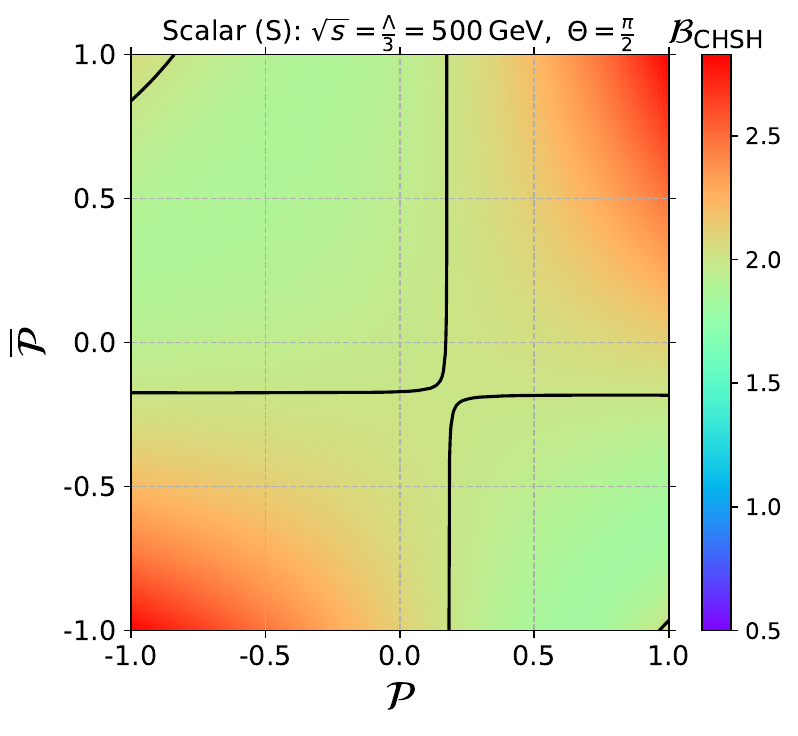}
\includegraphics[scale=0.335]{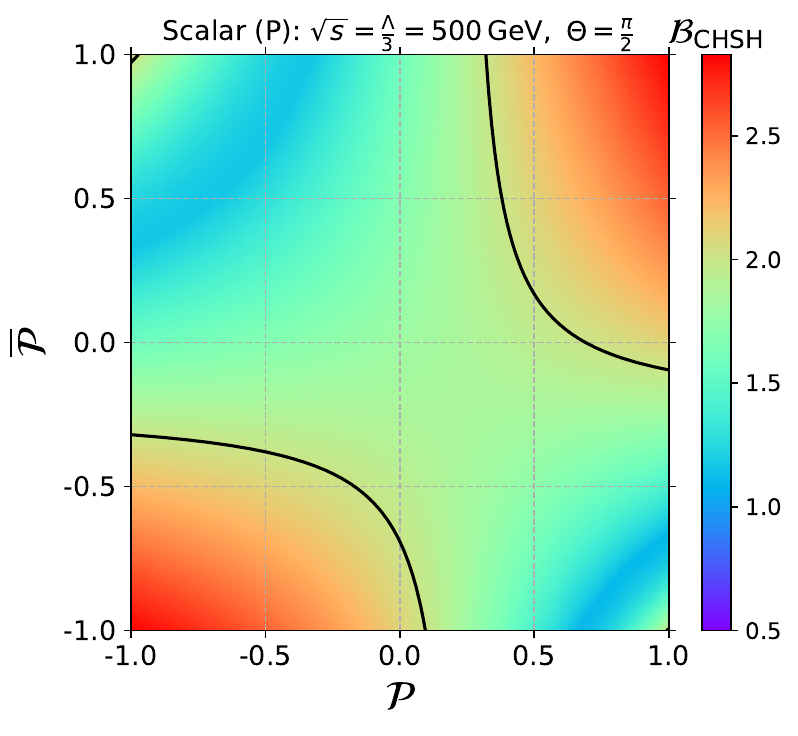}
\includegraphics[scale=0.335]{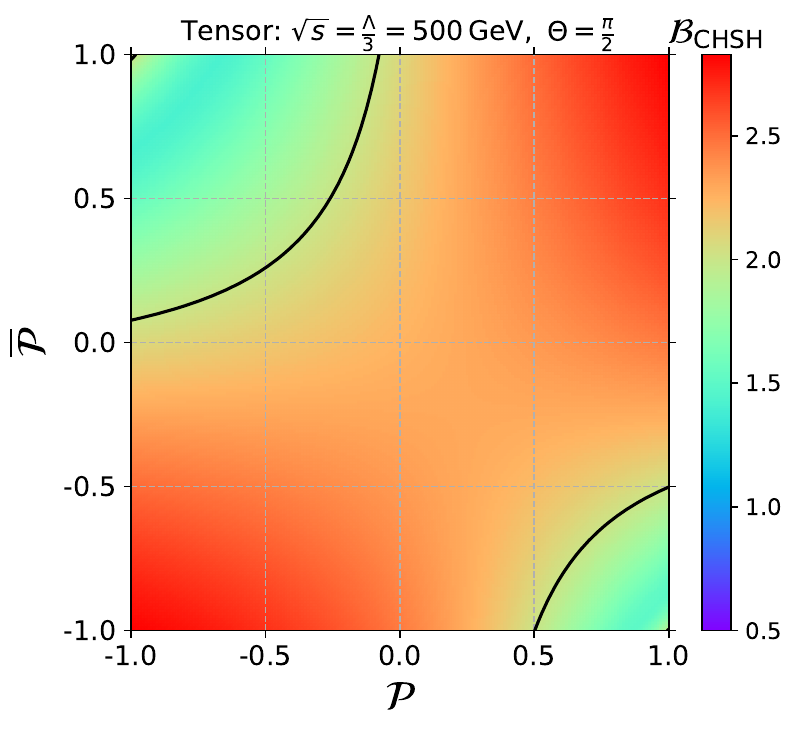}
\caption{\label{fig:chsh_500gev}
\small Bell-CHSH observable ${\cal B}_{\rm CHSH}$ over the $(\cal P,\overline {\cal P})$ plane. 
Contours of ${\cal B}_{\rm CHSH}=2$ are overlaid in black.
The centre-of-mass energy is $\sqrt{s}=500$ GeV and the production angle is fixed to $\Theta=\pi/2$. 
Top row: SM (left; outlined in black), vector (V,V) (middle), and axial–vector (A,A) (right). 
Bottom row: scalar (S) (left), scalar (P) (middle) and tensor (right).
}
\end{figure}

\begin{figure}[t!]
\centering
\fbox{\includegraphics[scale=0.335]{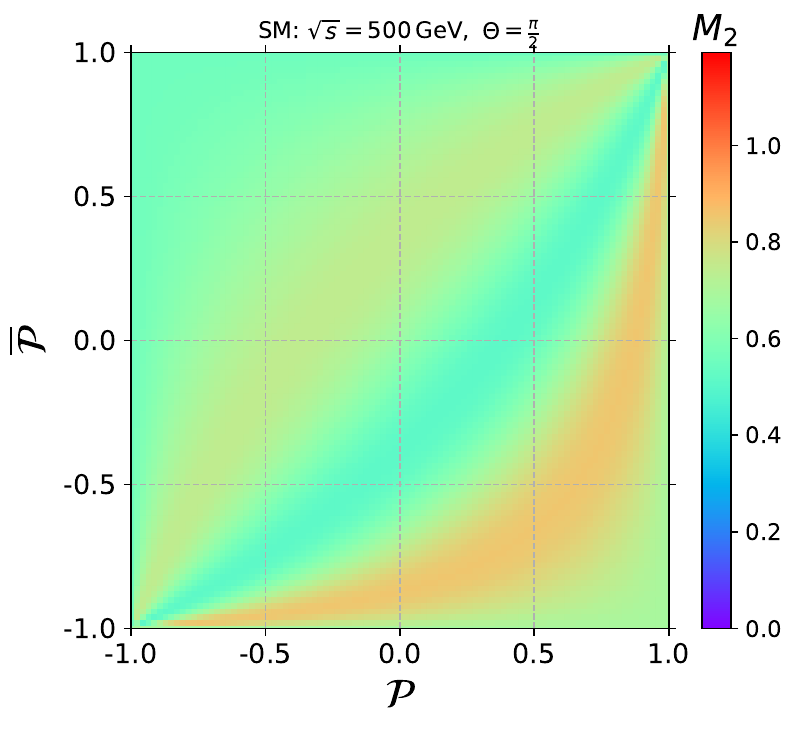}}
\includegraphics[scale=0.335]{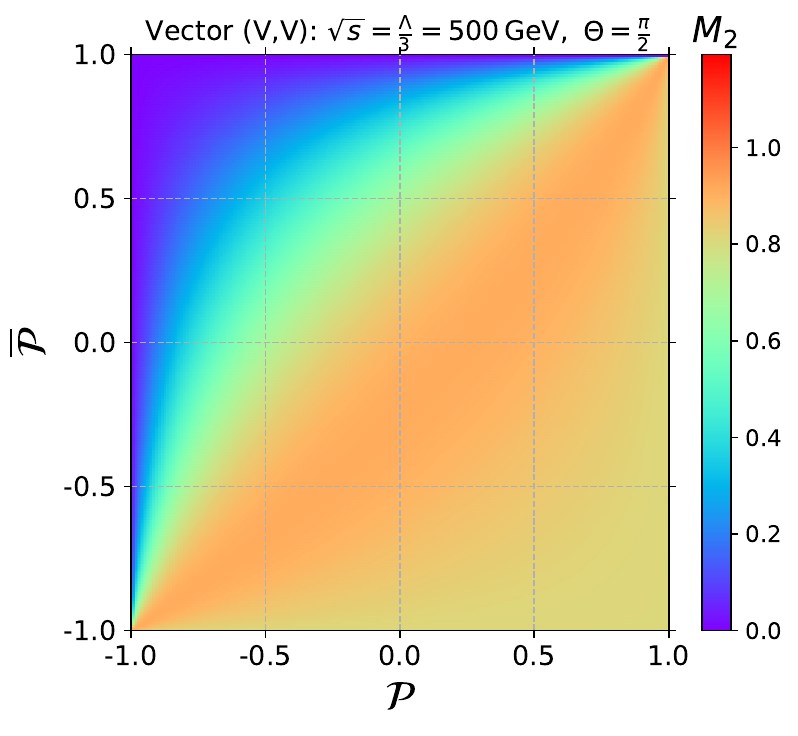}
\includegraphics[scale=0.335]{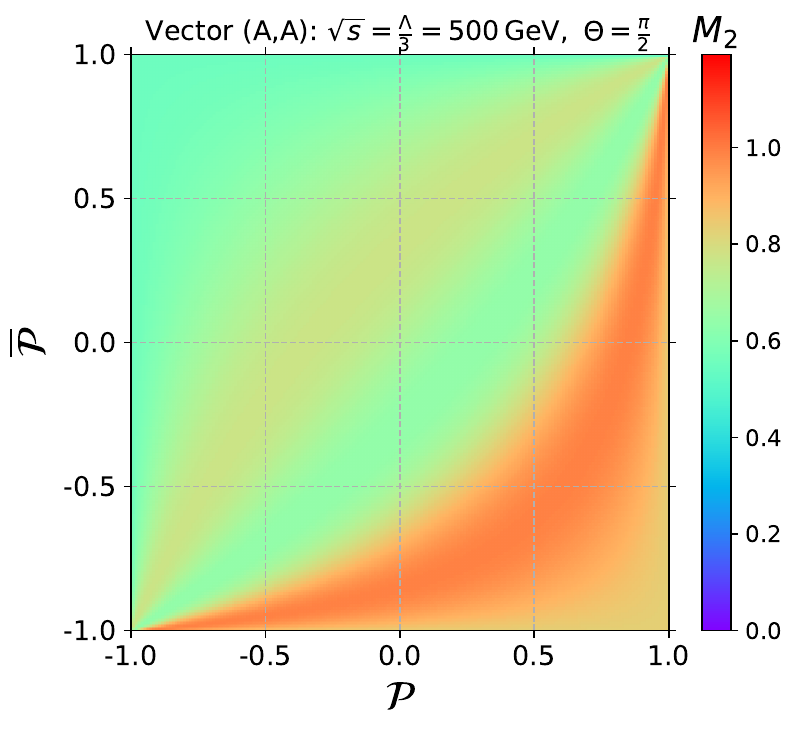}
\\
\includegraphics[scale=0.335]{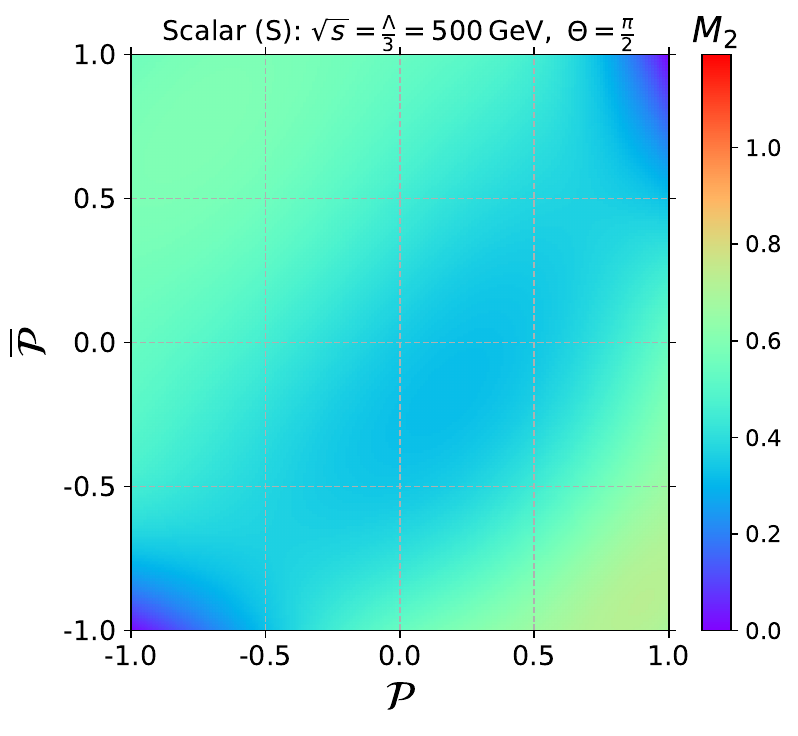}
\includegraphics[scale=0.335]{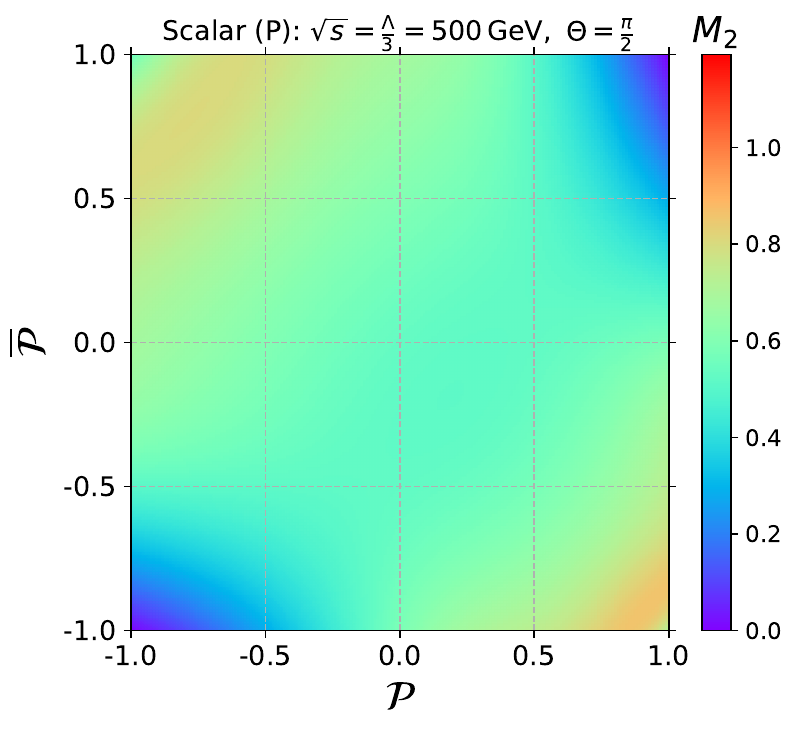}
\includegraphics[scale=0.335]{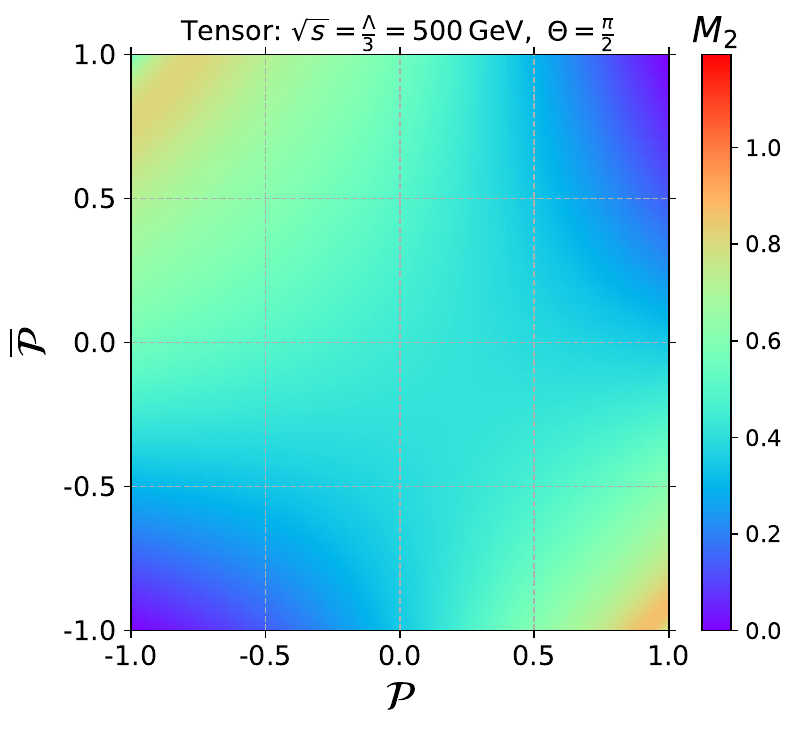}
\caption{\label{fig:M2_500gev}
\small Helicity-basis SRE $M_2$ over the $(\cal P,\overline {\cal P})$ plane. 
The centre-of-mass energy is $\sqrt{s}=500$ GeV and the production angle is fixed to $\Theta=\pi/2$. 
Top row: SM (left; outlined in black), vector (V,V) (middle), and axial–vector (A,A) (right). 
Bottom row: scalar (S) (left), scalar (P) (middle) and tensor (right).
}
\end{figure}

Figs.\ \ref{fig:pure_500gev}--\ref{fig:M2_500gev} present the purity $\Gamma$, the concurrence ${\cal C}$, the Bell-CHSH observable ${\cal B}_{\rm CHSH}$ and the helicity-basis SRE $M_2$ over the $({\cal P}, \overline {\cal P})$ plane. 
The production angle is fixed to $\Theta = \pi/2$.
In all figures, the upper-left plot (highlighted with a black frame) displays the SM prediction. 
In the upper panel, the middle and right plots correspond to the vector (V,V) and (A,A) scenarios. 
The lower-left, -middle and -right panels present the predictions of the scalar (S), the scalar (P) and the tensor scenarios, respectively.

Fig.\ \ref{fig:pure_500gev} shows the dependence of the purity $\Gamma$ on the beam polarisations. 
In the SM and in the vector (V,V) and (A,A) scenarios, the purity is highest when the polarisations are large and opposite in sign, $\overline {\cal P} = - {\cal P}$.
This is because the beams are effectively unpolarised on the $\overline {\cal P} = {\cal P}$ line for the vector-type interaction.  
For the scalar (S), scalar (P), and tensor scenarios, the purity is significantly reduced compared to the SM case. 
For the scalar (P) benchmark in particular, the purity can go as low as $\Gamma \simeq 0.464$ around $({\cal P}, \overline {\cal P}) \simeq (0.634, -0.634)$ and $(-0.308,  0.308)$.
Note that the final state purity below $\frac{1}{2}$ is never observed in the single operator study in section \ref{sec:EFT} and in the SM in section \ref{sec:sm}. 
For the benchmark scenarios other than the SM and vector-type, however, the $F \bar F$ state becomes pure for the extreme polarisations, $({\cal P}, \overline {\cal P}) = (\pm 1, \pm 1)$ and $(\pm 1, \mp 1)$.

Fig.\ \ref{fig:conc_500gev} shows the concurrence ${\cal C}$ over the $(\mathcal P,\overline{\mathcal P})$ plane. 
In the SM, ${\cal C}$ is nearly uniform across the plane, with $0.188 \lesssim {\cal C} \lesssim 0.244$. 
The EFT scenarios exhibit strikingly different patterns. 
In the vector (V,V) case, the concurrence is strongly enhanced near $({\cal P},\overline{\cal P})\simeq(-1,1)$ and suppressed near $(1,-1)$. 
The vector (A,A) case shares the suppression around $(1,-1)$ but lacks the corresponding enhancement near $(-1,1)$. 
Consequently, the concurrence at a polarised lepton collider provides strong discriminating power between the SM and the vector benchmarks (V,V) and (A,A).

The concurrence patterns for the scalar benchmarks (S) and (P) and for the tensor scenario are similar to one another but distinct from the SM and vector cases. 
All three exhibit two lobes of strong suppression located at large positive ${\cal P}$ with large negative $\overline{\cal P}$ and at large negative ${\cal P}$ with large positive $\overline{\cal P}$. 
The black curves in these panels mark the ${\cal C}=10^{-4}$ contours. 
For the tensor benchmark, these low-concurrence regions become even more pronounced at higher collision energy (see Fig.~\ref{fig:conc_1000gev} in Appendix~\ref{app:eft_1tev_plots} for $\sqrt{s}=1~\mathrm{TeV}$). 
In the opposite corners, ${\cal P}\simeq\overline{\cal P}\simeq\pm1$, the concurrence is strongly enhanced; maximally entangled states are reached at the extreme polarisations ${\cal P}=\overline{\cal P}=\pm1$.

Fig.\ \ref{fig:chsh_500gev} displays the Bell-CHSH observable ${\cal B}_{\rm CHSH}$, which broadly tracks the concurrence. 
In the SM and in the vector (A,A) benchmark, ${\cal B}_{\rm CHSH}$ varies only mildly across the $(\mathcal P,\overline{\mathcal P})$ plane: it lies in the range $2.03$--$2.06$ for the SM and $1.98$--$2.06$ for (A,A). 
Thus, in the SM the CHSH inequality is only weakly violated over the plane.

The other benchmarks show far more distinctive structures. 
Black contours denote ${\cal B}_{\rm CHSH}=2$, separating regions with and without CHSH violation. 
In the vector (V,V) case, violation occurs predominantly where $\mathcal P$ is large negative and $\overline{\mathcal P}$ is large positive, with the notable exception of the vicinity of $(\mathcal P,\overline{\mathcal P})=(1,-1)$. 
In particular, ${\cal B}_{\rm CHSH}$ becomes large near $(-1,1)$, approaching the Tsirelson bound.

For the scalar (S) and (P) benchmarks and for the tensor scenario, the trend is reversed: Bell-inequality violation is found mainly for large, same-sign polarisations. 
For these cases the Tsirelson bound is reached at $(\mathcal P,\overline{\mathcal P})=\pm(1,1)$. 
A distinctive feature of the scalar (P) benchmark is a strong suppression of ${\cal B}_{\rm CHSH}$ along bands near $(\mathcal P,\overline{\mathcal P})\!\sim\!(0.6,-0.6)$ and $(-0.8,0.8)$, where ${\cal B}_{\rm CHSH}\!\sim\!1.1$ (i.e.\ pronounced non-violation). 
This pattern sharply discriminates between the scalar (P) and the scalar (S) scenarios.

At higher energy, the CHSH–nonviolating region contracts for the vector benchmarks but expands for the tensor case (see Fig.~\ref{fig:chsh_1000gev} in Appendix~\ref{app:eft_1tev_plots} for $\sqrt{s}=1~\mathrm{TeV}$).

Fig.\ \ref{fig:M2_500gev} shows the helicity-basis SRE, $M_2$. 
Unlike the concurrence and the Bell-CHSH observable, the SM and the vector (A,A) scenario exhibit a visible dependence of $M_2$ on the beam polarisations. 
In these two benchmarks, $M_2$ is minimised along the diagonal ${\cal P}\simeq\overline{\cal P}$; moving away from this line, $M_2$ first rises and then falls toward the opposite corners $({\cal P},\overline{\cal P})\simeq(1,-1)$ and $(-1,1)$. 
A particularly striking feature of the vector (A,A) scenario is a pronounced peak with $M_2\simeq1$ along the strip near $({\cal P},\overline{\cal P})\simeq(0.5,-0.5)$, which serves as a distinctive discriminator for this benchmark.
The vector (V,V) scenario also has a large dependence on beam polarisations; the $M_2$ values can go from nearly zero (around $\overline {\cal P} \simeq - {\cal P} \simeq 1$) to 0.9 (along diagonal ${\cal P} \simeq \overline {\cal P}$). 

For the other benchmarks, scalar (S) and (P), and tensor, the trend is broadly opposite to that of ${\cal B}_{\rm CHSH}$: $M_2$ is enhanced (suppressed) where ${\cal B}_{\rm CHSH}$ is suppressed (enhanced). 
Moreover, for (S), (P), and tensor, one finds $M_2=0$ at the extreme, same-sign polarisations $({\cal P},\overline{\cal P})=\pm(1,1)$.

The beam-basis SRE $M_2^{(\hat z)}$ follows a closely similar pattern; representative maps over the $({\cal P},\overline{\cal P})$ plane for the various benchmarks are provided in Fig.~\ref{fig:M2_z_500gev} (Appendix~\ref{app:eft_500gev_plots}).

The above study clearly demonstrates that beam polarisation makes it possible to isolate specific $t \bar t$ final states, e.g.\ with maximal or vanishing entanglement, Bell-inequality violation, and stabiliser versus high-magic states, depending on the underlying interaction. 
This capability not only enhances sensitivity to new physics but also provides diagnostic power to infer the operator structure of the new interaction.

\medskip

In Figs.\ \ref{fig:pure_500gev}--\ref{fig:M2_500gev}, the production angle was fixed at $\Theta = \pi/2$ and the beam polarisations were varied.  
Below, we instead fix the polarisations and study the quantum observables as functions of $\cos \Theta$.
This mirrors the experimental procedure: beam polarisations are constant within a data-taking run, and the $t \bar t$ spin density matrix is reconstructed from post-selected events in different $\cos \Theta$ bins by quantum state tomography~\cite{Afik:2020onf,Ashby-Pickering:2022umy}.

At the ILC, the design values for the beam polarisations are 80\% (electron) and 30\% (positron)~\cite{Adolphsen:2013kya}.
Motivated by this, we consider the beam polarisation configurations listed in Table \ref{tab:pol} below.
\begin{table}[h!]
\centering
\renewcommand{\arraystretch}{1.1}
\begin{tabular}[h!]{ c | c }
beam configuration & $({\cal P},\, \overline {\cal P})$ \\ \hline
unpolarised & $(0,\,0)$ \\ 
polarised-PP & $(0.8,\,0.3)$ \\ 
polarised-PN & $(0.8,\,-0.3)$ \\ 
polarised-NP & $(-0.8,\,0.3)$ \\ 
polarised-NN & $(-0.8,\,-0.3)$ \\ 
\end{tabular}
\caption{\label{tab:pol}
\small The benchmark beam polarisation configurations. }
\end{table}

\begin{figure}[t!]
\centering
\includegraphics[scale=0.6]{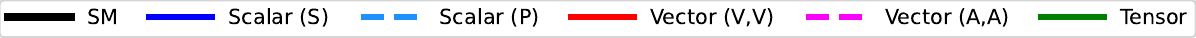}
\\
\includegraphics[scale=0.20]{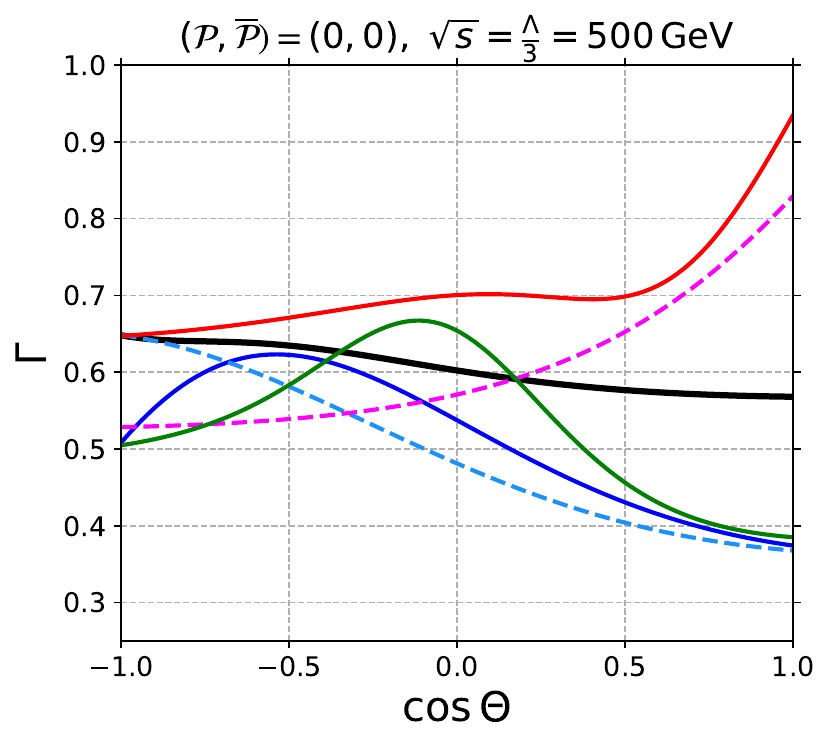}
\includegraphics[scale=0.20]{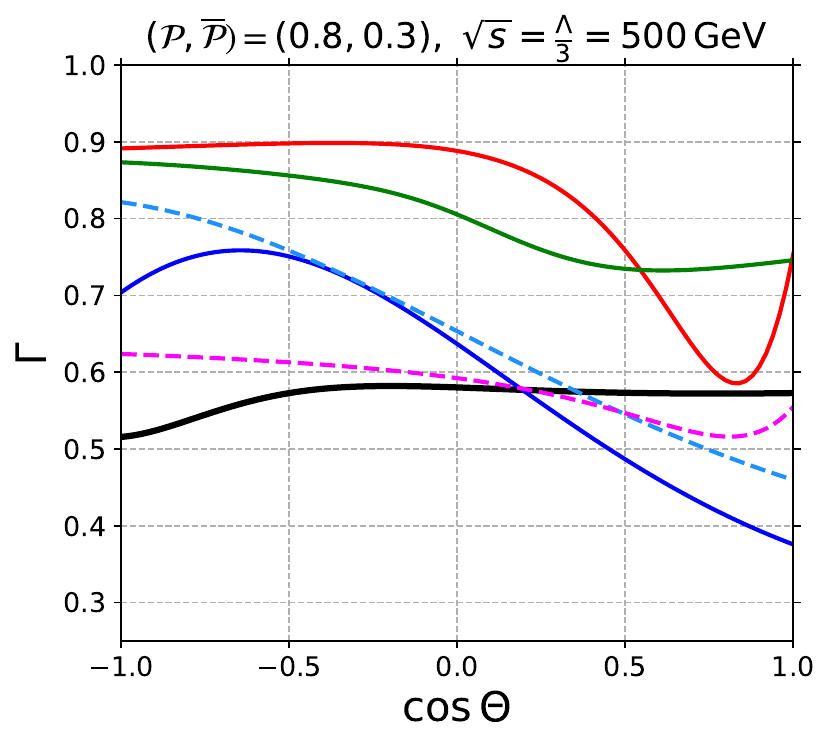}
\includegraphics[scale=0.20]{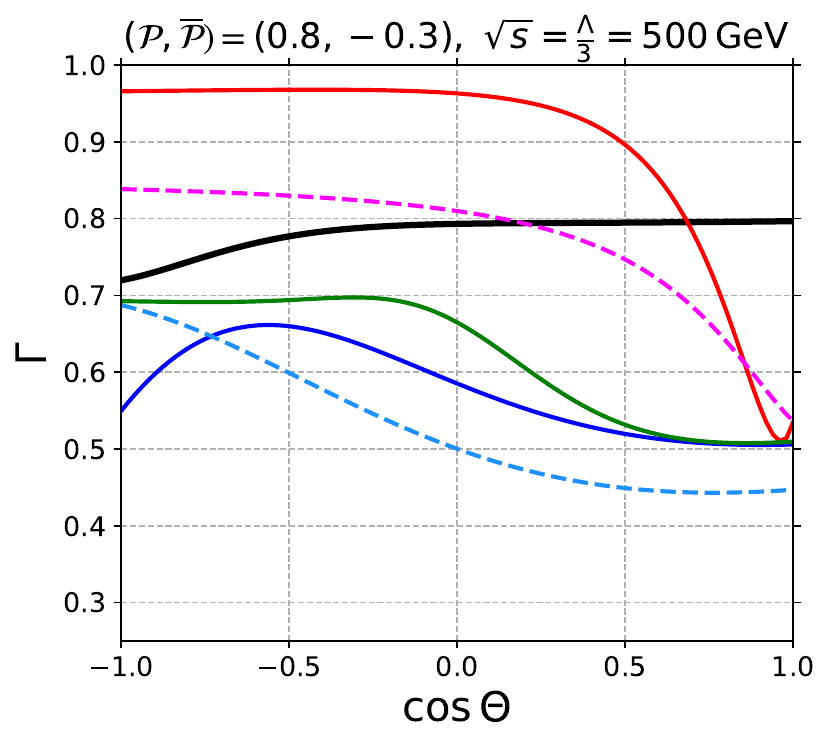}
\includegraphics[scale=0.20]{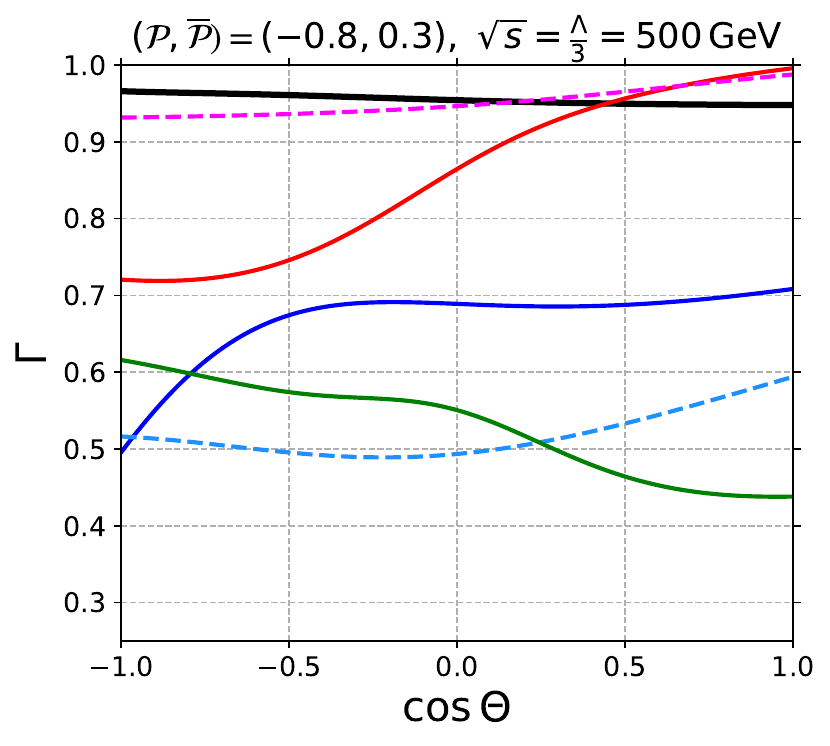}
\includegraphics[scale=0.20]{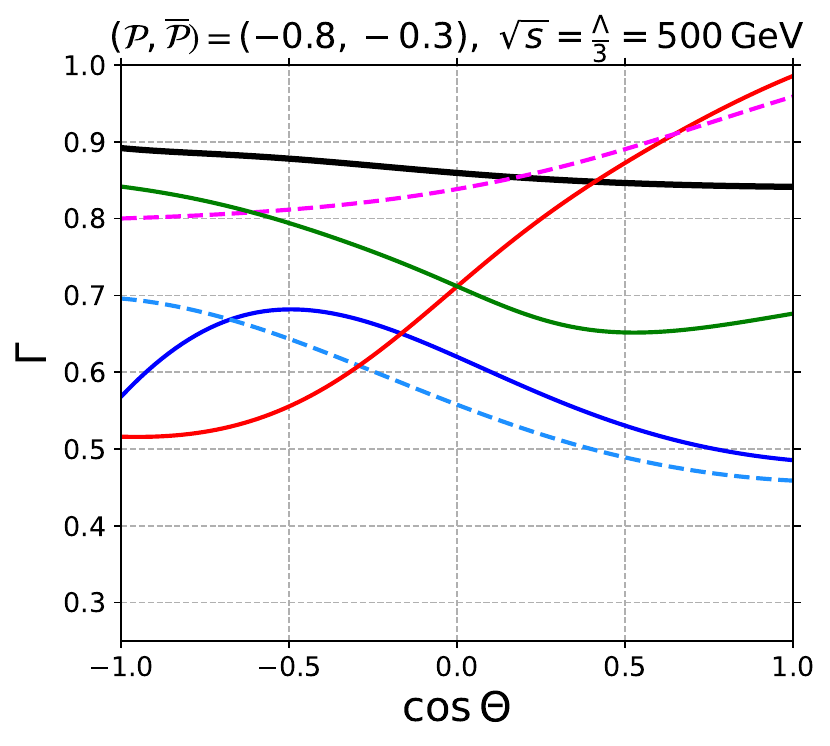}
\includegraphics[scale=0.20]{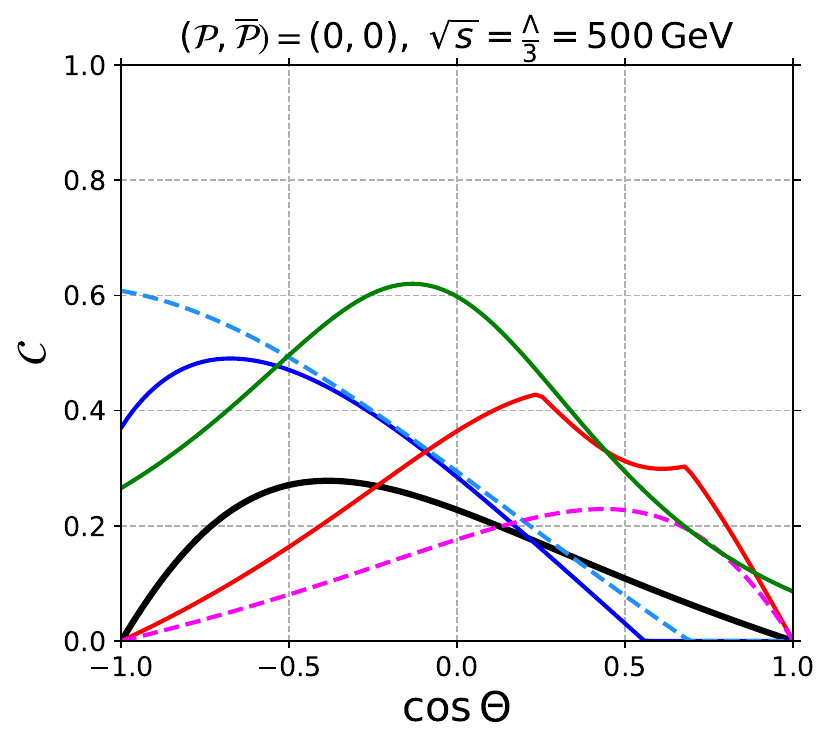}
\includegraphics[scale=0.20]{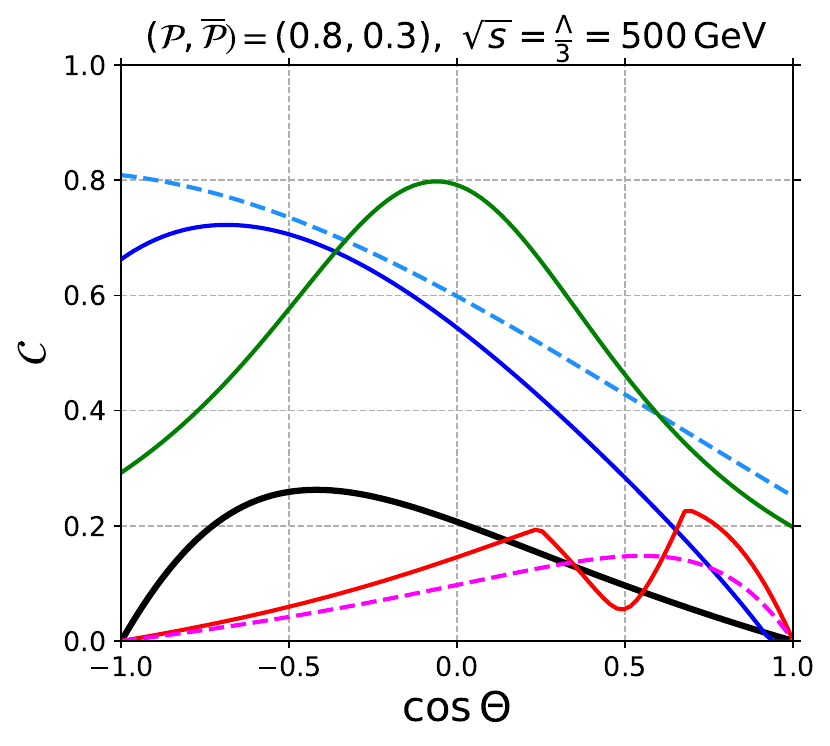}
\includegraphics[scale=0.20]{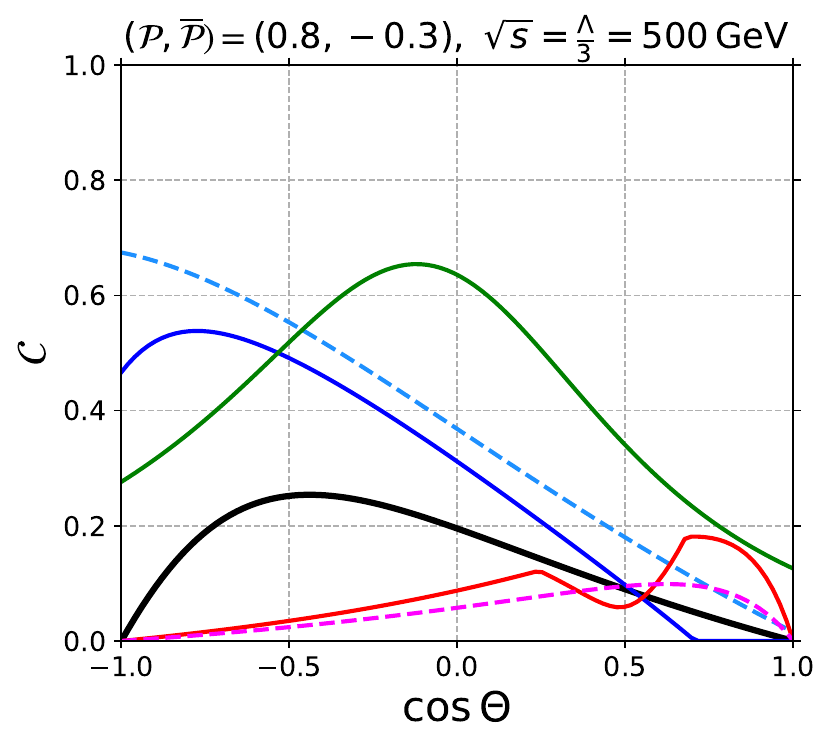}
\includegraphics[scale=0.20]{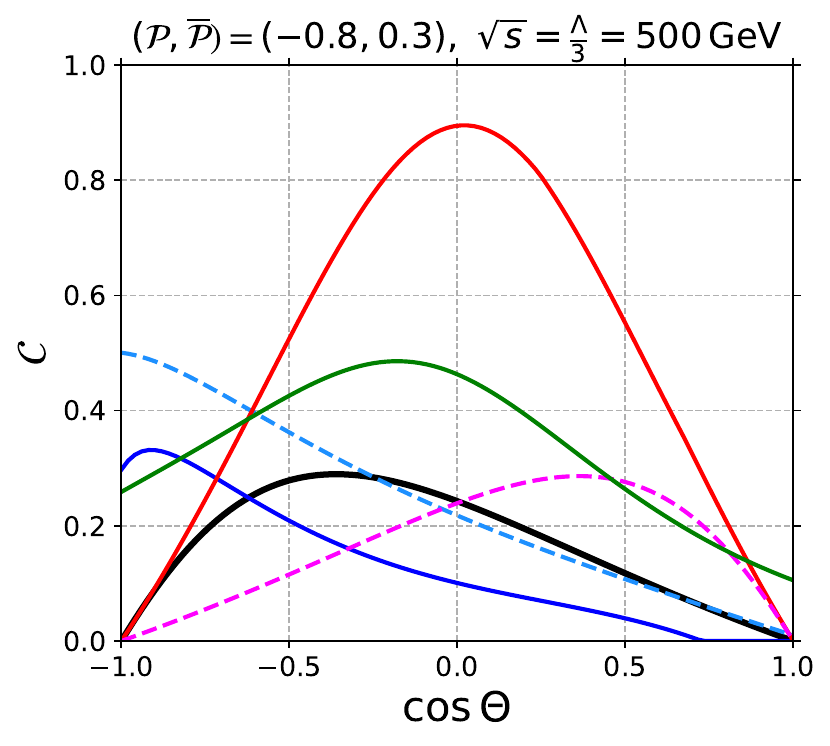}
\includegraphics[scale=0.20]{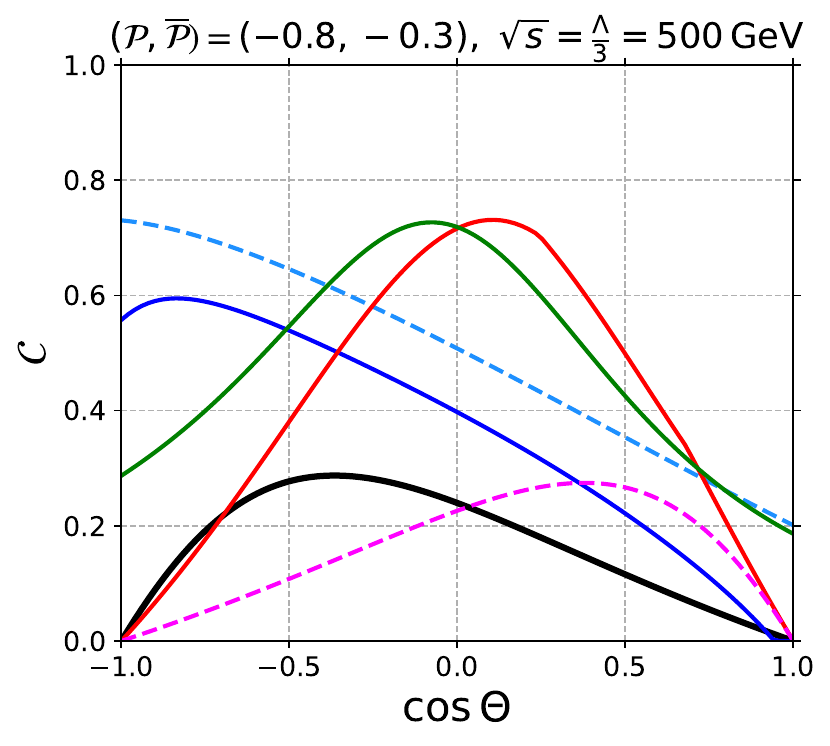}
\includegraphics[scale=0.20]{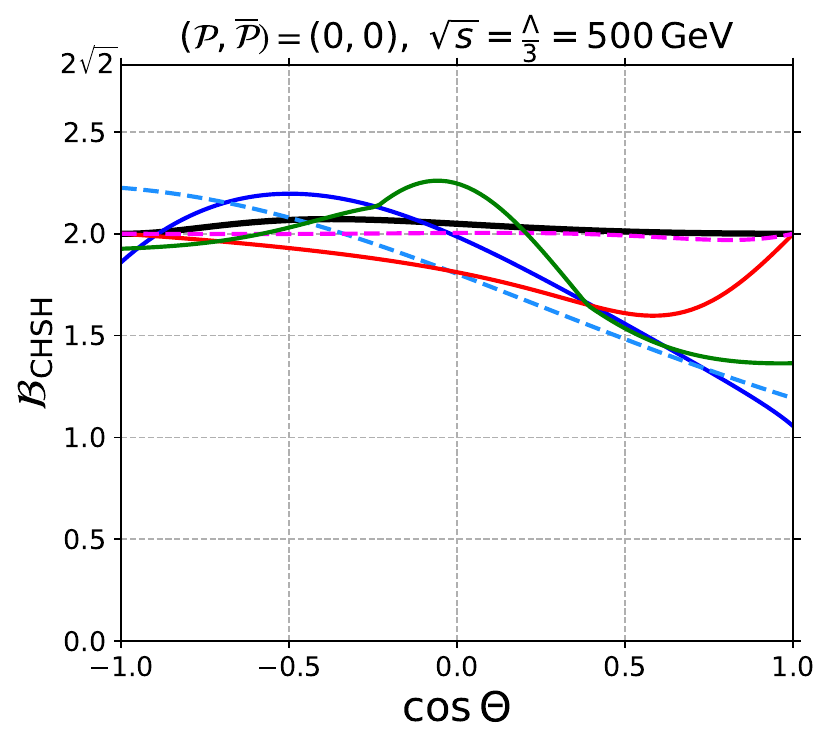}
\includegraphics[scale=0.20]{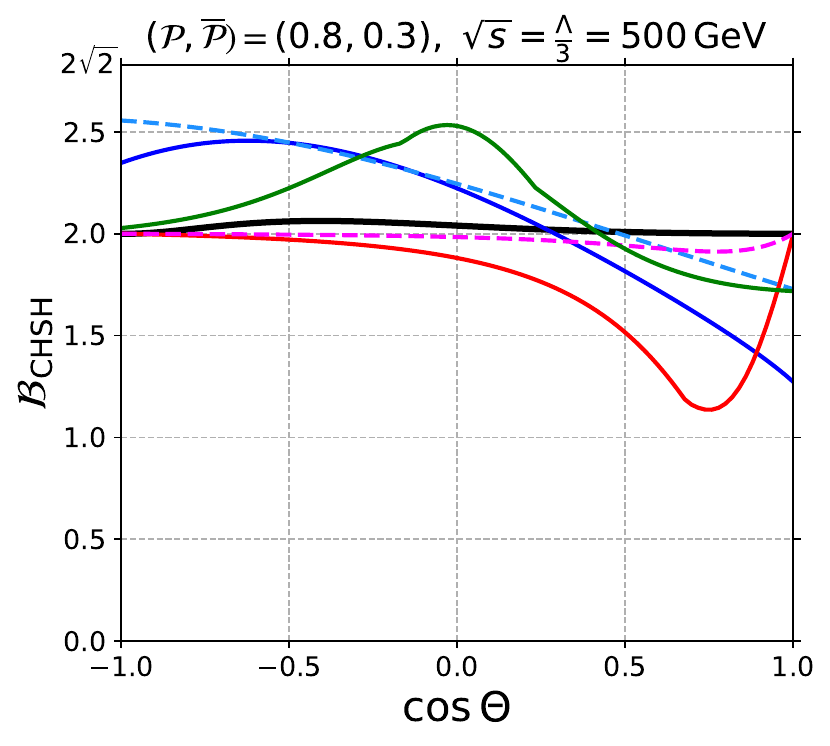}
\includegraphics[scale=0.20]{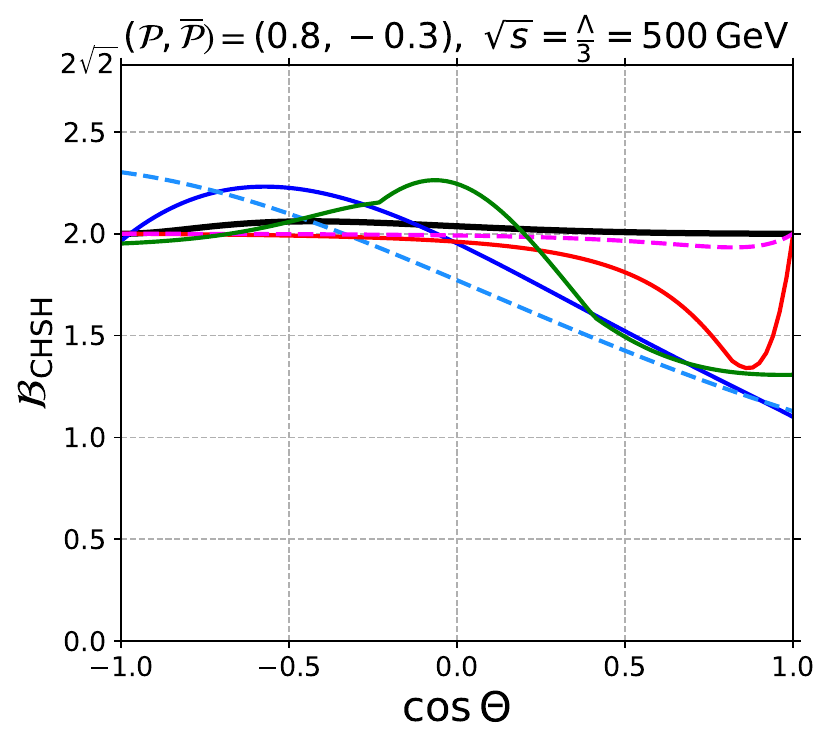}
\includegraphics[scale=0.20]{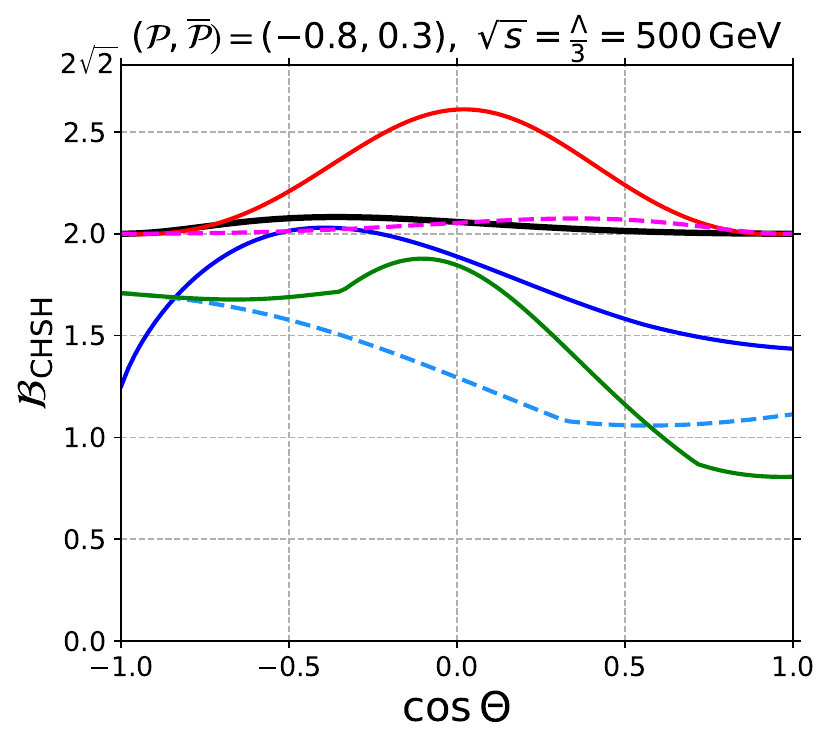}
\includegraphics[scale=0.20]{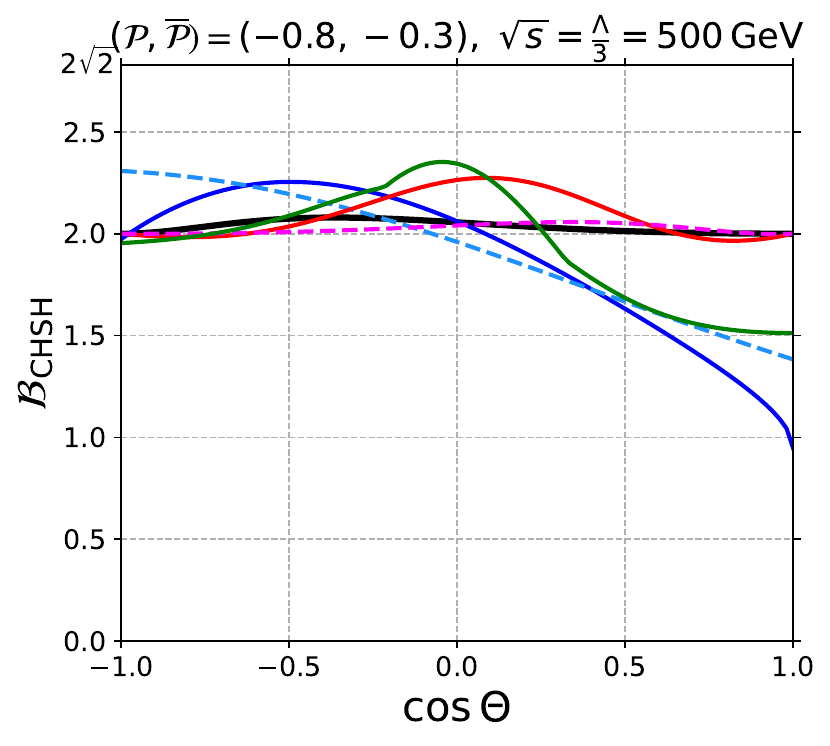}
\includegraphics[scale=0.20]{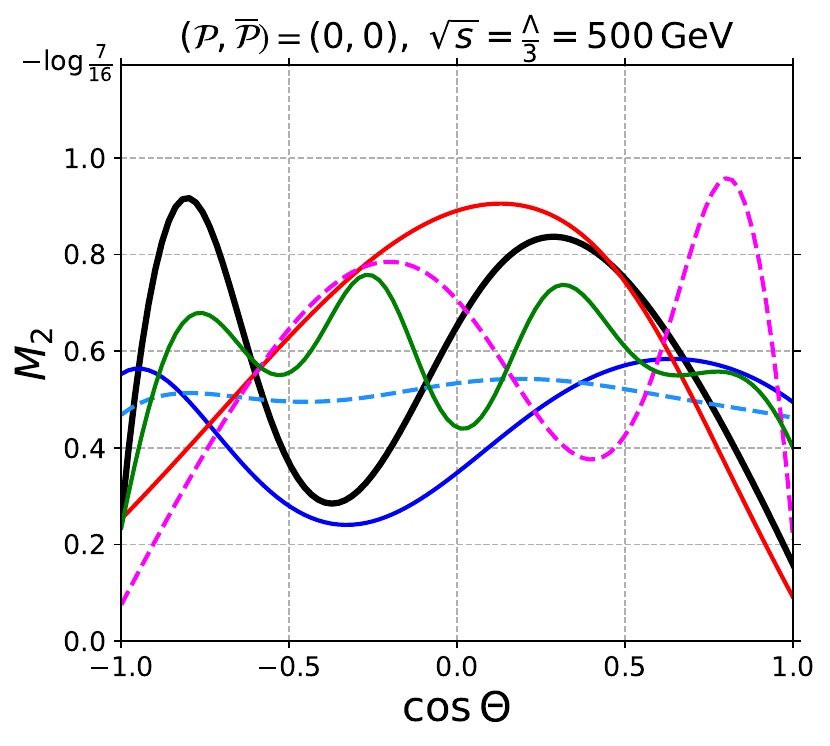}
\includegraphics[scale=0.20]{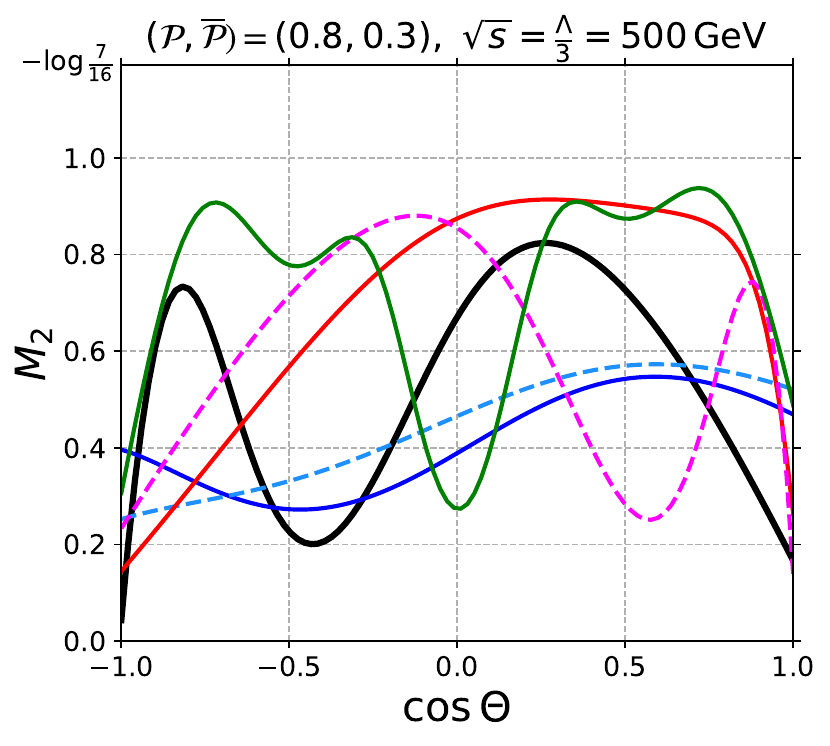}
\includegraphics[scale=0.20]{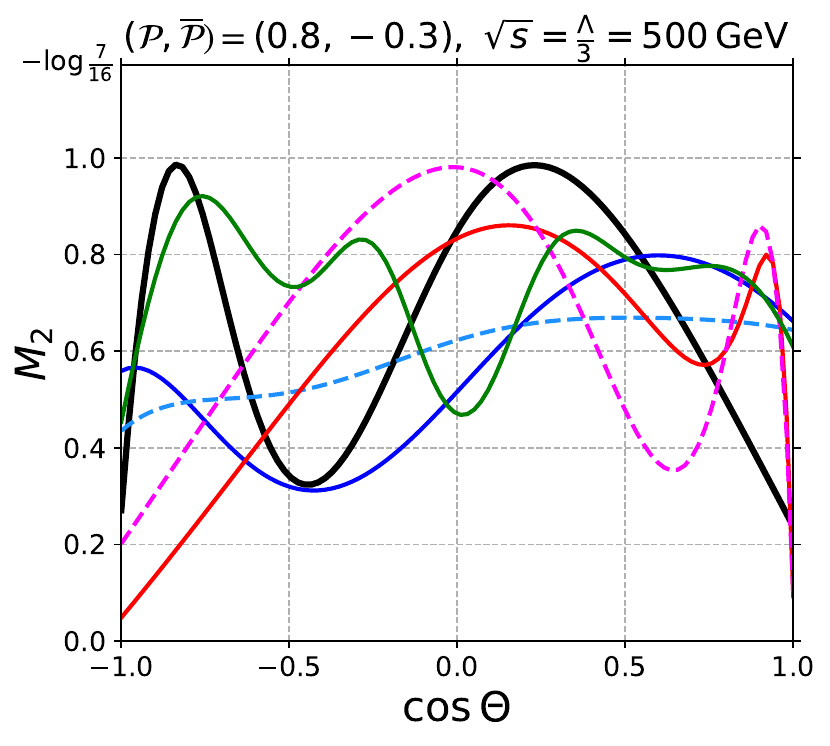}
\includegraphics[scale=0.20]{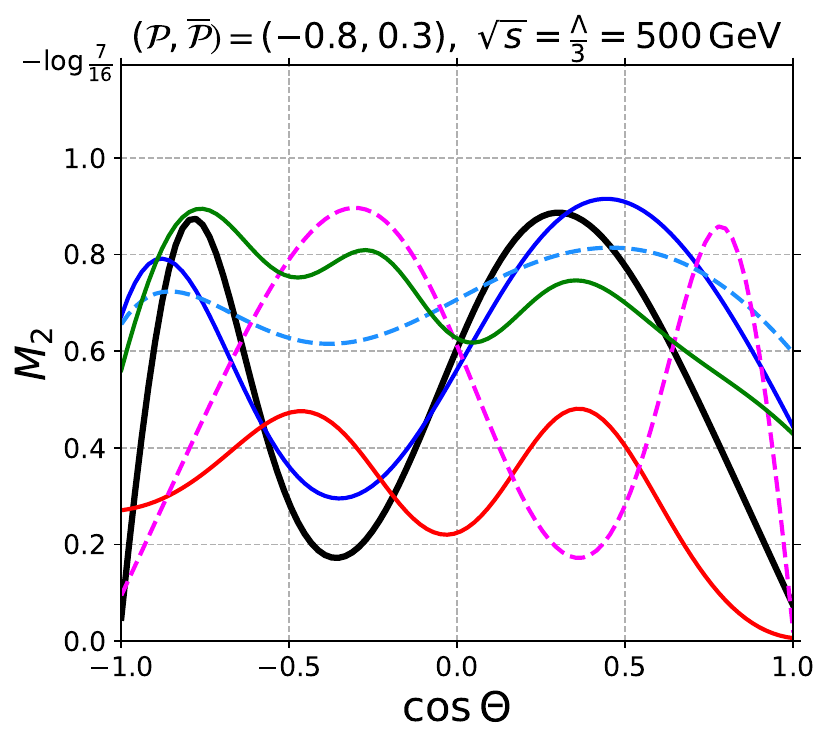}
\includegraphics[scale=0.20]{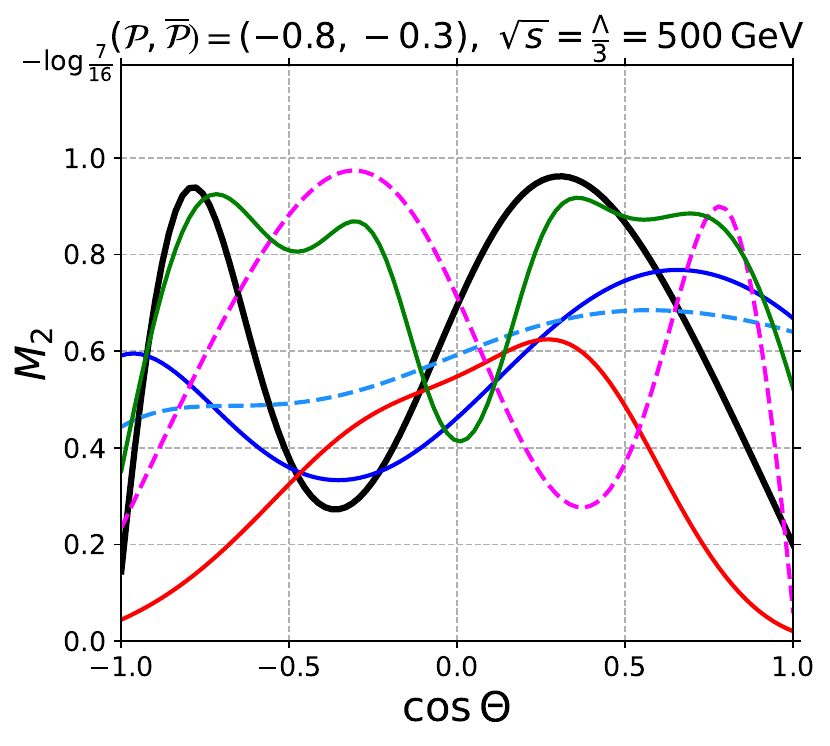}
\includegraphics[scale=0.20]{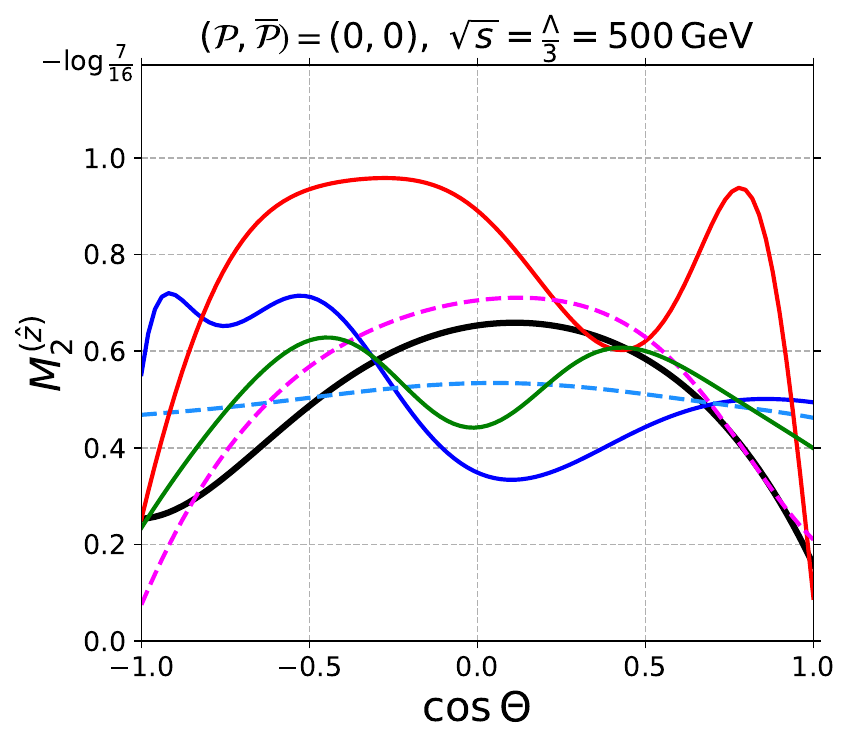}
\includegraphics[scale=0.20]{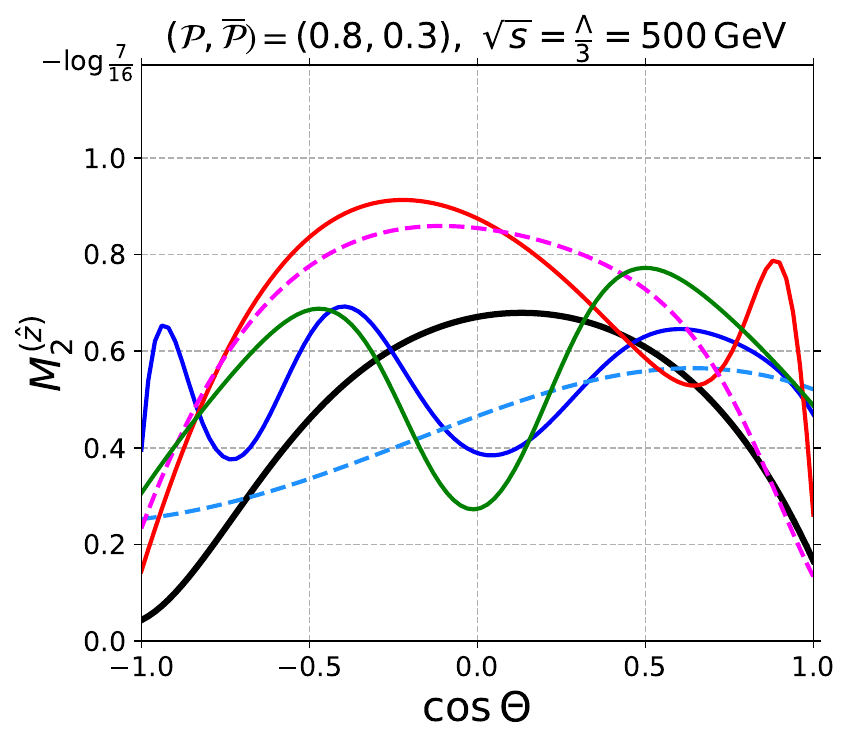}
\includegraphics[scale=0.20]{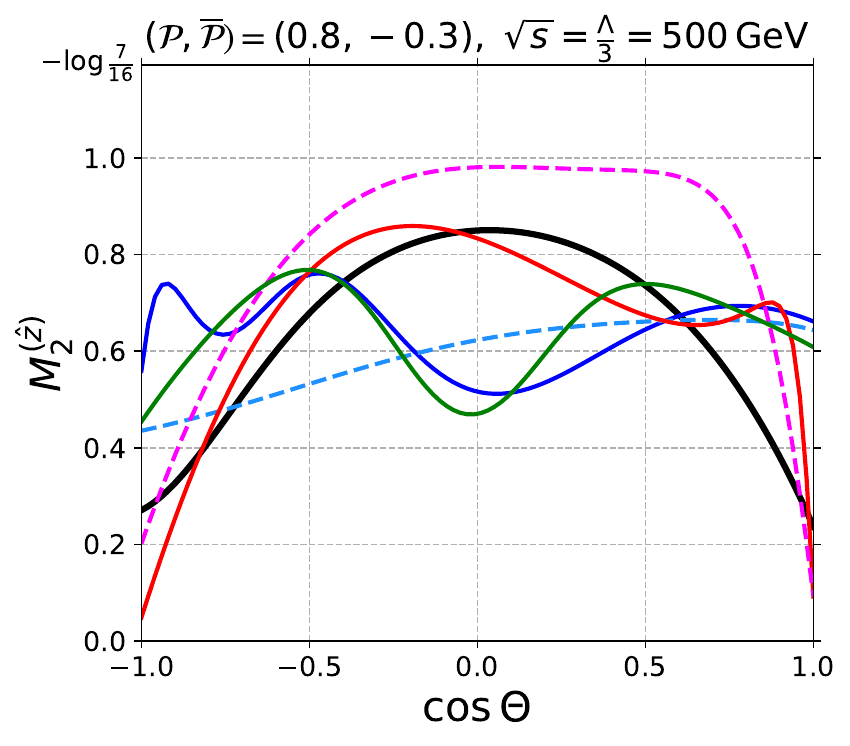}
\includegraphics[scale=0.20]{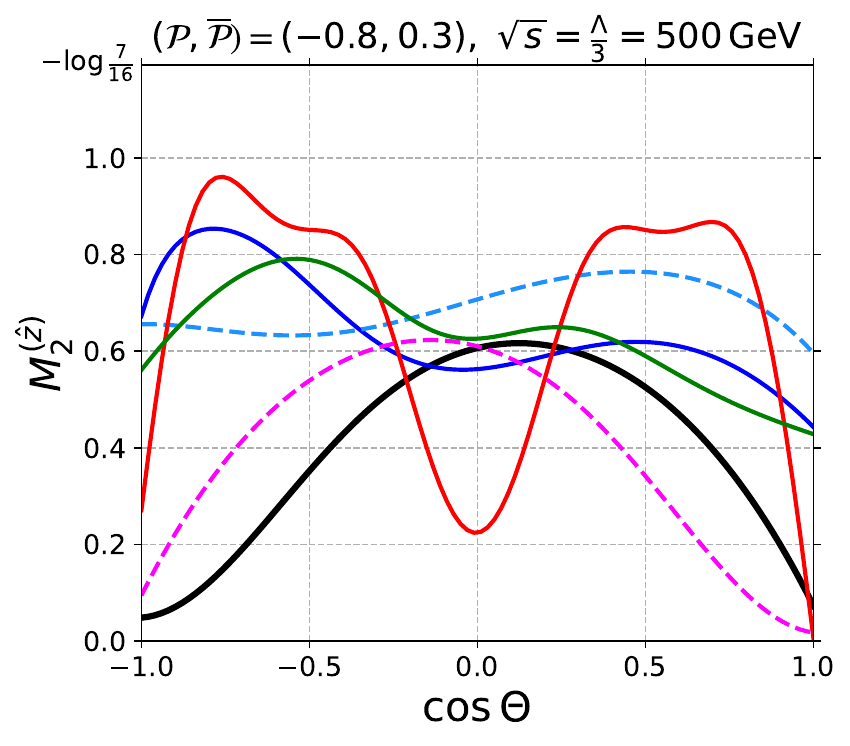}
\includegraphics[scale=0.20]{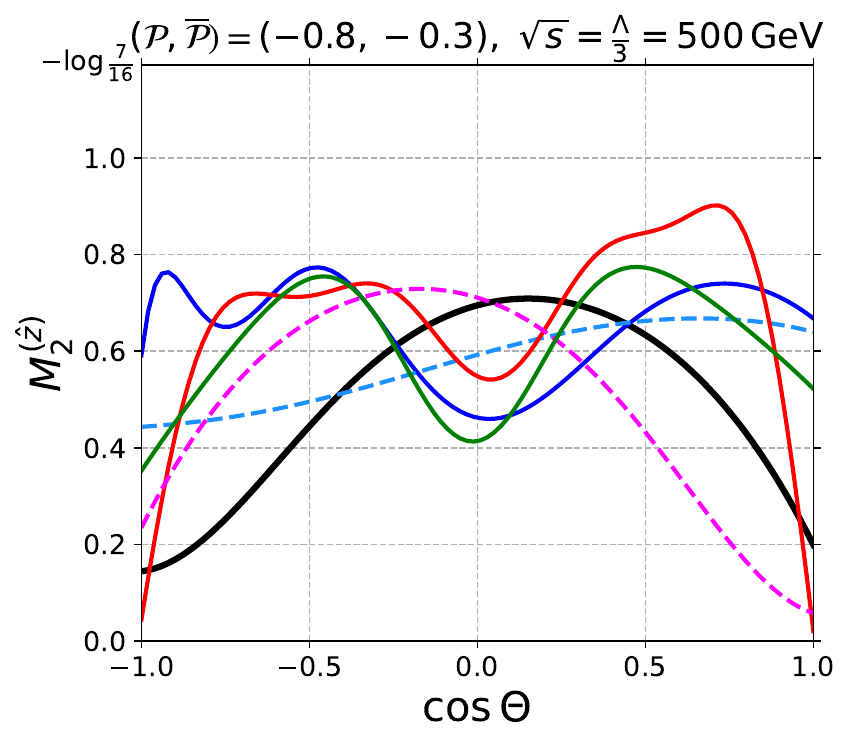}
\caption{\label{fig:1d_500}
\small 
Purity $\Gamma$ (top row), concurrence ${\cal C}$ (second row), Bell-CHSH observable ${\cal B}_{\rm CHSH}$ (third row), helicity-basis SRE $M_2$ (fourth row) and beam-basis SRE $M_2^{(\hat z)}$ (fifth row) as functions of $\cos \Theta$.
The centre-of-mass energy and the EFT cut-off scale are fixed as
$\sqrt{s} = \Lambda/3 = 500$ GeV.
Six benchmark scenarios are shown: 
the SM (solid black), 
scalar (S) (solid blue), 
scalar (P) (dashed light blue), 
vector (V,V) (solid red),
vector (A,A) (dashed magenta)
and tensor (solid green).
The leftmost column shows the unpolarised configuration; from left to right, the remaining columns display the polarised settings PP, PN, NP and NN, respectively.
}
\end{figure}

Fig.\ \ref{fig:1d_500} shows, as functions of $\cos \Theta$, 
the purity $\Gamma$ (first row), 
the concurrence ${\cal C}$ (second row), 
the Bell-CHSH observable ${\cal B}_{\rm CHSH}$ (third row), 
the helicity-basis $M_2$ (fourth row) and
the beam-basis $M_2^{(\hat z)}$ (fifth row).
The leftmost column corresponds to the unpolarised configuration. 
From left to right, the remaining columns show the polarised settings PP, PN, NP and NN, respectively.
We examine six benchmark scenarios: 
the SM (solid black), 
scalar (S) (solid blue), 
scalar (P) (dashed light blue), 
vector (V,V) (solid red),
vector (A,A) (dashed magenta)
and tensor (solid green).
The collision energy is set to $\sqrt{s} = 500$ GeV; the corresponding $\sqrt{s} = 1$ TeV results are presented in Fig.\ \ref{fig:1d_1000} (Appendix \ref{app:eft_1tev_plots}).

In the first row of Fig.\ \ref{fig:1d_500}, the SM purity is nearly flat in $\cos\Theta$. 
For the unpolarised and polarised-PP settings one finds relatively low values, $\Gamma\simeq 0.5$--$0.65$, whereas the NP and NN configurations yield high purities, $\Gamma\simeq 0.85$--$0.95$. 
By contrast, the EFT scenarios exhibit a pronounced $\cos\Theta$ dependence. 
For example, in the unpolarised and NN configurations the vector benchmarks (V,V) and (A,A) show an increase of $\Gamma$ toward the forward region, $\cos\Theta\gtrsim 0.5$, while the scalar (S) and (P) and the tensor cases predict smaller $\Gamma$ there. 
The NP configuration is particularly discriminating for separating the scalar and tensor scenarios from the SM and the vector ones: thanks to the high purity of the SM and vectors in this setting, the purity difference can approach $\Delta\Gamma\!\sim\!0.5$ over certain ranges of $\cos\Theta$.

In the second row of Fig.\ \ref{fig:1d_500}, the SM concurrence ${\cal C}$ vanishes in the forward/backward limits $\cos\Theta\to\pm1$ and attains a maximum near $\cos\Theta\simeq-0.5$ for all polarisation settings. 
The vector benchmarks (V,V) and (A,A) follow the same qualitative pattern, but the peak height and location vary markedly with the beam configuration. 
Whereas the SM peak lies in the backward region (negative $\cos\Theta$), the vector benchmarks typically peak in the forward region (positive $\cos\Theta$); notable exceptions are the (A,A) case with NP and NN beams, where the maxima occur near central production, $\cos\Theta\simeq0$, and are strongly enhanced; the concurrence there is larger than the SM by a factor of $\sim3$--$4$.

For the scalar benchmarks (S) and (P), ${\cal C}$ is generally large in the backward region and small in the forward region; for some cases vanishing exactly in the extreme forward region. 
The SM--EFT separation is therefore maximised for scalars at negative $\cos\Theta$. 
The tensor case, by contrast, yields moderately large concurrence across all beam settings, with a broad maximum near central production; values often exceed $0.6$ around $\cos\Theta\simeq 0$.

The third row of Fig.\ \ref{fig:1d_500} shows the Bell-CHSH observable ${\cal B}_{\rm CHSH}$. 
Across all beam settings, the SM and the vector (A,A) benchmark remain close to the local–hidden–variable bound, ${\cal B}_{\rm CHSH}\simeq 2$, over the full $\Theta$ range. 
The vector (V,V) case also yields ${\cal B}_{\rm CHSH}\simeq 2$ at the forward/backward limits $\cos\Theta=\pm1$, but develops a dip (CHSH non-violation) in the forward region for the unpolarised, PP, and PN configurations, while exhibiting a peak (CHSH violation) around $\cos \Theta \simeq 0$ for the NP and NN beam settings.
By contrast, the tensor and scalar benchmarks (S) and (P) yield systematically smaller ${\cal B}_{\rm CHSH}$, with clear non-violation in the forward region. 
Conversely, for these scenarios a pronounced CHSH violation is predicted in the backward region under the PP beam configuration.

The fourth and fifth rows of Fig.\ \ref{fig:1d_500} display $M_2$ and $M_2^{(\hat z)}$, respectively.  
For all polarisation settings, the SM $M_2$ and $M_2^{(\hat z)}$ are both suppressed near the forward/backward limits $\cos\Theta\simeq \pm1$.
In between, $M_2^{(\hat z)}$ exhibits a broad peak, while $M_2$ oscillates in such a way that it develops two peaks around $\cos \Theta \simeq -0.25$ and $0.3$.

All EFT benchmarks exhibit distinctive and often intricate $\cos\Theta$ dependences in each beam polarisation setting. 
In particular, $M_2$ provides strong discriminating power for the vector (A,A) case against the SM: its peaks typically coincide with SM dips, and vice versa. 
Conversely, for the NP configuration, $M_2^{(\hat z)}$ discriminates well between the SM and all EFT benchmarks other than (A,A).

Before concluding, we comment on the validity of the EFT expansion. As discussed above, our analysis includes both the SM--EFT interference and the squared EFT contributions. One may therefore question whether this choice is legitimate, and what would change if one were to strictly truncate the expansion at $\mathcal{O}(1/\Lambda^{2})$. A detailed discussion is presented in Appendix~E; here we summarise the main arguments and results.

A physically admissible spin density matrix must be Hermitian, properly normalised, and positive semi-definite, with non-negative eigenvalues admitting a probabilistic interpretation. When truncated at $\mathcal{O}(1/\Lambda^{2})$, these requirements impose non-trivial consistency conditions on the size of EFT corrections, which we find to be parametrically stronger than standard criteria based on cross-section positivity or on the dominance of linear over quadratic terms \cite{Liu:2016idz}.
In particular, density-matrix positivity enforces inequalities analogous to those of a mixed quantum state, bounding off-diagonal ``coherence'' terms by the geometric mean of the corresponding diagonal populations. As a result, the allowed size of $\mathcal{O}(1/\Lambda^{2})$ effects is more tightly constrained than suggested by rate-based arguments alone, providing a well-motivated general criterion for assessing EFT validity.

Overall, the study shows that each quantum observable exhibits a distinctive and often intricate dependence on the production angle $\Theta$ and on the beam polarisation settings. 
This demonstrates that quantum measurements at a polarised lepton collider offer enhanced discovery potential and a constrained framework in which complementary observables jointly probe, and cross-check, both the presence of new physics and the structure of the underlying interaction.

\section{Conclusions} 
\label{sec:concl}

We have presented a comprehensive study of the quantum properties of $F \bar F$ pairs produced in lepton collisions with polarised beams, focusing on quantum information observables: purity $\Gamma$, concurrence $\mathcal C$, the Bell-CHSH observable $\mathcal B_{\rm CHSH}$ and the stabiliser Rényi entropies (SREs) in both the helicity and beam bases, $M_2$ and $M_2^{(\hat z)}$, respectively.  

Section \ref{sec:EFT} presented such a study within a general QFT framework by analysing the spin state of a heavy fermion pair $F \bar F$ produced by the collision of polarised light fermions $l$ and $\bar l$.
In this analysis, the interaction is described by a single four-fermion operator of scalar, vector or tensor type.
In the scalar interaction, we find the $F \bar F$ spin state is characterised by the maximally entangled pure state (see Eq.\ \eqref{scalar_state}), which depends only on the effective CP angle $\delta_S$, defined in Eq.\ \eqref{dS}. 
Consequently, $\Gamma = 1$, ${\cal C} = 1$ and ${\cal B}_{\rm CHSH} = 2 \sqrt{2}$, independent of any kinetic, experimental and theoretical parameters. 
We find rather simple closed-form expressions for $M_2$ (as a function of $\d_S$) and $M_2^{(\hat z)}$ (as a function of $\d_S$ and $\Theta$) (see Eqs. \eqref{M2_scalar} and \eqref{M2z_scalar}).

In the vector interaction, the effect of beam polarisations is described by a single parameter
$\PhiV$, defined in Eq.\ \eqref{PhiV}.
It reveals that the beam polarisation has no effect as long as ${\cal P} = \overline {\cal P}$, independent of their value.  
The analytical expressions for the quantum observables were obtained for the vector-like coupling ($\sin \eta_A = 0$) (see Eqs. \eqref{pure_v}--\eqref{magic_v}).
The concurrence is a monotonic function of $\beta \sin \Theta$, interpolating between the vanishing entanglement at $\beta \sin \Theta = 0$ and the maximal entanglement at $\beta \sin \Theta = 1$.
Furthermore, the Bell-CHSH observable is one-to-one related to the concurrence as ${\cal B}_{\rm CHSH} = 2 \sqrt{ 1 + {\cal C}^2 }$.
The stabiliser R{\' e}nyi entropies $M_2$ and $M_2^{(\hat z)}$ show peculiar patterns in the $( \sin^2 \PhiV, \cos \Theta )$ plane as illustrated in Fig.\ \ref{fig:m2_v}.
The quantum observables for the axial-vector-like coupling ($\cos \eta_A = 0$) can be obtained by taking $\b \to 1$ limit in the corresponding formulae for the vector-like case, except for $M_2^{(\hat z)}$, whose formula was separately derived in Eq.\ \eqref{magic_a}.   

In the tensor interaction, the beam polarisation effect is effectively described by the angle $\PhiT$ defined in Eq.\ \eqref{PhiT}.
Contrary to the vector interaction, physics becomes insensitive to the  polarisation  for ${\cal P} = - \overline {\cal P}$, independently of their values. 
We derived the analytical formulae for the quantum observables for the tensor interaction (see Eqs.\ \eqref{pure_t}--\eqref{M2_t}).  
Remarkably, the concurrence and the Bell-CHSH observable are independent of the beam polarisations and one-to-one related to each other in the same way as in the vector interaction case with $\sin \eta_A = 0$ or 1.
The SREs $M_2$ and $M_2^{(\hat z)}$ for the tensor interaction exhibit peculiar patterns in the $(\sin^2 \PhiT, \cos \Theta)$ plane (see Fig.\ \ref{fig:M2tens}). 
Both SREs are strongly suppressed at $\cos \Theta \simeq 0$ for finite velocities.  

In section \ref{sec:sm+eft}, we investigated how the quantum observables and beam polarisations help us probe new physics beyond the SM in the $e^+ e^- \to F \bar F$ process. 
This was done by comparing the SM predictions with those obtained by various new physics benchmarks constructed by augmenting the SM Lagrangian with a single four-fermion operator with specified couplings (see Table \ref{tab:scenarios}). 

Figs.\ \ref{fig:pure_500gev}--\ref{fig:M2_500gev} compared quantum observables in the $({\cal P}, \overline {\cal P})$ plane and showed that 
they respond differently to new physics scenarios as well as to beam polarisations.  
In particular, the purity in the scalar and tensor scenarios is significantly lower than that in the SM and vector scenarios. 
While the SM concurrence and Bell-CHSH observable are almost constant across the polarisation plane, the EFT benchmarks predict these observables sensitive to polarisations and exhibit their own characteristic patterns. 
The stabiliser Rényi entropies exhibit distinctive patterns across all benchmarks, enabling discrimination between the SM and new physics as well as among different new physics scenarios.

This study demonstrates that beam polarisation not only enhances sensitivity to new physics, but also enables the isolation of specific quantum states, including maximally and minimally entangled states (see Fig.~\ref{fig:conc_500gev}), Bell-inequality violation (see Fig.~\ref{fig:chsh_500gev}) and stabiliser or high-magic states (see Fig.~\ref{fig:M2_500gev}). In Fig.~\ref{fig:1d_500}, we compared the $\cos\Theta$ distributions of the quantum observables for unpolarised and polarised beam configurations. 
These results reaffirm the power of quantum observables \emph{and} beam polarisation for new physics searches and for discriminating among competing scenarios. 
Each benchmark and polarisation choice yields a distinctive pattern, providing an overconstrained framework in which complementary observables jointly probe (and cross-check) potential new physics effects.

In summary, two main messages emerge. First, by controlling beam polarisations one can engineer the initial two-qubit state so as to selectively enhance or suppress to zero entanglement, Bell-inequality violation, and magic in the final $F\bar F$ system, thereby exposing operator structures that would remain hidden with unpolarised beams. Second, the combined use of $(\Gamma,\mathcal C,\mathcal B_{\rm CHSH},M_2,M_2^{(\hat z)})$, measured differentially in $\cos\Theta$ and across polarisation runs, provides a robust and overconstrained probe of new-physics scenarios, with distinct benchmarks yielding orthogonal signatures in this space. Together, these results strengthen the case for polarised lepton colliders as precision ``quantum laboratories'' for the top quark and as powerful instruments for indirect new-physics searches.

Finally, as a by-product of this study, we find that enforcing the EFT-truncated spin density matrix at $\mathcal{O}(1/\Lambda^{2})$ to define a valid quantum state leads to constraints that are parametrically stronger than those inferred from rate-level considerations, highlighting the importance of analysing EFT effects directly at the level of the spin density matrix.

Looking ahead, it will be natural to incorporate full SM predictions including NLO QCD and electroweak corrections as well as top decays, assess the detector-level feasibility of quantum-state tomography in $\cos\Theta$ bins, and extend the analysis to global EFT fits. We anticipate that the strategies outlined here will straightforwardly generalise to other fermion final states, such as $\tau^+\tau^-$, and motivate dedicated investigations of $VV(V')$ and $t\bar t+X$ ($X=h,W,Z$) final states.  In this spirit, it will also be interesting to further investigate whether quantum observables can more generally provide intrinsic validity conditions for EFTs that are stronger than those obtained from conventional rate-based criteria.

\section*{Acknowledgements}

KS thanks Daniel Jeans for helpful discussions.
FM acknowledges support by the FRS-FNRS (Belgian National Scientific Research Fund) IISN
projects 4.4503.16 (MaxLHC).  FM and PL are partially supported by the European Union (EU) COMETA COST Action (CA22130) and the Italian Ministry of University and Research (MUR) through the PRIN2022 grant Nr.~2022RXEZCJ,
and by the project ``QIHEP--Exploring the foundations of quantum information
in particle physics'', which is financed through the PNRR with NextGenerationEU funds, in the context of the extended
partnership PE00000023 NQSTI (CUP J33C24001210007). MMA was partially supported by the French Agence Nationale de la Recherche (ANR) under grant ANR-21-CE31-0023.

\bibliographystyle{JHEP}
\bibliography{ref}

\providecommand{\href}[2]{#2}\begingroup\raggedright\begin{thebibliography}{10}

\bibitem{Afik:2025ejh}
Y.~Afik et~al., \emph{{Quantum information meets high-energy physics: input to the update of the European strategy for particle physics}}, \href{http://dx.doi.org/10.1140/epjp/s13360-025-06752-9}{\emph{Eur. Phys. J. Plus} {\bf 140} (2025) 855}, [\href{http://arxiv.org/abs/2504.00086}{{\tt 2504.00086}}].

\bibitem{Durupt:2025wuk}
V.~Durupt, F.~Maltoni and O.~Mattelaer, \emph{{Automated computation of spin-density matrices and quantum observables for collider physics}},  \href{http://arxiv.org/abs/2510.17730}{{\tt 2510.17730}}.

\bibitem{2003.02280}
Y.~Afik and J.~R. M.~n. de~Nova, \emph{{Entanglement and quantum tomography with top quarks at the LHC}}, \href{http://dx.doi.org/10.1140/epjp/s13360-021-01902-1}{\emph{Eur. Phys. J. Plus} {\bf 136} (2021) 907}, [\href{http://arxiv.org/abs/2003.02280}{{\tt 2003.02280}}].

\bibitem{Fabbrichesi:2021npl}
M.~Fabbrichesi, R.~Floreanini and G.~Panizzo, \emph{{Testing Bell Inequalities at the LHC with Top-Quark Pairs}}, \href{http://dx.doi.org/10.1103/PhysRevLett.127.161801}{\emph{Phys. Rev. Lett.} {\bf 127} (2021) 161801}, [\href{http://arxiv.org/abs/2102.11883}{{\tt 2102.11883}}].

\bibitem{Severi:2021cnj}
C.~Severi, C.~D.~E. Boschi, F.~Maltoni and M.~Sioli, \emph{{Quantum tops at the LHC: from entanglement to Bell inequalities}}, \href{http://dx.doi.org/10.1140/epjc/s10052-022-10245-9}{\emph{Eur. Phys. J. C} {\bf 82} (2022) 285}, [\href{http://arxiv.org/abs/2110.10112}{{\tt 2110.10112}}].

\bibitem{Severi:2022qjy}
C.~Severi and E.~Vryonidou, \emph{{Quantum entanglement and top spin correlations in SMEFT at higher orders}}, \href{http://dx.doi.org/10.1007/JHEP01(2023)148}{\emph{JHEP} {\bf 01} (2023) 148}, [\href{http://arxiv.org/abs/2210.09330}{{\tt 2210.09330}}].

\bibitem{Aoude:2022imd}
R.~Aoude, E.~Madge, F.~Maltoni and L.~Mantani, \emph{{Quantum SMEFT tomography: Top quark pair production at the LHC}}, \href{http://dx.doi.org/10.1103/PhysRevD.106.055007}{\emph{Phys. Rev. D} {\bf 106} (2022) 055007}, [\href{http://arxiv.org/abs/2203.05619}{{\tt 2203.05619}}].

\bibitem{Afik:2022kwm}
Y.~Afik and J.~R. M.~n. de~Nova, \emph{{Quantum information with top quarks in QCD}}, \href{http://dx.doi.org/10.22331/q-2022-09-29-820}{\emph{Quantum} {\bf 6} (2022) 820}, [\href{http://arxiv.org/abs/2203.05582}{{\tt 2203.05582}}].

\bibitem{Aguilar-Saavedra:2022uye}
J.~A. Aguilar-Saavedra and J.~A. Casas, \emph{{Improved tests of entanglement and Bell inequalities with LHC tops}}, \href{http://dx.doi.org/10.1140/epjc/s10052-022-10630-4}{\emph{Eur. Phys. J. C} {\bf 82} (2022) 666}, [\href{http://arxiv.org/abs/2205.00542}{{\tt 2205.00542}}].

\bibitem{Afik:2022dgh}
Y.~Afik and J.~R. M.~n. de~Nova, \emph{{Quantum Discord and Steering in Top Quarks at the LHC}}, \href{http://dx.doi.org/10.1103/PhysRevLett.130.221801}{\emph{Phys. Rev. Lett.} {\bf 130} (2023) 221801}, [\href{http://arxiv.org/abs/2209.03969}{{\tt 2209.03969}}].

\bibitem{Cheng:2023qmz}
K.~Cheng, T.~Han and M.~Low, \emph{{Optimizing fictitious states for Bell inequality violation in bipartite qubit systems with applications to the tt\textasciimacron{} system}}, \href{http://dx.doi.org/10.1103/PhysRevD.109.116005}{\emph{Phys. Rev. D} {\bf 109} (2024) 116005}, [\href{http://arxiv.org/abs/2311.09166}{{\tt 2311.09166}}].

\bibitem{Han:2023fci}
T.~Han, M.~Low and T.~A. Wu, \emph{{Quantum entanglement and Bell inequality violation in semi-leptonic top decays}}, \href{http://dx.doi.org/10.1007/JHEP07(2024)192}{\emph{JHEP} {\bf 07} (2024) 192}, [\href{http://arxiv.org/abs/2310.17696}{{\tt 2310.17696}}].

\bibitem{Dong:2023xiw}
Z.~Dong, D.~Goncalves, K.~Kong and A.~Navarro, \emph{{Entanglement and Bell inequalities with boosted $t \bar t$}}, \href{http://dx.doi.org/10.1103/PhysRevD.109.115023}{\emph{Phys. Rev. D} {\bf 109} (2024) 115023}, [\href{http://arxiv.org/abs/2305.07075}{{\tt 2305.07075}}].

\bibitem{Cheng:2024btk}
K.~Cheng, T.~Han and M.~Low, \emph{{Optimizing Entanglement and Bell Inequality Violation in Top Anti-Top Events}},  \href{http://arxiv.org/abs/2407.01672}{{\tt 2407.01672}}.

\bibitem{Aguilar-Saavedra:2024hwd}
J.~A. Aguilar-Saavedra, \emph{{A closer look at post-decay $t \bar t$ entanglement}}, \href{http://dx.doi.org/10.1103/PhysRevD.109.096027}{\emph{Phys. Rev. D} {\bf 109} (2024) 096027}, [\href{http://arxiv.org/abs/2401.10988}{{\tt 2401.10988}}].

\bibitem{Aguilar-Saavedra:2023lwb}
J.~A. Aguilar-Saavedra, \emph{{Decay of entangled fermion pairs with post-selection}}, \href{http://dx.doi.org/10.1016/j.physletb.2023.138409}{\emph{Phys. Lett. B} {\bf 848} (2024) 138409}, [\href{http://arxiv.org/abs/2308.07412}{{\tt 2308.07412}}].

\bibitem{Aguilar-Saavedra:2024fig}
J.~A. Aguilar-Saavedra and J.~A. Casas, \emph{{Entanglement Autodistillation from Particle Decays}}, \href{http://dx.doi.org/10.1103/PhysRevLett.133.111801}{\emph{Phys. Rev. Lett.} {\bf 133} (2024) 111801}, [\href{http://arxiv.org/abs/2401.06854}{{\tt 2401.06854}}].

\bibitem{Cheng:2024rxi}
K.~Cheng, T.~Han and M.~Low, \emph{{Quantum Tomography at Colliders: With or Without Decays}},  \href{http://arxiv.org/abs/2410.08303}{{\tt 2410.08303}}.

\bibitem{Aguilar-Saavedra:2024vpd}
J.~A. Aguilar-Saavedra, \emph{{Full quantum tomography of top quark decays}}, \href{http://dx.doi.org/10.1016/j.physletb.2024.138849}{\emph{Phys. Lett. B} {\bf 855} (2024) 138849}, [\href{http://arxiv.org/abs/2402.14725}{{\tt 2402.14725}}].

\bibitem{Han:2024ugl}
T.~Han, M.~Low, N.~McGinnis and S.~Su, \emph{{Measuring quantum discord at the LHC}}, \href{http://dx.doi.org/10.1007/JHEP05(2025)081}{\emph{JHEP} {\bf 05} (2025) 081}, [\href{http://arxiv.org/abs/2412.21158}{{\tt 2412.21158}}].

\bibitem{Maltoni:2024tul}
F.~Maltoni, C.~Severi, S.~Tentori and E.~Vryonidou, \emph{{Quantum detection of new physics in top-quark pair production at the LHC}}, \href{http://dx.doi.org/10.1007/JHEP03(2024)099}{\emph{JHEP} {\bf 03} (2024) 099}, [\href{http://arxiv.org/abs/2401.08751}{{\tt 2401.08751}}].

\bibitem{Maltoni:2024csn}
F.~Maltoni, C.~Severi, S.~Tentori and E.~Vryonidou, \emph{{Quantum tops at circular lepton colliders}}, \href{http://dx.doi.org/10.1007/JHEP09(2024)001}{\emph{JHEP} {\bf 09} (2024) 001}, [\href{http://arxiv.org/abs/2404.08049}{{\tt 2404.08049}}].

\bibitem{Aoude:2025jzc}
R.~Aoude, H.~Banks, C.~D. White and M.~J. White, \emph{{Probing new physics in the top sector using quantum information}},  \href{http://arxiv.org/abs/2505.12522}{{\tt 2505.12522}}.

\bibitem{Fabbrichesi:2025psr}
M.~Fabbrichesi, R.~Floreanini and L.~Marzola, \emph{{Local vs. nonlocal entanglement in top-quark pairs at the LHC}},  \href{http://arxiv.org/abs/2505.02902}{{\tt 2505.02902}}.

\bibitem{ATLAS:2014aus}
{\scshape ATLAS} collaboration, G.~Aad et~al., \emph{{Measurements of spin correlation in top-antitop quark events from proton-proton collisions at $\sqrt{s}=7$ TeV using the ATLAS detector}}, \href{http://dx.doi.org/10.1103/PhysRevD.90.112016}{\emph{Phys. Rev. D} {\bf 90} (2014) 112016}, [\href{http://arxiv.org/abs/1407.4314}{{\tt 1407.4314}}].

\bibitem{CMS:2015cal}
{\scshape CMS} collaboration, V.~Khachatryan et~al., \emph{{Measurement of Spin Correlations in $t\bar{t}$ Production using the Matrix Element Method in the Muon+Jets Final State in $pp$ Collisions at $\sqrt{s} =$ 8 TeV}}, \href{http://dx.doi.org/10.1016/j.physletb.2016.05.005}{\emph{Phys. Lett. B} {\bf 758} (2016) 321--346}, [\href{http://arxiv.org/abs/1511.06170}{{\tt 1511.06170}}].

\bibitem{ATLAS:2016bac}
{\scshape ATLAS} collaboration, M.~Aaboud et~al., \emph{{Measurements of top quark spin observables in $ t\overline{t} $ events using dilepton final states in $ \sqrt{s}=8 $ TeV pp collisions with the ATLAS detector}}, \href{http://dx.doi.org/10.1007/JHEP03(2017)113}{\emph{JHEP} {\bf 03} (2017) 113}, [\href{http://arxiv.org/abs/1612.07004}{{\tt 1612.07004}}].

\bibitem{CMS:2016piu}
{\scshape CMS} collaboration, V.~Khachatryan et~al., \emph{{Measurements of t t-bar spin correlations and top quark polarization using dilepton final states in pp collisions at sqrt(s) = 8 TeV}}, \href{http://dx.doi.org/10.1103/PhysRevD.93.052007}{\emph{Phys. Rev. D} {\bf 93} (2016) 052007}, [\href{http://arxiv.org/abs/1601.01107}{{\tt 1601.01107}}].

\bibitem{CMS:2019nrx}
{\scshape CMS} collaboration, A.~M. Sirunyan et~al., \emph{{Measurement of the top quark polarization and $\mathrm{t\bar{t}}$ spin correlations using dilepton final states in proton-proton collisions at $\sqrt{s} =$ 13 TeV}}, \href{http://dx.doi.org/10.1103/PhysRevD.100.072002}{\emph{Phys. Rev. D} {\bf 100} (2019) 072002}, [\href{http://arxiv.org/abs/1907.03729}{{\tt 1907.03729}}].

\bibitem{ATLAS:2019zrq}
{\scshape ATLAS} collaboration, M.~Aaboud et~al., \emph{{Measurements of top-quark pair spin correlations in the $e\mu$ channel at $\sqrt{s} = 13$ TeV using $pp$ collisions in the ATLAS detector}}, \href{http://dx.doi.org/10.1140/epjc/s10052-020-8181-6}{\emph{Eur. Phys. J. C} {\bf 80} (2020) 754}, [\href{http://arxiv.org/abs/1903.07570}{{\tt 1903.07570}}].

\bibitem{CMS:2024pts}
{\scshape CMS} collaboration, A.~Hayrapetyan et~al., \emph{{Observation of quantum entanglement in top quark pair production in proton{\textendash}proton collisions at $\sqrt{s} = 13$ TeV}}, \href{http://dx.doi.org/10.1088/1361-6633/ad7e4d}{\emph{Rept. Prog. Phys.} {\bf 87} (2024) 117801}, [\href{http://arxiv.org/abs/2406.03976}{{\tt 2406.03976}}].

\bibitem{CMS:2024zkc}
{\scshape CMS} collaboration, \emph{{Measurements of polarization and spin correlation and observation of entanglement in top quark pairs using lepton+jets events from proton-proton collisions at $\sqrt{s}$ = 13 TeV}},  \href{http://arxiv.org/abs/2409.11067}{{\tt 2409.11067}}.

\bibitem{Altakach:2022ywa}
M.~M. Altakach, P.~Lamba, F.~Maltoni, K.~Mawatari and K.~Sakurai, \emph{{Quantum information and CP measurement in H\textrightarrow{}\ensuremath{\tau}+\ensuremath{\tau}- at future lepton colliders}}, \href{http://dx.doi.org/10.1103/PhysRevD.107.093002}{\emph{Phys. Rev. D} {\bf 107} (2023) 093002}, [\href{http://arxiv.org/abs/2211.10513}{{\tt 2211.10513}}].

\bibitem{Fabbrichesi:2022ovb}
M.~Fabbrichesi, R.~Floreanini and E.~Gabrielli, \emph{{Constraining new physics in entangled two-qubit systems: top-quark, tau-lepton and photon pairs}}, \href{http://dx.doi.org/10.1140/epjc/s10052-023-11307-2}{\emph{Eur. Phys. J. C} {\bf 83} (2023) 162}, [\href{http://arxiv.org/abs/2208.11723}{{\tt 2208.11723}}].

\bibitem{Ehataht:2023zzt}
K.~Ehat\"aht, M.~Fabbrichesi, L.~Marzola and C.~Veelken, \emph{{Probing entanglement and testing Bell inequality violation with e+e-\textrightarrow{}\ensuremath{\tau}+\ensuremath{\tau}- at Belle II}}, \href{http://dx.doi.org/10.1103/PhysRevD.109.032005}{\emph{Phys. Rev. D} {\bf 109} (2024) 032005}, [\href{http://arxiv.org/abs/2311.17555}{{\tt 2311.17555}}].

\bibitem{Fabbrichesi:2024xtq}
M.~Fabbrichesi and L.~Marzola, \emph{{Dipole momenta and compositeness of the $\tau$ lepton at Belle II}},  \href{http://arxiv.org/abs/2401.04449}{{\tt 2401.04449}}.

\bibitem{Fabbrichesi:2024wcd}
M.~Fabbrichesi and L.~Marzola, \emph{{Quantum tomography with \ensuremath{\tau} leptons at the FCC-ee: Entanglement, Bell inequality violation, sin\ensuremath{\theta}W, and anomalous couplings}}, \href{http://dx.doi.org/10.1103/PhysRevD.110.076004}{\emph{Phys. Rev. D} {\bf 110} (2024) 076004}, [\href{http://arxiv.org/abs/2405.09201}{{\tt 2405.09201}}].

\bibitem{Zhang:2025mmm}
Y.~Zhang, B.-H. Zhou, Q.-B. Liu, S.~Li, S.-C. Hsu, T.~Han et~al., \emph{{Entanglement and Bell Nonlocality in $\tau^+ \tau^-$ at the LHC using Machine Learning for Neutrino Reconstruction}},  \href{http://arxiv.org/abs/2504.01496}{{\tt 2504.01496}}.

\bibitem{Han:2025ewp}
T.~Han, M.~Low and Y.~Su, \emph{{Entanglement and Bell Nonlocality in $\tau^+ \tau^-$ at the BEPC}},  \href{http://arxiv.org/abs/2501.04801}{{\tt 2501.04801}}.

\bibitem{Rao:2019hsp}
K.~Rao, S.~D. Rindani and P.~Sarmah, \emph{{Probing anomalous gauge-Higgs couplings using Z boson polarization at e+e{\ensuremath{-}} colliders}}, \href{http://dx.doi.org/10.1016/j.nuclphysb.2019.114840}{\emph{Nucl. Phys. B} {\bf 950} (2020) 114840}, [\href{http://arxiv.org/abs/1904.06663}{{\tt 1904.06663}}].

\bibitem{Rao:2021eer}
K.~Rao, S.~D. Rindani, P.~Sarmah and B.~Singh, \emph{{Use of Z polarization in e+e{\ensuremath{-}} {\textrightarrow} ZH to measure the triple-Higgs coupling}}, \href{http://dx.doi.org/10.1016/j.nuclphysb.2021.115649}{\emph{Nucl. Phys. B} {\bf 975} (2022) 115649}, [\href{http://arxiv.org/abs/2109.11134}{{\tt 2109.11134}}].

\bibitem{Cheng:2025zaw}
K.~Cheng, T.~Han and S.~Trifinopoulos, \emph{{Quantum Information at the Electron-Ion Collider}},  \href{http://arxiv.org/abs/2510.23773}{{\tt 2510.23773}}.

\bibitem{Adolphsen:2013kya}
\emph{{The International Linear Collider Technical Design Report - Volume 3.II: Accelerator Baseline Design}},  \href{http://arxiv.org/abs/1306.6328}{{\tt 1306.6328}}.

\bibitem{Wootters:1997id}
W.~K. Wootters, \emph{{Entanglement of formation of an arbitrary state of two qubits}}, \href{http://dx.doi.org/10.1103/PhysRevLett.80.2245}{\emph{Phys. Rev. Lett.} {\bf 80} (1998) 2245--2248}, [\href{http://arxiv.org/abs/quant-ph/9709029}{{\tt quant-ph/9709029}}].

\bibitem{Hill:1997pfa}
S.~Hill and W.~K. Wootters, \emph{{Entanglement of a pair of quantum bits}}, \href{http://dx.doi.org/10.1103/PhysRevLett.78.5022}{\emph{Phys. Rev. Lett.} {\bf 78} (1997) 5022--5025}, [\href{http://arxiv.org/abs/quant-ph/9703041}{{\tt quant-ph/9703041}}].

\bibitem{PhysicsPhysiqueFizika.1.195}
J.~S. Bell, \emph{On the einstein podolsky rosen paradox}, \href{http://dx.doi.org/10.1103/PhysicsPhysiqueFizika.1.195}{\emph{Physics Physique Fizika} {\bf 1} (Nov, 1964) 195--200}.

\bibitem{PhysRevLett.23.880}
J.~F. Clauser, M.~A. Horne, A.~Shimony and R.~A. Holt, \emph{Proposed experiment to test local hidden-variable theories}, \href{http://dx.doi.org/10.1103/PhysRevLett.23.880}{\emph{Phys. Rev. Lett.} {\bf 23} (Oct, 1969) 880--884}.

\bibitem{Horodecki:1995nsk}
R.~Horodecki, P.~Horodecki and M.~Horodecki, \emph{{Violating Bell inequality by mixed spin- 1 2 states: necessary and sufficient condition }}, \href{http://dx.doi.org/10.1016/0375-9601(95)00214-N}{\emph{Phys. Lett. A} {\bf 200} (1995) 340--344}.

\bibitem{Cirelson:1980ry}
B.~S. Cirelson, \emph{{QUANTUM GENERALIZATIONS OF BELL'S INEQUALITY}}, \href{http://dx.doi.org/10.1007/BF00417500}{\emph{Lett. Math. Phys.} {\bf 4} (1980) 93--100}.

\bibitem{Gottesman:1998hu}
D.~Gottesman, \emph{{The Heisenberg representation of quantum computers}},  in \emph{{22nd International Colloquium on Group Theoretical Methods in Physics}}, pp.~32--43, 7, 1998.
\newblock \href{http://arxiv.org/abs/quant-ph/9807006}{{\tt quant-ph/9807006}}.

\bibitem{Gottesman:1999tea}
D.~Gottesman and I.~L. Chuang, \emph{{Demonstrating the viability of universal quantum computation using teleportation and single-qubit operations}}, \href{http://dx.doi.org/10.1038/46503}{\emph{Nature} {\bf 402} (1999) 390--393}, [\href{http://arxiv.org/abs/quant-ph/9908010}{{\tt quant-ph/9908010}}].

\bibitem{Aaronson:2004xuh}
S.~Aaronson and D.~Gottesman, \emph{{Improved simulation of stabilizer circuits}}, \href{http://dx.doi.org/10.1103/PhysRevA.70.052328}{\emph{Phys. Rev. A} {\bf 70} (2004) 052328}, [\href{http://arxiv.org/abs/quant-ph/0406196}{{\tt quant-ph/0406196}}].

\bibitem{Knill:1996dv}
E.~Knill, \emph{{Nonbinary unitary error bases and quantum codes}},  \href{http://arxiv.org/abs/quant-ph/9608048}{{\tt quant-ph/9608048}}.

\bibitem{Gottesman:1997zz}
D.~Gottesman, \emph{{Stabilizer codes and quantum error correction}}, {\emph{Phys. Rev. A} (5, 1997) }, [\href{http://arxiv.org/abs/quant-ph/9705052}{{\tt quant-ph/9705052}}].

\bibitem{Emerson:2013zse}
J.~Emerson, D.~Gottesman, S.~A.~H. Mousavian and V.~Veitch, \emph{{The resource theory of stabilizer quantum computation}}, \href{http://dx.doi.org/10.1088/1367-2630/16/1/013009}{\emph{New J. Phys.} {\bf 16} (2014) 013009}, [\href{http://arxiv.org/abs/1307.7171}{{\tt 1307.7171}}].

\bibitem{Howard:2014zwm}
M.~Howard, J.~J. Wallman, V.~Veitch and J.~Emerson, \emph{{Contextuality supplies the {\textquoteleft}magic{\textquoteright} for quantum computation}}, \href{http://dx.doi.org/10.1038/nature13460}{\emph{Nature} {\bf 510} (2014) 351--355}, [\href{http://arxiv.org/abs/1401.4174}{{\tt 1401.4174}}].

\bibitem{Leone:2021rzd}
L.~Leone, S.~F.~E. Oliviero and A.~Hamma, \emph{{Stabilizer R{\'e}nyi Entropy}}, \href{http://dx.doi.org/10.1103/PhysRevLett.128.050402}{\emph{Phys. Rev. Lett.} {\bf 128} (2022) 050402}, [\href{http://arxiv.org/abs/2106.12587}{{\tt 2106.12587}}].

\bibitem{Wang:2023sre}
Y.~Wang and Y.~Li, \emph{Stabilizer rényi entropy on qudits}, \href{http://dx.doi.org/10.1007/s11128-023-04186-9}{\emph{Quantum Information Processing} {\bf 22} (2023) 444}.

\bibitem{Liu:2025frx}
Q.~Liu, I.~Low and Z.~Yin, \emph{{Maximal Magic for Two-qubit States}},  \href{http://arxiv.org/abs/2502.17550}{{\tt 2502.17550}}.

\bibitem{Ohta:2025utz}
M.~Ohta and K.~Sakurai, \emph{{Extremal Magic States from Symmetric Lattices}},  \href{http://arxiv.org/abs/2506.11725}{{\tt 2506.11725}}.

\bibitem{Baumgart:2012ay}
M.~Baumgart and B.~Tweedie, \emph{{A New Twist on Top Quark Spin Correlations}}, \href{http://dx.doi.org/10.1007/JHEP03(2013)117}{\emph{JHEP} {\bf 03} (2013) 117}, [\href{http://arxiv.org/abs/1212.4888}{{\tt 1212.4888}}].

\bibitem{Afik:2020onf}
Y.~Afik and J.~R.~M. de~Nova, \emph{{Entanglement and quantum tomography with top quarks at the LHC}}, \href{http://dx.doi.org/10.1140/epjp/s13360-021-01902-1}{\emph{Eur. Phys. J. Plus} {\bf 136} (2021) 907}, [\href{http://arxiv.org/abs/2003.02280}{{\tt 2003.02280}}].

\bibitem{Ashby-Pickering:2022umy}
R.~Ashby-Pickering, A.~J. Barr and A.~Wierzchucka, \emph{{Quantum state tomography, entanglement detection and Bell violation prospects in weak decays of massive particles}}, \href{http://dx.doi.org/10.1007/JHEP05(2023)020}{\emph{JHEP} {\bf 05} (2023) 020}, [\href{http://arxiv.org/abs/2209.13990}{{\tt 2209.13990}}].

\bibitem{Liu:2016idz}
D.~Liu, A.~Pomarol, R.~Rattazzi and F.~Riva, \emph{{Patterns of Strong Coupling for LHC Searches}}, \href{http://dx.doi.org/10.1007/JHEP11(2016)141}{\emph{JHEP} {\bf 11} (2016) 141}, [\href{http://arxiv.org/abs/1603.03064}{{\tt 1603.03064}}].

\bibitem{Altomonte:2024upf}
C.~Altomonte, A.~J. Barr, M.~Eckstein, P.~Horodecki and K.~Sakurai, \emph{{Prospects for quantum process tomography at high energies}}, \href{http://dx.doi.org/10.1088/2058-9565/ae0af1}{\emph{Quantum Sci. Technol.} {\bf 10} (2025) 045060}, [\href{http://arxiv.org/abs/2412.01892}{{\tt 2412.01892}}].

\bibitem{Aguilar-Saavedra:2018ksv}
D.~Barducci et~al., \emph{{Interpreting top-quark LHC measurements in the standard-model effective field theory}},  \href{http://arxiv.org/abs/1802.07237}{{\tt 1802.07237}}.

\bibitem{Brivio:2022pyi}
I.~Brivio et~al., \emph{{Truncation, validity, uncertainties}},  \href{http://arxiv.org/abs/2201.04974}{{\tt 2201.04974}}.

\bibitem{Azatov:2016sqh}
A.~Azatov, R.~Contino, C.~S. Machado and F.~Riva, \emph{{Helicity selection rules and noninterference for BSM amplitudes}}, \href{http://dx.doi.org/10.1103/PhysRevD.95.065014}{\emph{Phys. Rev. D} {\bf 95} (2017) 065014}, [\href{http://arxiv.org/abs/1607.05236}{{\tt 1607.05236}}].

\bibitem{Falkowski:2016cxu}
A.~Falkowski, M.~Gonzalez-Alonso, A.~Greljo, D.~Marzocca and M.~Son, \emph{{Anomalous Triple Gauge Couplings in the Effective Field Theory Approach at the LHC}}, \href{http://dx.doi.org/10.1007/JHEP02(2017)115}{\emph{JHEP} {\bf 02} (2017) 115}, [\href{http://arxiv.org/abs/1609.06312}{{\tt 1609.06312}}].

\bibitem{Helset:2017mlf}
A.~Helset and M.~Trott, \emph{{On interference and non-interference in the SMEFT}}, \href{http://dx.doi.org/10.1007/JHEP04(2018)038}{\emph{JHEP} {\bf 04} (2018) 038}, [\href{http://arxiv.org/abs/1711.07954}{{\tt 1711.07954}}].

\bibitem{Azatov:2019xxn}
A.~Azatov, D.~Barducci and E.~Venturini, \emph{{Precision diboson measurements at hadron colliders}}, \href{http://dx.doi.org/10.1007/JHEP04(2019)075}{\emph{JHEP} {\bf 04} (2019) 075}, [\href{http://arxiv.org/abs/1901.04821}{{\tt 1901.04821}}].

\bibitem{ElFaham:2024uop}
H.~El~Faham, G.~Pelliccioli and E.~Vryonidou, \emph{{Triple-gauge couplings in LHC diboson production: a SMEFT view from every angle}}, \href{http://dx.doi.org/10.1007/JHEP08(2024)087}{\emph{JHEP} {\bf 08} (2024) 087}, [\href{http://arxiv.org/abs/2405.19083}{{\tt 2405.19083}}].

\bibitem{LoChiatto:2024dmx}
P.~Lo~Chiatto, \emph{{Interference resurrection of the {\ensuremath{\tau}} dipole through quantum tomography}}, \href{http://dx.doi.org/10.1103/8gtq-twfc}{\emph{Phys. Rev. D} {\bf 112} (2025) 015017}, [\href{http://arxiv.org/abs/2408.04553}{{\tt 2408.04553}}].

\end{thebibliography}\endgroup

\newpage

\appendix

\section{Derivation of the final state spin density matrix}
\label{derivation}

The goal of this section is to derive the formula \eqref{rho_formula}.
The procedure to obtain the final state spin density matrix $\rho^{\rm f}$ from the initial state spin density matrix $\rho^{\rm in}$ in the $l \bar l \to F \bar F$ process is outlined in Appendix B of Ref.\ \cite{Altomonte:2024upf}.
Our starting point is the initial spin state 
\ba
\hat \rho_{\rm in} 
\,=\,
\sum_{ \{ \l_{l \bar l} \}  }
\rho^{\rm in}_{ (\l'_l, \l'_{\bar l}),(\l_l, \l_{\bar l})  }
| \l'_l, \l'_{\bar l} \ketbra \l_l, \l_{\bar l} |\,,
\label{hat_rho}
\ea
In the scattering theory the unitary time evolution is governed by the $S$-matrix operator acting on the Fock space ${\cal F}_{\rm in}$.  
By complementing the momentum part of the $l \bar l$, $\hat \rho_{\rm in}$ can be promoted to an element of ${\cal F}_{\rm in}$ 
\be
| p_{l, \bar l} \ketbra p_{l, \bar l} | \otimes 
\hat \rho_{\rm in}
\,=\, 
\sum_{ \{ \l_{l \bar l} \}  }
\rho^{\rm in}_{ (\l'_l, \l'_{\bar l}),(\l_l, \l_{\bar l})  }
| \l'_l, \l'_{\bar l} ; p_{l, \bar l} \ketbra \l_l, \l_{\bar l} ; p_{l, \bar l}|
\,\in\, {\cal F}_{\rm in}
\,.
\label{full_in}
\ee
We assume that the momenta of $l$ and $\bar l$ are described by the momentum wave functions $\phi_l(k_1)$ and $\phi_{\bar l}(k_2)$, respectively, which are sharply peaked at ${\bf k}_i = {\bf p}_{X_i}$ for $i=1,2$ and $(X_1, X_2) = (l, \bar l)$.
Concretely, we write
\be
\phi_{X_i}(k_i) = \frac{(2 \pi)^3 \delta^3({\bf k}_i - {\bf p}_{X_i} ) }{ \sqrt{V} },
\label{pwave}
\ee
where $V$ is chosen to be the volume of all space 
$
V = \left[ \int d^3 x \, e^{i {\bf p} \cdot {\bf x}}  \right]_{ {\bf p} = {\bf 0} } = (2 \pi)^3 \delta^3( {\bf 0} )\,
$
to ensure the canonical normalisation 
$
\int \frac{d^3 k_i}{(2 \pi)^3} \left| \phi_{X_i}(k_i) \right|^2 = 1 \,.
$
Using these wave functions, and the Lorentz-covariantly normalised kets
\be
\bra k_1,k_2 | k_1',k_2' \ket = (2 \pi)^6 4 E_{k_1} E_{k_2} \delta^3( {\bf k}_1 - {\bf k}_1' )
\delta^3( {\bf k}_2 - {\bf k}_2' ),
\ee
one can construct the $l \bar l$ momentum state as
\be
| p_{l, \bar l} \ket = 
\int \frac{d^3 k_1}{(2 \pi)^3 } \frac{1}{\sqrt{2 E_{k_1}}} \phi_l( k_1 ) 
\frac{d^3 k_2}{(2 \pi)^3 } \frac{1}{\sqrt{2 E_{k_2}}} \phi_{\bar l}( k_2 )
| k_1, k_2 \ket
\label{ptilde}
\ee
with correct normalisation $\bra p_{l, \bar l} | p_{l, \bar l} \ket = 1$.

The next step is to evolve the full initial state \eqref{full_in} by the $S$-matrix
\be
\widetilde \rho \,=\,
S (| p_{l, \bar l} \ketbra p_{l, \bar l} | \otimes 
\hat \rho_{\rm in}) S^\dagger \,.
\label{rhotilde}
\ee
Prior to measurement, $\widetilde \rho$ is a coherent superposition of all kinematically allowed final state channels and momenta.
We condition on the $F\bar F$ final state with a restricted momentum region $x$ implemented by the projection operator 
\be
\Pi_x = \int_x 
\frac{d^3 k_F d^3 k_{\bar F} }{(2 \pi)^6 2 E_F 2 E_{\bar F} }
| k_F, k_{\bar F} \ketbra k_F, k_{\bar F} |\,.
\ee
The unnormalised state of $F \bar F$ spin subsystem is obtained after tracing out the momentum degrees of freedom:\footnote{$\hat \varrho_{\rm out}$ is not normalised due to the projection $\Pi_x$.}
\bea
&&
\hat \varrho_{\rm f}
\,=\,
{\rm Tr}_P \left[ \,
{\Pi}_{x} \widetilde \rho \,
\right]
\,=\,
\int_x 
\frac{d^3 k_F d^3 k_{\bar F} }{(2 \pi)^6 2 E_{F} 2 E_{\bar F}}
\bra k_F, k_{\bar F} | \widetilde \rho |  k_F, k_{\bar F} \ket
\nonumber \\
&&
~~~~~~=
\sum_{ \{ \l_{F \bar F} \} }
\int_x 
\frac{d^3 k_F d^3 k_{\bar F} }{(2 \pi)^6 2 E_{F} 2 E_{\bar F}}
\bra \l'_F, \l'_{\bar F} ; k_F, k_{\bar F} | \widetilde \rho |  \l_F, \l_{\bar F} ; k_F, k_{\bar F} \ket
\cdot 
|  \l'_F, \l'_{\bar F} \ketbra 
\l_F, \l_{\bar F} |.
\label{rhohatout}
\eea
In the second line, we inserted complete sets of the $F \bar F$ spin states, acting from the left and right of $\bra k_F, k_{\bar F} | \widetilde \rho |  k_F, k_{\bar F} \ket$, respectively, to obtain Fock space elements $|\l_F, \l_{\bar F}; k_F,k_{\bar F} \ket, |\l'_F, \l'_{\bar F}; k_F,k_{\bar F} \ket \in {\cal F}_{\rm out}$.

Using Eqs.\ \eqref{full_in}, \eqref{pwave}, \eqref{ptilde} and \eqref{rhotilde},
we find
\bea
&&
\bra \l'_F, \l'_{\bar F} ; k_F, k_{\bar F} | \widetilde \rho |  \l_F, \l_{\bar F} ; k_F, k_{\bar F} \ket
\,=\,
\frac{T}{V} \frac{1}{s} (2 \pi)^4 \delta^4( p_l+p_{\bar l} - k_F - k_{\bar F}  )
\times
\nonumber \\
&&
~~~~~~~~~~~~~~~~~~~~~~~~~~~~~~~~~~~~
\sum_{ \{ \l_{l \bar l}\} }
\rho^{\rm in}_{ (\l'_l, \l'_{\bar l}),(\l_l, \l_{\bar l})  }
{\cal M}^{\l'_l, \l'_{\bar l}}_{\l'_F, \l'_{\bar F}}(k_F,k_{\bar F})
\left[ {\cal M}^{\l_l, \l_{\bar l}}_{\l_F, \l_{\bar F}}(k_F, k_{\bar F}) \right]^*.
\eea
Here, we also used 
$s = 4 E_l E_{\bar l}$ at the centre-of-mass frame, 
\bea
\bra \l'_F, \l'_{\bar F} ; k_F, k_{\bar F} |  S | \l'_l, \l'_{\bar l} ; p_{l,\bar l} \ket
&\equiv& 
\bra \l'_F, \l'_{\bar F} ; k_F, k_{\bar F} | i {\cal T} | \l'_l, \l'_{\bar l} ; p_{l,\bar l} \ket
\nonumber \\
&\equiv&
(2 \pi)^4 
\delta^4( p_l+p_{\bar l} - k_F - k_{\bar F}  )
\cdot {\cal M}^{\l'_l, \l'_{\bar l}}_{\l'_F, \l'_{\bar F}}(k_F,k_{\bar F})\,,
\eea
and
$(2 \pi)^4 \delta^4(0) = T V$ with $TV$ being the volume of all space-time.
Putting everything together, we have
\be
\hat \varrho_{\rm f}
\,=\,
\frac{T}{V}
\frac{1}{s}
\sum_{ \{ \l_{F \bar F} \} }
\int_x 
d \Pi_{\rm LIPS}
\sum_{ \{ \l_{l \bar l}\} }
\rho^{\rm in}_{ (\l'_l, \l'_{\bar l}),(\l_l, \l_{\bar l})  }
{\cal M}^{\l'_l, \l'_{\bar l}}_{\l'_F, \l'_{\bar F}}(k_F,k_{\bar F})
\left[ {\cal M}^{\l_l, \l_{\bar l}}_{\l_F, \l_{\bar F}}(k_F, k_{\bar F}) \right]^*
\cdot 
|  \l'_F, \l'_{\bar F} \ketbra 
\l_F, \l_{\bar F} |,
\ee
where 
\be
\int_x d \Pi_{\rm LIPS}
\,=\,
\int_x \frac{d^3 k_F d^3 k_{\bar F}}{(2 \pi)^6 2 E_F 2 E_{\bar F} }
(2 \pi)^4 \delta^4( p_l+p_{\bar l} - k_F - k_{\bar F}  )
\,=\,
\int_x \frac{d \cos \Theta}{8 \pi}
\frac{|{\bf p}_F|}{\sqrt{s}}
\ee
is the Lorentz-invariant phase space factor. 
On the right-hand side, we have exploited the cylindrical symmetry of the system and integrated over the azimuthal angle $\phi$ of the $F \bar F$ production.  

In each $\cos \Theta$ bin, we normalise $\hat \varrho_{\rm f}$ and compare it with the canonical form of the spin density matrix  
\be
\hat \rho_{\rm f} \,=\, 
\sum_{ \{ \l_{F \bar F} \}  }
\rho^{\rm f}_{ (\l'_F, \l'_{\bar F}),(\l_F, \l_{\bar F})} \cdot 
| \l'_F, \l'_{\bar F} \ketbra \l_F, \l_{\bar F} |\,.
\ee
This procedure allows us to extract the formula 
\be
\rho^{\rm f}_{ (\l'_F, \l'_{\bar F}),(\l_F, \l_{\bar F})}
\,=\,
\frac{1}{\cal N} 
\sum_{ \{ \l_{l \bar l}\} }
\rho^{\rm in}_{ (\l'_l, \l'_{\bar l}),(\l_l, \l_{\bar l})  }
{\cal M}^{\l'_l, \l'_{\bar l}}_{\l'_F, \l'_{\bar F}}(\sqrt{s},\Theta)
\left[ {\cal M}^{\l_l, \l_{\bar l}}_{\l_F, \l_{\bar F}}(\sqrt{s},\Theta) \right]^*,
\label{rho_formula2}
\ee
where the normalisation factor ${\cal N}$ is given by
\be
{\cal N} = 
\sum_{\l_F, \l_{\bar F}}
\sum_{ \{ \l_{l \bar l}\} }
\rho^{\rm in}_{ (\l'_l, \l'_{\bar l}),(\l_l, \l_{\bar l})  }
{\cal M}^{\l'_l, \l'_{\bar l}}_{\l_F, \l_{\bar F}}(\sqrt{s},\Theta)
\left[ {\cal M}^{\l_l, \l_{\bar l}}_{\l_F, \l_{\bar F}}(\sqrt{s},\Theta) \right]^*.
\ee
Demanding that the initial spin density matrix is diagonal (Eq.\ \eqref{rho^in}), we arrive at the promised formula \eqref{rho_formula}.

\newpage

\section{The density matrices}
\label{app:rho}

Below we provide the analytical expressions of the $F \bar F$ density matrix for the vector and tensor interactions.
In each case, the effective polarisation angle is defined in Eqs.\ \eqref{PhiV} and \eqref{PhiT}, respectively.

\begin{description}

\item[Vector interaction]

\be
\rho^{\rm f} \,=\,
\frac{1}{a_+ + a_- + 2d}
\bmat
a_+ & b_- & b_- & c \\
b_- & d & d & b_+ \\
b_- & d & d & b_+ \\
c & b_+ & b_+ & a_- 
\emat
\label{eq:rho_vec}
\ee
with
\ba
a_{\pm} &=&
(\cos\eta_A \pm \beta\sin\eta_A)^2
\bigl(3+\cos 2\Theta \mp 4\cos\Theta\cos 2\PhiV\bigr)
\,,
\nn \\
b_{\pm} &=&  
\pm 2 \gamma^{-1} \cos\eta_A\,(\cos\eta_A \mp \beta\sin\eta_A)\,\sin\Theta\,(\cos\Theta \pm \cos 2\PhiV)
\,,
\nn \\
c_{~~} &=& 2 
 \bigl(\cos^2\eta_A-\beta^2\sin^2\eta_A\bigr)\sin^2\Theta
 \,,
 \nn \\
d_{~~} &=& 2 \gamma^{-2}
 \cos^2\eta_A\,\sin^2\Theta
\,.
\ea

\item[Tensor interaction]

\be
\rho^{\rm f} \,=\,
\frac{1}{2 \bar a + \bar c_+ + \bar c_-}
\bmat
~\bar a & ~\bar b_+ & ~\bar b_- & - \bar a \\
~\bar b_+ & ~\bar c_+ & ~\bar d & - \bar b_+ \\
~\bar b_- & ~\bar d & ~\bar c_- & - \bar b_- \\
- \bar a & - \bar b_+ & - \bar b_-  & ~ \bar a 
\emat
\label{eq:rho_tens}
\ee
with
\ba
\bar a_{~~} &=& \gamma^{-2}\sin^2 \Theta  
 \,,
\nn \\
\bar b_{\pm} &=&  
\frac{1}{2}\gamma^{-1} \sin 2 \Theta (1 \pm \beta \cos 2 \PhiT) 
\,,
\nn \\
\bar c_\pm &=& 
\cos^2\Theta
(1 + \beta^2 \pm 2 \beta \cos 2 \PhiT)
\,, \nn \\
\bar d_{~~} &=& 
\gamma^{-2}\cos^2\Theta
\,.
\ea

\end{description}

\newpage

\section{Supplemental plots for $\sqrt{s} = 500$ GeV}
\label{app:eft_500gev_plots}

We provide the supplemental plots for $\sqrt{s} = 500$ GeV. 

\begin{figure}[b!]
\centering
\fbox{\includegraphics[scale=0.335]{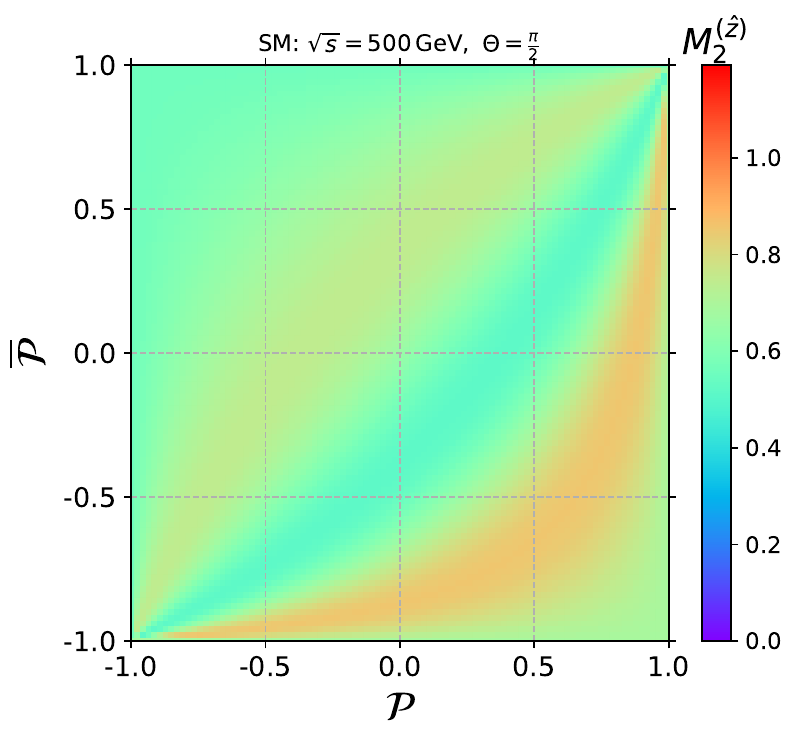}}
\includegraphics[scale=0.335]{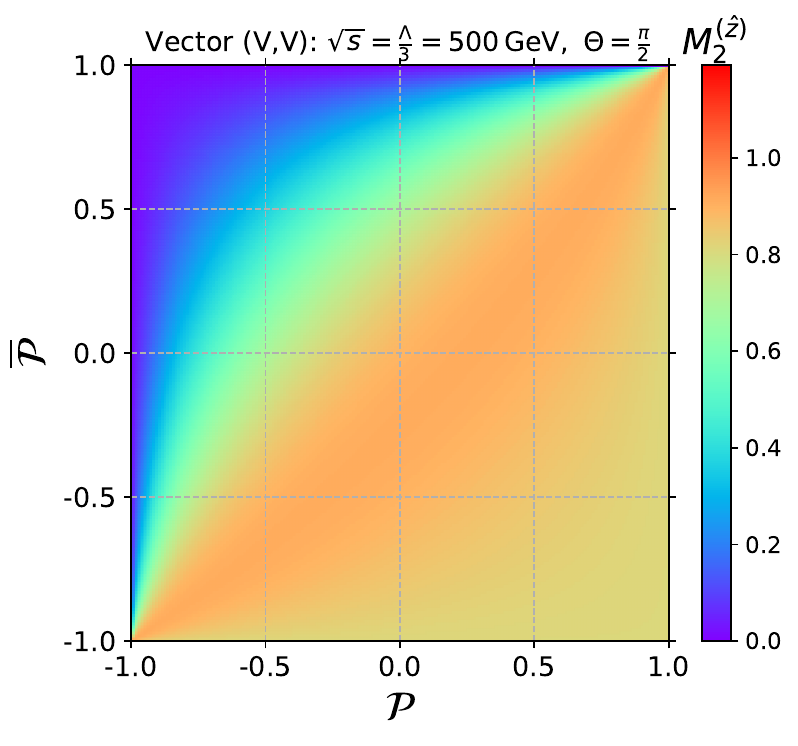}
\includegraphics[scale=0.335]{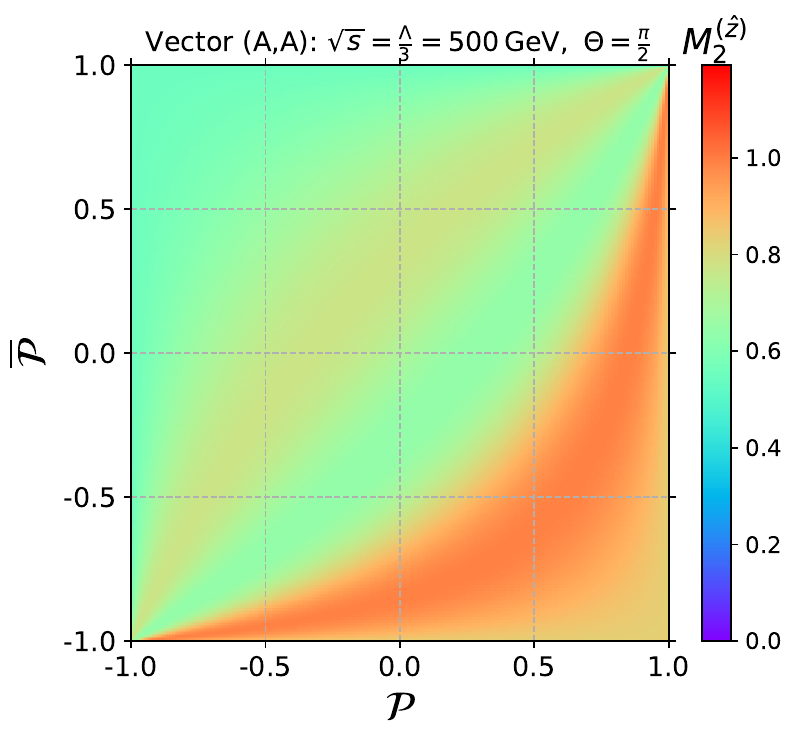}
\\
\includegraphics[scale=0.335]{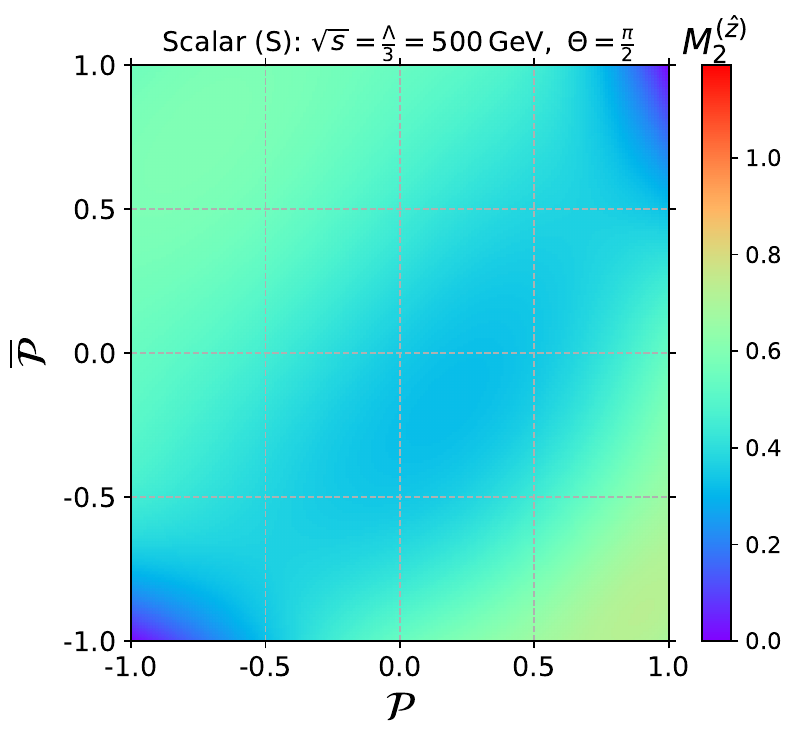}
\includegraphics[scale=0.335]{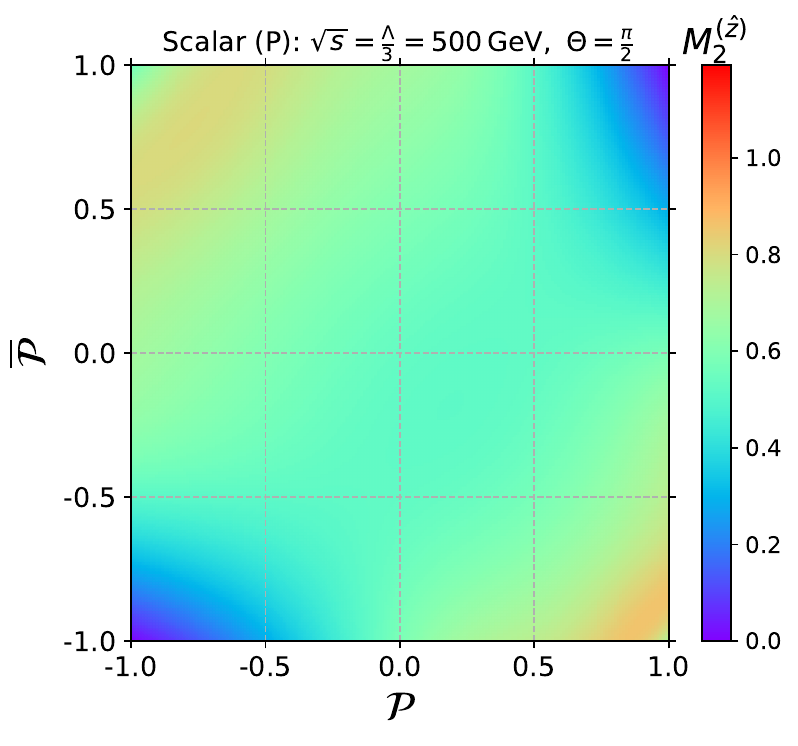}
\includegraphics[scale=0.335]{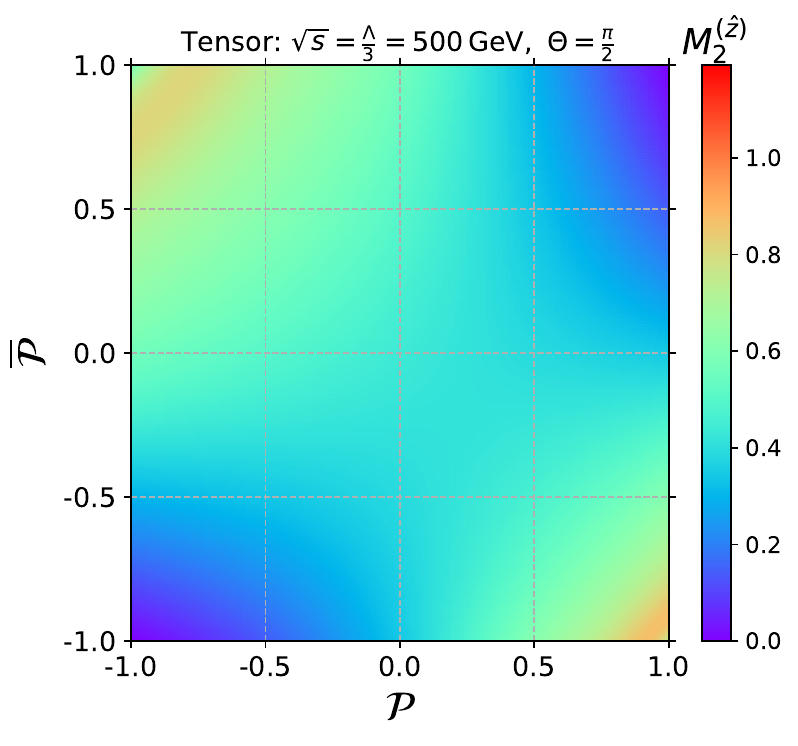}
\caption{\label{fig:M2_z_500gev}
\small Beam-basis SRE $M_2^{(\hat z)}$ over the $(\cal P,\overline {\cal P})$ plane. 
The centre-of-mass energy is $\sqrt{s}=500$ GeV and the production angle is fixed to $\Theta=\pi/2$. 
Top row: SM (left; outlined in black), vector (V,V) (middle), and axial–vector (A,A) (right). 
Bottom row: scalar (S) (left), scalar (P) (middle) and tensor (right).
}
\end{figure}

\begin{figure}[h!]
\centering
\includegraphics[scale=0.30]{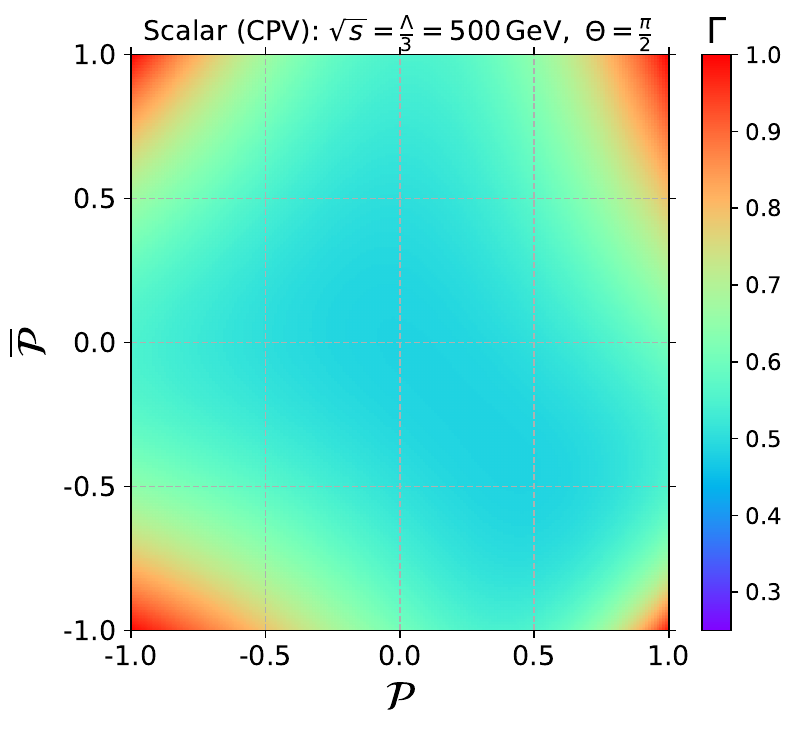}
\includegraphics[scale=0.30]{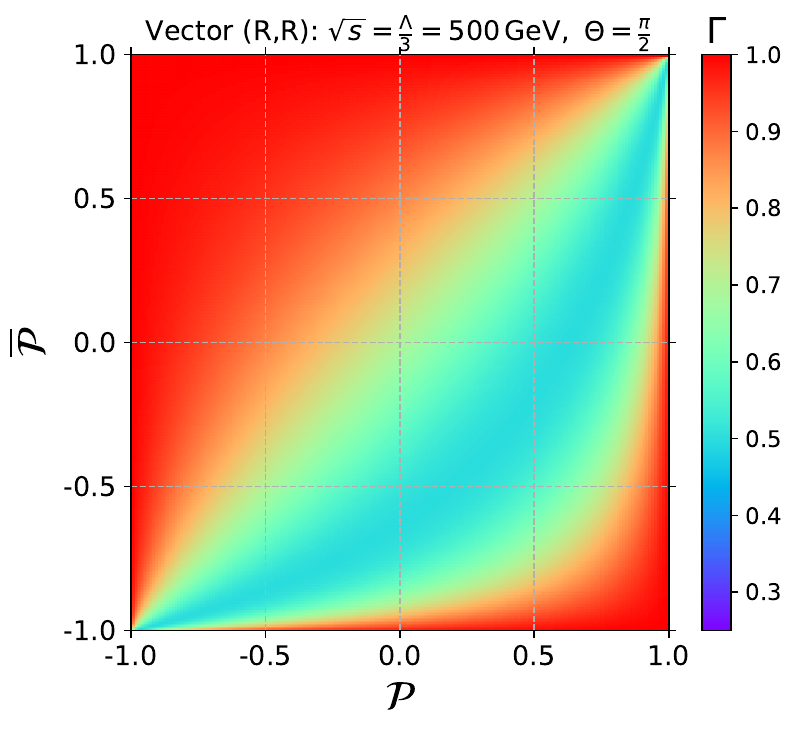}
\\
\includegraphics[scale=0.30]{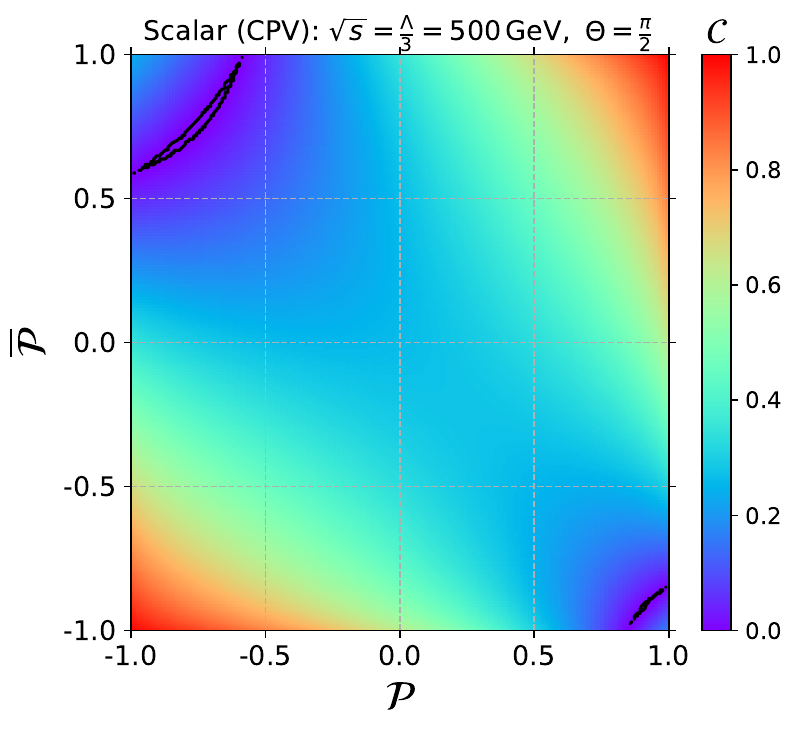}
\includegraphics[scale=0.30]{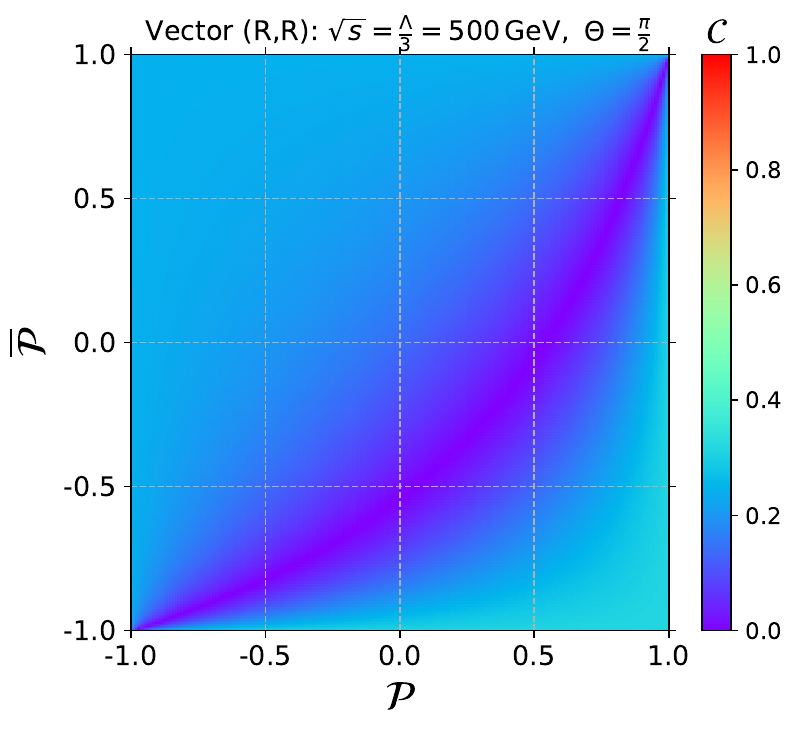}
\\
\includegraphics[scale=0.30]{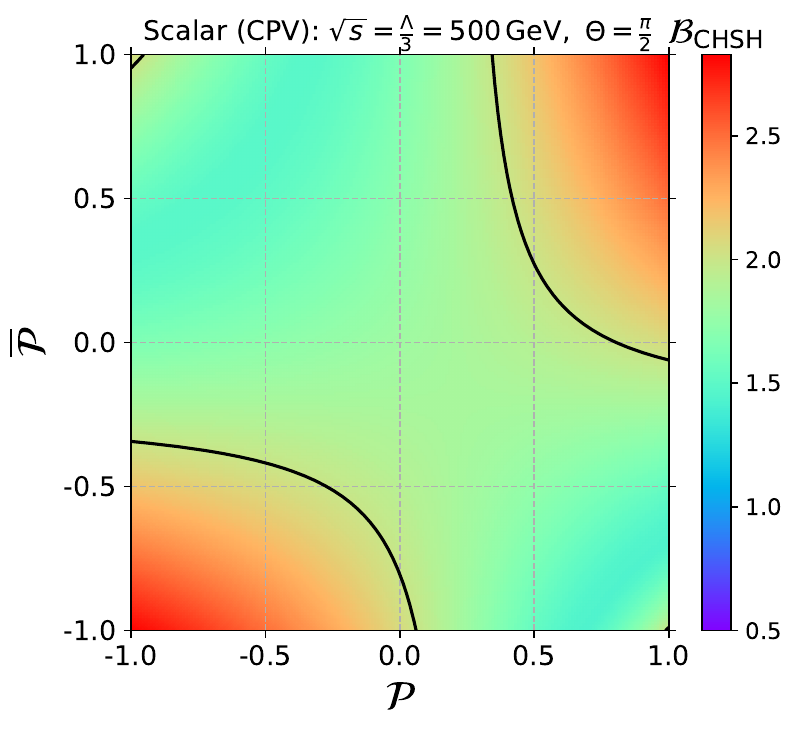}
\includegraphics[scale=0.30]{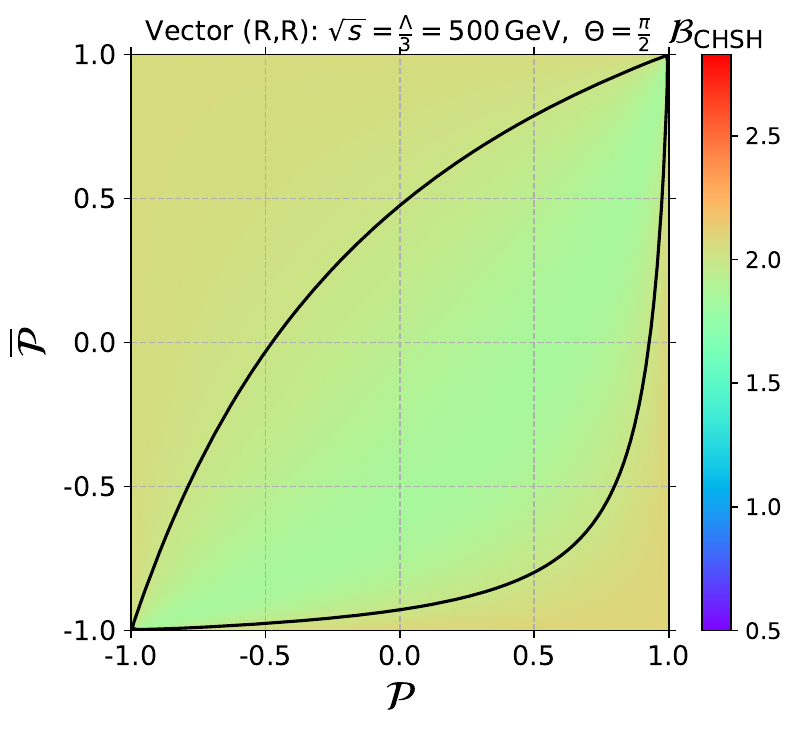}
\\
\includegraphics[scale=0.30]{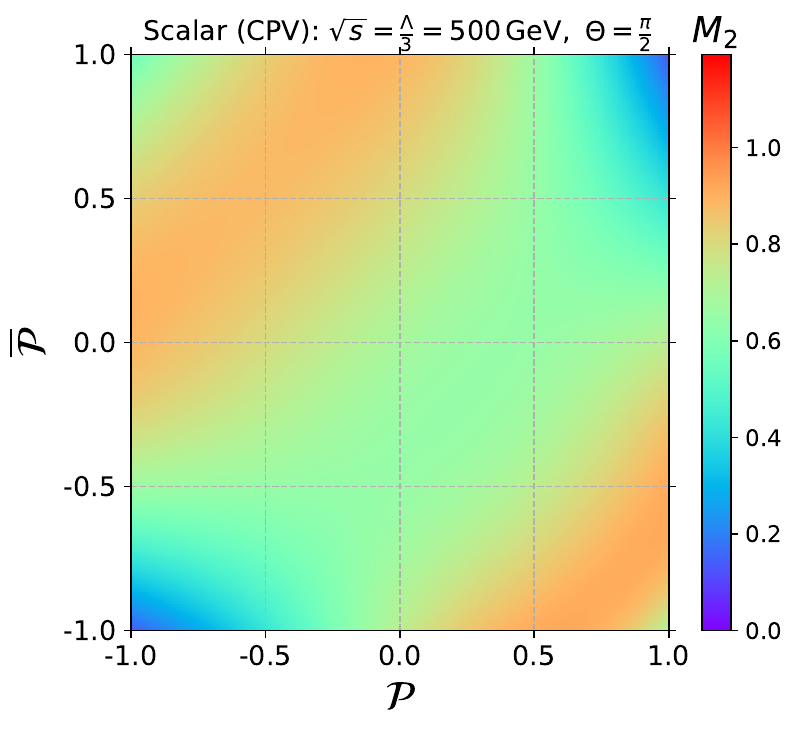}
\includegraphics[scale=0.30]{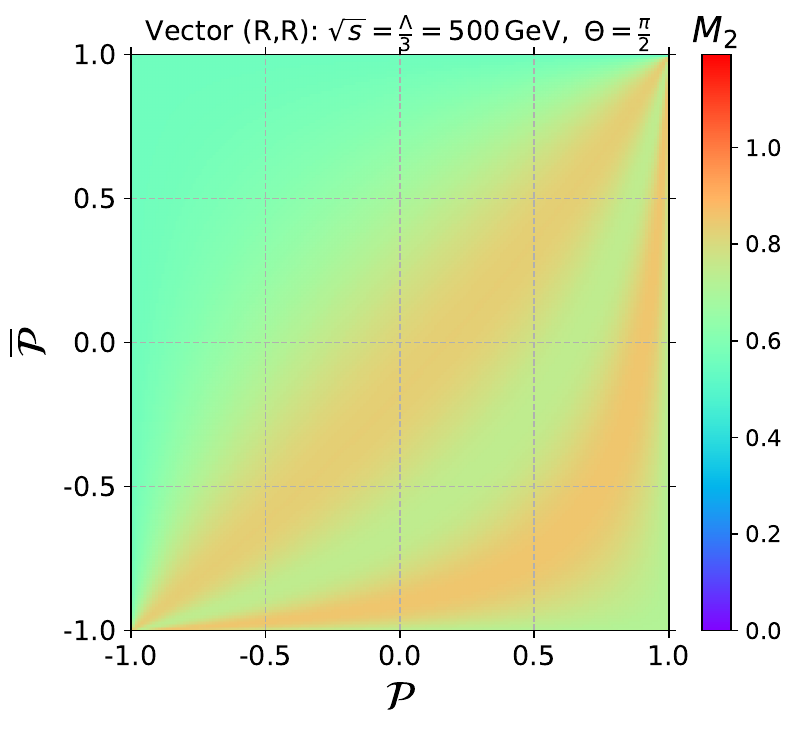}
\\
\includegraphics[scale=0.30]{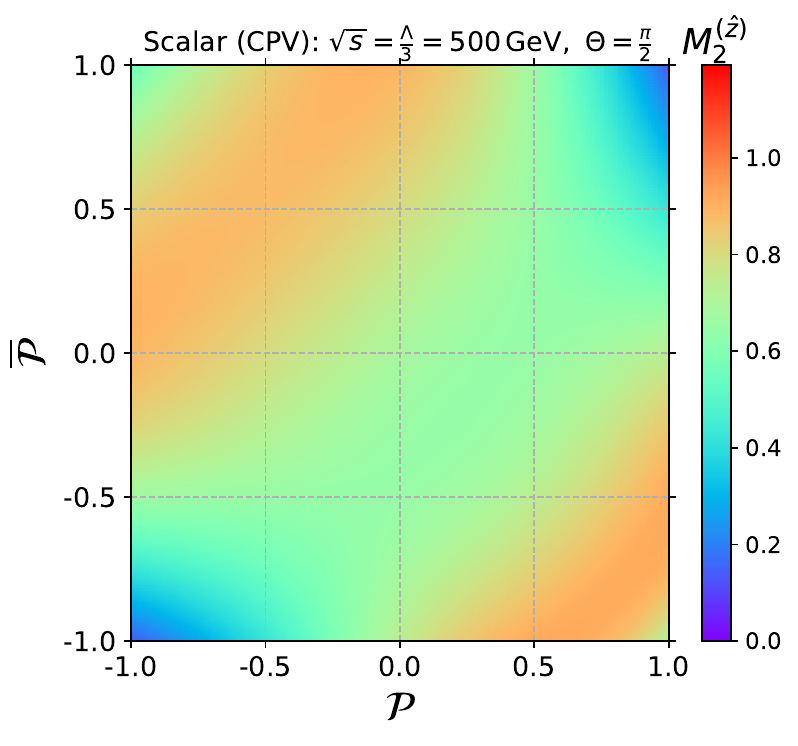}
\includegraphics[scale=0.30]{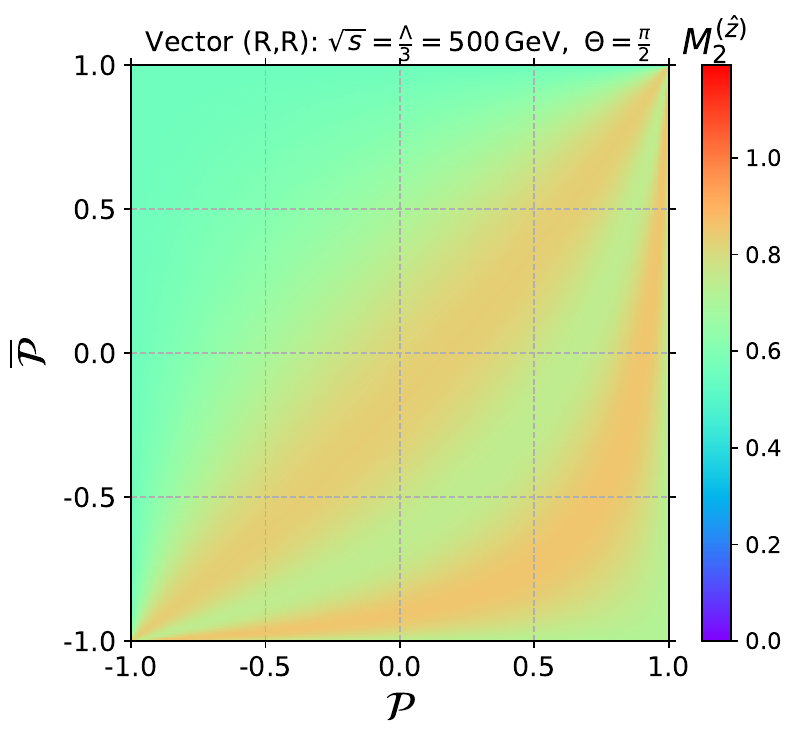}
\caption{\label{fig:500_others}
\small 
Purity $\Gamma$ (top row), concurrence ${\cal C}$ (second row), Bell-CHSH observable ${\cal B}_{\rm CHSH}$ (third row) [Contours of ${\cal B}_{\rm CHSH}=2$ are overlaid in black.], helicity-basis SRE $M_2$ (fourth row) and beam-basis SRE $M_2^{(\hat z)}$ (fifth row) over the $(\mathcal P,\overline{\mathcal P})$ plane. 
The centre-of-mass energy is $\sqrt{s}=500~\mathrm{GeV}$ and the production angle is fixed to $\Theta=\pi/2$. 
Shown are the scalar (CPV) (left column), and vector (R,R) (right column) benchmarks.
}
\end{figure}

\begin{itemize}

\item
Fig.\ \ref{fig:M2_z_500gev}: the beam-basis SRE $M_2^{(\hat z)}$ over the $(\cal P,\overline {\cal P})$ plane ($\Theta = \frac{\pi}{2}$).
The SM, vector (V,V), (A,A), scalar (S) and tensor benchmarks are shown.

\item
Fig.\ \ref{fig:500_others}: Purity $\Gamma$, concurrence ${\cal C}$, Bell-CHSH observable ${\cal B}_{\rm CHSH}$, helicity-basis SRE $M_2$ and beam-basis SRE $M_2^{(\hat z)}$ over the $(\mathcal P,\overline{\mathcal P})$ plane ($\Theta = \frac{\pi}{2}$). 
The scalar (CPV) and vector (R,R) scenarios are shown.

\end{itemize}

\newpage

\section{Supplemental plots for $\sqrt{s} = 1$ TeV}
\label{app:eft_1tev_plots}

We provide the supplemental plots for $\sqrt{s} = 1$ TeV. 

\begin{figure}[b!]
\centering
\fbox{\includegraphics[scale=0.335]{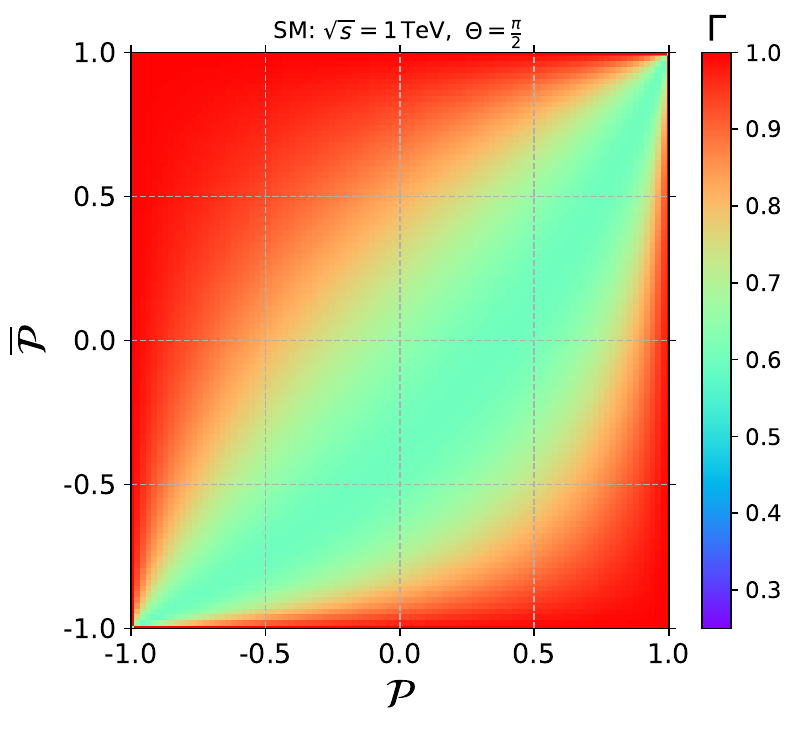}}
\includegraphics[scale=0.335]{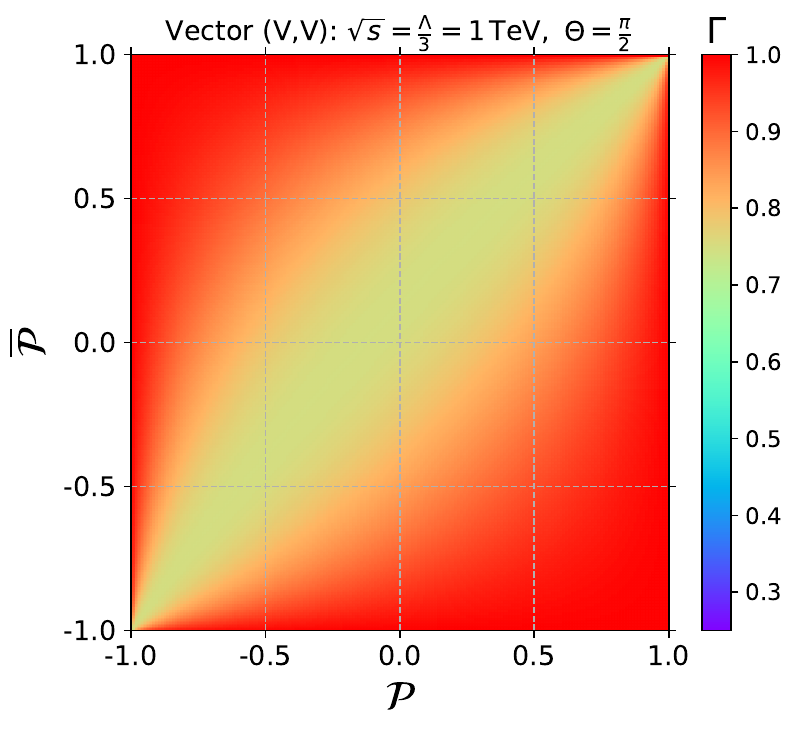}
\includegraphics[scale=0.335]{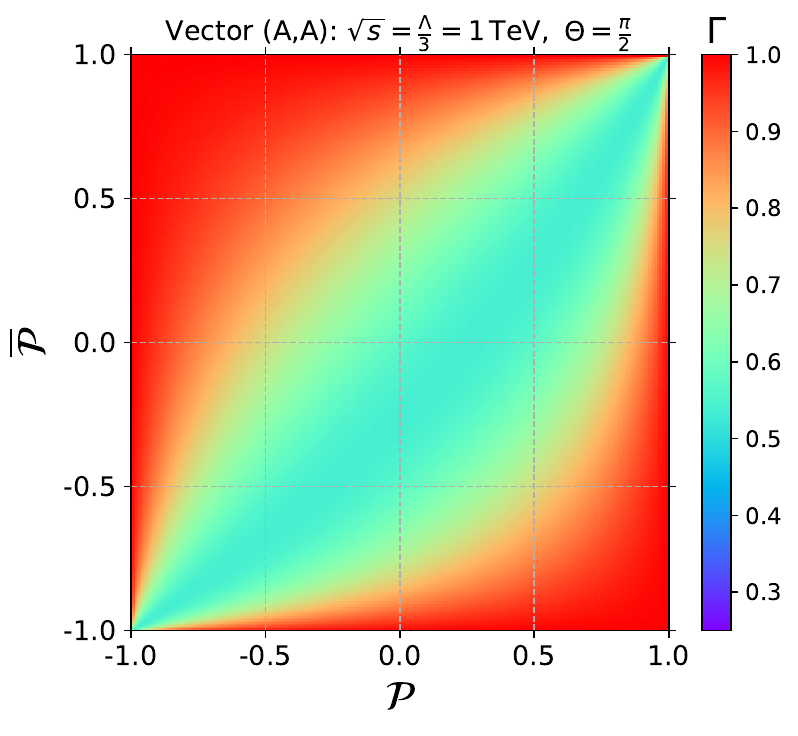}
\\
\includegraphics[scale=0.335]{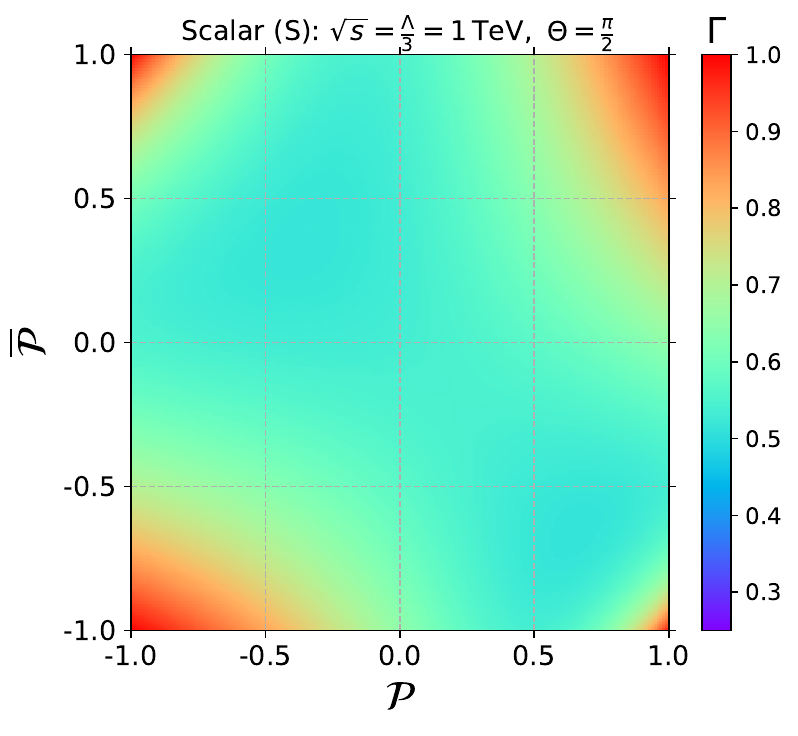}
\includegraphics[scale=0.335]{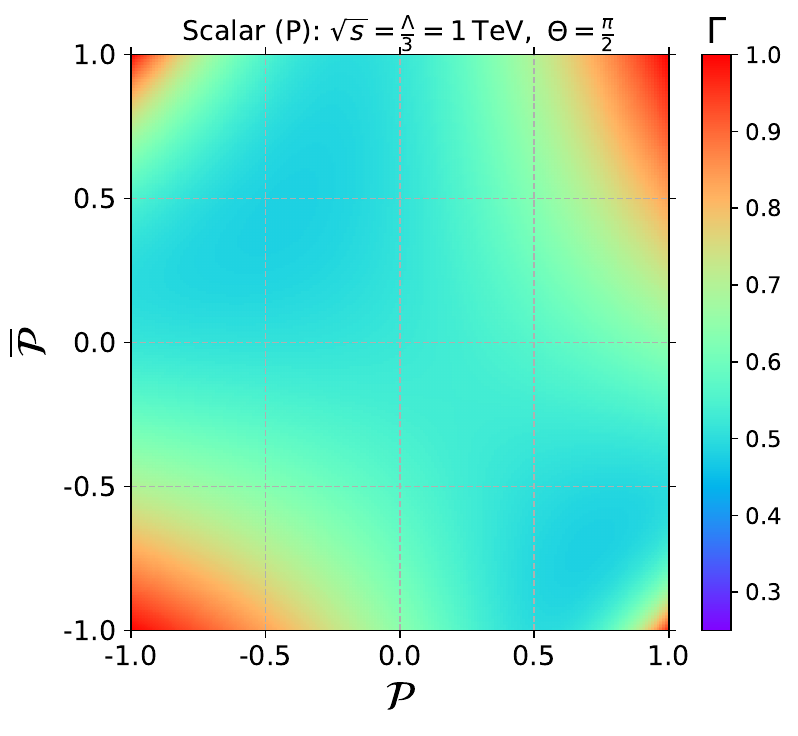}
\includegraphics[scale=0.335]{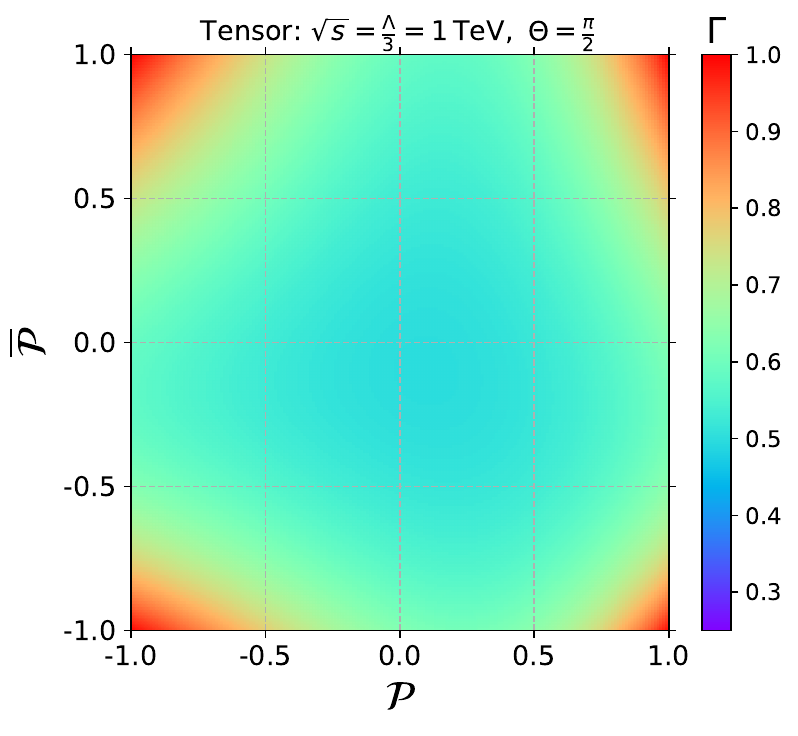}
\caption{\label{fig:pure_1000gev}
\small Purity $\Gamma$ over the $(\cal P,\overline {\cal P})$ plane. 
The centre-of-mass energy is $\sqrt{s}=1$ TeV and the production angle is fixed to $\Theta=\pi/2$. 
Top row: SM (left; outlined in black), vector (V,V) (middle), and axial–vector (A,A) (right). 
Bottom row: scalar (S) (left), scalar (P) (middle) and tensor (right).
}
\end{figure}

\begin{figure}[h!]
\centering
\fbox{\includegraphics[scale=0.335]{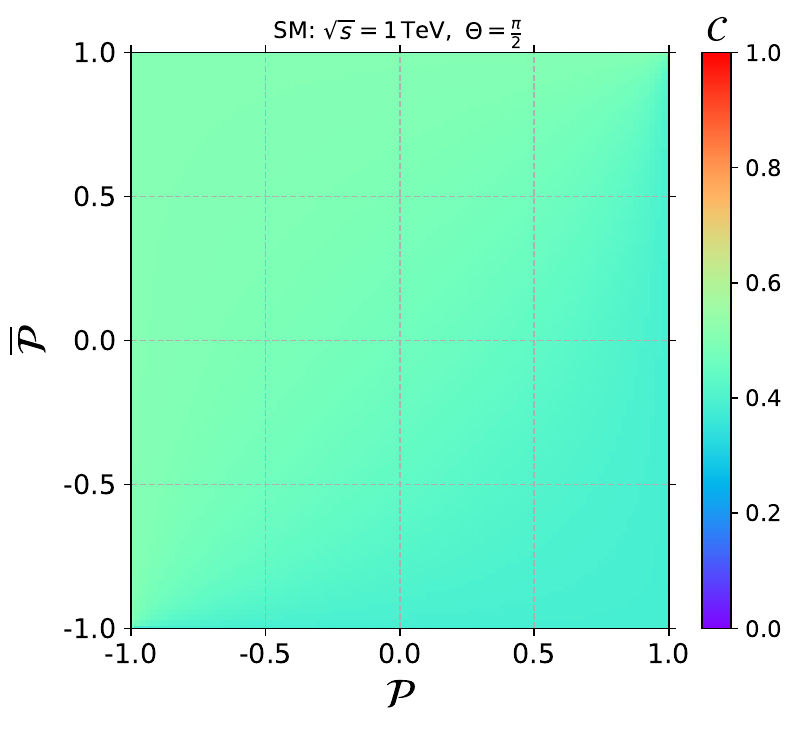}}
\includegraphics[scale=0.335]{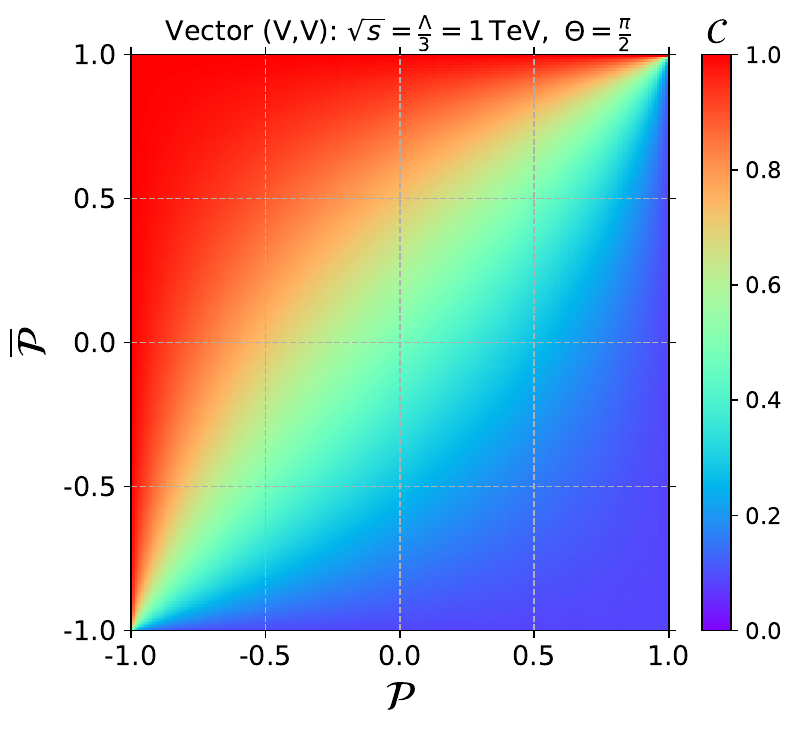}
\includegraphics[scale=0.335]{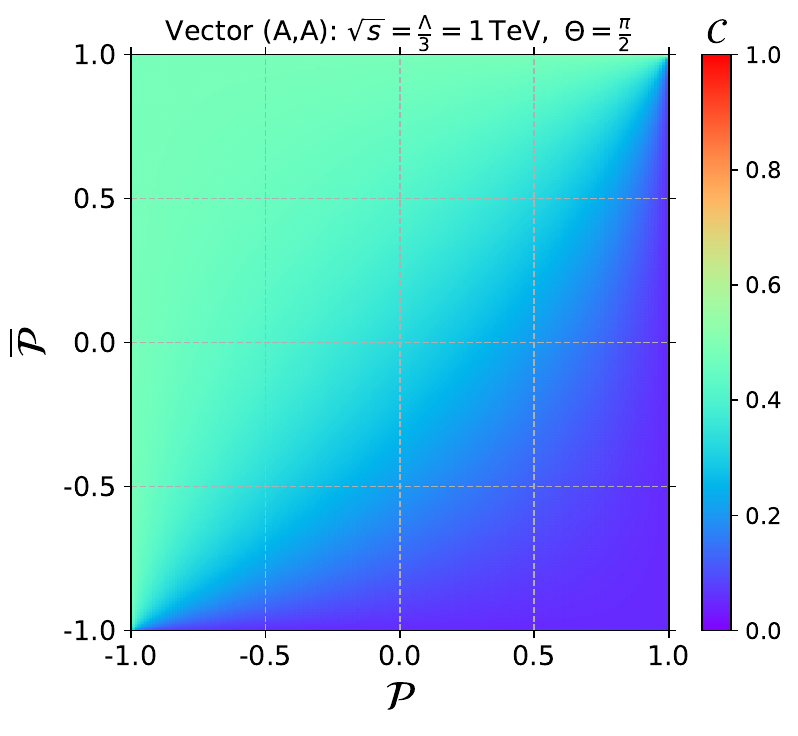}
\\
\includegraphics[scale=0.335]{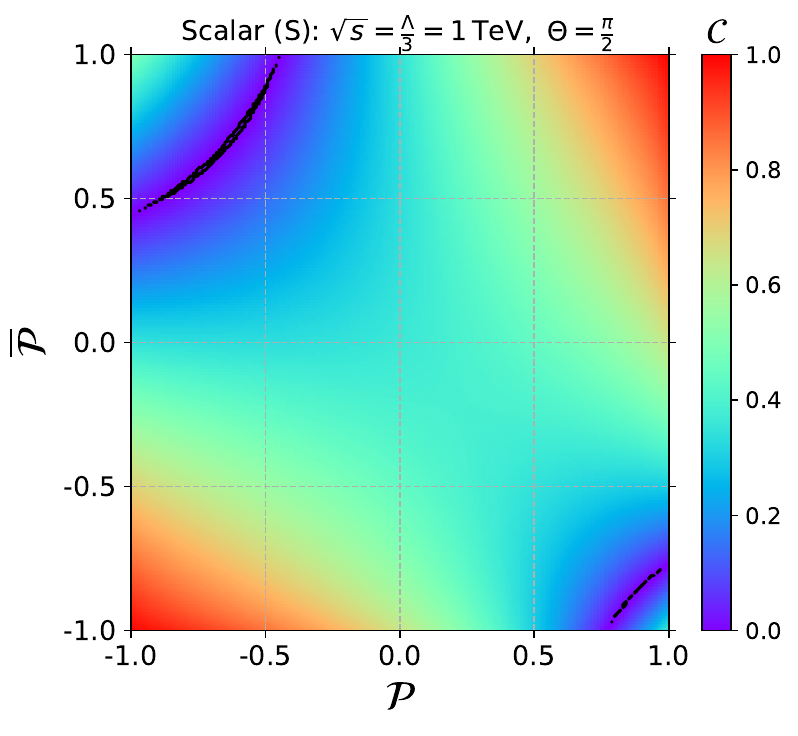}
\includegraphics[scale=0.335]{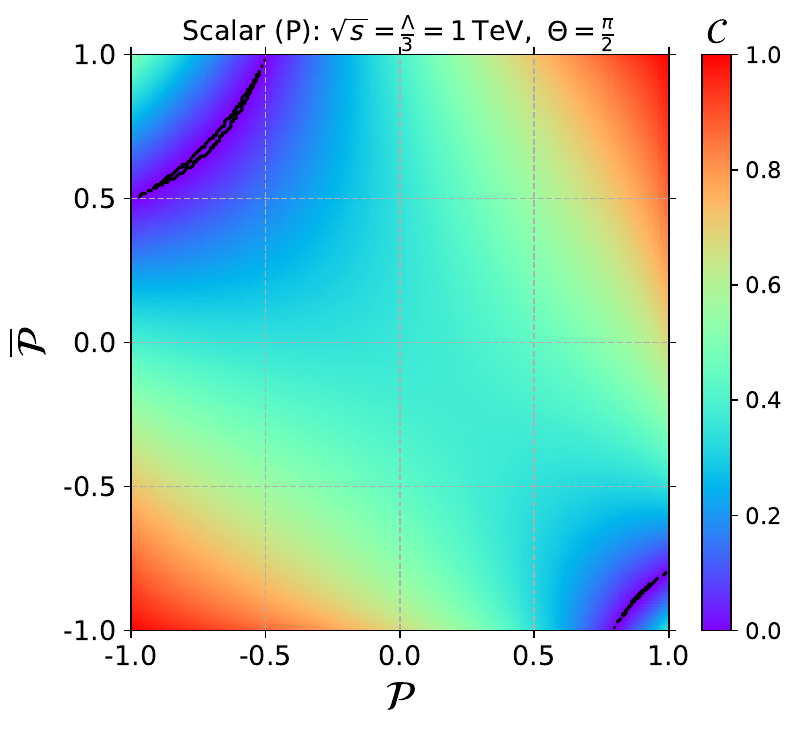}
\includegraphics[scale=0.335]{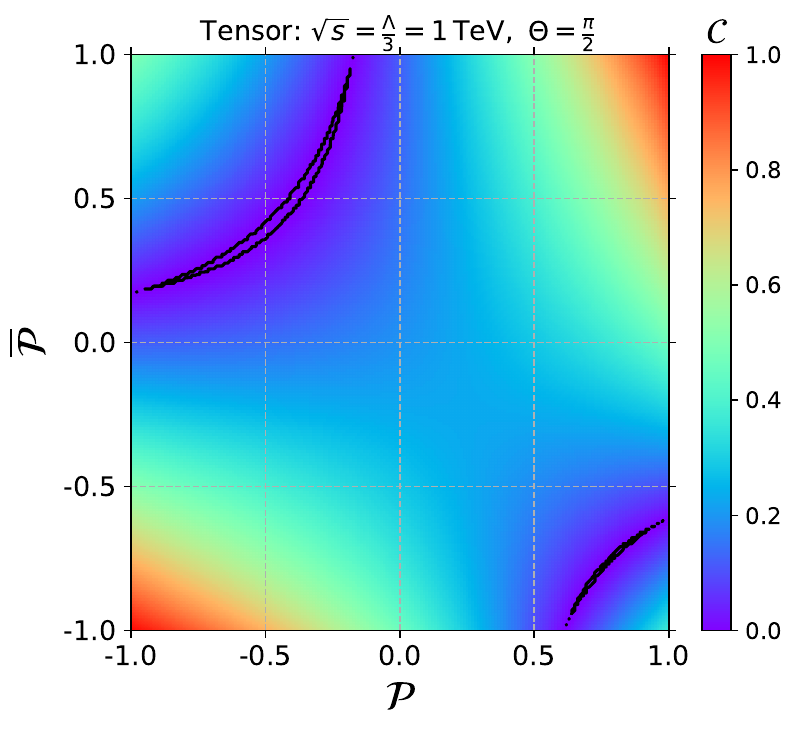}
\caption{\label{fig:conc_1000gev}
\small Concurrence ${\cal C}$ over the $(\cal P,\overline {\cal P})$ plane. 
Contours of ${\cal C}=10^{-4}$ are overlaid in black.
The centre-of-mass energy is $\sqrt{s}=1$ TeV and the production angle is fixed to $\Theta=\pi/2$. 
Top row: SM (left; outlined in black), vector (V,V) (middle), and axial–vector (A,A) (right). 
Bottom row: scalar (S) (left), scalar (P) (middle) and tensor (right).
}
\end{figure}

\begin{figure}[h!]
\centering
\fbox{\includegraphics[scale=0.335]{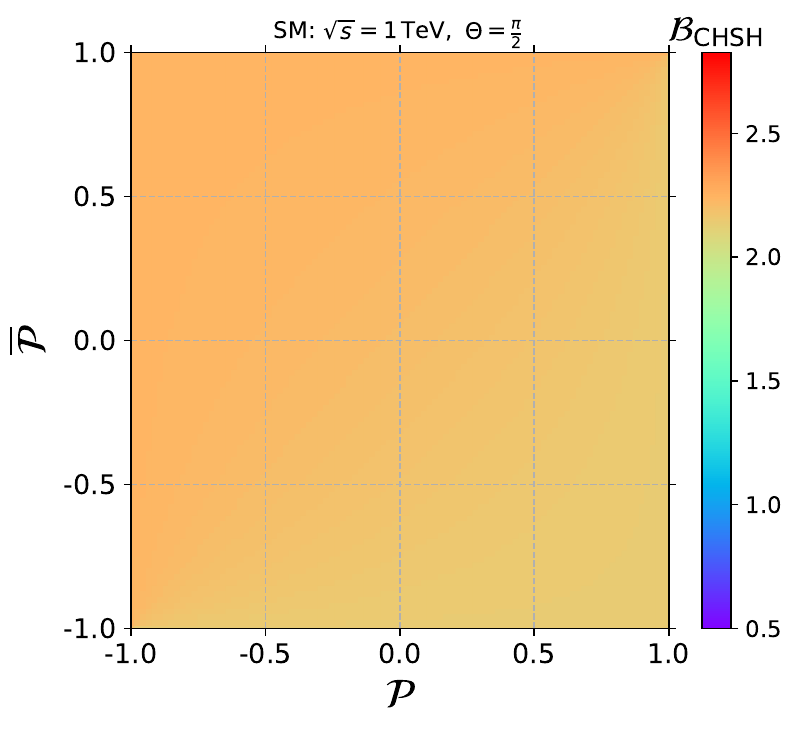}}
\includegraphics[scale=0.335]{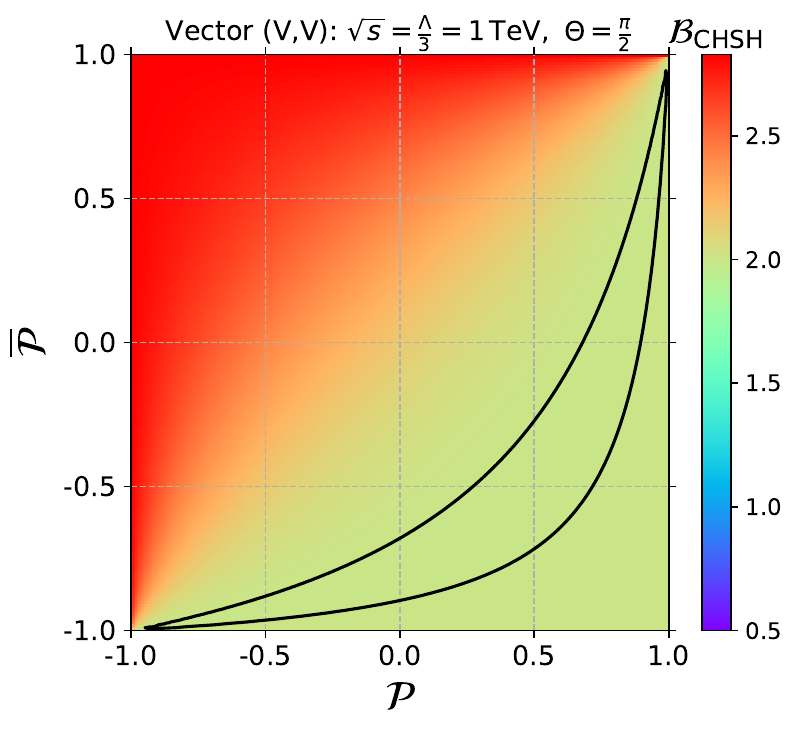}
\includegraphics[scale=0.335]{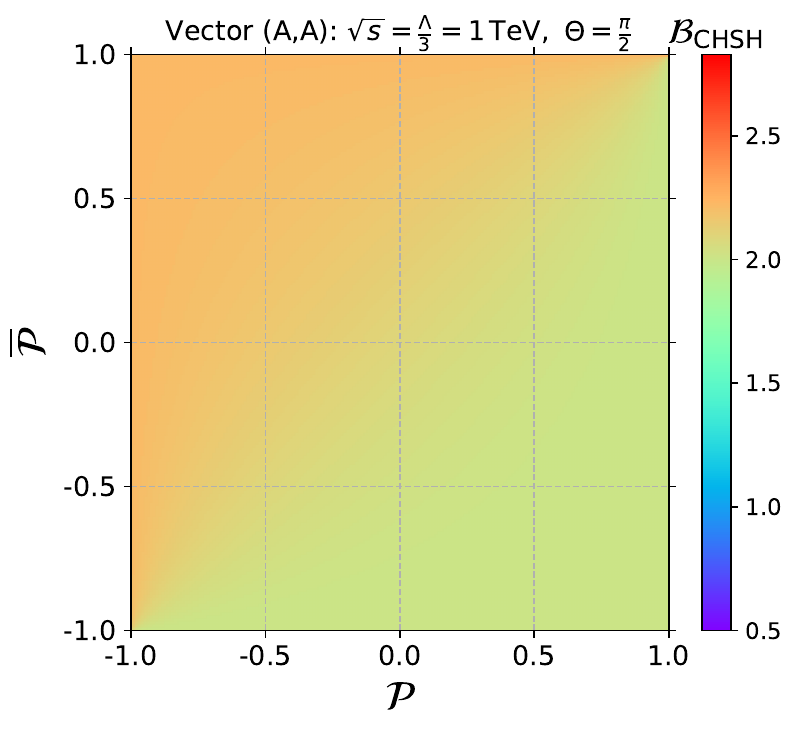}
\\
\includegraphics[scale=0.335]{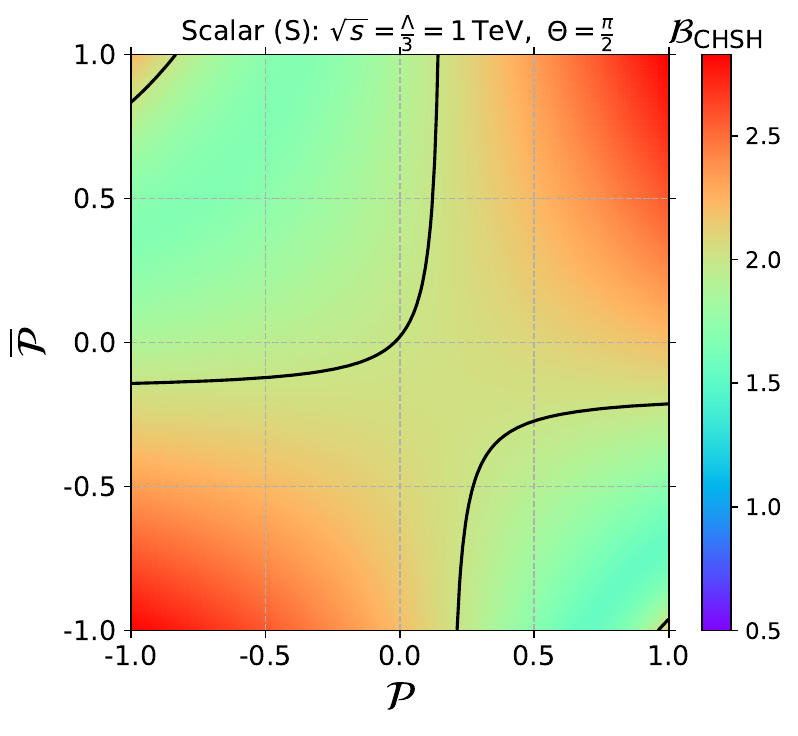}
\includegraphics[scale=0.335]{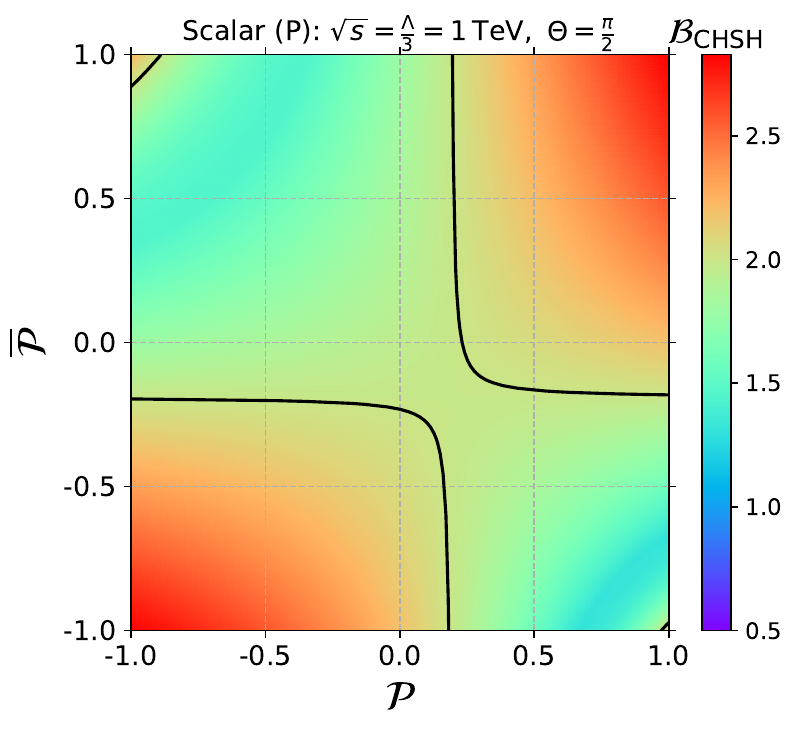}
\includegraphics[scale=0.335]{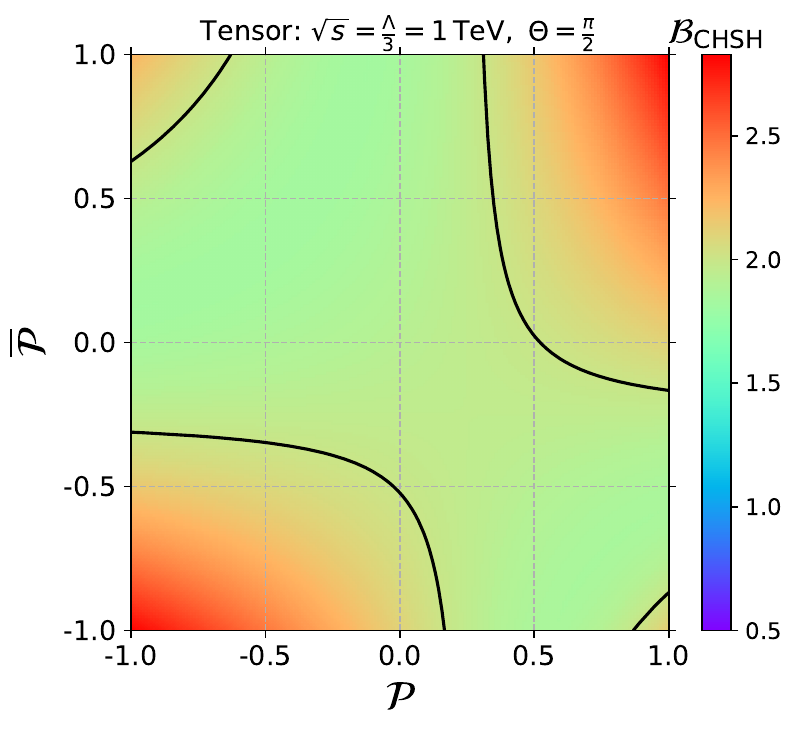}
\caption{\label{fig:chsh_1000gev}
\small Bell-CHSH observable ${\cal B}_{\rm CHSH}$ over the $(\cal P,\overline {\cal P})$ plane. 
Contours of ${\cal B}_{\rm CHSH}=2$ are overlaid in black.
The centre-of-mass energy is $\sqrt{s}=1$ TeV and the production angle is fixed to $\Theta=\pi/2$. 
Top row: SM (left; outlined in black), vector (V,V) (middle), and axial–vector (A,A) (right). 
Bottom row: scalar (S) (left), scalar (P) (middle) and tensor (right).
}
\end{figure}

\begin{figure}[h!]
\centering
\fbox{\includegraphics[scale=0.335]{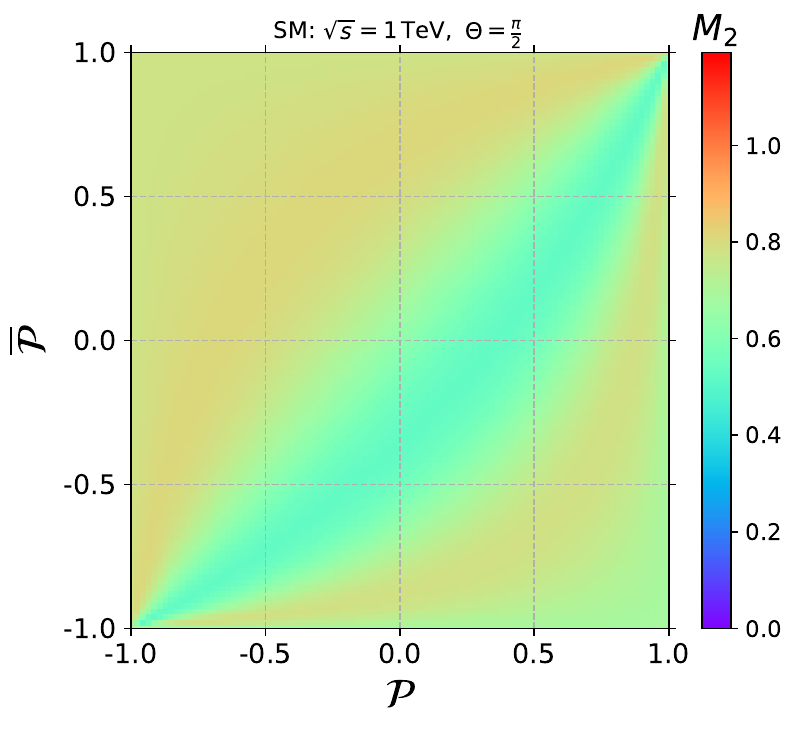}}
\includegraphics[scale=0.335]{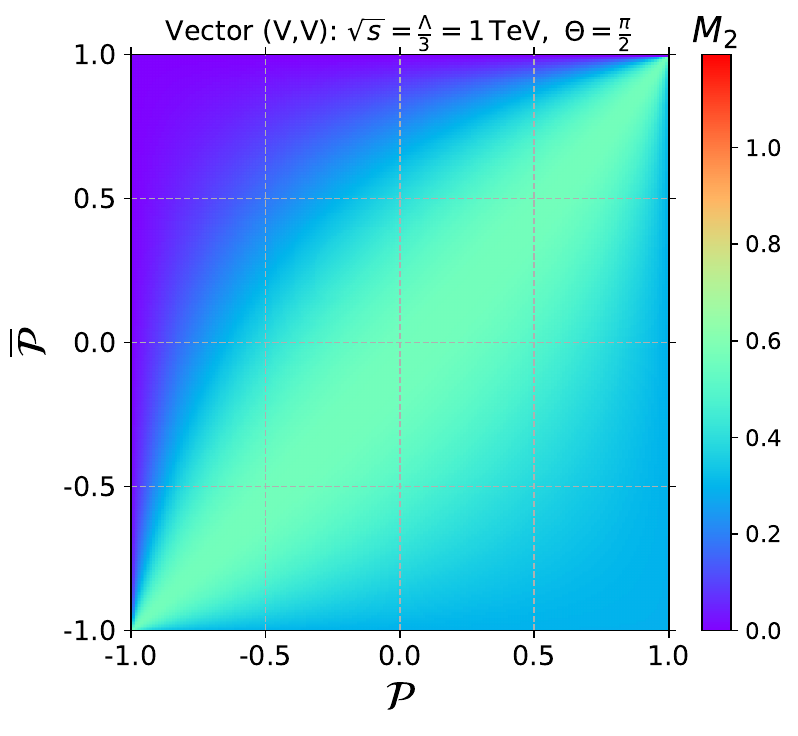}
\includegraphics[scale=0.335]{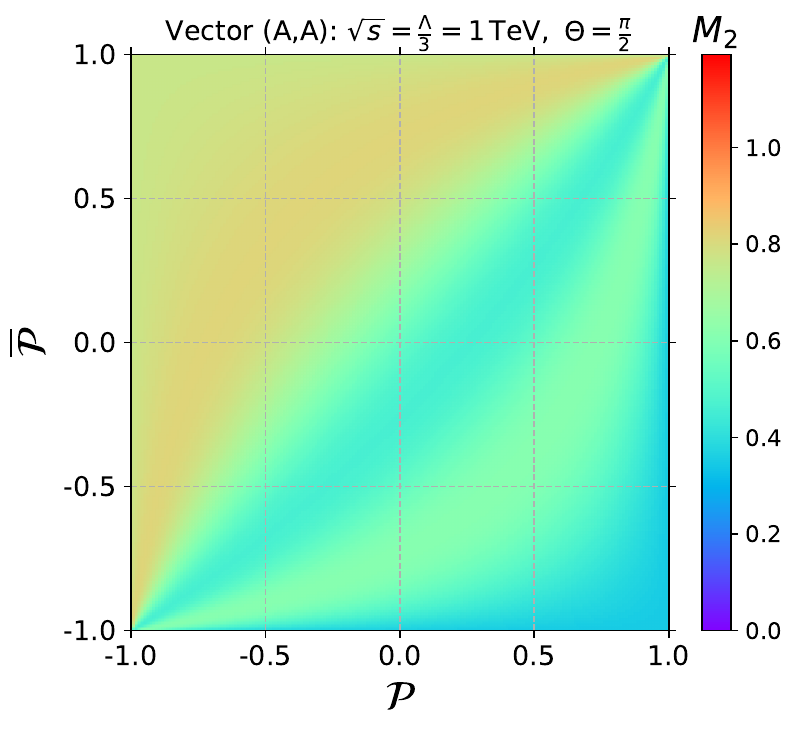}
\\
\includegraphics[scale=0.335]{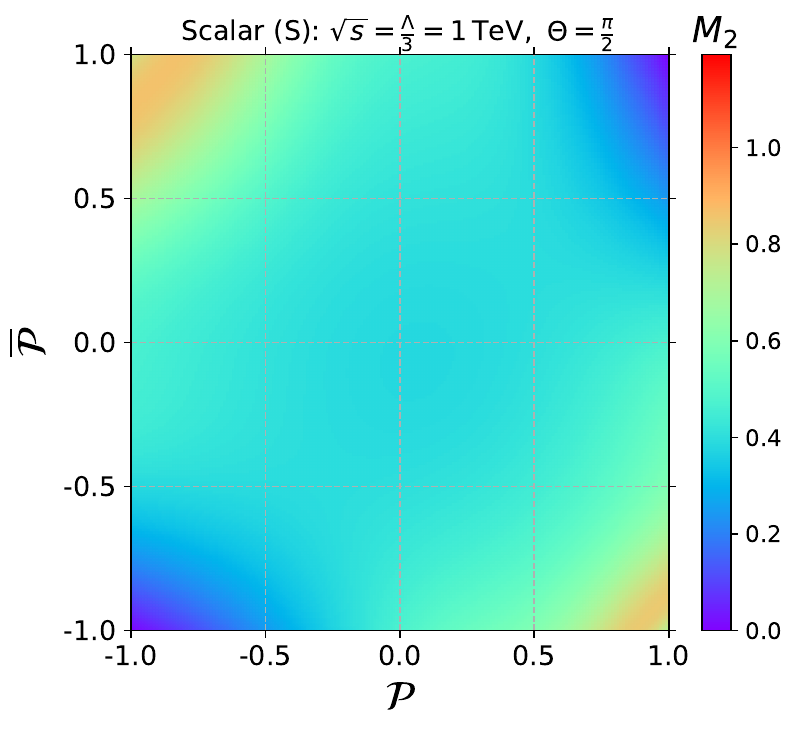}
\includegraphics[scale=0.335]{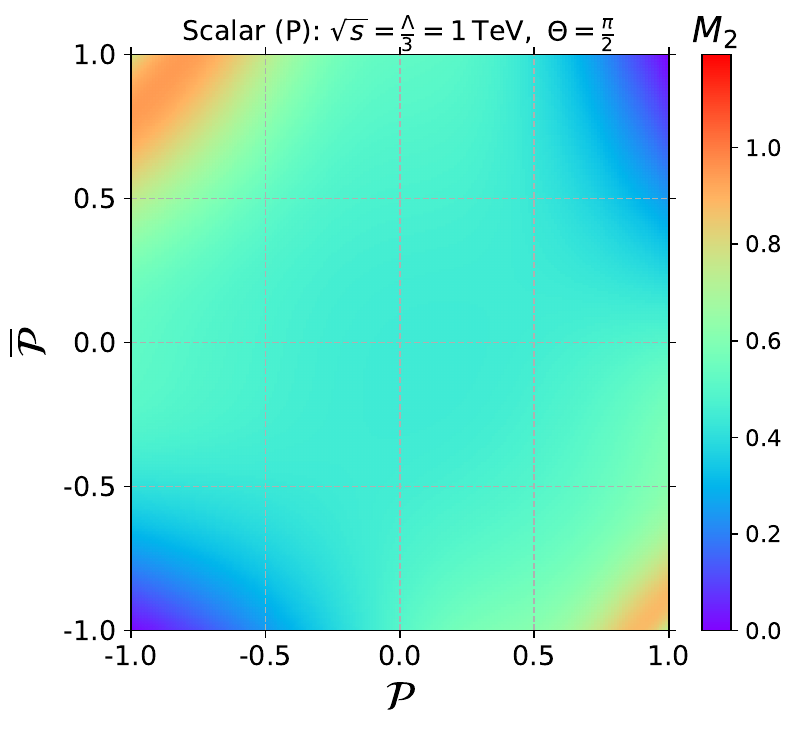}
\includegraphics[scale=0.335]{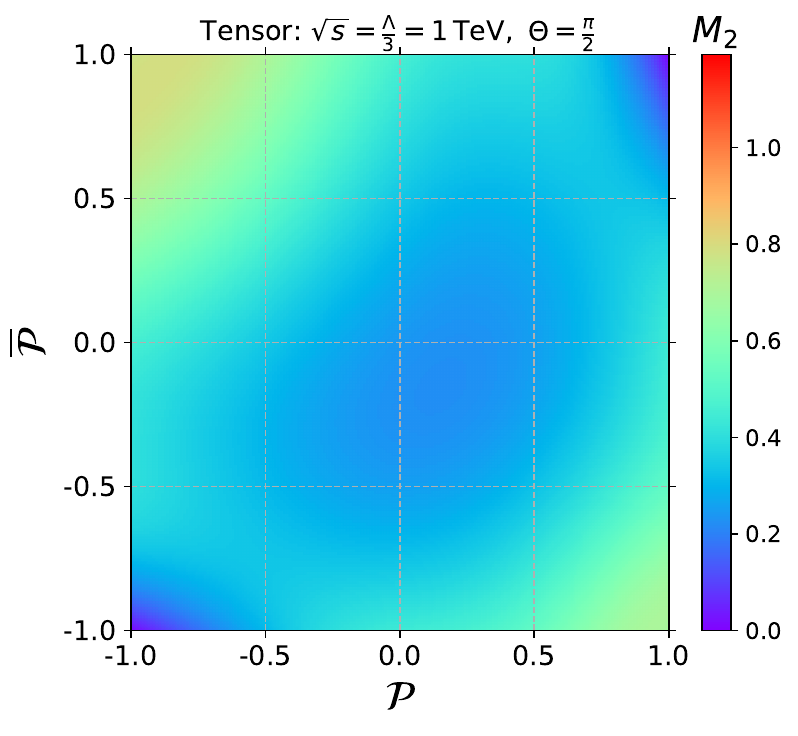}
\caption{\label{fig:M2_1000gev}
\small Helicity-basis SRE $M_2$ over the $(\cal P,\overline {\cal P})$ plane. 
The centre-of-mass energy is $\sqrt{s}=1$ TeV and the production angle is fixed to $\Theta=\pi/2$. 
Top row: SM (left; outlined in black), vector (V,V) (middle), and axial–vector (A,A) (right). 
Bottom row: scalar (S) (left), scalar (P) (middle) and tensor (right).
}
\end{figure}

\begin{figure}[h!]
\centering
\fbox{\includegraphics[scale=0.335]{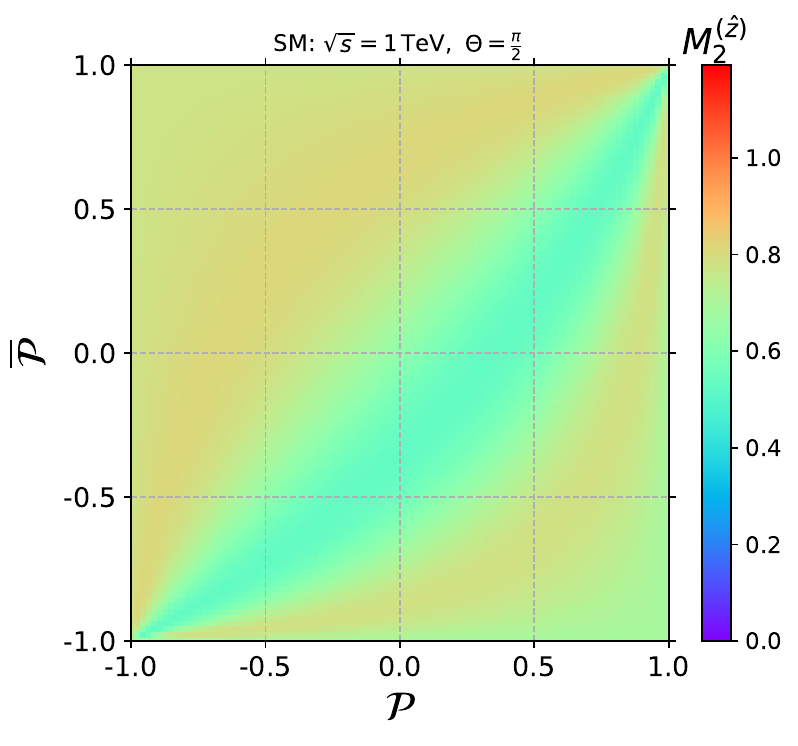}}
\includegraphics[scale=0.335]{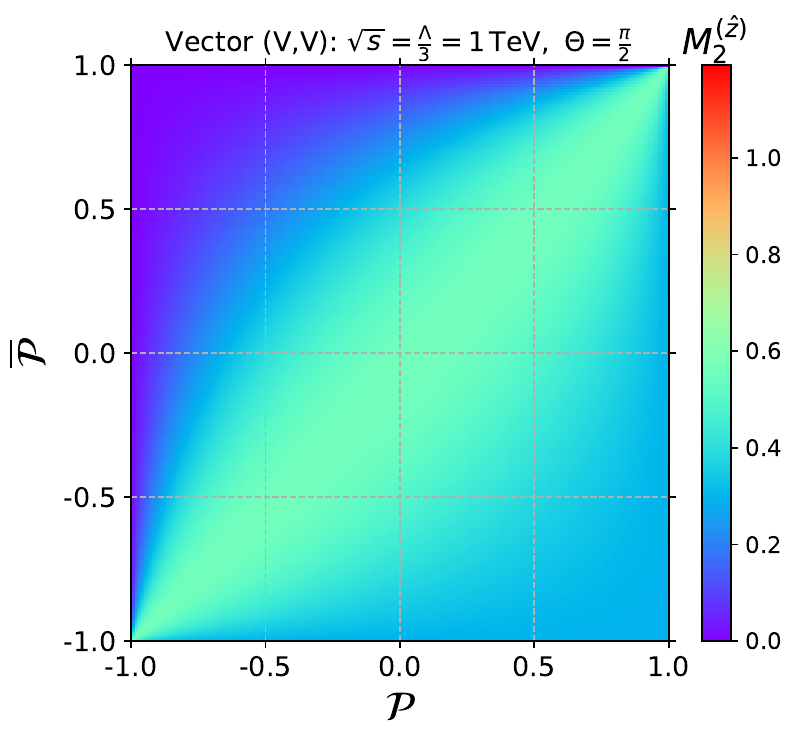}
\includegraphics[scale=0.335]{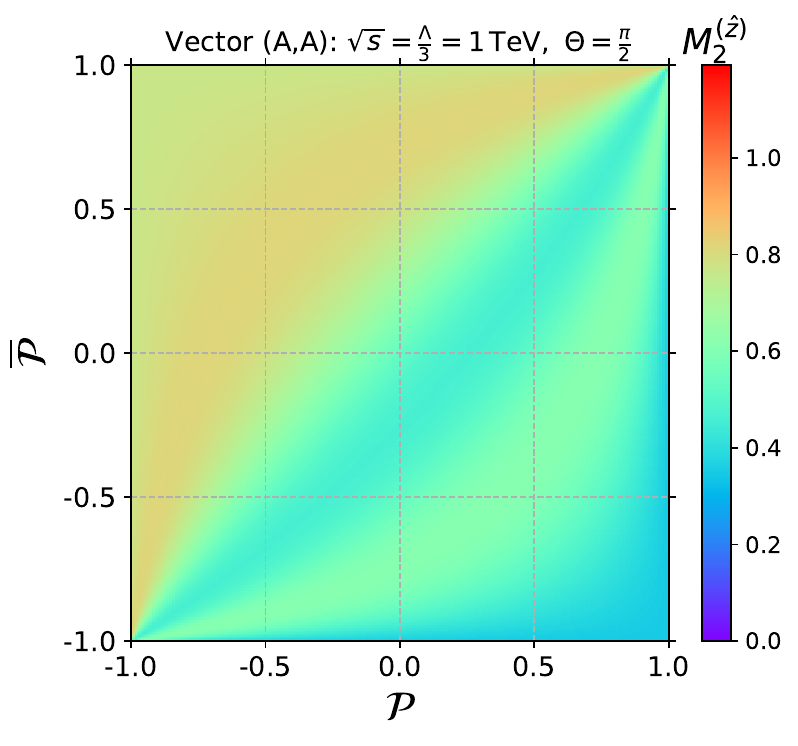}
\\
\includegraphics[scale=0.335]{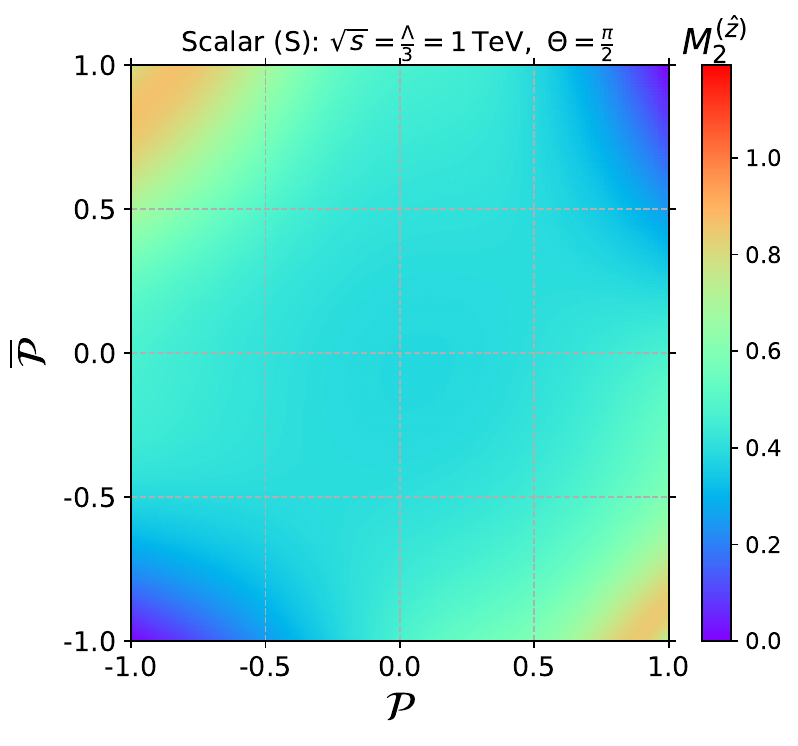}
\includegraphics[scale=0.335]{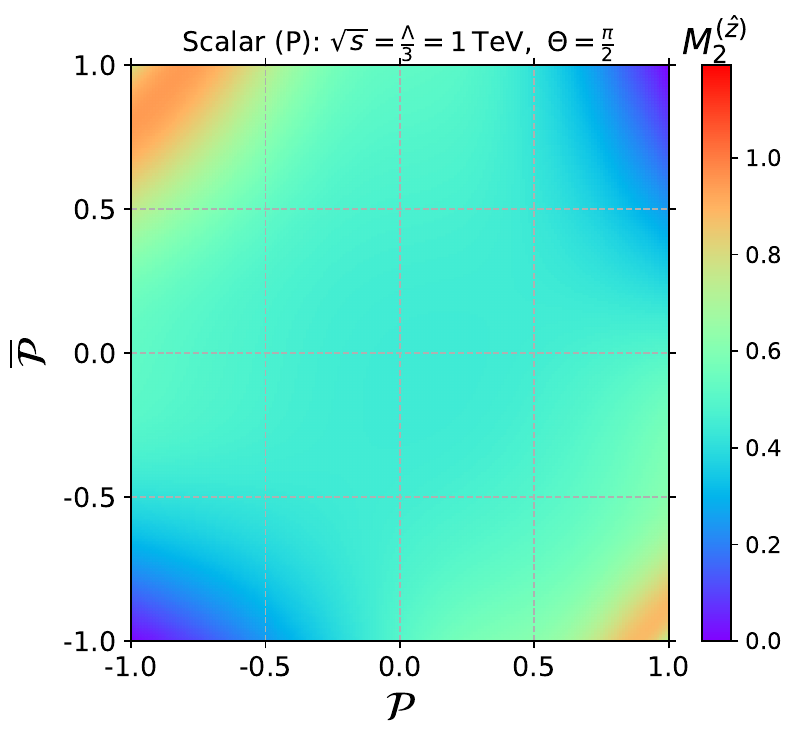}
\includegraphics[scale=0.335]{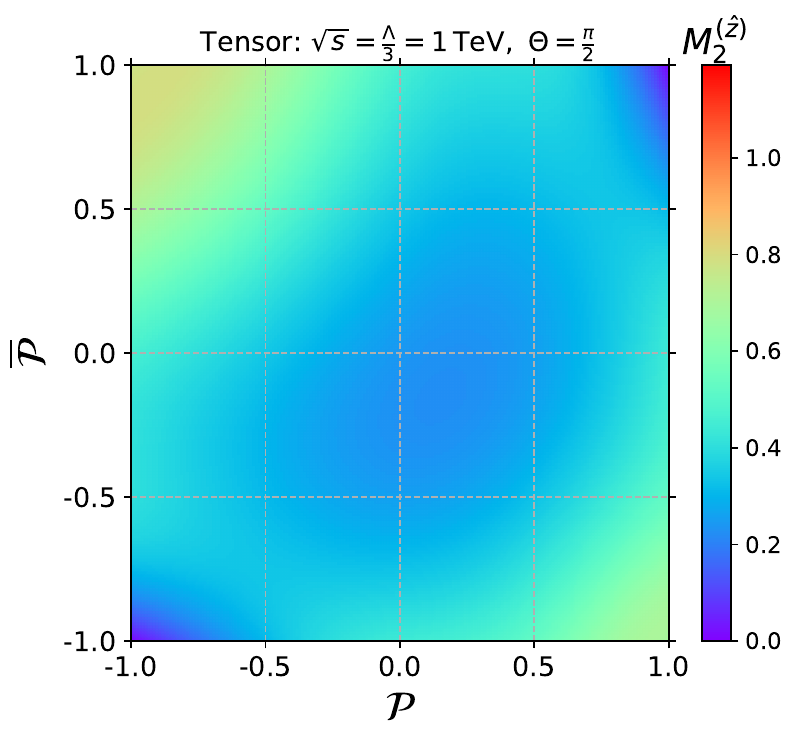}
\caption{\label{fig:M2_z_1000gev}
\small Beam-basis SRE $M_2^{(\hat z)}$ over the $(\cal P,\overline {\cal P})$ plane. 
The centre-of-mass energy is $\sqrt{s}=1$ TeV and the production angle is fixed to $\Theta=\pi/2$. 
Top row: SM (left; outlined in black), vector (V,V) (middle), and axial–vector (A,A) (right). 
Bottom row: scalar (S) (left), scalar (P) (middle) and tensor (right).
}
\end{figure}

\begin{figure}[h!]
\centering
\includegraphics[scale=0.7]{figs/legend_only}
\\
\includegraphics[scale=0.20]{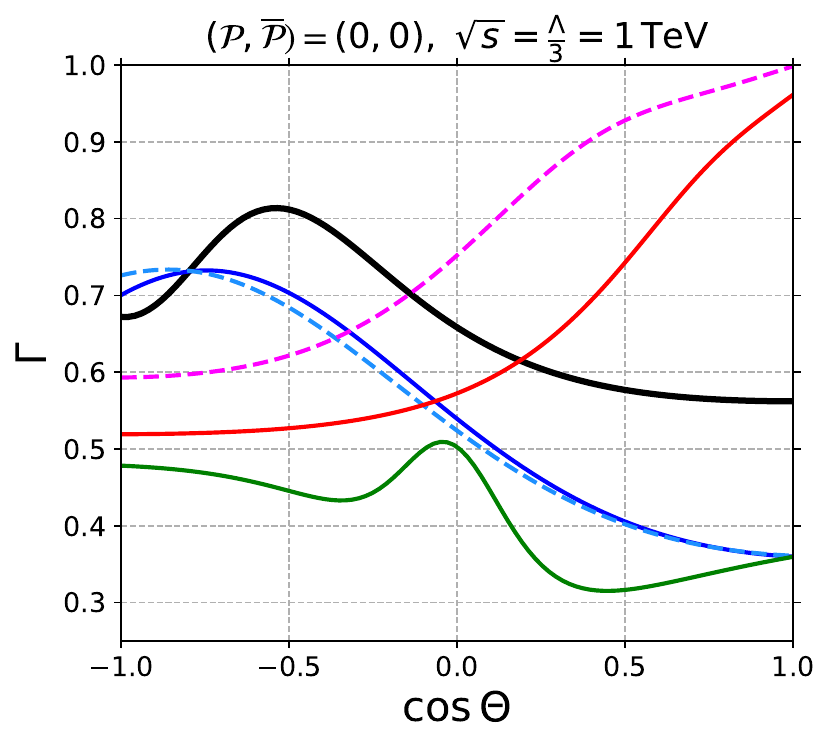}
\includegraphics[scale=0.20]{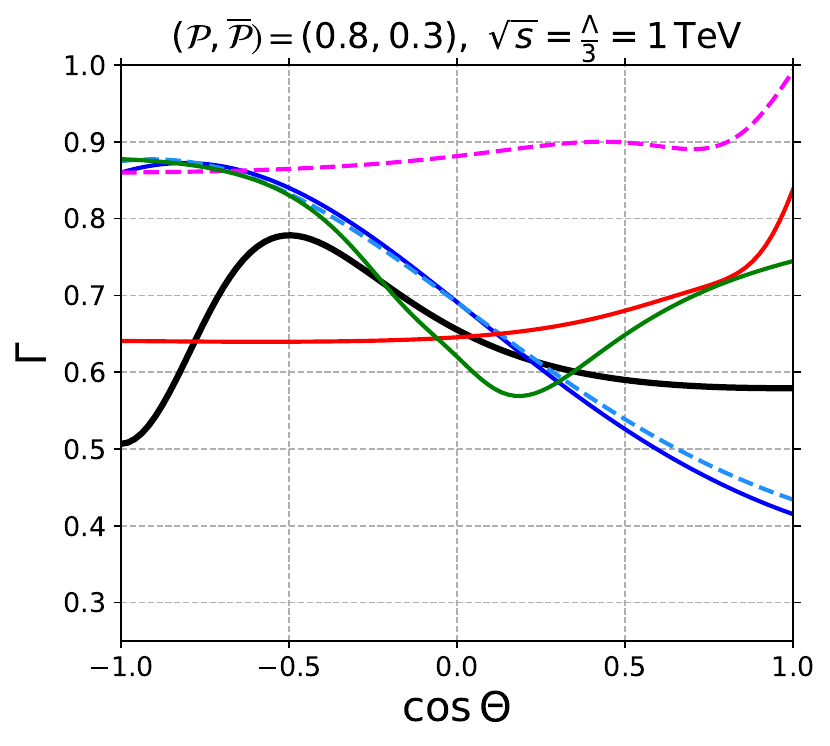}
\includegraphics[scale=0.20]{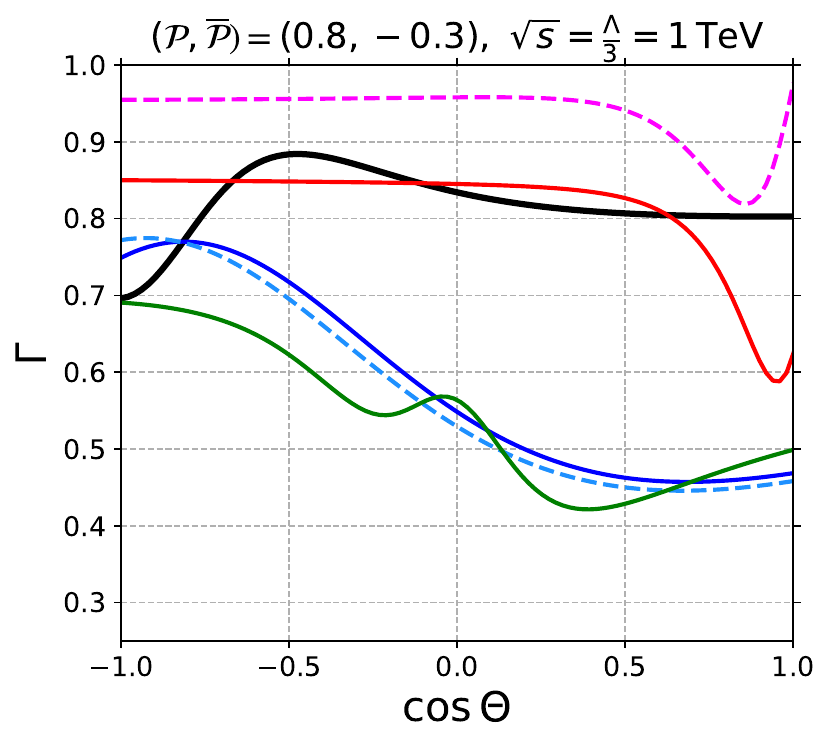}
\includegraphics[scale=0.20]{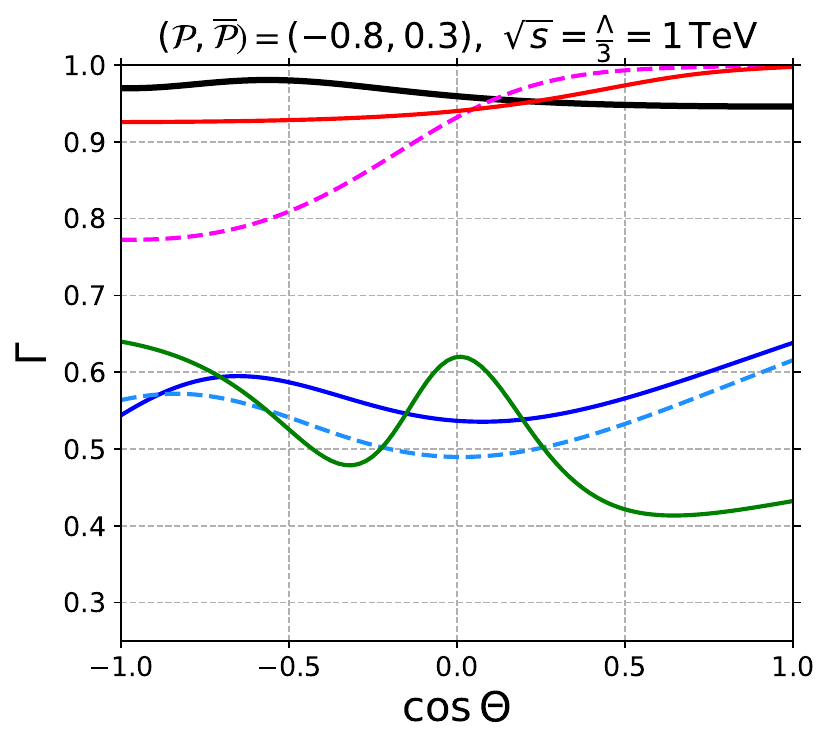}
\includegraphics[scale=0.20]{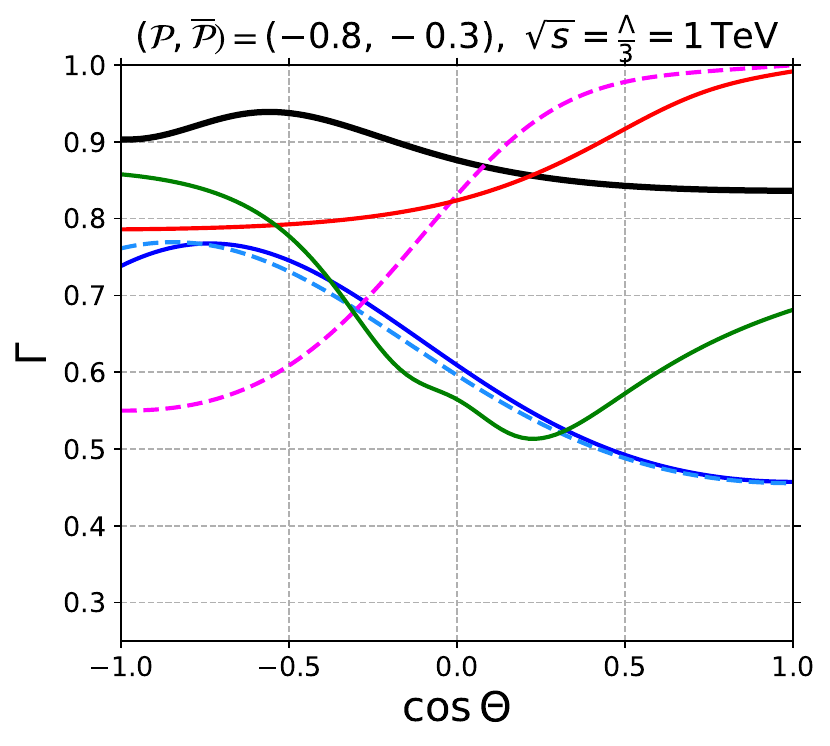}
\includegraphics[scale=0.20]{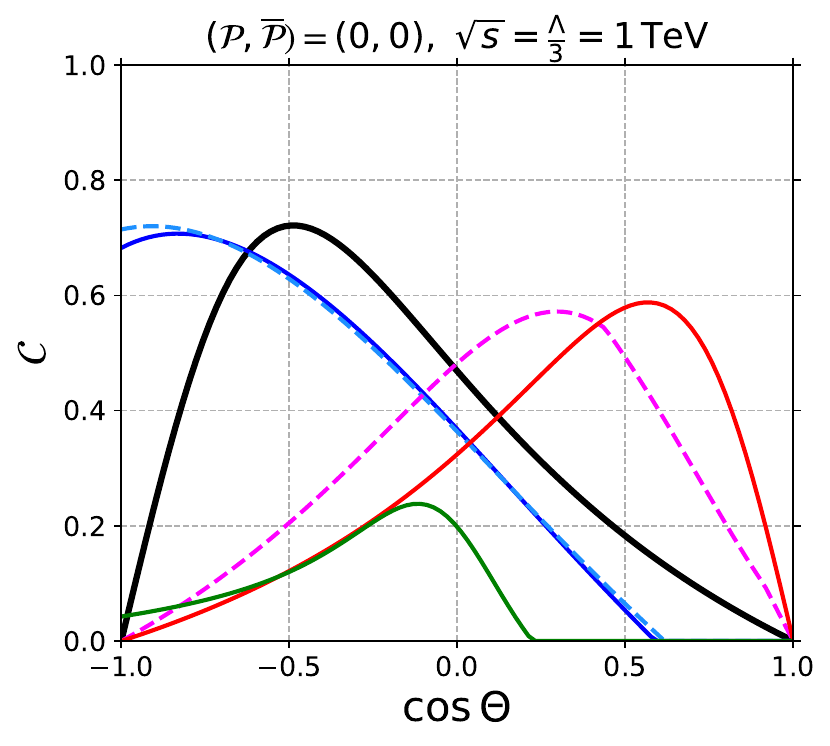}
\includegraphics[scale=0.20]{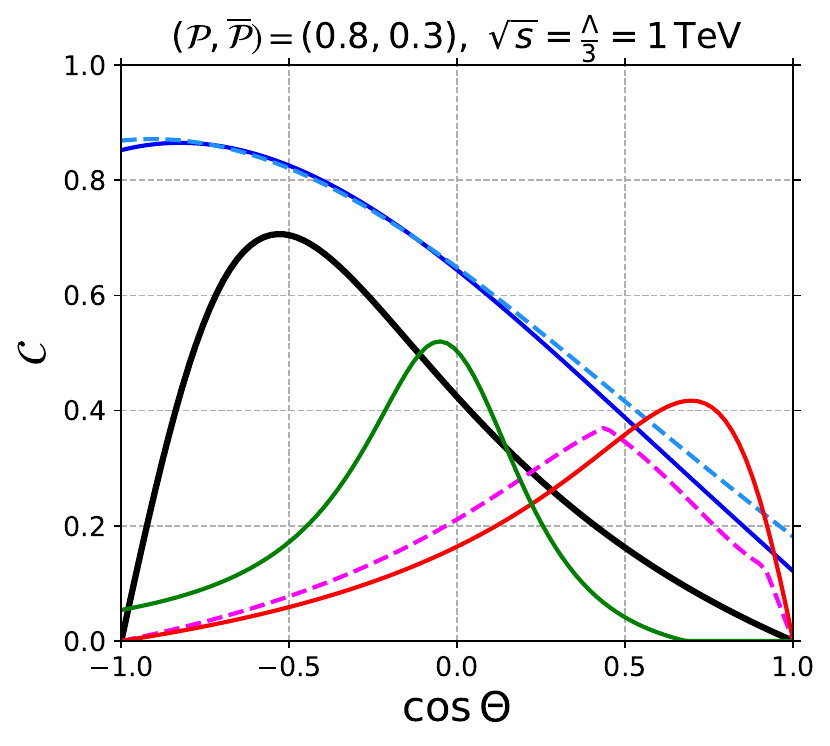}
\includegraphics[scale=0.20]{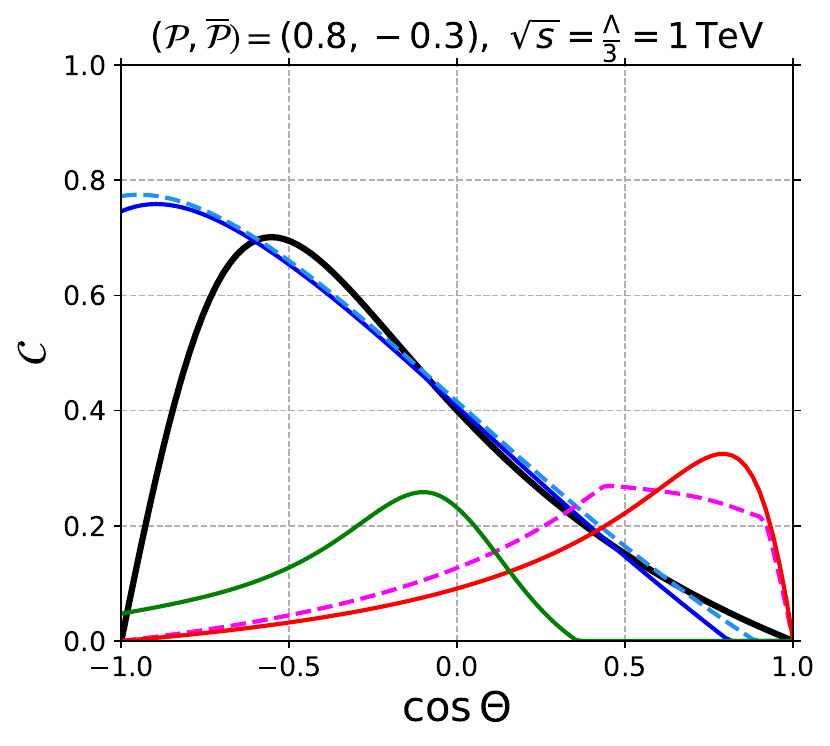}
\includegraphics[scale=0.20]{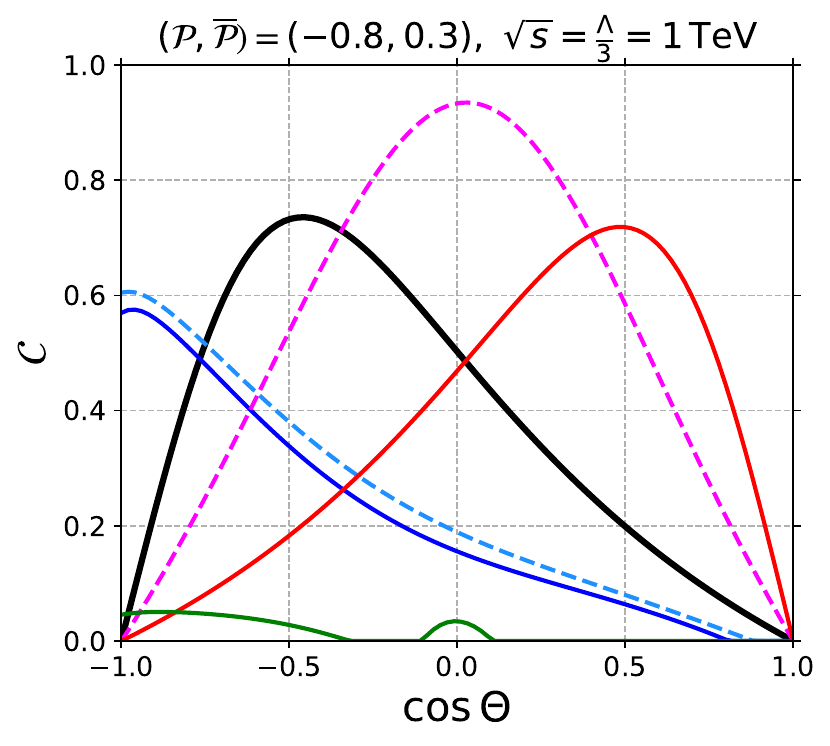}
\includegraphics[scale=0.20]{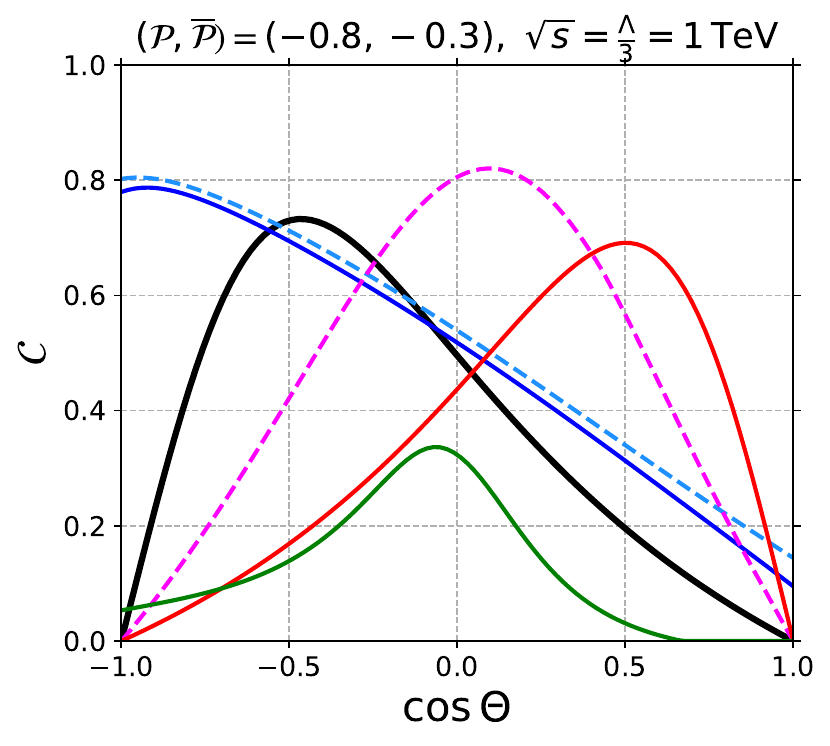}
\includegraphics[scale=0.20]{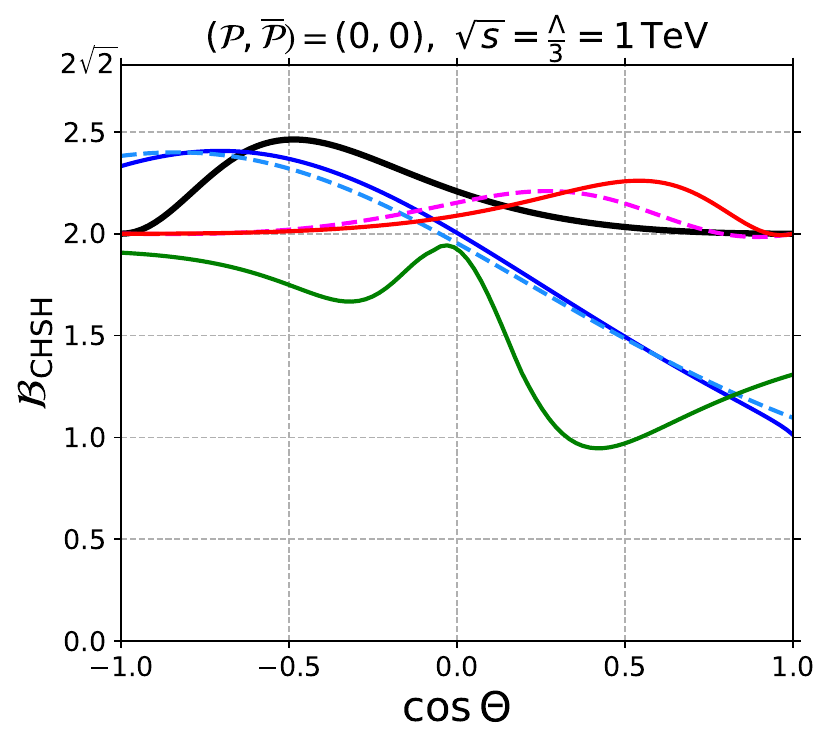}
\includegraphics[scale=0.20]{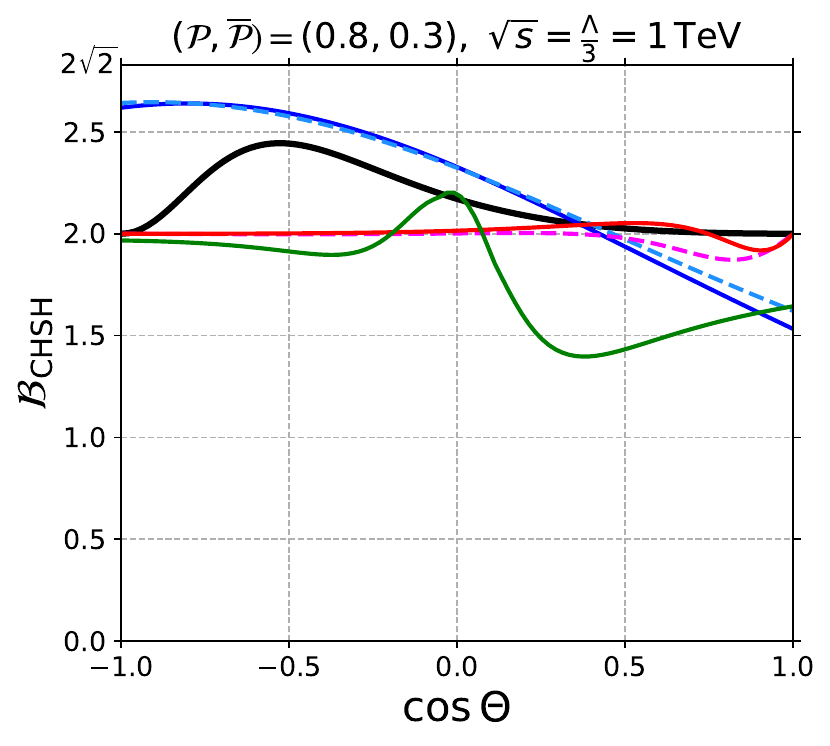}
\includegraphics[scale=0.20]{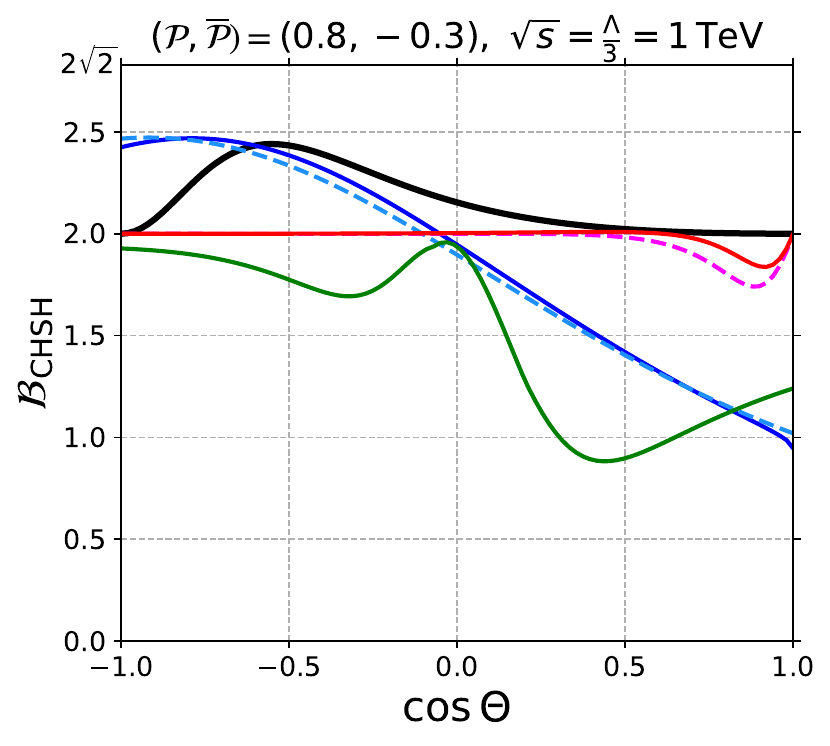}
\includegraphics[scale=0.20]{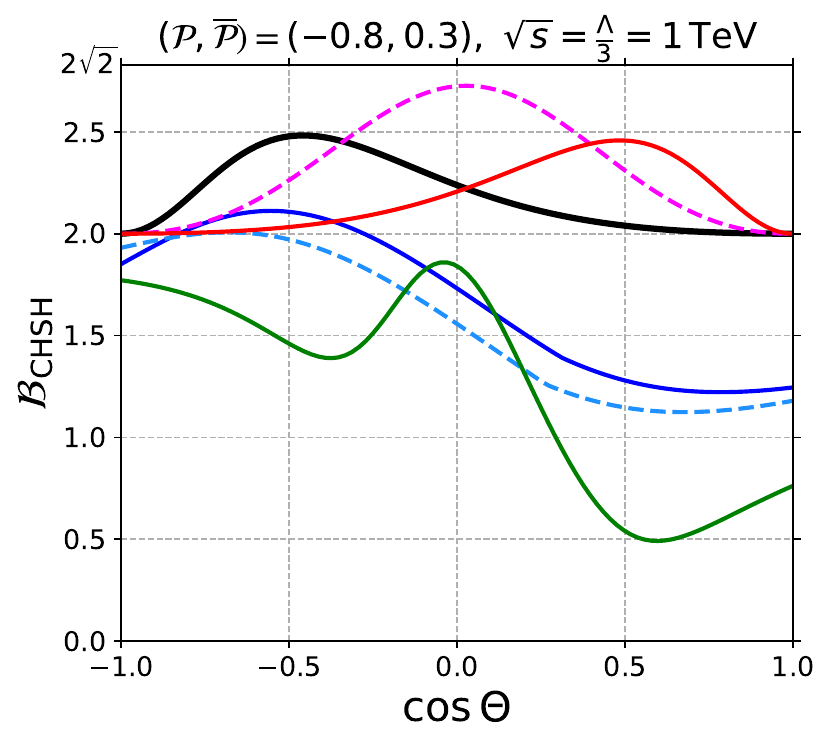}
\includegraphics[scale=0.20]{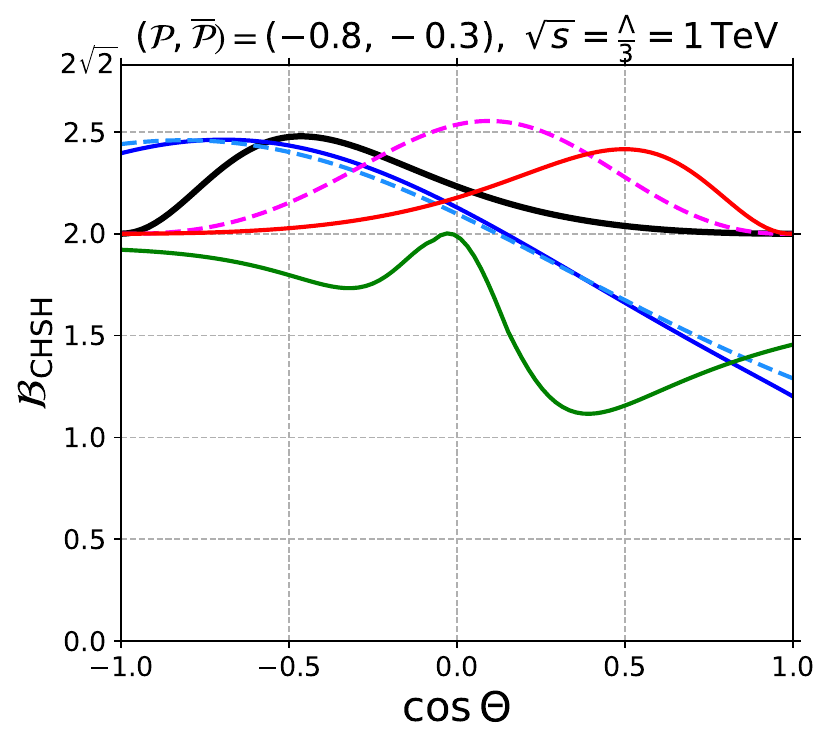}
\includegraphics[scale=0.20]{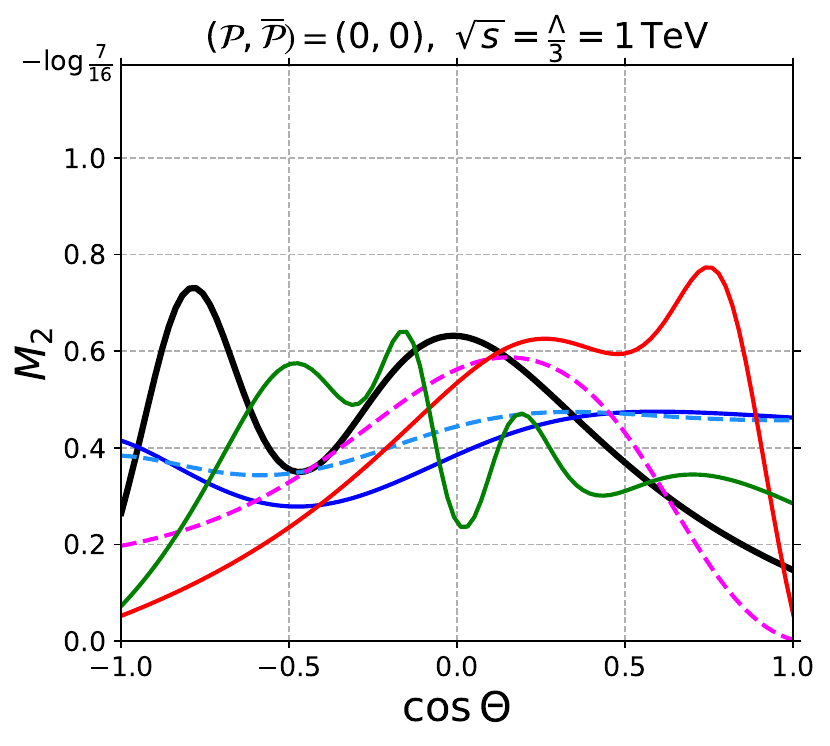}
\includegraphics[scale=0.20]{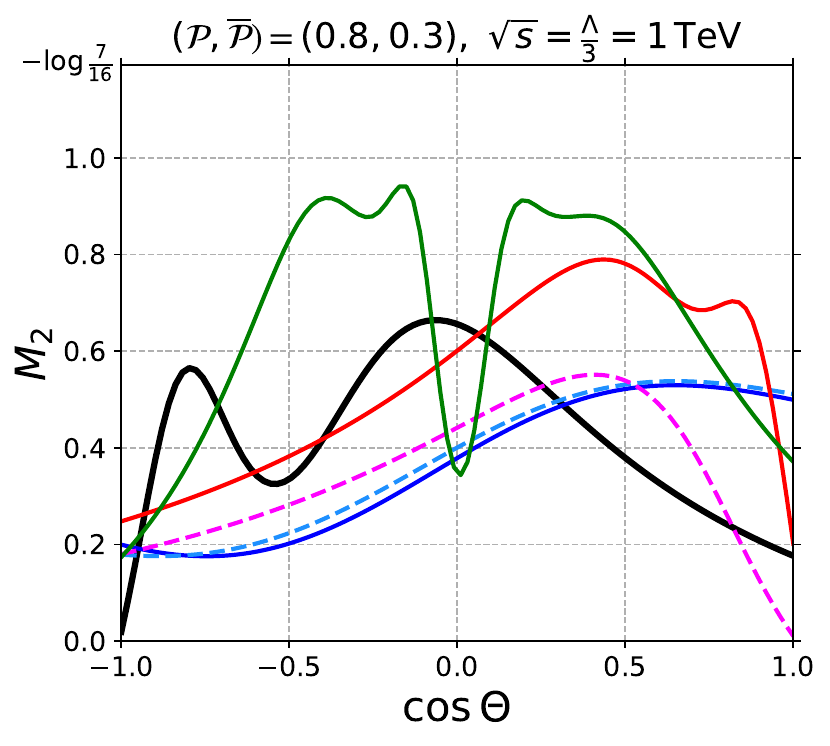}
\includegraphics[scale=0.20]{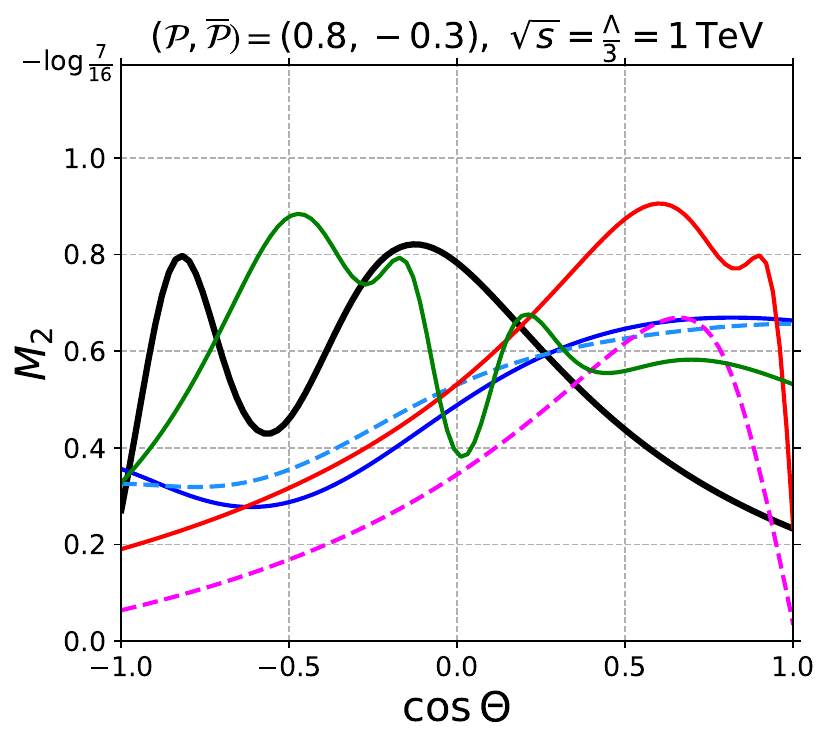}
\includegraphics[scale=0.20]{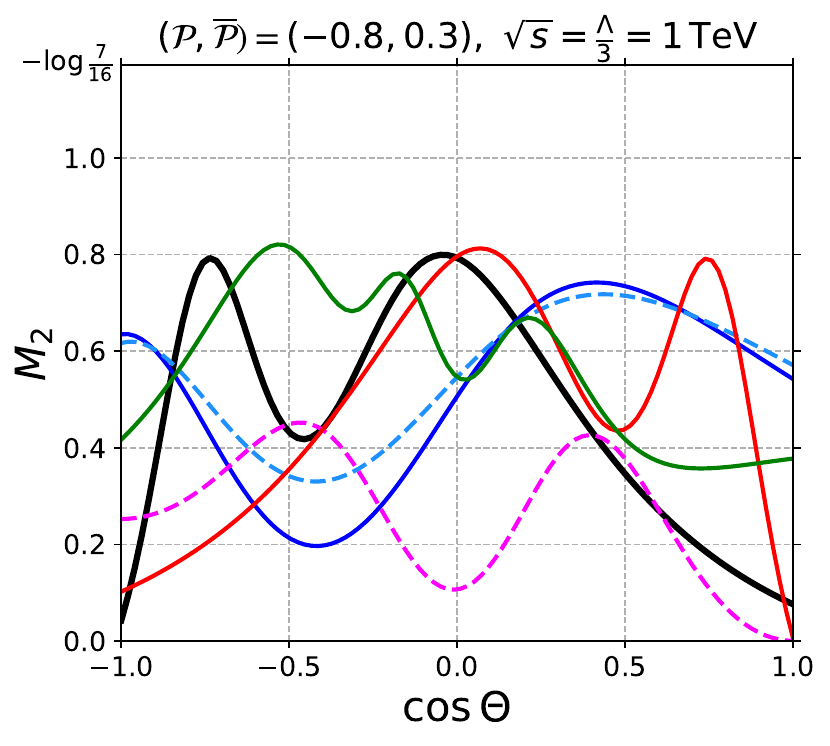}
\includegraphics[scale=0.20]{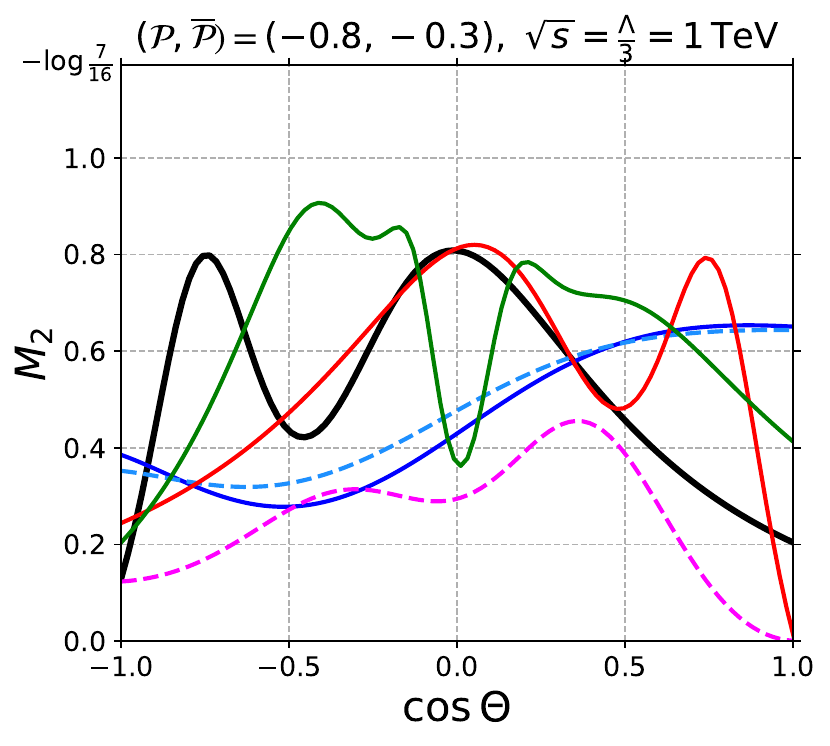}
\includegraphics[scale=0.20]{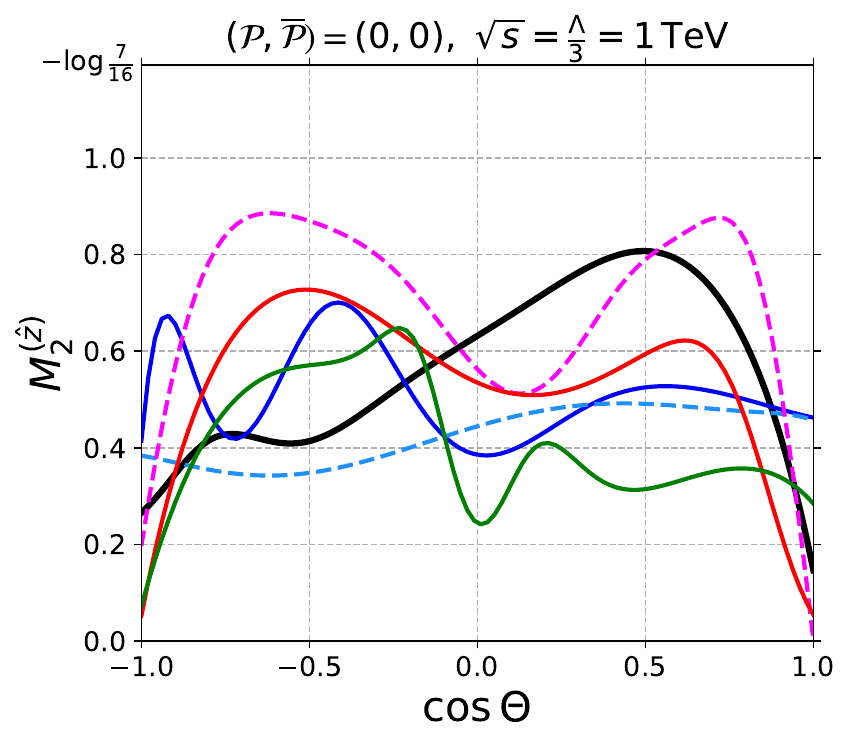}
\includegraphics[scale=0.20]{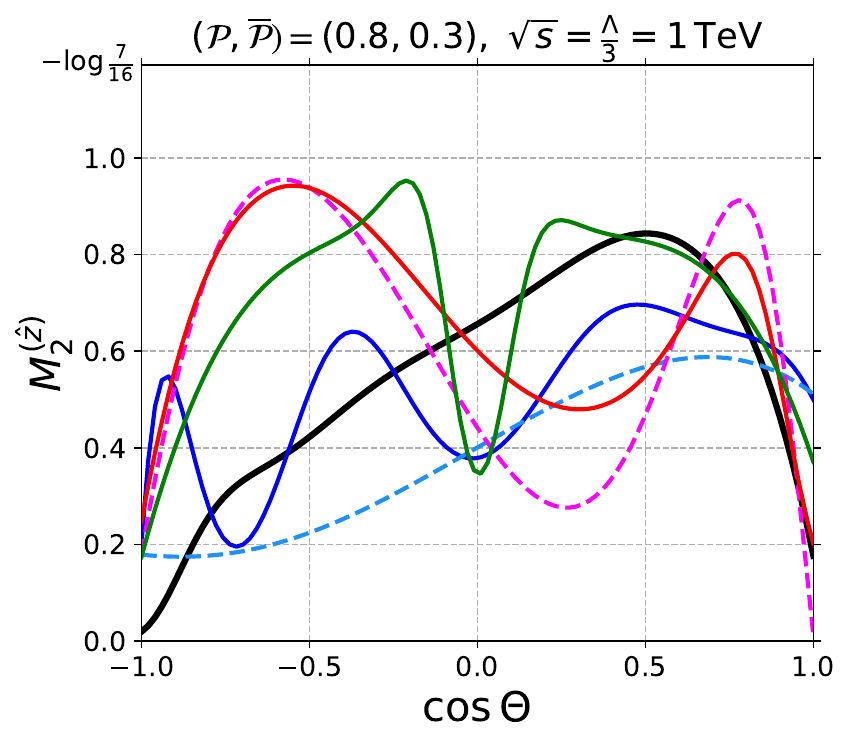}
\includegraphics[scale=0.20]{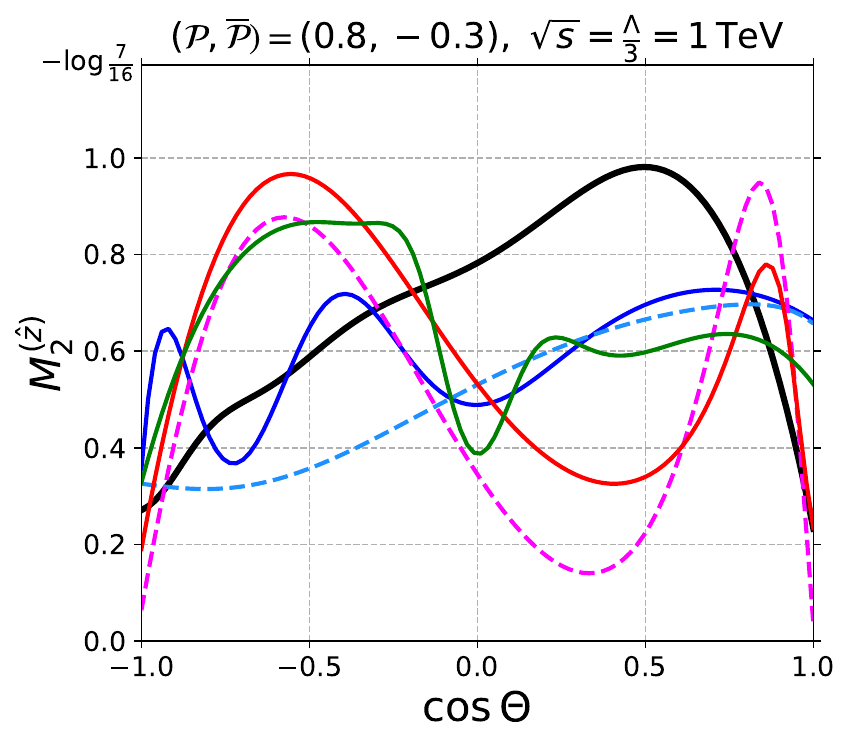}
\includegraphics[scale=0.20]{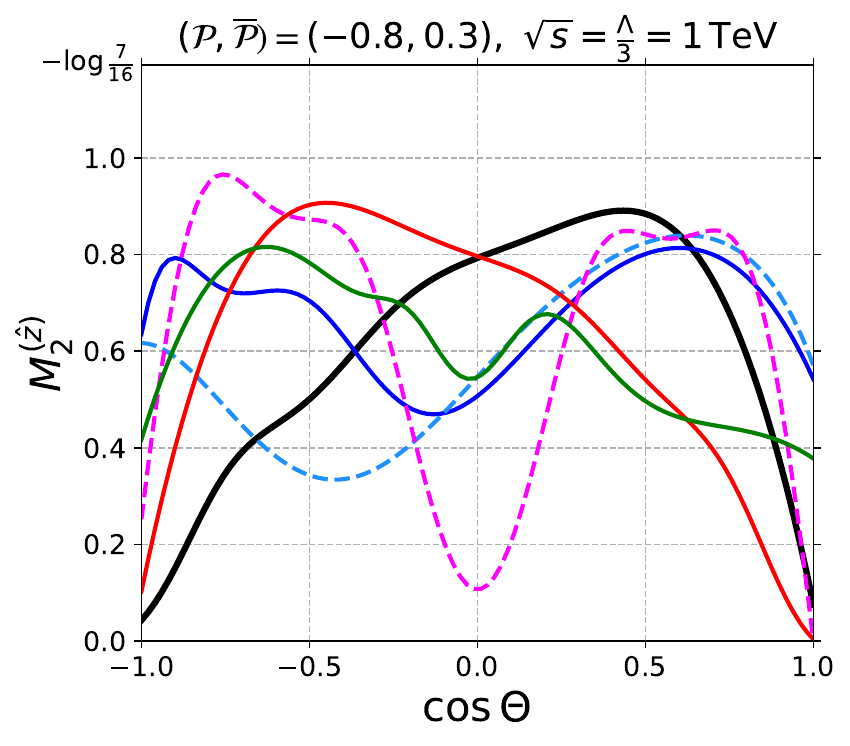}
\includegraphics[scale=0.20]{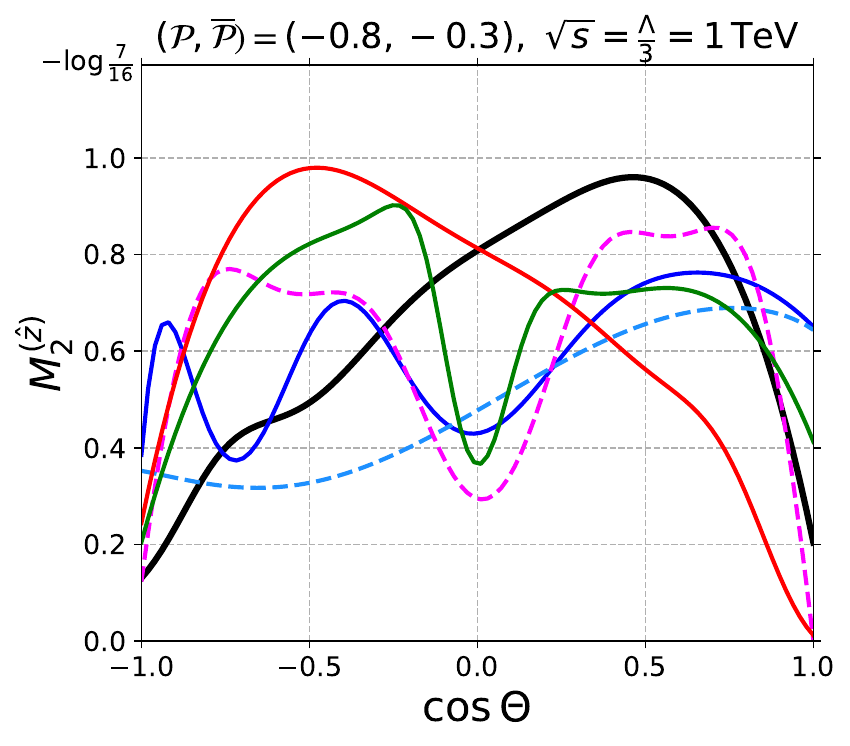}
\caption{\label{fig:1d_1000}
\small 
Purity $\Gamma$ (top row), concurrence ${\cal C}$ (second row), Bell-CHSH observable ${\cal B}_{\rm CHSH}$ (third row), helicity-basis SRE $M_2$ (fourth row) and beam-basis SRE $M_2^{(\hat z)}$ (fifth row) as functions of $\cos \Theta$.
The centre-of-mass energy and the EFT cut-off scale are fixed as
$\sqrt{s} = \Lambda/3 = 1$ TeV.
Six benchmark scenarios are shown: 
the SM (solid black), 
scalar (S) (solid blue), 
scalar (P) (dashed light blue), 
vector (V,V) (solid red),
vector (A,A) (dashed magenta)
and tensor (solid green).
The leftmost column shows the unpolarised configuration; from left to right, the remaining columns display the polarised settings PP, PN, NP and NN, respectively.
}
\end{figure}

\begin{itemize}

\item

Fig.\ \ref{fig:pure_1000gev}:
Purity $\Gamma$ over the $(\cal P,\overline {\cal P})$ plane ($\Theta = \frac{\pi}{2}$).

Fig.\ \ref{fig:conc_1000gev}:
Concurrence ${\cal C}$ over the $(\cal P,\overline {\cal P})$ plane ($\Theta = \frac{\pi}{2}$).

Fig.\ \ref{fig:chsh_1000gev}:
Bell-CHSH observable ${\cal B}_{\rm CHSH}$ over the $(\cal P,\overline {\cal P})$ plane ($\Theta = \frac{\pi}{2}$).

Fig.\ \ref{fig:M2_1000gev}: Helicity-basis SRE $M_2$ over the $(\cal P,\overline {\cal P})$ plane ($\Theta = \frac{\pi}{2}$).

Fig.\ \ref{fig:M2_z_1000gev}: Beam-basis SRE $M_2^{(\hat z)}$ over the $(\cal P,\overline {\cal P})$ plane ($\Theta = \frac{\pi}{2}$).

\item
Fig.\ \ref{fig:1d_1000}:
Purity $\Gamma$, concurrence ${\cal C}$, Bell-CHSH observable ${\cal B}_{\rm CHSH}$, helicity-basis SRE $M_2$ and beam-basis SRE $M_2^{(\hat z)}$ as functions of $\cos \Theta$.
Six benchmark scenarios are shown: 
the SM, 
scalar (S) and (P), 
vector (V,V) and (A,A), 
and tensor.
The left, middle and right columns represent the unpolarised, polarised-NP and polarised-NN beam configurations, respectively. 

\end{itemize}

\newpage

\section{Validity and truncation of the EFT}
\label{app:truncation}

In this appendix we discuss the domain of validity of the EFT expansion employed throughout the paper. Our discussion is intentionally brief, and we refer the reader to Refs.~\cite{Aguilar-Saavedra:2018ksv, Brivio:2022pyi} for a more comprehensive treatment.

For the process $e^{+}e^{-}\rightarrow t\overline{t}$ at tree level, scalar and tensor four-fermion operators do not interfere with the SM amplitudes, while vector and axial-vector operators generate non-vanishing interference terms. For this reason, the discussion below focuses on vector-type interactions.

The unnormalised spin density matrix of the $t\overline{t}$ pair can be expanded as
\begin{equation}
    R\,=\,R^{(0)}+\Lambda^{-2}R^{(1)}+\Lambda^{-4}R^{(2)},
    \label{eq:rho_expansion}
\end{equation}
where $R^{(0)}$ denotes the pure SM contribution, $R^{(1)}$ the SM--EFT interference term, and $R^{(2)}$ the contribution quadratic in the dimension-6 EFT amplitudes. For notational simplicity, in this appendix we set the relevant Wilson coefficient(s) to one; restoring them amounts to the replacement $R^{(1)}\to C\,R^{(1)}$ and $R^{(2)}\to C^{2}R^{(2)}$ (or the corresponding generalisation for multiple coefficients).

In the phenomenological studies presented in this paper we have retained all terms in Eq.~\eqref{eq:rho_expansion}. This guarantees a well-defined density matrix for any $\Lambda$, but it also keeps $\mathcal{O}(1/\Lambda^{4})$ contributions which are formally beyond the accuracy of a strict truncation at $\mathcal{O}(1/\Lambda^{2})$. It is therefore customary to assess the practical impact of these terms by inspecting the behaviour of the expansion in representative observables. The guiding idea is that, if the EFT series is well behaved and there is no reason for the interference term to be parametrically suppressed,\footnote{This may happen in notable cases. For example, selection rules can suppress the interference in double vector-boson production~\cite{Azatov:2016sqh}, and in processes where the SM amplitudes are themselves suppressed (e.g.\ Flavour Changing Neutral Currents (FCNC) or Charged Lepton Flavour Violation (CLFV)) the squared EFT contribution can dominate without implying a breakdown of the expansion: in such situations the SM--dimension-8 interference is also suppressed, so the dimension-6 squared term can remain the leading beyond-SM effect.} then including or dropping the quadratic terms should not qualitatively change the conclusions.

\paragraph{Linear truncation and positivity of $\rho$.}
As a first diagnostic we construct the linearly truncated density matrix by discarding the $\mathcal{O}(1/\Lambda^{4})$ term and normalising by the trace to enforce the unit-trace condition. The resulting $\mathcal{O}(1/\Lambda^{2})$ density matrix is
\begin{equation}
    \rho^{lin}=\frac{R^{(0)}+\Lambda^{-2}R^{(1)}}{\sigma^{(0)}+\Lambda^{-2}\sigma^{(1)}},
    \label{eq:rho_lin}
\end{equation}
where $\sigma^{(i)}\equiv \text{Tr}[R^{(i)}]$. The quantities $\sigma^{(0)}$ and $\Lambda^{-2}\sigma^{(1)}$ correspond to the SM and SM--EFT interference contributions to the total cross section, respectively.

While Eq.~\eqref{eq:rho_lin} is properly normalised by construction, it is not guaranteed to remain positive semidefinite for arbitrary choices of $\Lambda$ (and, more generally, of the Wilson coefficients). For sufficiently small $\Lambda$, the truncated matrix may develop negative eigenvalues, signalling that the linear truncation is no longer a consistent approximation at the level of the density matrix. To quantify this effect, we determine for each scattering angle $\Theta$ the value of $\Lambda$ at which the first negative eigenvalue appears, and then average this threshold over $\Theta$. We denote the resulting conservative lower bound by $\Lambda_{\rho}$.

\begin{table}[t!]
\centering
\renewcommand{\arraystretch}{1.1}
\begin{tabular}{ c | c | c  | c }
Lower Bound (TeV) & $\Lambda_\rho$ & $\Lambda_\sigma^{(1)}$ & $\Lambda_\sigma^{(2)}$  \\ \hline
Vector (V,V) & 9.7  & 2.3& 1.2  \\ \hline
Vector (A,A)  & 12.4 & 1.6 & 1.4 \\ \hline
Vector (R,R) & 7.0 &1.4 & 1.3 \\ \hline
\end{tabular}
\caption{\label{tab:linear}
\small
Comparison between the lower bounds $\Lambda_\rho$, $\Lambda_\sigma^{(1)}$ and $\Lambda_\sigma^{(2)}$ as defined in the text, on the new-physics scale $\Lambda$ appearing in the effective Lagrangian. The bounds are obtained for $e^+ e^- \to  t \bar t$ with unpolarised beams at $\sqrt{s} = 500$~GeV by imposing, respectively, positivity of the linearly truncated density matrix, positivity of the linearly expanded cross section, and a hierarchy criterion between the linear and quadratic terms. Values of $\Lambda$ are given in TeV and the (dimensionless) Wilson coefficients are set to one. Three vector-like EFT benchmark scenarios, (V,V), (A,A) and (R,R), are considered.
}
\end{table}

\paragraph{Cross-section based criteria.}
We next consider two commonly used constraints on the EFT scale based on the behaviour of the cross section.

A first requirement is that the (differential) cross section obtained from the \emph{linearly} truncated expansion remain non-negative. Since the interference contribution is not positive definite, one imposes
\begin{equation}
    \sigma^{(0)}\ge \Lambda^{-2}|\sigma^{(1)}| \,.
    \label{eq:sigma_linear_condition}
\end{equation}
We denote by $\Lambda_{\sigma}^{(1)}$ the corresponding lower bound on $\Lambda$ ensuring positivity of the cross section at linear order.

A second diagnostic is obtained by comparing the size of the quadratic contribution to the linear one. In terms of the total rates this can be expressed as
\begin{equation}
    |\Lambda^{-4}\sigma^{(2)}|\le|\Lambda^{-2}\sigma^{(1)}|.
    \label{eq:lambda_sigma}
\end{equation}
We define $\Lambda_{\sigma}^{(2)}$ as the value of $\Lambda$ at which the quadratic term becomes comparable to (and, for smaller $\Lambda$, larger than) the linear interference contribution. In other words, $\Lambda_{\sigma}^{(2)}$ marks the onset of a regime where quadratic effects can no longer be treated as a subleading correction to the interference term.

\paragraph{Comparison and interpretation.}
Table~\ref{tab:linear} reports the extracted values of $\Lambda_{\rho}$ and $\Lambda_{\sigma}^{(1,2)}$ (in TeV) for $e^{+}e^{-}\rightarrow t\overline{t}$ with unpolarised beams at $\sqrt{s}=500$~GeV, for the three vector scenarios (V,V), (A,A) and (R,R). One observes that $\Lambda_{\rho}$ is systematically larger than $\Lambda_{\sigma}^{(1,2)}$, in some cases by a sizeable factor. This indicates that demanding positivity of the \emph{density matrix} can yield a substantially more stringent consistency requirement than cross-section based criteria. We find this outcome noteworthy and plan to investigate its generality and its dependence on collider energy, beam polarisation, and operator choices in future work.

Two additional comments are in order. First, in $e^{+}e^{-}\rightarrow t\overline{t}$ the tree-level SM amplitudes are suppressed by the $1/s$ behaviour of the gauge-boson propagator. In such a situation, cross-section based criteria expressed solely in terms of $\sigma^{(0,1,2)}$ can be numerically conservative and may not directly translate into an optimal estimate of the EFT validity domain for all spin-sensitive observables. Second, when the SM and EFT amplitudes have different chiral structures the interference can vanish at tree level due to helicity selection rules \cite{Azatov:2016sqh,Falkowski:2016cxu}. 
This occurs for the scalar and tensor benchmark scenarios considered in this paper: due to the $s$-channel nature and negligible initial particle masses, the spin-1 (SM) and spin-0 and -2 (EFT) amplitudes select different initial spin confabulations. 
Therefore, the SM--EFT interference term vanishes at tree level, and the leading EFT contribution to the corresponding (vanishing) entries of the SM spin density matrix arises from the dimension-6 squared term.\footnote{
As the SM and EFT select different $e^+ e^-$ spins, ``interference resurrection'' \cite{Helset:2017mlf,Azatov:2019xxn,ElFaham:2024uop,LoChiatto:2024dmx} does not occur in the $t \bar t$ spin density matrix. 
}
In this case, dimension-8 contributions are subleading for both non-vanishing and vanishing SM entries. For non-vanishing SM entries, the SM–dimension-8 interference provides a correction to the existing SM contribution at order $\Lambda^{-4}$. For vanishing SM entries, dimension-8 effects first appear at order $\Lambda^{-8}$, and therefore only give subleading corrections to the dimension-6–squared contributions, which scale as $\Lambda^{-4}$. 

\end{document}